\documentclass[12pt]{article}
\usepackage{amsmath,epsfig,amssymb,bbm,amssymb,psfrag}

\newcommand{\jtheta}[1]{\vartheta \begin{bmatrix} #1 \end{bmatrix}}
\newcommand{\jstheta}[2]{\vartheta\big[^{#1}_{#2}\big]}
\newcommand{\Z}{{\bf Z}}
\newcommand{\1}{{\mathbbm{1}}}

\renewcommand{\Im}{{\rm Im}}

\begin{document}

\title{
\vbox{
\baselineskip 14pt
}
\vskip 2cm
\bf 
Chiral four dimensional field theory \\ from superstring and
higher dimensional super Yang-Mills theory
\vskip 0.5cm
}
\author{
 Hiroshi~Ohki\footnote{email: ohki@scphys.kyoto-u.ac.jp}\\*[20pt]
{\it \normalsize
Department of Physics, Kyoto University,
Kyoto 606-8502, Japan} 
\\
}

\date{}

\maketitle
\thispagestyle{empty}

\newpage
\begin{abstract}
We study four dimensional field theory from higher dimensional super
Yang-Mills theory based on the low-energy effective theory of
Type I, II or heterotic string theories.
Chiral fermions in four
dimensions are obtained by several mechanisms.
Especially, the background flux is one of the most
interesting mechanisms for obtaining four dimensional chiral
theories. Compactified extra dimensions with magnetic flux cause the
gauge symmetry breaking and non-trivial boundary conditions for
charged fields.  
Chiral matter fields have localized wavefunctions on
extra dimensions. We discuss about the relations between background
flux and low-energy spectra which are counted by their zero-mode.
We also study the low-energy constants and these moduli dependence.
They are calculated by usual dimensional reductions of super Yang-Mills
theory or supergravity theory. 
Yukawa couplings are free parameters in the standard model
and may be related to the underlying physics. 
In the string theory or its
low-energy limit, they are determined
by overlap integral of wavefunctions on extra dimensions.
We specify the simple compactifications, i.e. torus and compute the
overlap integral of three wavefunctions which correspond to
Yukawa interactions. We also study the higher order
couplings based on the field theoretical approach.  
From the analysis of generic n-point couplings, we can discuss about
flavor structures. We find that in such a construction some discrete
flavor symmetries appear in the four dimensional effective
theory. Their phenomenological implications are discussed.
Furthermore we extend these constructions to orbifold background.
Magnetic flux still plays an important role in
this background and leads to various types of low-energy spectra
different from that of toroidal compactifications. 
The orbifold models with heterotic string are also investigated. 
There are some discrete symmetries on orbifolds which reflect certain
geometrical symmetries of internal spaces. 
We use path integral methods to derive the anomaly in discrete
symmetries and show the anomaly coefficients for mixed gauge or
gravitational anomaly in heterotic orbifold models and 
higher dimensional field theory. 
We apply these mechanisms to realize the semi realistic model and
study phenomenological implications of these models.
\end{abstract}

\newpage

\tableofcontents
\newpage

\section{Introduction}
The theoretical particle physics has succeeded in explaining 
the physics of the elementary particles based on the quantum mechanism 
and its extension to the relativistic field theory.
It reached to the so-called standard model~(SM) of particle physics,
which contains the $SU(3)$$\times$ $SU(2)$$\times$$U(1)$ gauge groups
coupled to fundamental particles i.e. three generations of the quarks,
leptons and Higgs scalar. Although the SM can explain a lot of
independent high energy experimental data, it may not be accepted as
the fundamental physics of the world. 
This is because there are many free parameters which we only fix
posteriorly by the experiments, that is, the matter fields with
three replica and the masses of the matters in each of generations and
mixing in the quark and  lepton sectors. 
That is the so-called flavor problem. Indeed, most of free parameters in
the SM are originated from the flavor sector, that is, Yukawa couplings.
In addition, the reason why the strong interactions do not break CP
but weak interactions do is not explained. These issues are
accomplished by one specific choice of the infinite classes of
possible quantum filed theory of the SM. 
Beyond these issues, there are more intrinsic problems in the SM. 
We refer to the naturalness problems of Higgs scalar and quantum field
theory of the gravity. 
The former problem is related to the quantum corrections to the Higgs
scalar mass and may be solved by introducing the supersymmetry. 
The supersymmetric quantum field theory may become the extention of
the SM. In its minimal extension~(MSSM), it has a superpartner for
each of elementary particles with different spin-statistics.
The corresponding scalar partners of quarks and leptons are squarks
and sleptons and the Higgs scalar also has the partner i.e. Higgisino
and gauge boson has its fermionic partner as a gaugino.   
However the latter issue still remains.

The Superstring theory successfully unifies the concepts of
quantum field theory and general relativity. 
This is the most promising approach to overcome these issues.
In order to claim that
it incorporates a unification of all forces observed in nature, 
one has to prove the existence of string models reproducing SM
particle physics. The best way to prove its existence consists in the
construction of explicit models since that allows also to investigate
phenomenological implications of string theory. 

Perhaps the most
traditional, attempt of identifying realistic string models is given
by heterotic orbifold constructions~\cite{Dixon,IMNQ}.
In more recent years, this line of research was boosted by the
observation that phenomenological properties can be connected to
geometrical properties of the orbifold
\cite{Kobayashi:2004ud,Forste:2004ie,Buchmuller:2004hv,Kobayashi:2004ya,
Lebedev:2006kn,Kim:2006hw,Buchmuller:2005jr}.
Examples for quantities which
are directly tied to geometry are the K\"ahler potential for twisted
sector states as well as Yukawa couplings
\cite{Hamidi:1986vh,Burwick:1990tu,Kobayashi:2003vi,Kobayashi:2003gf}.
(For interesting applications see e.g.~Ref.\ \cite{Ko:2004ic}.) 
The Calabi-Yau compactifications are also interesting as compactified
six dimensional spaces preserving $\mathcal{N}=1$ supersymmetry.
The background metric is non-trivial and there are many kinds of six
dimensional Calabi-Yau manifolds. The concrete models for heterotic
orbifold can be regarded as the singular limit of the smooth
Calabi-Yau compactifiactions.
In a parallel development, semi realistic models have been obtained in
the free fermionic formulation of heterotic strings
\cite{Antoniadis:1989zy,Faraggi:1991jr}. Although there are some
indications \cite{Faraggi:1993pr} that these models are related to
${\mathbb Z}_2 \times {\mathbb Z}_2$ orbifolds a precise connection
has not been worked out in general. Hence a geometric picture is
missing for many free fermionic models.

The possibilities of the model building from type II string theory 
have been enriched by the discovory of the
D-branes~\cite{Polchinski:1995mt}. 
The open strings have their end points on certain D-branes.
Their lowest modes give rise to massless gauge fields and their
fermionic partners. Then n-stack of the D-branes have naturally 
$n^2$ number of massless gauge bosons and they have $U(n)$ gauge
symmetry in low-energy.
It is shown that these D-brane backgrounds give rise to realistic
string compatifications.
The first attempt to obtain the chiral matter fields is considered by  
two D-branes which are intersected each others. 
The chiral matter fields can appear in their localized intersecting
points as bi-fundamental gauge representaions.
The number of zero-modes i.e. the generation number is given by the 
intersection number in internal spaces. 
This model contruction has an advantage that 
the geometrical interpretation is easy as well as heterotic orbifold
models. 
The specific examples for this type of models are discussed in type
IIA string models with
D6-branes~\cite{Berkooz:1996km,Blumenhagen:2000wh,Angelantonj:2000hi,  
Blumenhagen:2005mu,Aldazabal:2000dg,Blumenhagen:2000ea,Cvetic:2001tj}.   

Their T-dual models i.e. magnetized D-brane models also have been
investigated.  In the language of T-duality, the intersecting angle of
two D-branes in the type~IIA side is interpreted as the magnetic flux
inside two internal spaces in the type~IIB picture. There is no
localized mode in the internal spaces and their low-energy effective
theory can be discrebed higher dimensional super Yang-Mills theory
with magnetic flux. 
The background magnetic fluxes cause a breaking of gauge groups and 
chiral matter fields can be obtained by solving the zero-mode solutions.
From the T-dual of $T^6$ toroidal compactification of intersecting 
D6-brane models, corresponding chiral matter fields are calculated
explicitly from a simple factorizable $T^2\times T^2\times T^2$
toroidal compactifications with constant magnetic flux.
The resulting solutions are represented by the products of Jacobi
theta functions. The bosonic mode, for example Higgs scalar or scalar
partners, are also calculated by solving Laplace operator with 
flux background. These results can be applied for lower dimensions
$D=6, 8$ and lead to various types of models. 

Concerning about the flavor problems explained above, 
one has to know Yukawa couplings.
In the string theory computation, Yukawa couplings are calculated by 
string amplitude of corresponding three vertex operators by using CFT
technique. Taking into accout the classical contributions of the
amplitude, Yukawa couplings are represented by a sum over worldsheet
instanton effects.  The magnitude of the Yukawa couplings is affected 
by the localization points for three matter fields. 
When their localization points are far away each other, the
exponetially surppressed Yukawa couplings are obtained. Thus Yukawa
couplings are geometrically determined.
On the other hand, in the T-dual picture, the calculations of the
Yukawa couplings are purely field theoretical.  
Yukawa interactions can be calculated by overlap integrals over
internal spaces with three wavefunctions as the following forms 
\begin{align}
Y= \int dy^6 \psi_i(y) \psi_j(y) \phi(y)
\end{align}
where $\psi_{i,j}(y)$ correspond to the internal wavefunctions of
chiral matter fields and $\phi(y)$ is the internal wavefunctions of
Higgs scalar fields.
The explicit calculations of the overlap integrals can tell
us the form of the Yukawa couplings.
It is found that two different approaches of stringy and field theory
calculations lead to the consistent results of the Yukawa couplings 
after proper transformation of moduli
parameters~\cite{Cremades:2004wa}. 
Furthermore the method using the field theoretical approach can
tell us the other constants like the normalization constants or
higher order couplings. For example, the former contribution is
related to the Kahler moduli. In order to obtain those in the
intersecting D-brane side, it needs quantum effects of stringy
correlators. 
These results are also consistent each other up to higher order
corrections to the normalization factors.

A rather bottom-up approach to understand the realistic quark/lepton mass
hierarchy and mixing angles is the flavor symmetry.
Symmetries play an important role in particle physics.
As long time ago it was suggested that the $U(1)$ symmetries can be
applied to obtain the hierarchical quark mass structure as called by
Froggatt-Nielsen mechanism~\cite{Froggatt:1978nt}.
More recently non-abelian discrete symmetries are investigated to
address the above flavor issue in particular in explanation of the 
large mixing of lepton flavor.
It is plausible that such non-abelian discrete 
flavor symmetries are originated from extra dimensional 
theories, because non-abelian symmetries 
are symmetries of geometrical solids.
Indeed, 
it has been shown that certain types 
of non-abelian discrete flavor symmetries 
such as $D_4$ and $\Delta(54)$ can appear 
in four-dimensional effective field theories 
derived from heterotic string theory with 
orbifold background~\cite{Kobayashi:2004ya,Kobayashi:2006wq}.
(See also \cite{Altarelli:2006kg}.)
In those analyses, 
the important ingredients to derive 
the non-abelian discrete flavor symmetry 
are geometrical symmetries of the compact space 
and stringy coupling selection rules.
To investigate the flavor symmetry, it is important to investigate the
higher order couplings.
In the field theoretical approach, the higher order couplings are also 
calculable in principle. 
For toroidal compactification $T^2$ case, the generic n-point
couplings are also represented analytically and have same properties
as CFT calculations. The main ingredient is stressed that the generic
n-point couplings are given by the products of three point couplings.
Then one can analyze the flavor symmetries and find that 
there are several types of discrete flavor symmetries in the model with
magnetized/intersecting D-brane models and a certain relation between the
number of generations and the flavor symmetries. 
Although in the normalization factor there is a small discrepancy in the
stringy and field theoretical calculations, this does not affect 
in the structures of the flavor since it is only determined by the
number of generations i.e. the index number.
Therefore it would be helpful to understand the flavor symmetries for
considering the flavor problems.

Recently the wavefunction profiles have been studied in some of
non-trivial background geometry, e.g. 
orbifold compactifications, $P^1\times P^1$, $P^2$
geometry~\cite{Conlon:2008qi}, warped
compactifications~\cite{Marchesano:2008rg} and 
flux compactifications~\cite{Camara:2009xy}
in which it is succeeded to obtain the explicit solutions 
of wavefunctions.    
Such string constructions are classified to two classes of global and
local models. 
As mentioned above, 
the ten-dimensional compactification is naturally corresponding to the 
low-energy limit of the heterotic and type I string theory.
From the view point of the field theory, one may consider less than ten 
dimensional field theory with gauge interactions. 
Global models are defined in the total compact space with certain
choice of topological features. 
The simplest example is the heterotic string 
theory with Calabi-Yau compactifications.
On the other hand, local models can be considered  localized modes 
living in a part of extra dimensions.
The gauge and matter fields depend on the local internal spaces
and do not depend on the details of other bulk topological features.
Thus there are attractive features of local model construction which
drastically simplify the structures of the geometry and it is easier
to calculate the low-energy physics than that of global models.
Indeed in the general Calabi-Yau compactifications, explicit metric of
such a global compact space is not known. 
For example we can consider the possibility of
the singular point in the six dimensional compact space with orbifold
and then such a metric can be represented as $T^{2n}$ $\times$
$\mathcal{C}^{3-n}/{\bf Z}_N$. 
One may put the $N$ stack of D$(3+2n)$-brane wrapping the $T^{2n}$ on 
the singular point of ${\bf Z}$ orbifold. 
Thus the low-energy effective theory is described 
by $\mathcal{N}=1$ $D=4+2n$ super Yang-Mills theory with proper 
gauge groups. 
In both cases, one can apply the formula of the overlap integrals and
the method of Kluza-Klein decomposition.
Thus the field theoretical approach is powerful method to 
calculate the low-energy constant including proper stringy effects and
moduli fields dependence, which allow to construct the phenomenologically interesting
models. 
In addition, these constructions are also related to the 
phenomenological model building within the extra dimensions.
Suppose that the chiral matter fields have Gaussian profiles in the
extra dimensions and  each of generations is localized in different way,  
the hierarchical structures of the masses for generations may be
obtained. Therefore one sees that D-brane model constructions give some
of concrete examples for these phenomenological models.
We then study these features of extra dimensional field theory for the
case of exceptional gauge groups e.g.$E_6$, $E_7$ or $E_8$ which are
regarded as phenomenological model buildings.   

In recent years a renewal of the local model building has been
developing, for instance, F-theory model
buildings~\cite{Donagi:2008ca,Beasley:2008dc}. 
In F-theory models, it naturally includes exceptional gauge groups
beyond the type~IIB D-brane. The flavor structures are different from
that of D-branes models, there are a lot of development for
phenomenological studies. 

For the selection of the vacua, 
one should study the potential of the moduli field and supersymmetric
four dimensional vacua. In the string theory, there are many fields
beyond the SM particle. Some of them are called as moduli fields and  
their vacuum expectation values correspond to the size or shape 
of the compactification spaces and positions of D-brane and so on.
These values are also related to the parameters like gauge coupling
constant or masses for four dimensional fields.
In the general Calabi-Yau compactifications, 
these moduli are not determined by means of minimalizing the
potential of moduli. 
To solve this problem, several mechanisms have been
proposed~\cite{Curio:2000sc,Kachru:2002he}.
In the string theory, they contain the anti-symmetric tensor fields as
$C_p$. Their field strengths are also appearing as
$F_{p+1}=dC_p$ and have non-vanishing background expectation values so
called three form flux. This flux affects in the low-energy potential 
in the moduli sectors. Thus some of these moduli fields are stabilized
in a flux compactification. Furthermore in type IIB theory, Kahler
moduli $T$ are stabilized by non-perturbative effects such as gaugino
condensation. In the supergravity potential, this minimum is 
supersymmetric and anti-de Sitter vacuum. The anti-D3 brane is
introduced in order to uplift the vacuum energy and realize the de
Sitter or Minkowski vacuum. This shifts the position of the potential
minimum and breaks the supersymmetry in a controllable way.
This is so called KKLT scenario~\cite{Kachru:2003aw}.
In addition, a new calculable method of moduli stabilization was
proposed, using the internal magnetic fields. 
This method can be used in simple toroidal compactifications,
stabilizing the geometric moduli in a supersymmetric vacuum that is
within a perturbative string description.
Once if we have a mechanism to break low-energy supersymmetry, 
the relevant soft supersymmetry breaking terms are also related to the
magnetic flux. 
Thus these approaches using the magnetic flux or KKLT scenario can
obtain the moduli parameters in a certain level so that we can analyze
the low-energy spectrum including the super particle.

Discrete symmetries play an important role in 
model building of particle physics.
For example, abelian and non-abelian discrete flavor symmetries 
are useful to derive realistic quark/lepton masses and their 
mixing~\cite{Altarelli:2007cd}.
Discrete non-abelian flavor symmetries can also be used to 
suppress flavor changing neutral current processes in 
supersymmetric models~\cite{dbkaplan,babu}.
Furthermore, discrete symmetries can be introduced to forbid
unfavorable couplings 
such as those leading to fast proton decay \cite{murayama,kakizaki}.
It is widely assumed that
superstring theory leads to anomaly-free effective theories.
In fact the anomalous $U(1)$ symmetries
are restored by the Green-Schwarz (GS) 
mechanism \cite{Green:1984sg,Witten:1984dg,Ibanez:1998qp}.
For this mechanism to work,
the mixed anomalies between the anomalous $U(1)$ and other continuous 
gauge symmetries have to satisfy a certain set of conditions,
the GS conditions,
at the field theory level.
In particular, in heterotic string theory the mixed anomalies between 
the anomalous $U(1)$ symmetries and other continuous gauge 
symmetries must be universal for different gauge 
groups up to their Kac-Moody 
levels \cite{Schellekens:1986xh,Kobayashi:1996pb}.
A well-known discrete symmetry in heterotic string theory is
 T-duality symmetry, and 
its effective theory has T-duality anomalies
\cite{Derendinger:1991hq}.
It has been shown that
the mixed anomalies between T-duality symmetry and 
continuous gauge symmetries are universal except 
for the sector containing an $N=2$ subsector
 and are exactly canceled by the GS mechanism  \cite{Ibanez:1992hc}.
That has phenomenologically interesting 
consequences which  have been  studied in early 90's
\cite{Ibanez:1992hc,Ibanez:1991zv,Kawabe:1993pz}.

For the above purposes, in this thesis, we study the phenomenological
aspects of the higher dimensional super Yang-Mills theory with various
dimensions as the best motivated theory for effective field theory of
string theory and study the low-energy physics related to the
extension of the SM. 
The contents of this thesis are as follows.
In section 2  we first study the higher dimensional $U(N)$ super
Yang-Mills theory on the torus background with magnetic fluxes using
the usual Kluza-Klein dimensional reductions for obtaining the chiral
matter fields. 
We also solve the wavefunctions explicitly in the toroidal
compactifications with and without toron configurations of twisted
boundary conditions~\cite{Cremades:2004wa,Abe:2009uz,Abe:2010ii}. 
We see the mechanism for obtaining the
chiral fermion and the number of generations.
Furthermore we extend this analysis to the exceptional gauge groups in
a similar way~\cite{Choi:2009pv}.
In section 3 we consider about calculation of the three point couplings
which can appear in the off-diagonal components for different three
gauge groups~\cite{Cremades:2004wa}. These analysis can extend to the
generic n-point couplings in which we see the structures of the
n-point couplings are related to the product of three point
couplings~\cite{Abe:2009dr}. We also discuss about the T-dual picture,
i.e. results from intersecting D-brane model.  

In section 4 we analyze the flavor structures based on the field
theoretical approach. Then we show that the discrete flavor symmetries
can appear in these types of models~\cite{Abe:2009vi}, which are 
phenomenologically interesting for the realistic patterns of the
lepton mixings.  
The corresponding representations for each generation are
classified. The role of the Wilson line parameters in flavor symmetry 
are shown and its phenomenological implications with symmetry breaking
are also discussed.

In section 5 we study the orbifold compactifications with magnetic
background~\cite{Abe:2008fi,Abe:2008sx}. We will see that a possible
orbifold projection on $T^2$ is restricted only $Z_2$
orbifold. Therefore the remaining matter field after orbifold
projections are consist of either even or odd wave functions. 
They are obtained from linear combination of the
wavefunctions on the torus. 
The zero-mode spectra are different from that of toroidal
compactifications. We classify the phenomenologically interesting
models with three generations of chiral matters and analyze the
Yukawa couplings. We also discuss the phenomenological model building
with magnetized extra-dimensions.

The question is whether discrete anomalous
symmetries can appear in string-derived models. The discrete
symmetries on orbifolds reflect certain geometrical symmetries of
internal space. Since the geometrical operations 
are embedded into the gauge group, one might suspect
that the discrete anomalies are related to gauge anomalies.
In section 6 we use path integral methods to derive the anomaly of
discrete symmetries including non-Abelian discrete symmetries and show
the anomaly coefficient for mixed gauge or gravitational anomaly.
We also briefly review the discrete anomalies focusing on the discrete
flavor symmetries appearing in heterotic orbifold models and higher 
dimensional field theory. 
Next, we define discrete $R$-charges, which is defined in heterotic
orbifold models and calculate the mixed anomalies between 
the discrete $R$-symmetries and the continuous gauge symmetries
in concrete models.
We also study the relations of R-anomalies with 
one-loop beta-function coefficients and T-duality anomalies.
Phenomenological implications of our results are also discussed.

Finally section 7 is devoted to conclusion and discussion.
In order to obtain the low-energy Lagrangian, we specify the functions 
of Kahler potential and super potential. In appendix A we perform the
Kaluza-Klein reductions of ten dimensional super Yang-Mills theory and
give a procedure for obtaining the scalar component of the chiral
matter field and calculation of the Kahler potential.
For toroidal compactifications the moduli dependence for
Yukawa and Kahler potential is obtained. 
In appendix B, C, we classify the orbifold models with three generation
and show the results of Yukawa couplings.
In appendix D, we give a short introduction of the properties of
discrete symmetries used in the thesis.

\newpage

\section{Super Yang-Mills theory on higher dimensions}

Understanding the structure of the SM is 
one of the fundamental problems of theoretical particle physics.
In particular, one of most outstanding puzzles
of the SM of particle physics is the structure
of the Yukawa couplings between the Higgs field and the 
SM fermions.
A correct description of the observed masses and mixing of quarks 
and leptons seems to require very different values for the Yukawa
coupling constants for the different generations.

In recent years the idea that there could be more than four dimensions
has been pursued intensively, particularly due to the study of string
theory which is naturally defined in 10 or 11 dimensions. 
In fact extra dimensional field theories, in particular 
string-derived extra dimensional field theories, 
play an important role in particle physics as well as cosmology.
There is a possibility of computing the Yukawa coupling in terms of 
the extra-dimensional geography.
Starting from d+4 dimensional compactified theory one may obtain the 
massless modes with factorized wavefunctions.
Gauge bosons in the extra dimensional component become 
scalars at low-energy and its fermionic component may give rise to
the matters. 
Yukawa couplings can appear from the higher dimensional gauge
interactions $A_i \bar{\Psi}\Gamma_i \Psi$.
Yukawa coupling constant is calculated in principle 
from the overlap integrals over extra dimensions.
Our aim is to study such theories of potential phenomenological interest.
We consider our starting point ten dimensional super Yang-Mills theory 
since it appears in the low-energy limit of the Type I, IIB and 
heterotic string theories. 

However when we start with extra dimensional theories, 
how to realize chiral theory is one of important 
issues from the viewpoint of particle physics.
Introducing magnetic fluxes in extra dimensions is 
one way to realize chiral fermions in 
field theories and superstring theories~\cite{Manton:1981es,
Bachas:1995ik,Cremades:2004wa,Troost:1999xn}.
In particular, magnetized D-brane models are T-duals of 
intersecting D-brane models 
and several interesting models have been constructed 
within the framework of intersecting D-brane
models~\cite{Berkooz:1996km,Blumenhagen:2000wh,Angelantonj:2000hi,
Aldazabal:2000dg,Blumenhagen:2000ea,Cvetic:2001tj}.\footnote{
See for a review \cite{Blumenhagen:2005mu} and references therein.}

In this section we introduce the $\mathcal{N}=1$ supersymmetric 
Yang-Mills theory with various dimensions.
We use the Kluza-Klein dimensional reductions for obtaining the
internal wavefunctions. The internal wavefunctions are chosen to be
eigenstates of the internal Laplace and Dirac operators.
We show that the chiral fermion can be obtain from the non-trivial
solutions of wavefunctions due to the internal background flux. 
The low-energy physics is described by those of zero-mode wavefunctions. 
Zero-modes are quasi-localized on the torus with the magnetic flux.
The number of zero-modes, which corresponds to 
the generation number, is determined by 
the value of the magnetic flux in the same way as 
that the generation number is determined by the 
intersecting number in intersecting D-brane models.
Furthermore we extend this analysis to other backgrounds and 
other gauge groups. In such a case, we also obtain the explicit 
wavefunctions and calculate the spectra within a field theoretical 
way. 

\subsection{Toroidal wavefunctions 
}

Let us consider $N=1$ super Yang-Mills theory in $D=4+2n$ dimensions.
Its Lagrangian density is given by 
\begin{equation}
\label{eq:SYM-L}
{\cal L} = - \frac{1}{4g^2}{\rm Tr}\left( F^{MN}F_{MN}  \right) 
+\frac{i}{2g^2}{\rm Tr}\left(  \bar \lambda \Gamma^M D_M \lambda
\right),
\end{equation}
where $M,N=0,\cdots, (D-1)$.
Here, $\lambda$ denotes gaugino fields, $\Gamma^M$ is the 
gamma matrix for $D$ dimensions and 
the covariant derivative $D_M$ is given as 
\begin{equation}
D_M\lambda = \partial_M \lambda - i [A_M, \lambda],
\end{equation}
where $A_M$ is the vector field.
Furthermore, the field strength $F_{MN}$ is given by 
\begin{equation}
F_{MN} = \partial_M A_N - \partial_N A_M -i[A_M,A_N].
\end{equation}

We consider the torus  $(T^2)^n$ as the extra dimensional 
compact space, whose coordinates are denoted by $y_m$ 
$(m=4, \cdots, 2n+3)$, while the coordinates of 
four-dimensional uncompact space $R^{3,1}$
are denoted by $x_\mu$ $(\mu=0,\cdots, 3)$.
We use orthogonal coordinates and choose the torus metric 
such that $y_m$ is identified by $y_m+n_m$ with $n_m=$ integer.
The gaugino fields $\lambda$ and the vector fields $A_m$ 
corresponding to the compact directions are decomposed as 
\begin{eqnarray}
\label{eq:gaugino-decomp}
\lambda(x,y) &=& \sum_n \chi_n(x) \otimes \psi_n(y), \\
\label{eq:vector-decomp}
A_m(x,y) &=& \sum_n \varphi_{n,m}(x) \otimes \phi_{n,m}(y).
\end{eqnarray}

\subsubsection{$U(1)$ gauge theory on magnetized torus
$T^2$}\label{sec:integer-mf} 

First, let us consider $U(1)$ gauge theory on $T^2$ 
with the coordinates $(y_4,y_5)$.
We study the non-vanishing constant magnetic flux $F_{45} = 2\pi M$.
We use the following gauge,
\begin{equation}
\label{eq:gauge}
A_4=-2\pi My_5, \qquad A_5 =0.
\end{equation}
Then, their boundary conditions can be written as 
\begin{eqnarray}
\label{eq:BC-gauge}
A_m(y_4+1,y_5)&=&A_m(y_4,y_5)+\partial_m \chi_4, \qquad   
\chi_4 = 0, \nonumber \\
A_m(y_4,y_5+1)&=&A_m(y_4,y_5)+\partial_m \chi_5, \qquad 
\chi_5 = -2\pi My_4.
\end{eqnarray}

Now, we study the spinor field $\psi(y)$ with the $U(1)$ charge $q=\pm
1$ 
on $T^2$, which corresponds to the compact part in the 
decomposition (\ref{eq:gaugino-decomp}).
The zero-mode satisfies the following equation,
\begin{equation}\label{eq:Dirac-T2}
\tilde \Gamma^m(\partial_m -iqA_m)\psi(y) = 0,
\end{equation}
for $m=4,5$, where $\tilde \Gamma^m$ corresponds to 
the gamma matrix for the two-dimensional torus $T^2$, e.g.
\begin{equation}
\tilde \Gamma^4 = \left(
\begin{array}{cc}
0 & 1 \\
1 & 0 
\end{array}
\right), \qquad 
\tilde \Gamma^5 = \left(
\begin{array}{cc}
0 & -i \\
i & 0 
\end{array}
\right),
\end{equation}
and $\psi(y)$ is the two component spinor,
\begin{equation}\label{eq:two-spinor}
\psi = \left(
\begin{array}{c}
\psi_+ \\ \psi_-
\end{array}
\right).
\end{equation}
Because of (\ref{eq:BC-gauge}), the spinor field satisfies 
the following boundary condition,
\begin{eqnarray}
\label{eq:bc-1}
\psi(y_4+1,y_5) &=& e^{iq\chi_4}\psi(y_4,y_5) = 
\psi(y_4,y_5), \\
\psi(y_4,y_5+1) &=& e^{iq\chi_5}\psi(y_4,y_5) = 
e^{-2\pi i qMy_4}\psi(y_4,y_5).
\label{eq:bc-2}
\end{eqnarray}
The consistency for the contractible loop, i.e.
$(y_4,y_5) \rightarrow (y_4+1,y_5) \rightarrow (y_4+1,y_5+1)$,  
$\rightarrow (y_4,y_5+1) \rightarrow (y_4+1,y_5+1)$,  
requires $M=$ integer.

\begin{figure}
\begin{tabular}{ccc}
\includegraphics[width=5cm,clip]{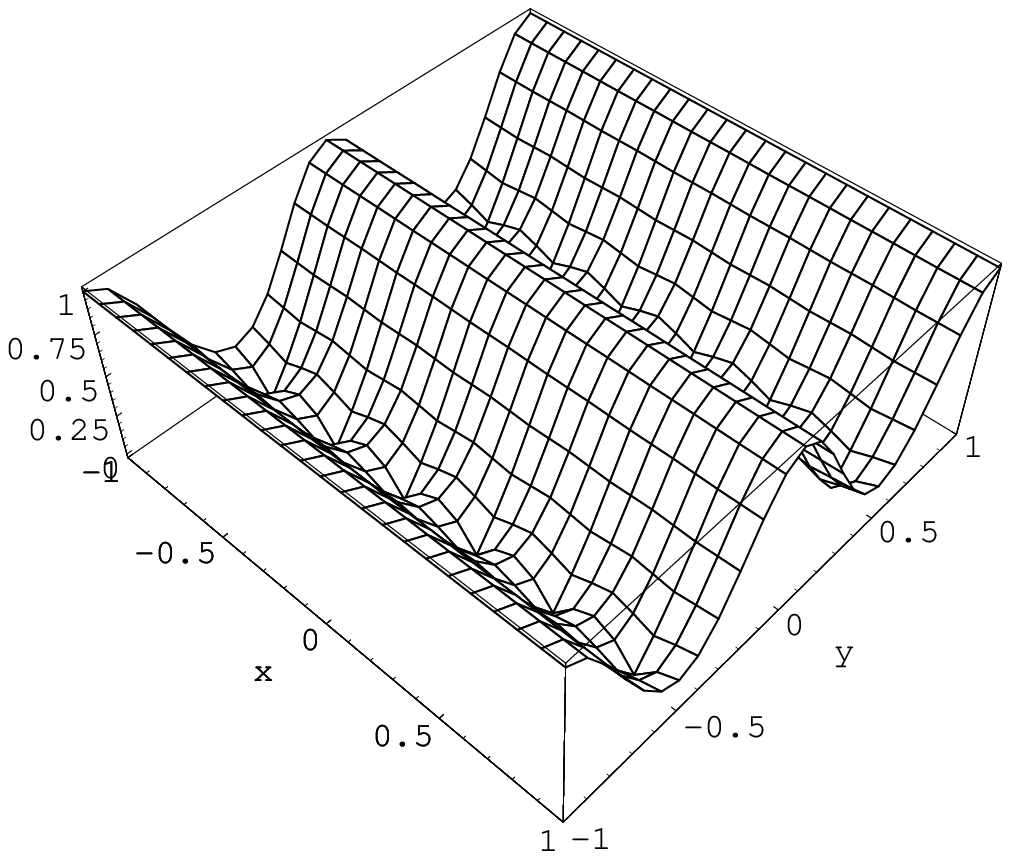}&
\includegraphics[width=5cm,clip]{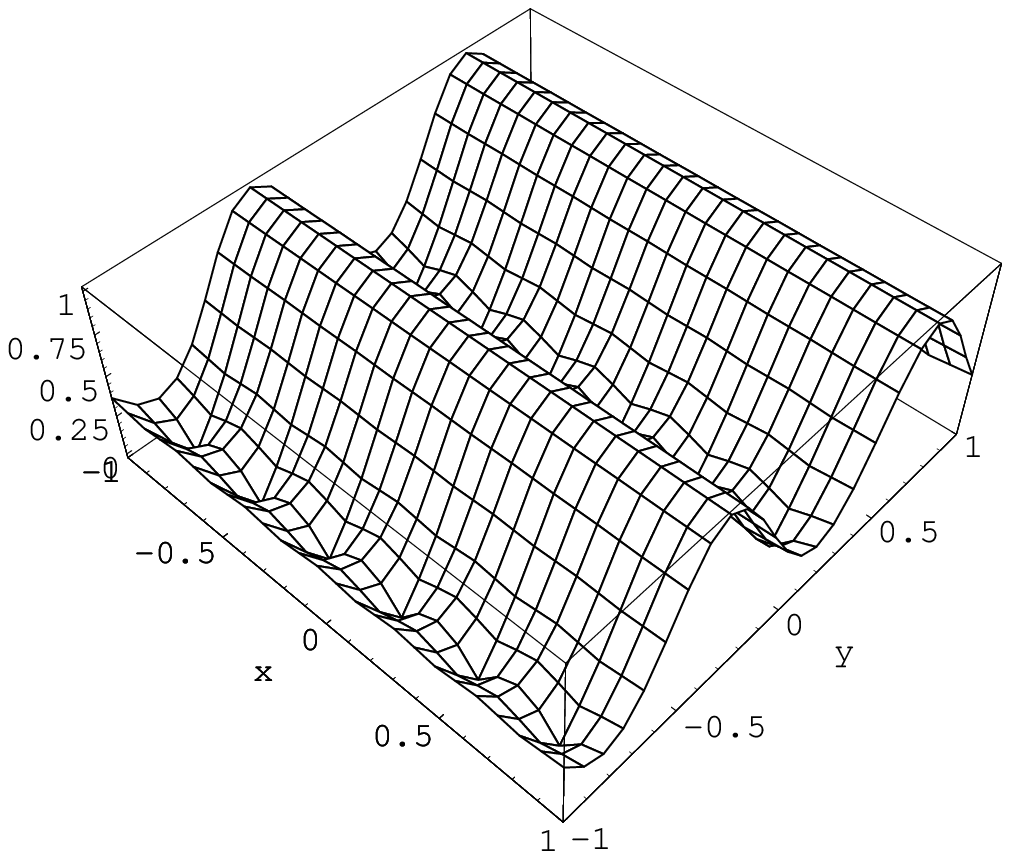}&
\includegraphics[width=5cm,clip]{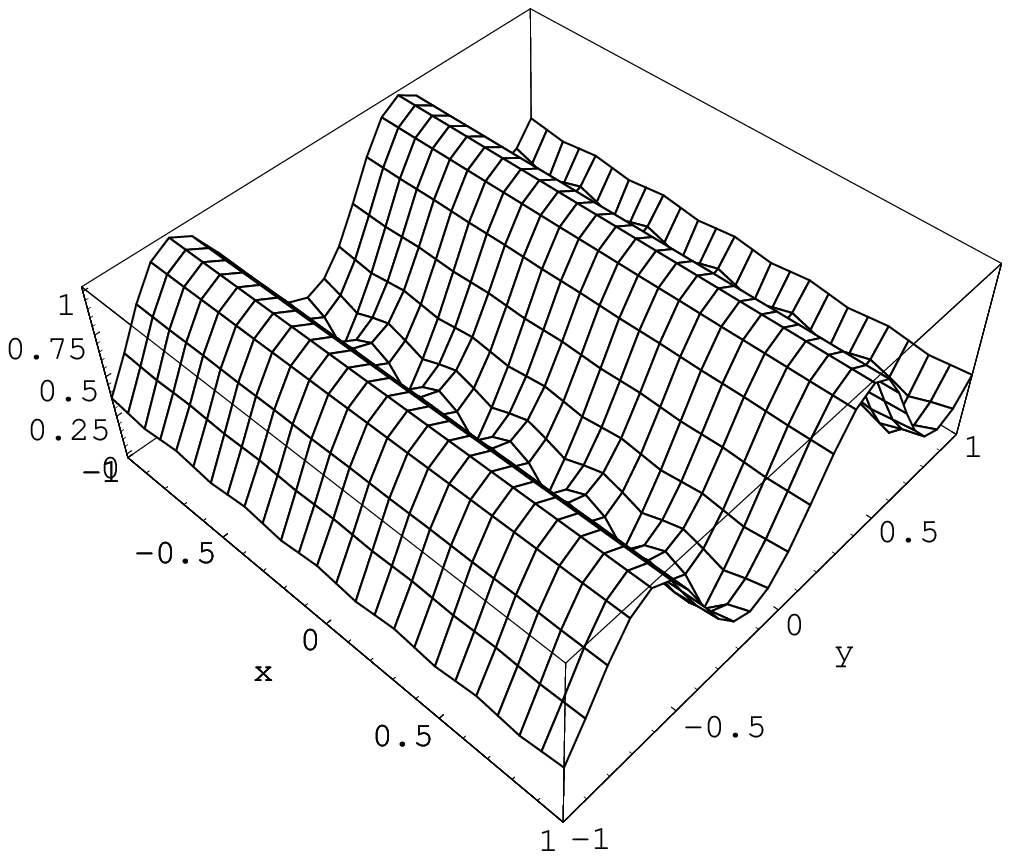} \\
j=0  & j=1 & j=2
\end{tabular}
\caption{The wavefunction profiles of $|\Theta^j(x,y)|$ for 
$M=3$.}
\label{fig:wf3}
\end{figure}

Because of the periodicity along $y_5$, 
$\psi_{\pm}$ can be written by 
\begin{equation}
\psi_{\pm}(y_4,y_5) = \sum_n c_{\pm,n}(y_4)e^{2\pi i ny_5}.
\end{equation}
Suppose that $qM >0$.
Then, the solution for the zero-mode equation of $\psi_+$ 
is given by 
\begin{equation}
c_{+,n}(y_4) = k_{+,n} e^{-\pi qMy_4^2-2\pi n y_4},
\end{equation}
where $k_{+,n}$ is a constant.
Furthermore the boundary condition requires 
\begin{equation}
k_{+,n} = a_n e^{-\pi n^2/(qM)},
\end{equation}
and $a_{n+qM}$ is equal to $a_n$, i.e. $a_{n+qM}=a_n$.
Thus, there are $|M|$ independent zero-modes of $\psi_+$, 
which have normalizable wavefunctions,
\begin{equation}\label{eq:zero-mode-wf}
\Theta^j(y_4,y_5) = N_je^{-M\pi y_5^2}\vartheta \left[
\begin{array}{c}
j/M \\ 0
\end{array} \right]\left( M(y_4+iy_5), Mi\right), 
\end{equation}
for $j=0,\cdots, M-1$, where $N_j$ is a normalization constant and 
\begin{equation}
\vartheta \left[
\begin{array}{c}
j/M \\ 0
\end{array} \right]
\left( M(y_4+iy_5), Mi\right) = \sum_n e^{-M\pi (n+j/M)^2 
+2\pi (n+j/M)M(y_4+iy_5)}, 
\end{equation}
that is, the Jacobi theta-function.
We can introduce the complex structure modulus $\tau$ by replacing 
the above Jacobi theta-function as 
\begin{equation}
\vartheta \left[
\begin{array}{c}
j/M \\ 0
\end{array} \right]
\left( M(y_4+iy_5), Mi\right) \rightarrow 
\vartheta \left[
\begin{array}{c}
j/M \\ 0
\end{array} \right]
\left( M(y_4+\tau y_5), M \tau \right).
\end{equation}
Thus, zero-mode wavefunctions depend on only 
the complex structure modulus, but not the overall 
size of $T^2$.
Furthermore, 
there is the degree of freedom to shift $y_m \rightarrow y_m + d_m$ 
with constants $d_m$.
They correspond to constant Wilson lines.
Their localization points of zero-mode profiles are different each
other and depend on the index $j$ and constants.  
Their wavefunction profiles for $M=3$ are shown 
in the Figure~\ref{fig:wf3}

On the other hand, the zero-mode equation for 
$\psi_-$ can be solved in a similar way, but their wavefunctions 
are unnormalizable.
Hence, we can derive chiral theory by introducing magnetic fluxes.
When $qM <0$, $\psi_-$ has $|M|$ independent zero-modes 
with normalizable wavefunctions, while zero-modes for 
$\psi_+$ have unnormalizable wavefunctions.
Bosonic fields are analyzed in a similar way.
(See e.g. \cite{Cremades:2004wa}.)

\subsubsection{$U(N)$ gauge theory on magnetized torus $T^2$}

Here, we study $U(N)$ gauge theory on $T^2$.
Let us consider the following form of (abelian) magnetic 
flux
\begin{equation}\label{eq:F45-UN}
F_{45} = 2 \pi \left(
\begin{array}{ccc}
M_1 {\bf 1}_{N_1\times N_1} & & 0 \\
 & \ddots & \\
0 & & M_n {\bf 1}_{N_n\times N_n}
\end{array}\right),
\end{equation}
where ${\bf 1}_{N_a \times N_a}$ denotes $(N_a \times N_a)$ 
identity matrix.
This abelian magnetic flux breaks the gauge group as 
$U(N) \rightarrow \prod_{a=1}^n U(N_a)$ with $N=\sum_a N_a$.
The rank is not reduced by the abelian magnetic flux.
When we consider non-abelian magnetic flux, 
i.e. the toron background \cite{'t Hooft:1979uj},  
the rank can be reduced.\footnote{
See e.g. \cite{Alfaro:2006is,vonGersdorff:2007uz} and references
therein.}
However, here we restrict ourselves to the abelian flux.

Now, let us study gaugino fields on this background.
We focus on the block including only $U(N_a) \times U(N_b)$ 
and such a block has the following magnetic flux,
\begin{equation}\label{eq:F-ab-block}
F_{45} = 2 \pi \left(
\begin{array}{cc}
M_a {\bf 1}_{N_a\times N_a}  & 0 \\
0 & M_b {\bf 1}_{N_b\times N_b}
\end{array}\right).
\end{equation}
We use the same gauge as (\ref{eq:gauge}), i.e.
\begin{equation}
A_4 = -F_{45}y_5, \qquad A_5 = 0.
\end{equation}

Similarly, the gaugino fields $\lambda$ in $R^{3,1}\times T^2$ are 
decomposed as 
\begin{equation}
\lambda(x,y) = \left(
\begin{array}{cc}
\lambda^{aa}(x,y)   & \lambda^{ab}(x,y)  \\
\lambda^{ba}(x,y) & \lambda^{bb}(x,y) 
\end{array}\right).
\end{equation}
Furthermore these gaugino fields are decomposed as 
(\ref{eq:gaugino-decomp}), 
\begin{eqnarray}
\lambda^{aa}(x,y)  &=& \sum_n \chi^{aa}_n(x) \otimes \psi^{aa}_n(y), 
\quad 
\lambda^{ab}(x,y) = \sum_n \chi^{ab}_n(x) \otimes \psi^{ab}_n(y), 
\nonumber \\
\lambda^{ba}(x,y)  &=& \sum_n \chi^{ba}_n(x) \otimes \psi^{ba}_n(y), 
\quad 
\lambda^{bb}(x,y) = \sum_n \chi^{bb}_n(x) \otimes \psi^{bb}_n(y). 
\end{eqnarray}
Each of $\psi^{aa}$, $\psi^{ab}$, $\psi^{ba}$ and $\psi^{bb}$ 
is a two-component spinor $(\psi_+,\psi_-)^T$.
Their zero-modes satisfy 
\begin{eqnarray}
\left(
\begin{array}{cc}\label{eq:Dirac-U(N)-T2-1}
\bar \partial \psi_+^{aa} & 
[\bar  \partial +2\pi i (M_a-M_b) y_5] \psi^{ab}_+ \cr \cr
  [\bar \partial +2\pi i (M_b-M_a)y_5] \psi^{ba}_+ & 
\bar \partial \psi_+^{bb}  \\
\end{array}
\right) &=& 0, \\
\left(
\begin{array}{cc}\label{eq:Dirac-U(N)-T2-2}
\partial \psi_-^{aa} & [\partial -2\pi i (M_a-M_b)y_5] \psi^{ab}_-  \cr
  \cr
[\partial -2\pi i (M_b-M_a)y_5] \psi^{ba}_- & \partial \psi_-^{bb}  \\
\end{array}
\right) &=& 0,
\end{eqnarray}
where $\bar \partial = \partial_4 +i \partial_5$ and 
$\partial = \partial_4  -i \partial_5$.

The zero-modes of $\psi^{aa}$ and $\psi^{bb}$ correspond to 
four-dimensional massless gauginos for the 
unbroken gauge group $U(N_a)\times U(N_b)$.
Dirac equations of $\psi^{aa}(y)$ and $\psi^{bb}(y)$ 
in (\ref{eq:Dirac-U(N)-T2-1}) and (\ref{eq:Dirac-U(N)-T2-2}) do not 
include any magnetic fluxes.
That is, both of $\psi_{\pm}$ have the same zero-modes as 
those on $T^2$ without magnetic fluxes.

Next, we study spinor fields, $\lambda^{ab}$ and $\lambda^{ba}$,  
which correspond to bi-fundamental matter fields, $(N_a,\bar N_b)$ 
and $(\bar N_a,N_b)$ for the unbroken gauge group  
$U(N_a)\times U(N_b)$.
When $M_a - M_b >0$, $\lambda^{ab}_+$ and $\lambda^{ba}_-$ 
have $(M_a-M_b)$ zero-modes with normalizable wavefunctions, 
i.e. $\Theta^j(y_4,y_5)$ for $j=0,\cdots, (M_a-M_b-1)$ as 
(\ref{eq:zero-mode-wf}), but 
zero-mode wavefunctions of $\lambda^{ab}_-$ and $\lambda^{ba}_+$ 
are unnormalizable.
On the other hand, when $M_a - M_b <0$, 
$\lambda^{ab}_-$ and $\lambda^{ba}_+$ have $(M_b - M_a)$ normalizable 
zero-modes.
Hence, we obtain chiral theory.

Similarly, we can analyze bosonic fields $A_m$.
In general, introduction of non-vanishing magnetic fluxes on $T^2$ 
breaks supersymmetry completely.

\subsubsection{$U(N)$ gauge theory on $(T^2)^3$}

Here, we extend the previous analysis to 
$U(N)$ gauge theory on $(T^2)^3$.
We consider the magnetic background, where 
only $F_{45}, F_{67}$ and $F_{89}$ are non-vanishing, 
but the others of $F_{mn}$ are vanishing.
Furthermore, $F_{45}, F_{67}$ and  $F_{89}$ are 
given by 
\begin{eqnarray}\label{eq:6D-flux}
F_{45} &=& 2 \pi \left(
\begin{array}{ccc}
M^{(1)}_1 {\bf 1}_{N_1\times N_1} & & 0  \\
 & \ddots & \\
0 & & M^{(1)}_n {\bf 1}_{N_n\times N_n}
\end{array}\right),  \nonumber \\
F_{67} &=& 2 \pi\left(
\begin{array}{ccc}
M^{(2)}_1 {\bf 1}_{N_1\times N_1} & & 0 \\
 & \ddots & \\
0 & & M^{(2)}_n {\bf 1}_{N_n\times N_n}
\end{array}\right),  \\
F_{89} &=& 2 \pi \left(
\begin{array}{ccc}
M^{(3)}_1 {\bf 1}_{N_1\times N_1} & & 0 \\
 & \ddots & \\
0 & & M^{(3)}_n {\bf 1}_{N_n\times N_n}
\end{array}\right). \nonumber 
\end{eqnarray}
This background breaks the gauge group $U(N)$ as 
$U(N) \rightarrow \prod_{a=1}^n U(N_a)$ with $N=\sum_a N_a$.

We can study gaugino fields on this background as 
a simple extension of the previous section 2.3.
That is, we focus on the block including only $U(N_a) \times U(N_b)$ 
and such a block has the following magnetic flux as 
(\ref{eq:F-ab-block}), 
\begin{equation}\label{eq:6D-F-ab-block}
F_{2i+2,2i+3} = 2 \pi\left(
\begin{array}{cc}
M^{(i)}_a {\bf 1}_{N_a\times N_a}  & 0 \\
0 & M^{(i)}_b {\bf 1}_{N_b\times N_b}
\end{array}\right),
\end{equation}
and  we use the following gauge  
\begin{equation}\label{eq:6D-gauge}
A_{2i+2} = -y_{2i+3}F_{2i+2,2i+3}, \qquad A_{2i+3} = 0,
\end{equation}
for $i=1,2,3$.
Then, we decompose the gaugino fields $\lambda(x,y)$ as 
(\ref{eq:gaugino-decomp}), i.e. the four-dimensional part $\chi(x)$ 
and the $i$-th $T^2$ part $\psi_{(i)}(y_{2i+2},y_{2i+3})$, 
whose zero-modes satisfy 
\begin{eqnarray}
\left(
\begin{array}{cc}
\bar \partial_i \psi_{(i)+}^{aa} & [\bar  \partial_i +2\pi i
(M^{(i)}_a-M^{(i)}_b)y_{2i+3} ] 
\psi^{ab}_{(i)+} \cr \cr
  [\bar \partial_i + 2\pi i (M^{(i)}_b-M^{(i)}_a)y_{2i+3}]
  \psi^{ba}_{(i)+} 
& \bar \partial_i \psi_{(i)+}^{bb}  \\
\end{array}
\right) &=& 0, \nonumber \\
 & & \\
\left(
\begin{array}{cc}
\partial_i \psi_{(i)-}^{aa} & 
[\partial_i - 2\pi i (M^{(i)}_a-M^{(i)}_b)y_{2i+3}] 
\psi^{ab}_{(i)-}  \cr \cr
[\partial_i -2 \pi i (M^{(i)}_b-M^{(i)}_a)y_{2i+3}] \psi^{ba}_{(i)-} & 

\partial_i \psi_{(i)-}^{bb}  \\
\end{array}
\right) &=& 0, \nonumber
\end{eqnarray}
where $\bar \partial_i = \partial_{2i+2}  +i \partial_{2i+3}$ and 
$\partial_i = \partial_{2i+2} -i \partial_{2i+3}$.

The gaugino fields, $\psi^{aa}$ and $\psi^{bb}$, for 
the unbroken gauge symmetry have no effect from magnetic 
fluxes in their Dirac equations.
Hence, they have the same zero-modes as those on 
$(T^{2})^3$ without magnetic fluxes.
On the other hand, $\psi^{ab}$ and $\psi^{ba}$ correspond 
to bi-fundamental matter fields, 
$(N_a, \bar N_b)$ and $(\bar N_a,N_b)$. 
For the $i$-th $T^2$ with $M^{(i)}_a -M^{(i)}_b >0$,
$\psi^{ab}_{(i)+}$ and $\psi^{ba}_{(i)-}$ have 
$|M^{(i)}_a - M^{(i)}_b |$ normalizable zero-modes, 
while $\psi^{ab}_{(i)-}$ and $\psi^{ba}_{(i)+}$ have 
no normalizable zero-modes.
When $M^{(i)}_a -M^{(i)}_b <0$, 
$\psi^{ab}_{(i)-}$ and $\psi^{ba}_{(i)+}$ have 
$|M^{(i)}_a - M^{(i)}_b |$ normalizable zero-modes.
Then, the total number of bi-fundamental zero-modes 
is given by $\prod_{i=1}^3|M_a^{(i)}-M_b^{(i)}|$ and 
all of them have the same six-dimensional chirality 
${\rm sign} \left[ \prod_{i=1}^3(M_a^{(i)}-M_b^{(i)})\right]$.
Since the ten-dimensional chirality of gaugino fields is fixed, 
bi-fundamental zero-modes for either $(N_a,\bar N_b)$ or 
$(\bar N_a,N_b)$ appear with a fixed four-dimensional chirality.
To summarize, the total number of bi-fundamental zero-modes 
for $(N_a,\bar N_b)$ is equal to 
\begin{equation}
I_{ab}=\prod_{i=1}^3(M_a^{(i)}-M_b^{(i)}),
\end{equation}
and their wavefunctions are given by a product
of two-dimensional parts, i.e. 
\begin{equation}\label{eq:zero-mode-6D}
\Theta^{i_1,i_2,i_3}(y) = \Theta^{i_1}(y_4,y_5) 
\Theta^{i_2}(y_6,y_7) \Theta^{i_3}(y_8,y_9),
\end{equation}
for $i_1 =0,\cdots, (M_a^{(1)}-M_b^{(1)}-1)$, 
$i_2 =0,\cdots, (M_a^{(2)}-M_b^{(2)}-1)$ and 
$i_3 = 0,\cdots, (M_a^{(3)}-M_b^{(3)}-1)$.
For $I_{ab} <0$, this means 
that there appear $|I_{ab}|$ independent zero-modes for 
$(\bar N_a,N_b)$.
It is also convenient to introduce the notation, 
$I^i_{ab} \equiv M_a^{(i)}-M_b^{(i)}$.

Similarly, we can analyze bosonic fields corresponding to $A_m$ 
for $m=4, \cdots, 9$.
For generic values of magnetic fluxes, supersymmetry 
is broken completely.
However, when they satisfy the following 
condition \cite{Troost:1999xn,Cremades:2004wa},
\begin{equation}\label{eq:SUSY-condition-0}
\sum_{i=1}^3\pm \frac{M_a^{(i)}-M_b^{(i)}}{{\cal A}^{(i)}} =0,
\end{equation}
for one combination of signs, 
where ${\cal A}^{(i)}$ denotes the area of the $i$-th torus,
there appear massless scalar modes as well as massive modes and 
four-dimensional N=1 supersymmetry remains unbroken at least in 
the $a-b$ sector.
When we consider ${\cal A}^{(i)}$ as free parameters, 
we can realize the above supersymmetric condition
(\ref{eq:SUSY-condition-0}) for 
most cases by choosing proper values of ${\cal A}^{(i)}$.
For the case with the universal area, 
${\cal A}^{(1)}={\cal A}^{(2)}={\cal A}^{(3)}$, the above condition 
(\ref{eq:SUSY-condition-0}) reduces to  
\begin{equation}\label{eq:SUSY-condition}
\sum_{i=1}^3\pm (M_a^{(i)}-M_b^{(i)}) =0.
\end{equation}
In addition to (\ref{eq:SUSY-condition-0}), 
when one of them is vanishing, i.e.
$(M_a^{(i)}-M_b^{(i)}) =0$ and 
\begin{equation}
\sum_{j \neq i} \pm \frac{M_a^{(j)}-M_b^{(j)}}{{\cal A}^{(j)}} 
=0,
\end{equation}
four-dimensional N=2 supersymmetry is unbroken.
In these supersymmetric models, 
zero-mode profiles of bosonic fields are the same as 
their superpartners, that is, zero-mode profiles 
of fermionic fields.

\subsection{General flux and non-abelian Wilson line}

Here, we consider $T^2$ of $(T^2)^n$, whose coordinates are 
denoted as $(y_4,y_5)$ with twisted boundary conditions.
As a $U(N)$ gauge background, we introduce the following form 
of (abelian) magnetic flux,
\begin{equation}\label{eq:F45-UN}
F_{45} = 2 \pi \left(
\begin{array}{cc}
f_a {\bf 1}_{N_a} & 0 \\
0 & 0
\end{array}\right),
\end{equation}
where ${\bf 1}_{N_a }$ denotes $(N_a \times N_a)$ 
identity matrix.
For example, we use the following gauge, 
\begin{equation}\label{eq:gauge2}
A_4 = -F_{45}y_5, \qquad A_5 = 0.
\end{equation}
Then, their boundary conditions can be written as 
\begin{eqnarray}
\label{eq:BC-gauge2}
A_m(y_4+1,y_5)&=&A_m(y_4,y_5)+\partial_m \chi_4, \qquad   
\chi_4 = 0, \nonumber \\
A_m(y_4,y_5+1)&=&A_m(y_4,y_5)+\partial_m \chi_5, \qquad 
\chi_5 = - 2 \pi f_a y_4.
\end{eqnarray}

This background breaks the gauge group $U(N)$ to 
$U(N_a) \times U(N-N_a)$.
The zero-mode $\psi(y)$ corresponding to the gaugino is also 
decomposed as 
\begin{equation}\label{eq:psi-ABCD} 
\psi = 
\begin{pmatrix}
A & B \\
C & D 
\end{pmatrix},
\end{equation}
depending on their $U(N_a) \times U(N-N_a)$ charges.
That is, $A$ and $D$ correspond to the gaugino fields of 
unbroken symmetries, $U(N_a)$ and $U(N-N_a)$, respectively, 
while $B$ and $C$ correspond to bi-fundamental representations, 
$(N_a,\overline{N-N_a})$ and $(\overline{N_a},{N-N_a})$, respectively.
The zero-mode satisfies the following equation,
\begin{equation}\label{eq:Dirac-T2}
\tilde \Gamma^m(\partial_m -iqA_m)\psi(y) = 0,
\end{equation}
for $m=4,5$, where $\tilde \Gamma^m$ corresponds to 
the gamma matrix for the two-dimensional torus $T^2$ and 
$\psi(y)$ is the two component spinor.
That is, $A, B, C$ and $D$ also have two components, 
$A_\pm, B_\pm, C_\pm$ and $D_\pm$.
Here, $q$ denotes the charge of $\psi$ under 
the gauge background $A_m$.
Since only the $U(1)$ part of $U(N_a)$  has the non-trivial
background, 
its charge is relevant, that is, $A, B, C$ and $D$ have 
charges $q=0,1,-1$ and $0$, respectively.

Because of (\ref{eq:BC-gauge2}), the spinor field satisfies 
the following boundary condition,
\begin{eqnarray}
\label{eq:bc-1}
\psi(y_4+1,y_5) &=& e^{iq\chi_4}\psi(y_4,y_5), \\
\psi(y_4,y_5+1) &=& e^{iq\chi_5}\psi(y_4,y_5) .
\label{eq:bc-2}
\end{eqnarray}
We write 
\begin{eqnarray}
\psi(y_4+1,y_5) &=& \Omega_4(y_4,y_5) \psi(y_4,y_5), \\
\psi(y_4,y_5+1) &=& \Omega_5(y_4,y_5) \psi(y_4,y_5) .
\end{eqnarray}
Then, the consistency for the contractible loop, i.e. 
$(y_4,y_5) \to (y_4+1,y_5) \to (y_4+1,y_5+1) \to (y_4,y_5+1) \to
(y_4,y_5)$ 
 requires 
\begin{eqnarray}\label{eq:loop}
\left( \Omega_5^{-1}(y_4,y_5+1)  \Omega_4^{-1}(y_4+1,y_5+1)
\Omega_5(y_4+1,y_5) 
\Omega_4(y_4,y_5) \right) \psi(y_4,y_5) = \psi(y_4,y_5),
\end{eqnarray}
for $\psi = A, B, C, D$.
The left hand side reduces to $e^{-2\pi i qf_a}\psi(y_4,y_5)$
in the above background.
Although that is trivial for $\psi = A$ and $D$, 
this condition for $\psi = B$ and $C$ leads to 
the quantization condition of the magnetic flux $f_a$. 
That is, the magnetic flux $f_a$ should be quantized such
that $f_a=$ integer.

When we introduce non-trivial background for 
the $SU(N_a)$ part of $U(N_a)$, the situation changes.
Now, let us impose the following boundary conditions for 
$\psi = B$,
\begin{eqnarray}
B(y_4+1,y_5) &=& \Omega_4(y_4,y_5) B(y_4,y_5) = 
e^{i\chi_4} \omega_4 B(y_4,y_5), \\
B(y_4,y_5+1) &=& \Omega_5(y_4,y_5) B(y_4,y_5) = 
e^{i\chi_5} \omega_5 B(y_4,y_5),
\end{eqnarray}
where $\omega_m$ are constant elements of $SU(N_a)$.
Then, the consistency condition (\ref{eq:loop}) reduces to 
\begin{eqnarray}\label{eq:loop-na}
\omega_5^{-1} \omega_4^{-1}  \omega_5 \omega_4  e^{-2\pi i f_a} = 
{\bf 1}_{N_a }.
\end{eqnarray}
If $\omega_4$ and $\omega_5$ commute each other, 
that would require gain  $e^{-2\pi i f_a} = 1$.
Thus, it is interesting that $\omega_4$ and $\omega_5$ do not 
commute each other, that is, non-Abelian Wilson lines.
In particular, we consider the case that 
$\omega_5^{-1} \omega_4^{-1}  \omega_5 \omega_4$ corresponds 
to the center of $SU(N_a)$, that is, 
\begin{eqnarray}\label{eq:twist-alg}
\omega_5^{-1} \omega_4^{-1}  \omega_5 \omega_4   = e^{ 2\pi i M_a/N_a}
{\bf 1}_{N_a },
\end{eqnarray}
where $M_a$ is an integer.
In this case, the consistency condition (\ref{eq:loop-na}) 
requires that the magnetic flux should satisfy $f_a= M_a/N_a$ 
(mod 1).

We denote $P_a = {\rm g.c.d.}(M_a,N_a)$, $m_a=M_a/P_a$ and 
$n_a=N_a/P_a$.\footnote{
Here, ${\rm g.c.d.}$ denotes the greatest common divisor.}
A solution of Eq.~(\ref{eq:twist-alg}) is given as 
\begin{eqnarray}\label{eq:nA-WL-1}
\omega_4 = P, \qquad \omega_5 = Q^{-m_a},
\end{eqnarray}
where
\begin{eqnarray}\label{eq:nA-WL-2}
P= 
\begin{pmatrix}
0 & {\bf 1}_{P_a } & 0 & 0 \\ 
0 & 0 & {\bf 1}_{P_a } & 0 \\
\cdots \\
{\bf 1}_{P_a } & 0 & 0 & 0 
\end{pmatrix}, \quad
Q=
\begin{pmatrix}
{\bf 1}_{P_a } & 0 & 0 & 0 \\ 
0 & \rho {\bf 1}_{P_a } & 0 & 0 \\
\cdots \\
0 & 0 & 0 & \rho^{n_a-1} {\bf 1}_{P_a }\\
\end{pmatrix},
\end{eqnarray}
with $\rho \equiv e^{{2\pi}i/{n_a}}$.

These non-Abelian Wilson lines break the gauge group $U(N_a)$ further.
The following condition on the $U(N_a)$ gauge field, 
\begin{eqnarray}
A_\mu = w_4 A_\mu \omega_4^{-1} = w_5 A_\mu \omega_5^{-1},
\end{eqnarray}
is required.
Then, the gauge group $U(N_a)$ breaks to $U(P_a)$.

\subsubsection{Matter fields}

Here, we consider the following form of $U(N)$ magnetic fluxes,
\begin{equation}\label{eq:F45-UN}
F_{45} = 2 \pi \left(
\begin{array}{ccc}
f_1 {\bf 1}_{N_1} & & 0 \\
 & \ddots & \\
0 & & f_n {\bf 1}_{N_n }
\end{array}\right).
\end{equation}
This form of magnetic fluxes breaks $U(N)$ to 
$\prod_i U(N_i)$ for $f_i=$ integer.
Furthermore, the gauge group is broken to 
$\prod_i U(P_i)$ when we choose $f_i=M_i/N_i$ with 
$P_i = {\rm g.c.d.}(M_i,N_i)$ and non-Abelian Wilson lines such that 
they satisfy the consistency condition like Eq.~(\ref{eq:loop}).

Now, let us focus on the $(N_a+N_b)\times (N_a+N_b)$ block 
in $U(N)$, which has the magnetic flux,
\begin{equation}\label{eq:mg-flux} 
F = 2 \pi
\begin{pmatrix}
\frac{m_a}{n_a} {\bf 1}_{N_a} &  \\
& \frac{m_b}{n_b} {\bf 1}_{N_b} \\
\end{pmatrix} .
\end{equation}
We use the same gauge as Eq.~(\ref{eq:gauge2}), i.e.
\begin{eqnarray}
A_4= - 2\pi
\begin{pmatrix}
\frac{ m_a}{n_a} {\bf 1}_{N_a} & \\ 
 & \frac{ m_b}{n_b} {\bf 1}_{N_b} \\ 
\end{pmatrix}y_5,
 \quad \quad 
A_5= 0. 
\end{eqnarray}
Similarly to Eq.~(\ref{eq:BC-gauge}), we denote 
their boundary conditions as 
\begin{eqnarray}
A_m(y_4+1,y_5)&=&A_m(y_4,y_5)+
\begin{pmatrix}
\partial_m \chi^a_4 & 0 \\
0& \partial_m \chi^b_4 
\end{pmatrix},  \nonumber \\
A_m(y_4,y_5+1)&=&A_m(y_4,y_5)+
\begin{pmatrix}
\partial_m \chi^a_5 & 0 \\
0& \partial_m \chi^b_5 
\end{pmatrix},
\end{eqnarray}
where 

\begin{eqnarray}
\chi_4^a=0, \ \ \ 
\chi_5^a=-2\pi\frac{m_a}{n_a}y_4, \ \ \   
\chi_4^b=0, \ \ \ 
\chi_5^b=-2\pi \frac{m_b}{n_b}y_4 .
\end{eqnarray}

We decompose the gaugino fields of this block 
in a way similar to Eq.~(\ref{eq:psi-ABCD}).
That is, $A$ and $D$ correspond to adjoint matter fields 
of $U(N_a)$ and $U(N_b)$, respectively, while 
$B$ and $C$ correspond to bi-fundamental representations, 
$(N_a,\overline{N_b})$ and $(\overline{N_a},N_b)$, respectively.
Among them, we concentrate on the field $B$, 
which satisfies the boundary conditions,
\begin{eqnarray}\label{eq:B-BC}
B(y_4+1,y_5) &=&
\Omega_4^a B(y_4,y_5)  (\Omega_4^b)^\dagger =
e^{i(\chi_4^a-\chi_4^b)} \omega_4^a B(y_4,y_5) (\omega^b_4)^\dagger, 
\nonumber \\
 B(y_4,y_5+1) &=&
\Omega_5^a B(y_4,y_5)  (\Omega_5^b)^\dagger =
e^{i(\chi_5^a-\chi_5^b)} \omega_5^a B(y_4,y_5) (\omega^b_5)^\dagger.
\end{eqnarray}
Here, $\omega_{4,5}^{a,b}$ are non-Abelian Wilson lines, 
which are given as Eqs.~(\ref{eq:nA-WL-1}) and (\ref{eq:nA-WL-2}).
Then, the gauge symmetry is broken to $U(P_a)$ and $U(P_b)$.
We study zero-mode profiles of $B$ fields in what follows.

Here, we study zero-mode profiles in the models with 
fractional magnetic fluxes and non-Abelian Wilson lines.

\subsubsection{$n_a = n_b$}

First, let us study the magnetic flux (\ref{eq:mg-flux}) 
for $n= n_a = n_b$.
In this case, the non-Abelian Wilson lines break
the gauge group $U(N_a)\times U(N_b)$ to 
$U(P_a) \times U(P_b)$, where $P_a=N_a/n$ and $P_b=N_b/n$.
Following this breaking pattern, we decompose the fields $B$ as 
\begin{eqnarray}
B= 
\begin{pmatrix}
B_{00} & B_{01} & \cdots & \\
B_{10} & B_{11} & \cdots & \\
\cdots \\
B_{n-1,0} & B_{n-1,1} & \cdots & B_{n-1,n-1} 
\end{pmatrix}.
\end{eqnarray}
Each of $B_{p,q}$ components is  
$(P_a \times P_b)$ matrix-valued fields, 
which correspond to bi-fundamental $(P_a,\bar P_b)$
fields under $U(P_a)\times U(P_b)$.
The boundary condition (\ref{eq:B-BC}) due to 
the non-Abelian Wilson lines is written as
\begin{eqnarray}\label{eq:B-BC-1}
B_{pq}(y_4+1,y_5) &=& B_{p+1,q+1}(y_4,y_5), \nonumber \\ 
B_{pq}(y_4,y_5+1) &=& \rho^{-(m_ap-m_bq)}e^{-\frac{2\pi im}{n}y_4} 
B_{p,q}(y_4,y_5),
\end{eqnarray}
where $m$ is used as $m=m_a-m_b$.
That leads to the boundary condition,
\begin{eqnarray}\label{eq:B-BC-n}
B_{pq}(y_4+n,y_5) &=& B_{pq}(y_4,y_5),  \nonumber \\ 
B_{pq}(y_4,y_5+n) &=& e^{-2\pi im y_4} B_{pq}(y_4,y_5).
\end{eqnarray}
Suppose that $mn>0$.
Then, similar to section \ref{sec:integer-mf}, 
the $B_+$ component for $B_{p,q}$ has 
$nm$ independent solutions for the zero-mode 
Dirac equation (\ref{eq:Dirac-T2}) with the above 
condition (\ref{eq:B-BC-n}).
These solutions are given by 
\begin{eqnarray}
\Theta^{j}(y_4,y_5)
&=&
\sum_l e^{- nm\pi(l+\frac{j}{nm})^2 + 2\pi i
m(l+\frac{j}{nm})y_4-\frac{\pi m}{n}y^2_5
-2\pi m(l+\frac{j}{nm})y_5  } \nonumber  \\
&=&
e^{-\frac{\pi m}{n}y_5^2} \ 
\jtheta{\frac{j}{nm} \\ 0}(mz,nm\tau) ,
\end{eqnarray}
where $j= 0, 1, \cdots, nm-1$ and $\tau=i$.
On the other hand, the $B_-$ component has no 
normalizable zero-modes.
One finds that these solutions satisfy 
the boundary conditions,
\begin{eqnarray}
\Theta^j(y_4+1,y_5) &=& e^{\frac{2\pi ij}{n}   } \Theta^{j}(y_4,y_5), 
\nonumber \\ 
\Theta^j(y_4,y_5+1) &=& e^{-\frac{2\pi im}{n} y_4}
\Theta^{j+m}(y_4,y_5).  
\end{eqnarray}
Thus, the zero-mode solutions with 
the boundary conditions  (\ref{eq:B-BC-1}) due to non-Abelian Wilson
lines 
can be written in terms of $\Theta^j$ as 
\begin{eqnarray} \label{eq:zeromode-1}
& & B^j_{pq}(y_4,y_5)=c_{pq}^j \sum_{r=0}^{n-1} e^{2 \pi
  i(m_ap-m_bq)\frac{r}{n}}\Theta^{j+mr},  
\end{eqnarray}
where $j=0,1,...,m-1$.
Here,  $c_{pq}^j$ is a constant normalization, which can be 
determined by the boundary conditions.

We have concentrated on the $B_{+}$ fields.
Similarly, when $mn >0$, 
the $C_-$ fields have the same solutions as $B_+$.
However, the $B_-$ and $C_+$ have no normalizable zero-modes 
for $mn >0$.
On the other hand, when $mn <0$
the $B_-$ and $C_+$ have normalizable zero-modes with the same 
wavefunctions as the above, 
while $B_+$ and $C_-$ have normalizable zero-modes.

We have considered the zero-modes profiles of 
fermionic fields.
If 4D N=1 supersymmetry is preserved, 
the scalar mode has the same zero-mode profiles as its 
fermionic superpartner.

\subsubsection{$n_a \neq n_b$}

Next, we study the model with $n_a \neq n_b$.
In this case, the non-Abelian Wilson lines break
the gauge group $U(N_a)\times U(N_b)$ to 
$U(P_a) \times U(P_b)$, where $P_a=N_a/n_a$ and $P_b=N_b/n_b$.
Similar to the previous subsection, 
we decompose the fields $B$ as 
\begin{eqnarray}
B= 
\begin{pmatrix}
B_{00} & B_{01} & \cdots & B_{0,n_b-1}\\
B_{10} & B_{11} & \cdots & \\
\cdots \\
B_{n_a-1,0} & B_{n_a-1,1} & \cdots & B_{n_a-1,n_b-1} 
\end{pmatrix}.
\end{eqnarray}
Each of $B_{p,q}$ components is  
$(P_a \times P_b)$ matrix-valued fields.
The boundary condition (\ref{eq:B-BC}) due to 
the non-Abelian Wilson lines is written as
\begin{eqnarray}\label{eq:B-BC-1-2}
B_{pq}(y_4+1,y_5) &=& B_{p+1,q+1}(y_4,y_5), \nonumber \\ 
B_{pq}(y_4,y_5+1) &=& e^{-2\pi i (\frac{m_a}{n_a}-\frac{m_b}{n_b})
y_4}
                  e^{2\pi i(\frac{m_a}{n_a}p-\frac{m_b}{n_a}q)} 
                  B_{p,q}(y_4,y_5).
\end{eqnarray}
That leads to the boundary condition,
\begin{eqnarray}\label{eq:B-BC-Q}
B_{pq}(y_4+Q_{ab},y_5) &=& B_{pq}(y_4,y_5), \nonumber \\ 
B_{pq}(y_4,y_5+Q_{ab}) &=& e^{-\frac{2\pi i}{k_{ab}} I_{ab} y_4}
B_{p,q}(y_4,y_5), 
\end{eqnarray}
where $I_{ab} = n_bm_a-n_am_b$ and $Q_{ab}$ is defined by 
$Q_{ab}={\rm l.c.m.}(n_a,n_b)$.\footnote{
Here, l.c.m. denotes the least common multiple.} 
In addition, we define $k_{ab}={\rm g.c.d.}(n_a,n_b)$, which is
related 
with $Q_{ab}$ as  $Q_{ab}=\frac{n_a n_b}{k_{ab}}$. 
There are $S_{ab}=\frac{n_an_b}{k^2_{ab}}I_{ab}$ independent zero-mode 
profiles, which satisfy the boundary condition (\ref{eq:B-BC-Q}).
Those functions are obtained as 
\begin{eqnarray}
\Theta^{j}(y_4,y_5)
&=&
\sum_n e^{- \pi S_{ab}(n+\frac{j}{S_{ab}})^2 + \frac{2\pi i
S_{ab}}{Q_{ab}}
(n+\frac{j}{S_{ab}})y_4-\frac{\pi S_{ab}}{Q_{ab}^2}y_5^2
-2\pi \frac{S_{ab}}{Q_{ab}}(n+\frac{j}{S_{ab}})y_5} \nonumber \\
&=&
e^{-\frac{\pi S_{ab}}{Q_{ab}^2}y_5^2}
\jtheta{\frac{j}{S_{ab}} \\
0}\left((S_{ab}/Q_{ab})z,S_{ab}\tau\right),
\end{eqnarray}
where $\tau=i$.
These wavefunctions satisfy the following boundary conditions,
\begin{eqnarray}
\Theta^j (y_4+1,y_5) &=& e^{2\pi i \frac{k_{ab}}{n_a n_b}j}
\Theta^j(y_4,y_5) , \nonumber \\
\Theta^j (y_4,y_5+1) &=&
e^{2\pi i(\frac{m_a}{n_a}-\frac{m_b}{n_b})y_4}
\Theta^{j-\frac{I_{ab}}{k_{ab}}}(y_4,y_5) .
\end{eqnarray}
Thus, the zero-mode wavefunctions, which satisfy the 
boundary conditions (\ref{eq:B-BC-1-2}), are obtained as 
\begin{eqnarray}\label{eq:B-wf-2}
B_{pq}^j(y_4,y_5)
= c_{pq}^j 
\sum_{r=0}^{Q_{ab}-1}
e^{2\pi i(\frac{m_a}{n_a} p-\frac{m_b}{n_b}q) r}
\Theta^{j+\frac{I_{ab}}{k_{ab}}r}(y_4,y_5) ,
\end{eqnarray}
where $j =0,1, \cdots, \frac{I_{ab}}{k_{ab}}-1$.

As an illustrating example, we consider the model with 
$n_a=2, n_b=4$ and $m_a=m_b=3$.
Then, we have $k_{ab}={\rm g.c.d.}(n_a,n_b)=2 \ne 1$ and $I_{ab}=6$.
We decompose the  bi-fundamental fields $B$ with  
the $2\times 4$ matrix entries as
\begin{eqnarray}
B = 
\begin{pmatrix}
B_{00} & B_{01} & B_{02} & B_{03} \\
B_{00} & B_{11} & B_{12} & B_{13}
\end{pmatrix}.
\end{eqnarray}
From the wavefunction formula in Eq.~(\ref{eq:B-wf-2}),
one obtains the three independent solutions labeled by $j=0,1,2$ for
each
component of $B_{pq}$ and 
these are represented by linear combination of $\Theta^i$, 
for example $B_{00}$ and $B_{01}$ are 
\begin{eqnarray}
B^j_{00} &=& 
\Theta^j +\Theta^{j+3}+\Theta^{j+6} + \Theta^{j+9}, \nonumber \\
B^j_{01} &=& 
\Theta^j + e^{-\frac{3\pi i}{2}} \Theta^{j+3}
+e^{-3\pi i} \Theta^{j+6} + e^{-\frac{9\pi i}{2}} \Theta^{j+9}.
\end{eqnarray}
Obviously, the $y_4$-direction boundary condition can connect some of 
components of $B$ as follows 
\begin{eqnarray}
B_{00} \to B_{11} \to B_{02} \to B_{13} \to B_{00} \\
B_{01} \to B_{12} \to B_{03} \to B_{14} \to B_{01}.
\end{eqnarray}
Hence, there are 6 zero-mode solutions in this background.

\begin{itemize}
\item Another representation of solutions
\end{itemize}

In the previous section, we have presented 
solutions in terms of the  $\Theta^j$ functions.
However, by using the properties of the theta function, 
one can represent the wavefunctions (\ref{eq:zeromode-1}) 
and (\ref{eq:B-wf-2}) as a single theta 
function as  
\begin{eqnarray}\label{eq:B-wf-3}
B_{pq}^j(y_4,y_5)
=
C_{p,q}^j
e^{-\pi \tilde{I}_{ab}y_5^2}
\times 
\jtheta{\frac{j}{M_{ab}} \\ 0}
\left(\tilde{I}_{ab} z+ 
\left(\frac{m_a}{n_a}p-\frac{m_b}{n_b}q \right),\tilde{I}_{ab} \tau
\right),
\end{eqnarray}
where $\tilde I_{ab} = I_{ab}/n_an_b$.
The constant $C_{p,q}^j$ can be determined by the 
boundary conditions.
The net number of zero-mode multiplicity is given by
$M_{ab}=I_{ab}/k_{ab}$.
Therefore the wavefunctions $B_{pq}^{j'}(y_4,y_5)$ with $j'=j+M_{ab}$
should be equal to $B_{pq}^j(y_4,y_5)$.
Furthermore we impose the $B_{p+n_a,q}^j, B_{p,q+n_b}^j=B_{p,q}^j$ and 
we have twist boundary condition
$B_{pq}^j(y_4+1,y_5)=B_{p+1,q+1}^j(y_4,y_5)$.
Then these conditions imply the following constraint for the
coefficients of $C_{pq}^j$ as
\begin{eqnarray}
e^{2\pi ij\frac{m_a}{M_{ab}} }C_{p+n_a,q}^{j}
=
e^{-2\pi ij\frac{m_b}{M_{ab}} }C_{p,q+n_b}^{j}
=
C_{pq}^j, \\
C_{p+1,q+1}^j=C_{p,q}^j, \quad
C_{p,q}^{j+M_{ab}}=C_{p,q}^j.  
\end{eqnarray}
In general, their solutions should not be determined uniquely.
We find that a simple solutions is
\begin{eqnarray}
C_{pq}^{j}=
e^{ 2\pi i j\frac{L}{M_{ab}}(p-q)}, 
\end{eqnarray}
where $L$ is a certain integer given by
\begin{eqnarray}
L= \frac{M_{ab}l_a-m_a}{n_a}=-\frac{M_{ab}l_b+m_b}{n_b},
\end{eqnarray}
where $l_a$ and $l_b$ are also integers.
Then the forms of wavefunctions become simple as
\begin{eqnarray}
B_{pq}^j(y_4,y_5)
&=&
N_j e^{ 2\pi i j\frac{L}{M_{ab}}(p-q)} 
e^{-\pi \tilde{I}_{ab} y_5^2} \times
\jtheta{\frac{j}{M_{ab}} \\ 0}
\left( \tilde{I}_{ab} z+ 
\left(\frac{m_a}{n_a}p-\frac{m_b}{n_b}q \right), \tilde{I}_{ab} \tau
\right).
\end{eqnarray}
However this expression is only valid if there exists such an integer
$L$ 
satisfying the relations.
Furthermore, when $\frac{m_a}{M_{ab}}=\frac{m_b}{M_{ab}}=\rm{integer}$, 
$C_{pq}^j$ is reduced to $C_{pq}^j=\rm{const.}$

\subsubsection{Continuous Wilson line}

So far, we have considered the simple $T^2$, 
where $y_4$ and $ y_5$ are identified as 
$y_4 \sim y_4+1$ and $ y_5 \sim y_5 +1$.
Similarly, we can study the torus compactification 
with arbitrary value of the complex structure modulus $\tau$, 
although we have fixed $\tau =i$ in the above analysis.
Then, we obtain the same zero-mode wavefunctions 
for arbitrary value of $\tau$ as Eq.~(\ref{eq:B-wf-3}) 
except replacing $z=y_4+iy_5$ in the theta function by $z=y_4+\tau
y_5$.
In this section, we also discuss about the effect of the constant gauge
potential called by Wilson line on the gauge group and wavefunctions.
It is useful to use the holomorphic basis of $z$ and 
gauge potential. 
In order for this reason, 
we take the following form of magnetic flux on $T^2$,
\begin{eqnarray}
F  ={\pi i \over \Im \tau} m \ (dz \wedge d \bar z), 
\end{eqnarray}
where $m$ is an integer~\cite{toron}.
We also take the following gauge of vector potential  
\begin{eqnarray}
A(z) ={\pi m \over \Im \tau} \Im (\bar z \ dz).
\end{eqnarray}
This form of the vector potential satisfies the 
following relations,
\begin{eqnarray}
A(z+1) &=& A(z) +{\pi m \over \Im \tau} \Im (dz), \\
A(z+\tau ) &=& A(z) +{\pi m \over \Im \tau} \Im (\bar \tau \ dz).
\end{eqnarray}
The Dirac equations of the zero-modes are modified 
by the Wilson line background, $\xi = \xi_4 + \tau \xi_5$ as 
\begin{eqnarray}\label{eq:zero-mode-WL}
& & \left( \bar \partial + \frac{\pi  q}{2 \Im (\tau)}  (mz +\xi)
\right) 
\psi_+(z,\bar z) =0, \\
& & \left( \partial - \frac{\pi  q}{2 \Im (\tau)}  (m\bar z+\bar \xi)
\right) 
\psi_- (z,\bar z) =0, 
\end{eqnarray}
where $\xi_4$ and $\xi_5$ are real constants.
That is, we can introduce the Wilson line background, 
$\xi = \xi_1 + \tau \xi_2$ by replacing $\chi_i$ in 
as~\cite{Cremades:2004wa}
\begin{eqnarray}\label{eq:chi-WL}
\chi_4 = {\pi  \over \Im \tau} \Im (mz +\xi), \qquad 
\chi_5 = {\pi  \over \Im \tau} \Im (\bar \tau (mz + \xi)).
\end{eqnarray}
Because of this Wilson line, the number of 
zero-modes does not change, but their wavefunctions 
are replaced as 
\begin{eqnarray}\label{eq:WL-shift}
\Theta^{j,M}(z) \rightarrow \Theta^{j,M}(z+\xi/m).
\end{eqnarray}

It would be useful to consider $U(1)_a \times U(1)_b$ theory from 
the phenomenological viewpoint.
We consider the fermion field $\lambda (x,z)$ 
with $U(1)_a \times U(1)_b$ charges, $(q_a,q_b)$.
We assume the following form of $U(1)_a$ magnetic flux on $T^2$,
\begin{eqnarray}\label{eq:magne-a}
F^a_{z \bar z} ={ \pi i \over \Im \tau} m_a, 
\end{eqnarray}
where $m_a$ is integer, but there is no magnetic flux in 
$U(1)_b$.
On top of that, we introduce Wilson lines $\xi^a$ and $\xi^b$ for 
$U(1)_a$ and $U(1)_b$, respectively.
The zero-mode equations are written as 
\begin{eqnarray}\label{eq:zero-mode-WL-ab}
& & \left( \bar \partial + \frac{\pi  }{2\Im (\tau)} 
 (q_a(m_az +\xi^a)+q_b\xi^b) \right) 
\psi_+(z,\bar z) =0, \\
& & \left( \partial - \frac{\pi  }{2\Im (\tau)} 
 (q_a(m_a\bar z +\bar \xi^a)+q_b\bar \xi^b) \right) 
\psi_- (z,\bar z) =0.
\end{eqnarray}
Then, the number of zero-modes is obtained as $M=q_a m_a $ 
and their wavefunctions are written as 
\begin{eqnarray}\label{eq:WL-shift-2}
\Theta^{j,M}(z+\xi/m_a),
\end{eqnarray}
where $\xi= \xi^a + \xi^bq_b/q_a$.
Here we give a few comments.
All of modes with $q_a=0$ become massive and there do not appear 
zero-modes with $q_a=0$.
For $q_a \neq 0$, zero-modes with $q_b=0$ appear 
and the number of zero-modes is independent of $q_b$.
Obviously, when we introduce Wilson lines $\xi^a$ and/or  $\xi^b$ 
without magnetic flux $F^a$, zero-modes do not appear.
The shift of wavefunctions depends on $1/m_a$ and 
the charge $q_b$.
Note that although $F^b=0$, 
Wilson lines $\xi_b$ and charges $q_b$ for $U(1)_b$ 
are also important \footnote{Wilson lines $\xi_b$ and charges $q_b$ for
  $U(1)_b$  are in a sense more important than 
Wilson lines $\xi_a$ and charges $q_a$ for $U(1)_a$, because  
the shift of wavefunctions (\ref{eq:WL-shift-2}) 
depends on $q_b$.}.

The above aspects of magnetic fluxes and Wilson lines 
are phenomenologically interesting.
We consider 6D super Yang-Mills theory with non-Abelian gauge group
$G$.
We introduce a magnetic flux $F^a$ along a Cartan direction of $G$.
Then, the gauge group breaks to $G'\times U(1)_a$ without reducing the
rank.
Furthermore, there appear the massless fermion fields $\lambda'$ , 
which correspond to the gaugino fields for the broken gauge group
part and have the fundamental representation 
of $G'$ and non-vanishing $U(1)_a$ charge.
Furthermore, we introduce a Wilson line along a Cartan direction of 
$G'$.
Then, the gauge group is broken to $G'' \times U(1)_a \times U(1)_b$ 
without reducing the rank.
The gaugino fields corresponding to the broken gauge part in $G'$ 
do not remain as massless modes, but they gain masses due to 
the Wilson line $U(1)_b$.
However, the fermion fields $\lambda'$ remain still massless 
with the same degeneracy.

\begin{figure}[t] \begin{center}
\includegraphics[height=3cm]{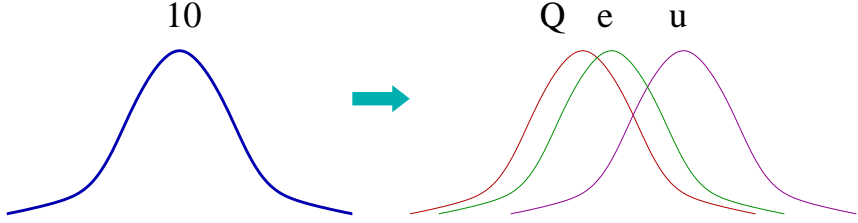}
\caption{Wavefunction splitting by Wilson lines}
\label{fig:WL-10}
\end{center} \end{figure}

Let us explain more on this aspect.
Suppose that we introduce magnetic fluxes in a model 
with a larger group $G$ such that they break 
$G$ to a GUT group like $SU(5)$ and 
this model includes three families of matter fields 
like $10$ and $\bar 5$.
Their Yukawa couplings are computed by the overlap integral of three 
zero-mode profiles. 
We obtain the GUT relation among Yukawa coupling 
matrices when wavefunction profiles of matter fields in 
$10$ ($\bar 5$) are degenerate.
Then, we introduce a Wilson line along $U(1)_Y$, which 
breaks $SU(5)$ to $SU(3) \times SU(2) \times U(1)_Y$.
Because of Wilson lines, $SU(5)$ gauge bosons except 
the $SU(3) \times SU(2) \times U(1)_Y$ gauge bosons become massive 
and the corresponding gaugino fields become massive.
However, three families of $10$ and $\bar 5$ matter fields remain
massless.
Importantly, this Wilson line resolves the degeneracy of 
wavefunction profiles of left-handed quarks, right-handed up-sector 
quarks and right-handed charged leptons in $10$ and 
right-handed down-sector quarks and left-handed charged leptons 
in $\bar 5$ as Figure \ref{fig:WL-10}.
That  is, the GUT relation among Yukawa coupling matrices is 
deformed.
As an illustrating model, we study  the Pati-Salam model in 
the next subsection.

Here we study effects due to discrete values of Wilson lines 
such as $\xi=k\tau$ with $k=$ integer.
We find 
\begin{eqnarray}\label{eq:WV}
\Theta^{j,M}(z+k\tau/M) = e^{\pi i k \Im (\bar \tau z)/\Im(\tau)}
\Theta^{j+k,M}(z).
\end{eqnarray}
Thus, the effect of discrete Wilson lines $\xi=k\tau$ is 
to replace the $j$-th zero-mode by the $(j+k)$-th zero-mode
up to $e^{\pi i k \Im (\bar \tau z)/\Im(\tau)}$.
However, when we consider 3-point and higher order couplings, 
the gauge invariance requires that the sum of Wilson lines 
of matter fields should vanish, that is, $\sum_i k_i=0$ for 
allowed n-point couplings.
Thus, the part $e^{\pi i k \Im (\bar \tau z)/\Im(\tau)}$ is irrelevant
to 
4D effective theory and 
the resultant 4D effective theory is the equivalent 
even when we introduce $\xi=k\tau$.
Similarly, introducing the Wilson lines $\xi=k$ with $k=$ integer leads 
to the equivalent 4D effective theory.

\subsection{Pati-Salam model}

As an illustrating model, we consider 
the Pati-Salam model.
We start with 10D N=1 $U(8)$ super Yang-Mills theory. 
We compactify the extra 6 dimensions on 
$T^2_1\times T^2_2 \times T^2_3$, and we denote the complex coordinate 
for the $d$-th $T^2_d$ by $z^d$, where $d=1,2,3$.
Then, we introduce the following form of magnetic fluxes,
\begin{equation} \label{toronbg} \begin{split}
 F_{z^d \bar z^d} = {\pi i \over \Im \tau_d}  \begin{pmatrix}
   m_1^{(d)} \1_{4} & & \\
  & m_2^{(d)} \1_{2}  & \\ & & m_3^{(d)} \1_{2} 
\end{pmatrix}, \quad d=1,2,3,
\end{split} \end{equation}
in the gauge space, 
where $\1_{N}$ are the unit matrices of rank $N$, $m_i^{(d)}$ are
integers.
We assume that the above background preserves 4D N=1 supersymmetry
(SUSY).
Here, we denote $M^{(d)}_{ij}=m^{(d)}_i - m^{(d)}_j$ and 
$M_{ij}=M^{(1)}_{ij}M^{(2)}_{ij}M^{(3)}_{ij}$.
This magnetic flux breaks the gauge group $U(8)$ to 
$U(4)\times U(2)_L \times U(2)_R $, that is the Pati-Salam gauge group 
up to $U(1)$ factors.
The gauge sector corresponds to 4D N=4 SUSY vector multiplet, that is, 
there are $U(4)\times U(2)_L \times U(2)_R $ N=1 vector multiplet and 
three adjoint chiral multiplets.
In addition, there appear bifundamental matter fields like  
$\lambda_{(4,2,1)}$, $\lambda_{(\bar 4,1,2)}$ and 
$\lambda_{(1,2,2)}$, and their numbers of zero-modes are equal to 
$M_{12}$, $M_{31}$ and $M_{23}$.
When $M_{ij}$ is negative, that implies their 
conjugate matter fields appear with the degeneracy $|M_{ij}|$.
The fields $\lambda_{(4,2,1)}$ and $\lambda_{(\bar 4,1,2)}$ 
correspond to left-handed and right-handed matter fields,
respectively, while $\lambda_{(1,2,2)}$ corresponds to 
up and down Higgs (higgsino) fields.
For example, we can realize three families by 
$M^{(d)}_{12}=(3,1,1)$ and $M^{(d)}_{31}=(3,1,1)$.
That leads to $|M_{23}|=0$ or 24.
At any rate, the flavor structure is determined by 
the first $T^2_1$ in such a model.
Explicitly, the zero-mode wavefunctions of both 
$\lambda_{(4,2,1)}$ and $\lambda_{(\bar 4,1,2)}$ 
are obtained as
\begin{eqnarray}
\Theta^{j,3}(z^1)\Theta^{1,1}(z^2)\Theta^{1,1}(z^3).
\end{eqnarray}
Their Yukawa matrices are constrained by the Pati-Salam 
gauge symmetry, that is, up-sector quarks, down-sector quarks,  
charged leptons and neutrinos have the same Yukawa matrices with 
Higgs fields.
Even with such a constraint, one could derive 
realistic quark/lepton masses and mixing angles, because 
this model has many Higgs fields and their vacuum expectation values 
generically break the up-down symmetry.

We introduce Wilson lines in $U(4)$ and $U(2)_R$ such that 
$U(4)$ breaks to $U(1)\times U(3)$ and $U(2)_R$ breaks 
$U(1) \times U(1)$.
Then, the gauge group becomes the standard gauge group 
up to $U(1)$ factors.
Furthermore, the profiles of left-handed quarks and 
leptons in $\lambda_{(4,2,1)}$ shift differently because of 
Wilson lines.
Similarly, right-handed up-sector quarks, down-sector quarks, 
charged leptons and neutrinos in $\lambda_{(\bar 4,1,2)}$ 
shift differently.
The flavor structure is determined by the first $T^2_1$.
Thus, when we introduce Wilson lines on the second or third torus,  
the resultant Yukawa matrices are constrained by the 
$SU(4) \times SU(2)_L \times SU(2)_R$.
For example, we introduce Wilson lines on $T^2_2$.
Then, zero-mode profiles of 
quarks, $(Q,u,d)$ and leptons $(L,e,\nu)$ split as 
\begin{eqnarray}
Q&:& \Theta^{j,3}(z^1)\Theta^{1,1}(z^2+\xi^a)\Theta^{1,1}(z^3),
\nonumber \\
L&:& \Theta^{j,3}(z^1)\Theta^{1,1}(z^2-3\xi^a)\Theta^{1,1}(z^3),
\nonumber 
\\
u^c&:& \Theta^{j,3}(z^1)\Theta^{1,1}(z^2-\xi^a+\xi^b)\Theta^{1,1}(z^3), 
\nonumber \\
d^c&:& \Theta^{j,3}(z^1)\Theta^{1,1}(z^2-\xi^a-\xi^b)\Theta^{1,1}(z^3), \\
e^c&:& \Theta^{j,3}(z^1)\Theta^{1,1}(z^2+3\xi^a-\xi^b)\Theta^{1,1}(z^3), 
\nonumber \\
\nu^c&:& \Theta^{j,3}(z^1)\Theta^{1,1}(z^2+3\xi^a+\xi^b)\Theta^{1,1}(z^3),
\nonumber  
\end{eqnarray}
where $\xi^a$ and $\xi^b$ are the Wilson lines 
to break $U(4) \rightarrow U(3) \times U(1)$ and 
$U(2)_R \rightarrow U(1) \times U(1)$, respectively. 
Those Wilson lines just change the overall factors 
of Yukawa matrices, but ratios among elements in one Yukawa 
matrix 
do not change.
Also we can introduce Wilson lines along the same $U(1)$ directions 
as the magnetic fluxes, but they do not deform 
the up-down symmetry of Yukawa matrices, either.

On the other hand, when we introduce Wilson lines 
on the first $T^2_1$, the zero-mode wavefunctions
 split as 
\begin{eqnarray}
Q&:& \Theta^{j,3}(z^1+\xi^a/3)\Theta^{1,1}(z^2)\Theta^{1,1}(z^3),
\nonumber \\
L&:& \Theta^{j,3}(z^1-\xi^a)\Theta^{1,1}(z^2)\Theta^{1,1}(z^3),
\nonumber 
\\
u^c&:&
\Theta^{j,3}(z^1-\xi^a/3+\xi^b/3)\Theta^{1,1}(z^2)\Theta^{1,1}(z^3), 
\nonumber \\
d^c&:&
\Theta^{j,3}(z^1-\xi^a/3-\xi^b/3)\Theta^{1,1}(z^2)\Theta^{1,1}(z^3), \\
e^c&:& \Theta^{j,3}(z^1+\xi^a-\xi^b/3)\Theta^{1,1}(z^2)\Theta^{1,1}(z^3), 
\nonumber \\
\nu^c&:&
\Theta^{j,3}(z^1+\xi^a+\xi^b/3)\Theta^{1,1}(z^2)\Theta^{1,1}(z^3). 
\nonumber  
\end{eqnarray}
In this case, the flavor structure is deviated from the 
$SU(4) \times SU(2)_L \times SU(2)_R$ relation, that is, 
mass ratios and mixing angles can change.
Also we can introduce Wilson lines $\xi^a$ to $T^2_2$ and $\xi^b$ to 
$T^2_1$.
Then we realize
\begin{eqnarray}
Q&:& \Theta^{j,3}(z^1)\Theta^{1,1}(z^2+\xi^a)\Theta^{1,1}(z^3),
\nonumber \\
L&:& \Theta^{j,3}(z^1)\Theta^{1,1}(z^2-3\xi^a)\Theta^{1,1}(z^3),
\nonumber 
\\
u^c&:& \Theta^{j,3}(z^1+\xi^b/3)\Theta^{1,1}(z^2-\xi^a)\Theta^{1,1}(z^3), 
\nonumber \\
d^c&:& \Theta^{j,3}(z^1-\xi^b/3)\Theta^{1,1}(z^2-\xi^a)\Theta^{1,1}(z^3),
\\
e^c&:& \Theta^{j,3}(z^1-\xi^b/3)\Theta^{1,1}(z^2+3\xi^a)\Theta^{1,1}(z^3), 
\nonumber \\
\nu^c&:&
\Theta^{j,3}(z^1+\xi^b/3)\Theta^{1,1}(z^2+3\xi^a)\Theta^{1,1}(z^3). 
\nonumber  
\end{eqnarray}

Indeed, this behavior is well-known in the intersecting 
D-brane models, which are T-duals of magnetized D-brane 
models.
In the intersecting D-brane side, introduction of 
Wilson lines corresponds to split D-branes.
By splitting D-branes, the gauge group breaks 
as $U(M+N) \rightarrow U(M)\times U(N)$, but 
the number of massless bi-fundamental modes does not change, 
although they decompose because of the gauge symmetry breaking.

\subsection{Exceptional gauge groups}

Here we extend these analysis to the exceptional gauge symmetry.
Gauge theories with the gauge groups $E_6$, $E_7$ and $E_8$, are 
quite interesting as grand unified theory in particle physics, 
which would lead to 
the standard model at low-energy.
All of quarks and leptons are involved in 
the ${\bf 16}$ representations of 
$SO(10)$ and such a ${\bf 16}$ representation 
appears from the adjoint representation and 
${\bf 27}$ representation of $E_6$.
Furthermore, these representations are included 
in adjoint representations of $E_7$ and $E_8$. 
These exceptional gauge theories can be derived 
in heterotic string theory, type IIB string theory with 
non-perturbative effects and F-theory.
Indeed, interesting models have been studied e.g. in
heterotic orbifold models \cite{Kobayashi:2004ud,Forste:2004ie,
Buchmuller:2005jr,Kim:2006hw,Lebedev:2006kn} 
and F-theory~\cite{Beasley:2008dc,Donagi:2008ca,Font:2008id,
Bourjaily:2009vf,Hayashi:2009ge,Marsano:2009ym,Blumenhagen:2009up}.

We start with 10 dimensional super Yang-Mills theory with gauge group
$G$. 
Introducing the magnetic flux on general gauge group $G$ is 
achieved by taking the background gauge potential along the Cartan
direction of gauge group $G$. We take this direction as $U(1)_a$. 

By the magnetic flux along the $U(1)_a$ direction,  
all of 4D gauge vector fields $A_\mu$, which have $U(1)_a$ charges, 
become massive, that is, the gauge group is broken from 
$G$ to $G'\times U(1)_a$ without reducing 
its rank,\footnote{For example, when 
$G=SU(N)$, $G'$ would correspond to $SU(N-1)$.} where 
4D gauge fields $A_\mu$ in $G' \times U(1)_a$ have vanishing 
$U(1)_a$ charges and their zero-modes $\phi_\mu(z)$ have 
a flat profile.
Since the magnetic flux has no effect on the unbroken 
gauge sector, 4D N=4 supersymmetry remains in the 
$G'\times U(1)_a$ sector, that is, there are massless 
four adjoint gaugino fields and six adjoint scalar fields.\footnote{
In string terminology, these adjoint scalar fields correspond to 
open string moduli, that is, D-brane position moduli.
How to stabilize these moduli is one of important issues.}

In addition, matter fields appear from  gaugino fields 
corresponding to the broken gauge part, that is, 
they have non-trivial representations under $G'$ and non-vanishing
$U(1)_a$ charges $q^a$.\footnote{For example, when $G=SU(N)$ and 
$G'\times U(1)_a= SU(N-1)\times U(1)_a $, these matter fields 
have $(N-1)$ fundamental representation under $SU(N-1)$ and 
$U(1)_a$ charge $q^a=1$ and their conjugates.}
Therefore we can obtain the chiral zero-modes 
by using same technique even for the exceptional gauge groups.

Now, let us introduce Wilson lines along the $U(1)_b$ direction of 
$G'$.
That breaks further the gauge group $G'$ to $G'' \times
U(1)_b$ without reducing its rank.\footnote{
For example, when $G'=SU(N-1)$, the Wilson line breaks it 
to $SU(N-2)\times U(1)_b$.}
All of the $U(1)_b$-charged fields including 4D vector, spinor and 
scalar fields become massive because of the Wilson line, when 
they are not charged under $U(1)_a$ and their zero-mode profiles 
are flat.
On the other hand, the matter fields with non-trivial profiles 
due to magnetic flux have different behavior. 
For matter fields with $U(1)_a$ charge $q^a$ and $U(1)_b$ charge
$q^b$, 
the Dirac equations of the zero-modes are modified 
by the Wilson line background, $\xi^b_d = \xi^b_{d,1} + \tau_d \xi^b_{d,2}$
.
That is, we can introduce Wilson lines along the $U(1)_b$ 
direction. 
Because of this Wilson line, the number of 
zero-modes does not change, but their wavefunctions 
are shifted as 
\begin{eqnarray}\label{eq:WL-shift}
\Theta^{j,M}(z_d) \rightarrow
\Theta^{j,M}(z_d+q^b\xi^b_d/(q^am^a_{(d)})).
\end{eqnarray}
Note that the shift of zero-mode profiles depend on 
$U(1)_b$ charges of matter fields.
Similarly, we can introduce the Wilson line $\xi^a_d$ 
along the $U(1)_a$ direction.
Then, the zero-mode wavefunctions shift as 
\begin{eqnarray}\label{eq:WL-shift-ab}
\Theta^{j,M}(z_d+q^b\xi^b_d/(q^am^a_{(d)})) \rightarrow \Theta^{j,M}
(z_d+\xi^a_d/m^a_{(d)}+q^b\xi^b_d/(q^am^a_{(d)})).
\end{eqnarray}
However, the shift due to $\xi^a_d$ is rather universal shift, but 
the shift by $\xi^b_d$ depends on the charges $q^b$ of matter fields.
Thus, the shift by $\xi^b_d$ would be much more important than 
one by $\xi^a_d$, in particular from the phenomenological viewpoint.

Suppose that we introduce magnetic fluxes in a model with 
a larger gauge group $G$ such that they break $G$ to a GUT 
group like $SO(10)$  and this model include 
three families of matter fields like the ${\bf 16}$ representation, 
corresponding to all of quarks and leptons.
Then, we assume that the $SO(10)$ gauge symmetry is broken 
to $SU(3) \times SU(2) \times U(1)_Y$  by some mechanism.
If zero-mode profiles of quarks and leptons are degenerate 
even after such $SO(10)$ breaking, 
couplings in 4D effective field theory are constrained 
(at the lowest level) by the $SO(10)$ symmetry.
That is, Yukawa matrices would be the same between 
the up-sector, the down-sector and the lepton sector.
However, when we break $SO(10)$ to 
$SU(3) \times SU(2) \times U(1)_Y \times U(1)$ 
by introducing Wilson lines along the $U(1)_Y \times U(1)$ direction, 
these Wilson lines resolve the degeneracy of zero-mode profiles 
among quarks and leptons.
Then, Yukawa matrices would become different from each other 
among the up-sector, the down-sector and the lepton sector.
Similarly we can analyze 4D massless scalar modes~\cite{Cremades:2004wa}.
We are assuming that 4D N=1 supersymmetry is 
preserved~\cite{Troost:1999xn,Cremades:2004wa}.
Thus, the number of zero-modes and the profiles for 
4D scalar fields are the same as those for their superpartners, 
i.e. the above spinor fields.
For example, for Higgs fields, we study 
zero-modes and their profiles of Higgsino fields.

\subsubsection{$E_6$ model}\label{sec:E6}

Here, we consider 10D super Yang-Mills theory with 
the $E_6$ gauge group.

We compactify extra six-dimensions on $T^6$.
We introduce magnetic fluxes along the $U(1)_a$ direction, which breaks 
the gauge group, $E_6 \rightarrow SO(10) \times U(1)_a$.
The $E_6$ adjoint representation is decomposed as
\begin{eqnarray}\label{eq:78-rep}
{\bf 78} = {\bf 45}_0 + {\bf 1}_0+ {\bf 16}_1 + \overline {\bf
16}_{-1},
\end{eqnarray}
for $SO(10) \times U(1)_a$.
Here, ${\bf 16}_1$ and $\overline {\bf 16}_{-1}$ correspond to 
the broken part and the corresponding 
gaugino fields appear as matter fields.

For example, we assume magnetic fluxes, 
\begin{eqnarray}\label{eq:mag-3-family} 
m^a_{(1)}=3, \qquad m^a_{(2)}=1, \qquad m^a_{(3)}=1.
\end{eqnarray}
Then, the chiral matter fields corresponding to 
${\bf 16}_1$ and $s_d=(+,+,+)$ have zero-modes, 
but there are no massless modes for $\overline {\bf 16}_{-1}$.
Furthermore, the number of ${\bf 16}_1$ is equal to 
$m^a_{(1)} m^a_{(2)} m^a_{(3)}=3$, that is, 
the model with three families of ${\bf 16}_1$.
Their wavefunctions are written as 
\begin{eqnarray}
\Theta^{j,3}(z_1) \Theta^{1,1}(z_2) \Theta^{1,1}(z_3).
\end{eqnarray}
The flavor structure is determined by the first torus $T^2_1$.
Thus, the massless matter spectrum is realistic, although 
there is no Higgs fields and the gauge sector has 
4D N=4 SUSY.

The $U(1)_a$ symmetry is anomalous.
We assume that its gauge boson become massive by 
the Green-Schwarz mechanism.
Hereafter, we also assume that if other $U(1)$ symmetries 
become anomalous they become massive by the 
Green-Schwarz mechanism.

Here, we break the $SO(10)$ gauge group further 
to the standard model gauge group 
up to $U(1)$ factors, i.e.
$SU(3) \times SU(2) \times U(1)_Y\times U(1)_b$, 
by introducing Wilson lines along $U(1)_Y$ and $U(1)_b$ directions.
The   ${\bf 16}$ representation of $SO(10)$ is decomposed 
under $SU(3) \times SU(2) \times U(1)_Y\times U(1)_b$ as 
\begin{eqnarray}
{\bf 16} = ({\bf 3},{\bf 2})_{1,-1} + (\bar {\bf 3},{\bf 1})_{-4,-1}
+ ({\bf 1},{\bf 1})_{6,-1} + (\bar {\bf 3},{\bf 1})_{2,3}
+ ({\bf 1},{\bf 2})_{-3,3} + ({\bf 1},{\bf 1})_{0,-5}, 
\end{eqnarray}
where we normalize $U(1)_Y$ and $U(1)_b$ charges, such that 
minimum charges satisfy $|q^Y|=1$ and $|q^b|=1$.

By introducing  Wilson lines along $U(1)_Y$ and $U(1)_b$ directions, 
the generation number does not change, but
the zero-mode profiles of three families of 
${\bf 16}$ split differently each other among 
quarks and leptons.
Furthermore, their splitting behaviors depend on 
which torus $T^2_d$ we introduce Wilson lines.
Recall that in this model 
the flavor structure is determined by the first torus $T^2_1$.
For example, when we introduce Wilson lines along $U(1)_Y$ and
$U(1)_b$
directions on the second torus $T^2_2$, 
the zero-mode profiles of quarks ($Q,u,d$) and leptons ($L,e,\nu$) 
split as 
\begin{eqnarray}\label{eq:ql-wl-2}
Q&:& \Theta^{j,3}(z^1)\Theta^{1,1}(z^2+\xi^Y-\xi^b)\Theta^{1,1}(z^3),
\nonumber \\
u^c&:& \Theta^{j,3}(z^1)\Theta^{1,1}(z^2-4\xi^Y-\xi^b)\Theta^{1,1}(z^3), 
\nonumber \\
d^c&:& \Theta^{j,3}(z^1)\Theta^{1,1}(z^2+2\xi^Y+3\xi^b)\Theta^{1,1}(z^3),
\\
L&:& \Theta^{j,3}(z^1)\Theta^{1,1}(z^2-3\xi^Y+3\xi^b)\Theta^{1,1}(z^3),
\nonumber 
\\
e^c&:& \Theta^{j,3}(z^1)\Theta^{1,1}(z^2+6\xi^Y-\xi^b)\Theta^{1,1}(z^3), 
\nonumber \\
\nu^c&:& \Theta^{j,3}(z^1)\Theta^{1,1}(z^2-5\xi^b)\Theta^{1,1}(z^3),
\nonumber  
\end{eqnarray}
where $\xi^Y$ and $\xi^b$ are the Wilson lines along 
$U(1)_Y$ and $U(1)_b$ directions.
On the other hand, when we introduce Wilson lines on 
the first torus $T^2_1$, the zero-mode profiles  
of quarks ($Q,u,d$) and leptons ($L,e,\nu$) 
split as 
\begin{eqnarray}\label{eq:ql-wl-1}
Q&:& \Theta^{j,3}(z^1+\xi^Y/3-\xi^b/3)\Theta^{1,1}(z^2)\Theta^{1,1}(z^3),
\nonumber \\
u^c&:&
\Theta^{j,3}(z^1-4\xi^Y/3-\xi^b/3)\Theta^{1,1}(z^2)\Theta^{1,1}(z^3), 
\nonumber \\
d^c&:& \Theta^{j,3}(z^1+2\xi^Y/3+\xi^b)\Theta^{1,1}(z^2)\Theta^{1,1}(z^3),
\\
L&:& \Theta^{j,3}(z^1-\xi^Y+\xi^b)\Theta^{1,1}(z^2)\Theta^{1,1}(z^3),
\nonumber 
\\
e^c&:& \Theta^{j,3}(z^1+2\xi^Y-\xi^b/3)\Theta^{1,1}(z^2)\Theta^{1,1}(z^3), 
\nonumber \\
\nu^c&:& \Theta^{j,3}(z^1-5\xi^b/3)\Theta^{1,1}(z^2)\Theta^{1,1}(z^3).
\nonumber  
\end{eqnarray}
Since the flavor structure is determined by the first torus  $T^2_1$, 
the first case (\ref{eq:ql-wl-2}) preserves the $SO(10)$ flavor 
structure.
However, such flavor structure is deformed in the second case 
(\ref{eq:ql-wl-1}) by Wilson lines.

Obviously, other configurations of Wilson lines are possible, 
e.g. $\xi^Y$ on $T^2_1$ and $\xi^b$ on $T^2_2$ and so on.
In any case, the flavor structure is determined by which 
Wilson lines we introduce on the first $T^2_1$.
For example, if we introduce only $\xi^b$ on $T^2_1$, 
the resultant Yukawa matrices would have the $SU(5)$ GUT 
relation.

Thus, the above model is interesting.
Its chiral matter spectrum is realistic and 
the model has the interesting flavor structure, 
although electro-weak Higgs fields do not appear 
and the gauge sector has 4D N=4 SUSY.

\subsubsection{$E_7$ and $E_8$ models}

Similarly, we can study $E_7$ and $E_8$ models.
Their ranks are larger than $E_6$ and their adjoint 
representations include several representations.
The $E_8$ adjoint representation ${\bf 248}$ is decomposed 
under $E_7 \times U(1)_{E8}$ as 
\begin{eqnarray}\label{eq:248-rep}
{\bf 248} = {\bf 133}_0 + {\bf 1}_0 +{\bf 56}_1 + 
{\bf 56}_{-1}+{\bf 1}_2 + {\bf 1}_{-2} .
\end{eqnarray}
Note that we are using $U(1)$ charge normalization such that 
the minimum charge except vanishing charge is equal to one, $|q|=1$. 
Then, the $E_7$ adjoint representation ${\bf 133}$ 
is decomposed under $E_6 \times U(1)_{E7}$ as 
\begin{eqnarray}\label{eq:133-rep}
{\bf 133} = {\bf 78}_0 + {\bf 1}_0 +{\bf 27}_{-2} + 
\overline {\bf 27}_{2},
\end{eqnarray}
 and the ${\bf 56}$ representation of $E_7$ is decomposed 
under $E_6 \times U(1)_{E7}$ as 
\begin{eqnarray}\label{eq:56-rep}
{\bf 56} = {\bf 27}_1 + 
\overline {\bf 27}_{-1} + {\bf 1}_2 + {\bf 1}_{-2}.
\end{eqnarray}
Furthermore, the ${\bf 27}$ representation of $E_6$ 
is decomposed under $SO(10)\times U(1)_{E6}$ as 
\begin{eqnarray}
{\bf 27} ={\bf 16}_1 + {\bf 10}_{-2} + {\bf 1}_4.
\end{eqnarray}
Thus, we can construct various models from 
$E_7$ and $E_8$ models.
Quark and lepton matter fields can be originated from 
several sectors, although such matter fields are originated from 
${\bf 16}$ of the $E_6$ adjoint sector in the models 
of the previous section.
In addition,  the $E_7$ and $E_8$ adjoint representations 
include exotic representations. 
Hence, exotic matter fields, in general, appear in 
4D massless spectra.
Instead of $U(1)_{E8} \times U(1)_{E7}$, we use the  
$U(1)_c \times U(1)_d$ basis, such that those charges are related as 
\begin{eqnarray}
q_c=\frac{1}{2}q_{E8} + \frac{1}{2}q_{E7}, \qquad 
q_d=-\frac{1}{2}q_{E8} + \frac{1}{2}q_{E7},
\end{eqnarray}
where $q_c$, $q_d$, $q_{E8}$ and $q_{E7}$ denote 
$U(1)_c$, $ U(1)_d$, $U(1)_{E8}$ and $U(1)_{E7}$ charges, 
respectively.
In addition, we denote $U(1)_{E6}$ by $U(1)_a$ as in section \ref{sec:E6}.
Also, as in section \ref{sec:E6}, we use the notation $U(1)_b$, which 
appears through the $SO(10)$ breaking as $SO(10) \rightarrow SU(5)
\times U(1)_b$.

Here, we show just simple illustrating models.
First of all, we can construct almost the same model as 
the $E_6$ models.
For example, we start with the 10D $E_7$ super Yang-Mills theory.
We can introduce magnetic fluxes with the same form in 
$U(1)_{E6}$ as (\ref{eq:mag-3-family}).
Furthermore, we introduce Wilson lines such that the 
gauge group is broken down to 
$SU(3) \times SU(2) \times U(1)_Y$ up to $U(1)$ factors.
Then, we realize three families of quarks and leptons under 
the standard model gauge group, 
that is, the same 4D massless spectrum as one in section \ref{sec:E6}, 
although the gauge sector has partly 4D N=4 SUSY and 
there is no Higgs fields.
Similarly, the same model can be derived from the 10D 
$E_8$ super Yang-Mills theory.

Now, let us consider another illustrating model with different
aspects.
We start with the 10D $E_8$ super Yang-Mills theory.
When $E_8$ is broken to the standard model gauge group, there are 
five $U(1)$'s including $U(1)_Y$, i.e., $U(1)_I$ $(I=a,b,c,d,Y)$.
We introduce magnetic fluxes $m^I_{(d)}$ 
along these five $U(1)_I$ directions.
Then,  the sum of magnetic fluxes $M=\sum_I q^I m^I_{(d)}$
appears in the zero-mode Dirac equation for the matter field with 
charges $q^I$.
We require that $\sum_I q^Im^I_{(d)}$  should be integer for 
all of matter fields, that is, the quantization condition of 
magnetic fluxes~\cite{toron}.

For example, five $({\bf 3},{\bf 2})_{1}$ representations 
under $SU(3) \times SU(2) \times U(1)_Y$ 
as well as their conjugates appear from the ${\bf 248}$ adjoint 
representation.
Three of them appear from three ${\bf 27}$ representations 
of ${\bf 248}$, i.e., Eqs.~(\ref{eq:248-rep}), 
(\ref{eq:133-rep}) and (\ref{eq:56-rep}).
In the zero-mode equations of such three $({\bf 3},{\bf 2})_{1}$
matter fields, 
the following sum of magnetic fluxes $\sum_I q^I m^I_{(d)}$ appears
\begin{eqnarray}
m^{Q1}_{(d)} &=& m^c_{(d)} + m^a_{(d)} - m^b_{(d)} + m^{Y}_{(d)}, 
\nonumber \\
m^{Q2}_{(d)} &=& m^d_{(d)} + m^a_{(d)} - m^b_{(d)} + m^{Y}_{(d)},\\
m^{Q3}_{(d)} &=& -m^c_{(d)} -m^d_{(d)} 
+ m^a_{(d)} - m^b_{(d)} + m^{Y}_{(d)}. \nonumber
\end{eqnarray}
In addition, one $({\bf 3},{\bf 2})_{1}$ representation appears from 
${\bf 16}$ of the $E_6$ adjoint ${\bf 78}$ representation 
(\ref{eq:78-rep}) as section 3.
In the zero-mode equation of such $({\bf 3},{\bf 2})_{1}$ matter
field, 
the following sum of magnetic fluxes $\sum_I q^I m^I_{(d)}$ appears 
\begin{eqnarray}
m^{Q4}_{(d)} &=& -3 m^a_{(d)} - m^b_{(d)} + m^{Y}_{(d)} .
\end{eqnarray}
Moreover, the $SO(10)$ adjoint ${\bf 45}$ representation also includes
a 
$({\bf 3},{\bf 2})_{1}$ representation and 
the corresponding matter field has the sum 
of magnetic fluxes $\sum_I q^I m^I_{(d)}$, \footnote{The 
$SO(10)$ adjoint ${\bf 45}$ representation includes another 
$({\bf 3},{\bf 2})$ representation but its $U(1)_Y$ charge is
different.}
\begin{eqnarray}
m^{Q5}_{(d)} = 4m^b_{(d)} + m^{Y}_{(d)} ,
\end{eqnarray}
in the zero-mode equation.
Here, we require that all of 
$m^{Q1}_{(d)}$, $m^{Q2}_{(d)}$, $m^{Q3}_{(d)}$, $m^{Q4}_{(d)}$
and $m^{Q5}_{(d)}$ should be integers.
Similarly, we require that $\sum_I q^Im^I_{(d)}$  should be integers
for 
all of matter fields with charges $q^I$, which appear from 
the $E_8$ adjoint ${\bf 248}$ representation.
By an explicit computation, it is found that 
the sum $\sum_I q^Im^I_{(d)}$ for any charge 
$q^I$ appearing from ${\bf 248}$ can be written as a linear 
combination of $m^{Q1}_{(d)}$, $m^{Q2}_{(d)}$, $m^{Q3}_{(d)}$,
$m^{Q4}_{(d)}$
and $m^{Q5}_{(d)}$ with integer coefficients.
Thus, when all of $m^{Q1}_{(d)}$, $m^{Q2}_{(d)}$, $m^{Q3}_{(d)}$,
$m^{Q4}_{(d)}$
and $m^{Q5}_{(d)}$ are integers, 
the sum $\sum_I q^Im^I_{(d)}$ for any charge $q^I$ of 
${\bf 248}$ is always integer.

Using the above notation, 
we introduce the  magnetic fluxes such as ,
\begin{eqnarray}
 & & m^{Q1}_{(1)} = 1, \qquad m^{Q1}_{(2)} = -1, \qquad m^{Q1}_{(3)} =
 -3, 
\nonumber  \\
 & & m^{Q2}_{(1)} = -1, \qquad m^{Q2}_{(2)} = 0, \qquad m^{Q2}_{(3)} =
 1,  
\nonumber \\
 & &  m^{Q3}_{(1)} = -1, \qquad m^{Q3}_{(2)} = 0, \qquad m^{Q3}_{(3)}
 = 1, \\
 & &  m^{Q4}_{(1)} = -1, \qquad m^{Q4}_{(2)} = 0, \qquad m^{Q4}_{(3)}
 = 1,  
\nonumber \\
 & &  m^{Q5}_{(1)} = -2, \qquad m^{Q5}_{(2)} = -1, \qquad m^{Q5}_{(3)}
 = 0. 
\nonumber
\end{eqnarray}
In addition, we also introduce all possible Wilson lines 
on each torus along five $U(1)$ directions.
Then, the gauge group is $SU(3) \times SU(2) \times U(1)_Y$ 
with $U(1)$ factors.

The 4D massless spectrum of this model includes 
the following matter fields under the standard gauge group, 
$SU(3) \times SU(2) \times U(1)_Y$,
\begin{eqnarray}
& & 3 \times \left[ ({\bf 3},{\bf 2})_1 + (\overline {\bf 3},{\bf
1})_{-4} 
+ (\overline {\bf 3},{\bf 1})_{2} + ({\bf 1},{\bf 2})_{-3} 
  + ({\bf 1},{\bf 1})_{6} \right]  \nonumber \\
& & + 8 \left[ ({\bf 1},{\bf 2})_{3}+ ({\bf 1},{\bf 2})_{-3} \right]
\\
& & + 15 \times  \left[  ({\bf 3},{\bf 1})_{4} + 
 (\overline {\bf 3},{\bf 1})_{-4} \right] 
+ 6 \times  \left[  ({\bf 3},{\bf 1})_{-2} + 
 (\overline {\bf 3},{\bf 1})_{2} \right] 
+ 27 \times  \left[ ({\bf 1},{\bf 1})_{6} 
+ ({\bf 1},{\bf 1})_{-6}\right],  \nonumber 
\end{eqnarray}
and $SU(3) \times SU(2)$ singlets with vanishing $U(1)_Y$ charges.
That is, this massless spectrum includes 
three families of quarks and leptons as well as 
eight pairs of up- and down-sectors of electroweak Higgs fields.
In addition, many vector-like matter fields appear, 
but exotic matter fields 
do not appear even in vector-like form.
Such exotic matter fields have (effectively) vanishing 
magnetic flux on one of $T^2_d$.
Then, such fields become massive when we switch on proper Wilson
lines.\footnote{In the limit of vanishing Wilson lines, colored 
Higgs fields appear in the vector-like form, but 
they become massive for finite values of Wilson lines.}
Thus, this model has semi-realistic massless spectrum, 
although the gauge sector still has 4D N=4 SUSY.
We can write the wavefunctions of these zero-modes.
For example, the zero-mode wavefunctions of 
left-handed quarks are written as 
\begin{eqnarray}
\Theta^{1,1}(z_1+\xi_1)\Theta^{1,1}(z_2+\xi_2)\Theta^{j,3}(z_3+\xi_3/3),
\end{eqnarray}
for $j=1,2,3$,
where $\xi_d$ denote Wilson lines along five $U(1)$ directions.
Thus, the flavor structure for the left-handed quarks is 
determined by the third torus.
Similarly, we can write zero-mode wavefunctions of the other 
matter fields.
The above massless spectrum includes several vector-like generations
of 
right-handed quarks as well as right-handed leptons.
These vector-like generations may gain mass terms.
Thus, the flavor structure of chiral right-handed quarks depends 
on mass matrices of vector-like generations.

Similarly, various models can be constructed within 
the framework of $E_7$ and $E_8$ models with 
magnetic flux and Wilson line backgrounds.

\newpage

\section{Calculation of Yukawa interaction and higher order couplings}

\subsection{Low-energy effective action} 

In this section we study the low-energy phenomenology based on 
the general set up of the string theory or supergravity theory.
In the low-energy limit of these theories can be 
described as the effective action for $N=1$ super Yang-Mills 
theory with chiral matter fields as far as 
low-energy supersymmetry exist.
Their action consists of only three functions as
Kahler potential, super potential  and gauge kinetic functions.
Furthermore such functions usually depend on the moduli fields
which are corresponding the background of the higher dimensional 
space or tensor fields.
To describe the realistic world, these moduli field
should be stabilized.
This can be achieved by some mechanism e.g. 
background flux induced super potential or non-perturbative super
potential which means that moduli fields have vacuum
expectation values. 
Therefore it is important and necessary to study the moduli field and
their stabilization for understanding 
the dynamics of string theory or quantum field theory.
For the phenomenological aspects understanding these moduli dependence  
is important. In such scenario the vacuum often breaks the
supersymmetry.
This affects on the soft supersymmetry breaking terms mediated 
by moduli fields. Actually we have seen the Yukawa coupling is
determined by the background of the compactified space and depend on
the complex structure moduli. Therefore we need to know the moduli 
parameters and dynamics of the mediation mechanism of supersymmetry
breaking to understand low-energy phenomena.

Since these three functions are dependent on the light matter fields
$C_\alpha$
and heavy moduli field $\mathcal{M}$, they may be represented by the 
expansion of the light matter fields and the general expressions for
super potential $W(\mathcal{M},C)$ and 
Kahler potential $K(\mathcal{M},\bar{\mathcal{M}},C,\bar{C})$ are
given by 
\begin{align}\label{eq:W}
W(\mathcal{M},C) =
\sum_\alpha \xi(\mathcal{M}) C_\alpha +
\sum_{\alpha,\beta} \mu_{\alpha\beta}(\mathcal{M}) C_\alpha C_\beta
+\sum_{\alpha,\beta,\gamma} 
Y_{\alpha\beta\gamma}(\mathcal{M}) C_\alpha C_\beta C_\gamma +\cdots,
\end{align}
and 
\begin{align}
K(\mathcal{M},\bar{\mathcal{M}},C,\bar{C})=
K_0(\mathcal{M},\bar{\mathcal{M}})
+\sum_{\alpha,\beta} K_{C_\alpha C_\beta}
(\mathcal{M},\bar{\mathcal{M}}) C_\alpha  C_{\bar{\beta}}+\cdots.
\end{align}
By giving these coefficients of the moduli parameters
we may obtain the low-energy constants up to higher order
corrections which is denoted by ellipsis. For example,
the second and third terms of the super potential give rise to the 
supersymmetric masses and the Yukawa couplings.  
We are interested in the information on the explicit form given in 
this expressions.
There are usually two ways to obtain the effective actions.
One method is to use the string S matrix calculation.
This enables to compute the amplitude for massless string states at
least perturbatively in $\alpha'$ and string coupling $g_s$.
From the expressions we can extract the interaction terms and
dependency of the moduli fields at arbitrary order of $\alpha'$ and
$g_s$ in principle. Therefore this approach gives the solid results  
including the stringy effects. For this calculations it needs 
the technically higher knowledge about string vertex operators and 
calculation of CFT.

The second method to construct the four dimensional
effective theory is easier way to start with higher dimensional 
field theory or DBI action which is an effective action of D-brane
models and take the ordinary dimensional reductions to
four dimensions. This also provides the low-energy interactions
including the moduli dependence at certain accuracy.
Indeed we will see such a discrepancy in the calculation of the
normalization constant and Yukawa couplings in field theory
which are discussed in section\ref{sec:Intersecting}
The explicit calculations for the dimensional reduction of toroidal
compactifications are studied in appendix\ref{dimensional}.

\subsection{General setup }

We consider dimensional reduction of ten-dimensional ${\cal N}=1$
super Yang--Mills theory with $U(N)$ gauge group \cite{SYM}, on a six
torus in Abelian magnetic flux background. 
We factorize the six-torus into two-tori  $(T^2)^3$, each of which is
specified by the complex structure $\tau_d$ and the area 
$A_d = (2 \pi R_d)^2~\Im \tau_d$ where $d=1,2,3$.
We shall focus on the case with trivial background which means 
torus without non-abelian Wilson line.
For the fractional flux case, the analysis of these couplings
will be discussed later.
From the periodicity of torus, the background magnetic flux 
is quantized as \cite{toron}
\begin{equation} \label{toronbg} \begin{split}
 F_{z^d \bar z^d} = {2 \pi i \over \Im \tau_d}  \begin{pmatrix}
   m_1^{(d)} \1_{N_1} & & \\
  & \ddots & \\ & & m_n^{(d)} \1_{N_n} \end{pmatrix}, \quad d=1,2,3,
\end{split} \end{equation}
where $\1_{N_a}$ are the unit matrices of rank $N_a$, $m_i^{(d)}$ are
integers and $z^d$ are the complex coordinates. 
This background  breaks the gauge symmetry 
$U(N) \to \prod_{a=1}^n U(N_a)$ where
$N=\sum_{a=1}^n N_a.$
We have the $|M^{(d)}|$ zero-modes labeled by the index $j$.
Note that the wavefunction for $j=k+M^{(d)}$ is identical to one 
for $j=k$.
They satisfy the orthonormal condition,
\begin{equation} \label{orthre2}
 \int d^2z^d\ \psi_d^{i,M^{(d)}}(z^d) \left( \psi_d^{j,M^{(d)}}(z^d)
 \right)^* =
  \delta_{ij}.
\end{equation}
The important part of zero-mode wavefunctions is written 
in terms of the Jacobi theta function
\begin{equation} \label{jacobitheta}
 \jtheta{a \\ b}(\nu,\tau) = \sum_{n=-\infty}^\infty \exp\left[ \pi i
   (n+a)^2 \tau + 2 \pi i (n+a)(\nu +b)\right].
\end{equation}
It transforms under the symmetry of torus lattice and
has several important properties \cite{Mu}. 
One of them is the following product rule
\begin{equation} \begin{split} \label{thetaprod}
 \jtheta{{i/M_1} \\ 0}&(z_1,\tau M_1) \cdot \jtheta{j/M_2 \\
 0}(z_2,\tau M_2)
 \\
=& \sum_{m \in \Z_{M_1+M_2}} \jtheta{{i+j+M_1 m \over
     M_1 + M_2} \\ 0 }(z_1 + z_2,\tau(M_1 +M_2)) \\
 & \times \jtheta{{M_2 i - M_1 j + M_1 M_2 m \over M_1 M_2(M_1 +M_2)}
 \\ 0}(z_1 M_2 - z_2 M_1,\tau M_1 M_2(M_1+M_2)). \\
\end{split} \end{equation}
Here $\Z_M$ is the cyclic group of order $|M|$, $\Z_M =
\{1,\dots,|M|\}$
where every number is defined modulo $M$.
Although this expression looks asymmetric under the exchange 
between $i$ and $j$, it is symmetric if we take into account the
summation.
By using the product property (\ref{thetaprod}), 
we can decompose a product of two zero-mode wavefunctions as follows,
\begin{equation}\begin{split}\label{wvprod}
\psi_d^{i,M_1}(z^d) \psi^{j,M_2}_d(z^d) = &  \frac {N_{M_1}
N_{M_2}}{N_{M_1+M_2}} 
\sum_{m \in \Z_{M_1+M_2}}  \psi_d^{i+j+M_1m,M_1+M_2}(z^d)   \\
& \times \jtheta{{M_2 i - M_1 j + M_1 M_2 m \over M_1 M_2(M_1 +M_2)}
 \\ 0}(0,\tau_d M_1 M_2(M_1+M_2)), \\
\end{split} \end{equation}
where the normalization factor $N_M$ is obtained 
as
\begin{equation} \label{normalization}
N_{M^{(d)}} = \left( {2 \Im \tau_{d} |M^{(d)}| \over
   A_{d}^2 } \right)^{1/4} .
\end{equation}

In this section, we calculate the generalization of Yukawa couplings to
arbitrary order $L$ couplings
\begin{equation}\label{eq:L-coupling}
 Y_{i_1 \dots i_{L_\chi} i_{L_\chi+1}\cdots i_L} \chi^{i_1}(x) \cdots 
\chi^{i_{L_\chi}}(x) \phi^{i_{L_{\chi}+1}}(x) \dots \phi^{i_L}(x),
\end{equation}
with $L=L_\chi + L_\phi$, where $\chi$ and $\phi$ collectively 
represent four-dimensional components of fermions and bosons,
respectively. The system under consideration can be
understood as low-energy effective field theory of open string theory.
The magnetic flux is provided by stacks of D-branes filling in the
internal dimension. The leading order terms in $\alpha'$ are identical
to
ten-dimensional super-Yang--Mills theory, whose covariantized gaugino
kinetic term gives the three-point coupling upon dimensional
reduction \cite{Cremades:2004wa,DiVecchia:2008tm}. The higher order couplings can be
read off from the effective Lagrangian of the Dirac--Born--Infeld
action with
supersymmetrization. The internal component of bosonic and 
fermionic wavefunctions is the same \cite{Cremades:2004wa}.
Therefore it suffices to calculate the wavefunction overlap in the
extra
dimensions
\begin{equation} \label{wavefnoverlap}
 Y_{i_1 i_2 \dots i_L} = g_L^{10}  \int_{T^6} d^6z \ 
\prod_{d=1}^3 \psi^{i_1,M_1}_d(z) 
\psi^{i_2,M_2}_d(z) \dots \psi^{i_L,M_L}_d(z),
\end{equation}
where $g_L^{10}$ denotes the coupling in ten dimensions.

\subsection{Three-point coupling}

In this section, we calculate the three-point coupling 
considering the coupling selection rule. 
As we see later, the three-point coupling provides a building block of
higher order couplings.

The gauge group dependent part is contracted by the gauge invariance, 
so that the choice of three blocks $m_a,m_b,m_c$ in (\ref{toronbg})
automatically fixes the relative
 magnetic fluxes
\begin{equation}
 (m_a - m_b) + (m_b - m_c) = (m_a - m_c), \quad \text{and} 
\quad M_1 + M_2 = M_3,
\end{equation}
where $M_1 = m_a -m_b$, $M_2=m_b-m_c$ and $M_3 = m_a -m_c$.
Here every $M_i$ is assumed to be a positive integer.
This relation is interpreted as the selection rule, in analogy of
intersecting brane case \cite{Cremades:2003qj,Higaki:2005ie}, to which we
come back
later. If it is not satisfied, there is no corresponding gauge
invariant operator in ten dimensions.
In terms of quantum numbers the coupling has the form $\bf (N_a,
\overline N_b,1) \cdot (1,N_b, \overline N_c) 
\cdot (\overline N_a,1,N_c)$ under $U(N_a) \times U(N_b)
\times U(N_c)$.

The internal part including the wavefunction integrals 
on the $d$-th $T^2$ gives
\begin{equation}\label{eq:Yukawa}
  y_{ij\bar k} = \int d^2 z \ 
\psi^{i,M_1}(z) \psi^{j,M_2}(z) \left( \psi^{k,M_3}(z) \right)^* .
\end{equation}
The complete three-point coupling is the direct product of those in
$d=1,2,3$ and $g^{10}_3$.
For the moment we neglect the normalization factors $N_M$,
and consider two-dimensional
wavefunctions, omitting the extra dimensional index $d$.
By using the relation (\ref{wvprod}), we can decompose 
the product of the first two wavefunctions 
$\psi^{i,M_1}(z) \psi^{j,M_2}(z)$ in terms of $\psi^{k,M_3}(z) $
and we apply the orthogonality relation (\ref{orthre2}). 
Then, we obtain 
\begin{equation}\label{yijk-1}
 y_{ij\bar k} = \sum_{m \in \Z_{M_3}}  \delta_{i+j+M_1m,k}
 ~\jtheta{{M_2 i - M_1 j + M_1 M_2 m \over M_1 M_2 M_3}
 \\ 0}(0,\tau M_1 M_2 M_3),
\end{equation}
where the numbers in the Kronecker delta is defined modulo $M_3$. 
This expression is symmetric under the exchange 
$(i,M_1) \leftrightarrow (j,M_2)$.

For $\gcd(M_1,M_2)=1$, we solve the constraint from the Kronecker
delta $\delta_{i+j+M_1m,k}$,
\begin{equation} \label{deltaconstraint}
 i+j-k = M_3 l - M_1 m, \quad m \in {\bf Z}_{M_3}, l \in {\bf
   Z}_{M_1}.
\end{equation}
Using Euclidean algorithm, it is easy to see that, in the
relatively prime case $\gcd(M_1,M_2)=1$, there is always a unique
solution
for given $i,j,k$. 
This situation is the same as one in intersecting D-brane models 
\cite{Cremades:2003qj,Higaki:2005ie}.
The argument of the theta function 
in eq.(\ref{yijk-1}) becomes
\begin{equation} \label{thetaarg}
 {M_2 i - M_1 j + M_1 M_2 m \over M_1 M_2 (M_1 + M_2)}
  = {M_2 k - M_3 j + M_2 M_3 l \over (M_3 - M_2) M_2 M_3}.
\end{equation}
Therefore, the three-point coupling is written as 
\begin{equation}  \label{3pt}
  y_{ij\bar k}(l) = \jtheta{{M_2 k - M_3 j + M_2 M_3 l \over M_2
  M_3(M_3-M_2)} \\ 0}
  (0,\tau (M_3-M_2) M_2 M_3),
\end{equation}
where $l$ is an integer related to $i,j,k$ through
(\ref{deltaconstraint}).
This is called the 2-3 picture, or the $j$-$k$ picture, 
where the dependence on $i$ and $M_1$
is only implicit.

In the case with a generic value of $\gcd(M_1,M_2)=g$, we can show
\begin{equation} \label{modyukawa}
 y_{ij \bar k} =  \sum_{n=1}^g \vartheta \left[ \begin{matrix}
    {M_2k - M_3j + M_2 M_3 l \over M_1 M_2 M_3 } + {n \over g} \\ 0
    \end{matrix}
  \right](0,\tau M_1 M_2 M_3).
\end{equation}
The point is that, for a given particular solution $(i,j,k)$, the
number of general solutions satisfying Eq.~(\ref{deltaconstraint})
is equal to $g$.  We can use a similar argument as above, now
considering
${\bf Z}_{M_1/g}$ and ${\bf Z}_{M_3/g}$ instead of the original
region.
There is a unique
pair $(l,m)$ in $({\bf Z}_{M_1/g},{\bf Z}_{M_3/g})$ satisfying the
constraint (\ref{deltaconstraint}), i.e. ,
\begin{equation}
 {i+j-k \over g} = {M_3 \over g} l - {M_1 \over g} m.
\end{equation}
Obviously, when $(l,m)$ is a particular solution, 
the following pairs,
\begin{equation} \label{argshift}
 \left(l+ \frac{M_1}{g},m+\frac{M_3}{g} \right) 
\in({\bf  Z}_{M_1},{\bf Z}_{M_3}) ,
\end{equation}
also satisfy the equation with the same right-hand side (RHS).
Since ${\bf Z}_{M_1}$ and ${\bf Z}_{M_3}$ are respectively unions of
$g$ identical
copies of ${\bf Z}_{M_1/g},{\bf Z}_{M_3/g}$, there are $g$ different
solutions. 
This situation is the same as one in intersecting D-brane models.
If we reflect the shift (\ref{argshift}) in
(\ref{thetaarg}), we obtain the desired result (\ref{modyukawa}).

There can be Wilson lines $\zeta \equiv \zeta_r + \tau \zeta_i$,
whose effect is just a translation of each wavefunction~\cite{Cremades:2004wa}
\begin{equation}
 \psi^{j,M}(z) \to \psi^{j,M}(z+\zeta), \quad \text{ for all } j.
\end{equation}
Thus the corresponding product for (\ref{thetaprod}) is 
obtained as 
\begin{equation} \begin{split} 
 \jtheta{{i/M_1} \\ 0}&((z+\zeta_1)M_1,\tau M_1) \cdot 
\jtheta{j/M_2 \\ 0}((z +\zeta_2)M_2,\tau M_2)
 \\
=& \sum_{m \in \Z_{M_1+M_2}} \jtheta{{i+j+M_1 m \over
     M_1 + M_2} \\ 0 }((M_1+M_2)(z+\zeta_3),\tau(M_1 +M_2)) \\
 & \times \jtheta{{M_2 i - M_1 j + M_1 M_2 m \over M_1 M_2(M_1 +M_2)}
 \\ 0}(M_1M_2(\zeta_1 -\zeta_2)),\tau M_1 M_2(M_1+M_2)), \\
\end{split} \end{equation}
where $M_3 = M_1 + M_2$ and $\zeta_3 M_3 = \zeta_1 M_1 + \zeta_2 M_2$.

\begin{figure}[h] \begin{center}
\psfrag{i M1}[c]{$i,M_1$}
\psfrag{j M2}[c]{$j,M_2$}
\psfrag{k M3}{$k,M_3$}
\includegraphics[height=3cm]{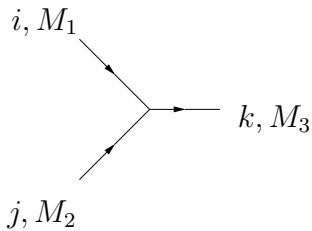}
\caption{A three-point coupling provides a building block of higher
  order couplings. 
This diagram corresponds to the three-point coupling
(\ref{Yukawa}). The direction of an arrow depends on the
holomorphicity
of the corresponding external state.}  
\label{f:3pt}
\end{center} \end{figure}

Finally, we take into account the six internal dimensions $T^2 \times
T^2 \times T^2$. Referring to (\ref{wavefnoverlap}), essentially the
full coupling is the direct product of the coupling on each
two-torus. The overall factor in (\ref{wavefnoverlap}) is the physical 
ten dimensional gauge coupling $g_3^{10} = g_{\rm YM}$, since this is
obtained by dimensional reduction of super Yang--Mills theory.
Collecting the normalization factors (\ref{normalization}) from
(\ref{wvprod}), the full three-point coupling becomes
\begin{equation} \label{Yukawa} \begin{split}
 Y_{ij \bar k} = & g_{\rm YM} \prod_{d=1}^3 \left({2 \Im \tau_d 
\over A^2_d}{M_1^{(d)} M_2^{(d)} \over M_3^{(d)}}\right)^{1/4} \\
&\times \exp\left(i \pi (M_1^{(d)} \zeta_1^{(d)} \Im \zeta_1^{(d)} +
   M_2^{(d)} \zeta_2^{(d)} \Im \zeta_2^{(d)} + M_3^{(d)}
   \zeta_3^{(d)} \Im \zeta_3^{(d)})/\Im \tau_d \right)\\ 
&\times \sum_{n_d=1}^{g_d} \vartheta \left[ \begin{matrix}
    {M_2^{(d)}k - M_3^{(d)}j + M_2^{(d)} M_3^{(d)} l \over M_1^{(d)}
      M_2^{(d)} M_3^{(d)} } + {n_d \over g_d} \\ 0 \end{matrix} 
  \right](M_2^{(d)} M_3^{(d)}(\zeta_2^{(d)} - \zeta_3^{(d)}),\tau_d
      M_1^{(d)} M_2^{(d)} M_3^{(d)}) .
\end{split} 
\end{equation}
Here the index $d$ indicates that the corresponding quantity is the
component in $d$-th direction. 
For later use, it is useful to visualize the three-point coupling like
Feynman diagram in Fig. \ref{f:3pt}.

\subsection{Higher order coupling}

\subsubsection{Four-point coupling}

We calculate the four-point coupling 
\begin{equation} \label{4pt}
 y_{ijk\bar l} \equiv \int d^2 z \ 
\psi^{i,M_1}(z) \psi^{j,M_2}(z) \psi^{k,M_3}(z) 
\left( \psi^{l,M_4}(z) \right)^* ,
\end{equation}
and represent it in various ways.
The main result is that the four-point coupling can be expanded by 
three-point couplings. Thus by iteration, we can generalize it to
higher
order couplings.

We consider the case without Wilson lines, since the generalization is
straightforward. The product of the first two wavefunctions 
$\psi^{i,M_1}(z) \psi^{j,M_2}(z) $ in (\ref{4pt}) is the
same as in (\ref{wvprod}). Again, we suppose $M_1 + M_2 + M_3 = M_4$.
Then the product of the first three wavefunctions 
$\psi^{i,M_1}(z) \psi^{j,M_2}(z) \psi^{k,M_3}(z) $
in (\ref{4pt}) gives
\begin{equation} \label{threeprod}
\begin{split}
 \sum_{m \in \Z_{M_1 + M_2}} & \sum_{n \in \Z_{M_4}}
\psi^{i+j+k+M_1 m + (M_1 + M_2) n , M_4}(z)   
  ~\jtheta{{M_2 i - M_1 j + M_1 M_2 m \over M_1 M_2(M_1+M_2)} \\
  0}(0,\tau
 M_1 M_2 (M_1 + M_2)) \\
 & \times \jtheta{{M_3 (i+j+M_1 m) - (M_1 + M_2)k + (M_1+M_2)M_3 n
   \over (M_1+M_2) M_3 M_4} \\ 0}(0,\tau(M_1+M_2)M_3 M_4).
\end{split} \end{equation}
Now, we product the last wavefunction $\left( \psi^{l,M_4}(z)
\right)^*$
 in (\ref{4pt}), acting on the first factor in (\ref{threeprod}),
 yielding
the Kronecker delta 
$ \delta_{i+j+k+M_1 m + (M_1 + M_2) n,l} $.
The relation is given modulo $M_4$, reflecting that $i,j,k,l$ are
defined modulo $M_1,M_2,M_3,M_4$, respectively. It is non-vanishing if
there is $r$ such that
\begin{equation} \label{selrule}
 i+j+k+M_1 m + (M_1 + M_2) n = l + M_4r.
\end{equation}
We solve the constraint equation in terms of $n$.

For $\gcd(M_1,M_2,M_3)=1$,
any coupling specified by $(i,j,k,l)$ satisfies the constraint.
For a coupling $y_{ijk\bar l}$ with fixed $(m,r)$ there is always a
unique $n$ satisfying the constraint.
This means that by solving the constraint equation in terms of $n$,
we can remove the summation over $n$ in (\ref{threeprod}). The result
is
\begin{equation} \label{sdecomp-1} \begin{split}
 y_{ijk\overline l} = \sum_{m \in \Z_{M_1+M_2}} & \jtheta{{M_2 i - M_1
     j + M_1 M_2 m \over M_1 M_2 M} \\ 0} (0, \tau M_1 M_2M)
 \cdot \jtheta{{ M_3 l - M_4 k + M_3 M_4 r \over M M_3
     M_4} \\ 0 }(0,\tau M_3 M_4 M ) ,
\end{split} \end{equation}
where $M
= M_1 + M_2 = -M_3 + M_4$. 
This form (\ref{sdecomp-1}) 
is expressed in terms of only `external lines', $i, j, k, l$, 
and in the
`internal line' $r$ is uniquely fixed by $m$ from the relation
(\ref{selrule}).
This is to be interpreted as expansion in terms of three-point
couplings (\ref{3pt}).
{}From the property of the theta function, we have relations like
$y_{ij\bar k} = y_{\bar \imath \bar \jmath k}^*$, etc.
Thus we can write 
\begin{equation} \label{4pt-4}
 y_{ijk\bar l} = \sum_{m \in \Z_{M_1+M_2}} y_{i j \bar m}(m)
 \cdot  y_{k   m \bar l}(r) ,
\end{equation}
where $m$ and $r$ are uniquely related by the relation
(\ref{selrule}). 
Recall that three-point coupling can be
expressed in terms of `two external lines' depending on the 2-3
`picture.'

The result (\ref{sdecomp-1}) can be written by arranging the summation
of 
quantum numbers as follows,
\begin{equation} \label{sdecomp-2} \begin{split}
y_{ijk\overline l}  = \sum_{s \in \Z_{M_1+M_2}} & \jtheta{{M_2
s-Mj+M_2 Mr \over
      M_1M_2M} \\ 0}(0,\tau M_1 M_2M)
 \cdot \jtheta{{-Ml+M_4 s + MM_4 n \over M_3 M_4 M} \\ 0}
       (0,\tau M_3 M_4 M).
\end{split} \end{equation}
Here, we rewrite (\ref{selrule}) 
\begin{align}
 i + j + M_1 m &= s + (M_1+M_2) r, \nonumber \\
 -k + l +M_3 r &= s + (M_1+M_2) n,
\end{align}
by introducing 
an auxiliary label $s$, defined modulo
$M=M_1+M_2=-M_3+M_4$. This is uniquely fixed by other numbers from
(\ref{selrule}) and it can be traded with $m$. Thus we arrive at the
second form (\ref{sdecomp-2}), which becomes 
\begin{equation} \label{4pt-4-2}
 y_{ijk\bar l} 
  = \sum_{s \in \Z_{M_1+M_2}}  y_{i j \bar s}
 \cdot  y_{k  s \bar l} .
\end{equation}
The second expression (\ref{sdecomp-2}), explicitly depends on the  
`internal line' $s$. 
It is useful to track the intermediate quantum number $s$. 

\begin{figure}[t] \begin{center}
\includegraphics[height=3cm]{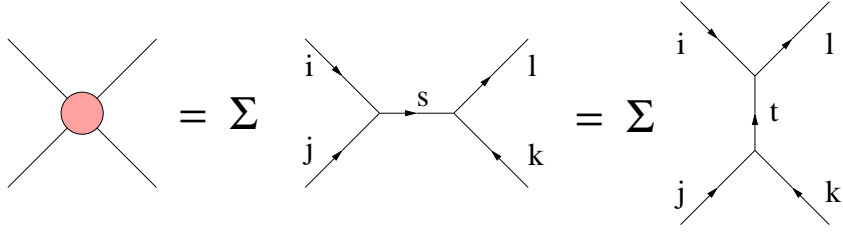}
\caption{A four-point coupling is decomposed into products of
  three-point couplings. It also has `worldsheet' duality. We have
  another `$u$-channel' diagram.}
\label{f:4ptdecomp}
\end{center} \end{figure}

We saw that in the case $\gcd(M_1,M_2)=1$, there is a unique solution.
Since we expand higher order coupling in terms of three-point
couplings, if any of them have degeneracies as in (\ref{modyukawa}),
i.e., $\gcd(M_i,M_j)=g_{ij}>1$, we should take into account their
effects.
It is interpreted that each three-point coupling contains a flavor
symmetry $\Z_{g_{ij}}$~\cite{Abe:2009vi}.
For the four-point coupling with 
$\gcd(M_1,M_2) = g_{12}$ and $\gcd(M_3,M_4)=g_{34}$ we have
also $\gcd(g_{12},g_{34})=g = \gcd(M_1,M_2,M_3,M_4)$, without loss of
generality (see below).
Employing the `intermediate state picture', or the
($j$-$s)\times(s$-$l)$
picture, in the last expression in
(\ref{sdecomp-2}), we have
\begin{equation} \begin{split}
 \sum_{p \in \Z_g} \sum_{s \in \Z_{M_1+M_2}} & \jtheta{{M_2s-Mj+M_2 Mr
 \over
      (M-M_2)M_2M} + {p \over g} \\ 0}(0,\tau (M-M_2)M_2M) \\
  & \times \jtheta{{-Ml+M_4s + MM_4 n \over M M_4 (M_4-M)} + {p \over
 g} \\ 0}
       (0,\tau M M_4 (M_4-M)).
\end{split} \end{equation}
It shows that the two symmetries $\Z_{g_{12}}$ and $\Z_{g_{34}}$ are
broken down to the largest common symmetry $\Z_g$, due to the
constraint. Otherwise we cannot put together the vertices with the
common intermediate state $s$.

Reminding that we are examining the overlap of four wavefunctions, and
it {\em does not depend on the order of product}.
If we change the order of the product in (\ref{4pt}), namely consider
the product of the second and the third wavefunctions
$\psi^{j,M_2}(z) \psi^{k,M_3}(z) $
first, we have
differently-looking constraint relation
which is equivalent to (\ref{selrule}) undergoing the decomposition,
\begin{align}
 j + k + M_2 m' &= t + (M_2+M_3) r', \nonumber \\
 -i + l +M_1 r' &= t + (M_2+M_3) n'.
\end{align}
This looks like the `$t$-channel' and we have
\begin{equation} \label{tdecomp} \begin{split}
 y_{ijk\overline l} = \sum_{t \in \Z_{M'}} & \jtheta{{M_3t-M' k+M_3 M'
 r' \over
      (M'-M_3)M_3M'} \\ 0}(0,\tau (M'-M_3)M_3M')) \\
  & \times \jtheta{{-M'l+M_1t + M'M_1 n \over M' M_1 (M_1-M')} \\ 0}
       (0,\tau M' M_1 (M_1-M')) \\
  = \sum_{t \in \Z_{M'}} & y_{i\bar l t} \cdot y_{jk\bar t},
\end{split} \end{equation}
with $M' = -M_1 + M_4 = M_2 +M_3$.
The result  has a behavior like `worldsheet' {\em duality} in those of
Veneziano and Virasoro--Shapiro \cite{VSV}. This means that, in
decomposing the
diagram, the position of an insertion does not matter. 

If we have Wilson lines, we just replace the three-point couplings
by those with Wilson lines (\ref{Yukawa}).

\subsubsection{Generic $L$-point coupling}
We have seen that the four point coupling is expanded in terms of
three-point couplings.
We can generalize the result to obtain arbitrary higher order
couplings.
The constraint relations and the higher order couplings are 
always decomposed into products of three-point couplings. 
It is easily calculated by Feynman-like diagram.

The decompositions (\ref{sdecomp-1}),(\ref{sdecomp-2}),(\ref{tdecomp}) 
are understood as
inserting the identity expanded by the complete set of orthonormal
eigenfunctions
$\{ \psi^{i,M}_n \}$  as follows.
For example, we split the integral  (\ref{4pt}) as 
\begin{equation} \label{split-1}
  y_{ijk\bar l} = \int d^2 z d^2z'\ \psi^{i,M_1}(z)
 \psi^{j,M_2}(z) \delta^2(z-z')
 \psi^{k,M_3} (z') \left(
 \psi^{l,M_4} (z') \right)^*.
\end{equation}
Then, we use the complete set of orthonormal eigenfunctions 
$\{ \psi^{i,M}_n \}$ of the Hamiltonian with a magnetic 
flux $M$.
That is, they satisfy 
\begin{equation}\label{complet-set}
\sum_{s,n} \left( \psi^{s,M}_n  (z) \right)^* \psi^{s,M}_n  (z') 
= \delta^2(z-z').
\end{equation}
We insert LHS instead of the delta function $\delta^2(z-z')$ in 
(\ref{split-1}).
Since $\psi^{i,M_1}(z) \psi^{j,M_2}(z)$ is decomposed 
in terms of $ \psi^{s,M_1+M_2}_n  (z)$, it is convenient to 
take $M = M_1 +M_2$ for inserted wavefunctions 
$\left( \psi^{s,M}_n  (z) \right)^* \psi^{s,M}_n  (z')$.
In such a case, only zero-modes of $\psi^{s,M}_n (z)$ appear 
in this decomposition.
If we take $M \neq M_1 + M_2$, higher modes of $\psi^{s,M}_n (z)$ 
would appear.
At any rate, when we take $M = M_1 +M_2$, 
we can lead to the result (\ref{sdecomp-2}) and 
(\ref{4pt-4}).
On the other hand, we can split 
\begin{equation} \label{split-2}
  y_{ijk\bar l} = \int d^2 z d^2z'\ \psi^{j,M_2}(z)
 \psi^{k,M_3}(z) \delta^2(z-z')
 \psi^{i,M_1} (z') \left(
 \psi^{l,M_4} (z') \right)^*,
\end{equation}
and insert (\ref{complet-set}) with $M=M_2 +M_3$.
Then, we can lead to (\ref{tdecomp}).
Furthermore, we can calculate the four-point coupling after 
splitting 
\begin{equation} \label{split-3}
  y_{ijk\bar l} = \int d^2 z d^2z'\ \psi^{i,M_1}(z)
 \psi^{k,M_3}(z) \delta^2(z-z')
 \psi^{j,M_2} (z') \left(
 \psi^{l,M_4} (z') \right)^*.
\end{equation}
How to split corresponds to `s-channel', `t-channel' and 
`u-channel'.
Note that only zero-modes appear in 
`intermediate states', when we take proper values of $M$ 
because of the product property.

We have considered the four-point couplings 
with $M_1+M_2+M_3 = M_4$ for $M_i >0$.
We may consider the case with 
$M_1+M_2 = M_3 + M_4$ for $M_i >0$, 
which corresponds to 
\begin{equation} \label{4pt-2}
 y_{ij\bar k \bar l} \equiv \int d^2 z \ 
\psi^{i,M_1}(z) \psi^{j,M_2}(z) \left( \psi^{k,M_3}(z) \right)^* 
\left( \psi^{l,M_4}(z) \right)^* .
\end{equation}
In order to consider both of this case and the previous case at the
same time, 
we would have more symmetric expression for the four-point coupling
\begin{equation} \label{4pt-3}
 y_{ijkl} = \int d^2 z \ \psi^{i_1,M_1}(\tilde z)
 \psi^{i_2,M_2}(\tilde z)
 \psi^{i_3,M_3} (\tilde z)
 \psi^{i_4,M_4} (\tilde  z) ,
\end{equation}
by defining
\begin{equation} \label{ext}
 \psi^{i,-M}(\bar z) \equiv  \left( \psi^{i,M} (z) \right)^*,
\end{equation}
with
$$M_1 + M_2 + M_3 + M_4 =0, $$
where some of $M_i$ are negative, 
and $\tilde z = z$ for $M >0$ and $\tilde z = \bar z$ for $M <0$.  

\begin{figure}[t] \begin{center}
\includegraphics[height=2cm]{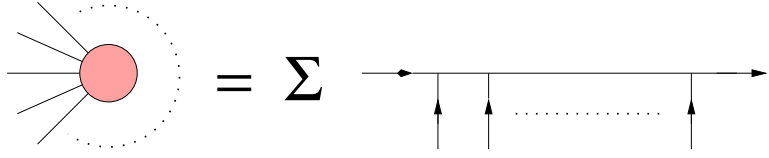}
\caption{Likewise, any amplitude with arbitrary external lines is
  decomposed into product of three-point amplitudes.}
\end{center} \end{figure}

We can extend the above calculation to the $L$-point coupling,
\begin{equation}
 y_{i_1 i_2 \dots i_L} \equiv \int d^2 z \prod_{j=1}^L \psi^{ i_j,M_j} 
(\tilde z), 
\end{equation}
with the extension as in (\ref{ext}). We have then the selection rule
\begin{equation}
 \sum_{j=1}^L M_j = 0,
\end{equation}
where some of $M_j$ are negative.
The constraint is given as
\begin{equation}
 \sum_{j=1}^L \left( i_j + \left(\sum_{l=1}^j M_l \right )r_j \right)
 = 0.
\end{equation}
Again, it shows the conservation of the total flavor number $i_j$,
reflecting the fact that each $i_j$ is defined modulo $M_j$.
We can decompose $L$-point coupling into $(L-1)$ and three-point
couplings
\begin{align}
 \sum_{j=1}^{L-3} \left( i_j + \left(\sum_{l=1}^{j} M_l \right )r_j
 \right) + i_{L-2}
 &= s -  K r_{L-1}, \nonumber  \\
  i_{L-1} + i_L + M_{L-1} r_{L-1} &= - s - K r_{L-2},
\end{align}
where
\begin{equation}
K = \sum_{k=1}^{L-2} M_i = - M_{L-1} - M_L ,
\end{equation}
is the intermediate quantum number.
Therefore if $\gcd(M_1,M_2,\dots,M_L)=1$, by induction we see that
there is a unique solution by Euclidean algorithm.
By iteration
\begin{equation} \label{iteration}
 y_{i_1 i_2 \dots i_L} = \sum_s y_{i_1 i_2 \dots i_{L-2} s} \cdot
 y_{\bar s
   i_{L-1} i_{L}} ,
\end{equation}
we can obtain the coupling including the normalization.
Thus, we can obtain $L$-point coupling out of $(L-1)$-point coupling.
Due to the independence of ordering, we can insert (or cut and glue)
any node.

\begin{figure}[t] 
\psfrag{a}[r]{$i_1,M_1$}
\psfrag{b}[c]{$i_2,M_2$}
\psfrag{c}[c]{$i_3,M_3$}
\psfrag{d}[c]{$i_4,M_4$}
\psfrag{e}{$i_5,M_5$}
\psfrag{f}[c]{$s_1$,$M_1$+$M_2$}
\psfrag{g}[c]{$s_2$,$M_1$+$M_2$+$M_3$}
\begin{center}
\includegraphics[height=3cm]{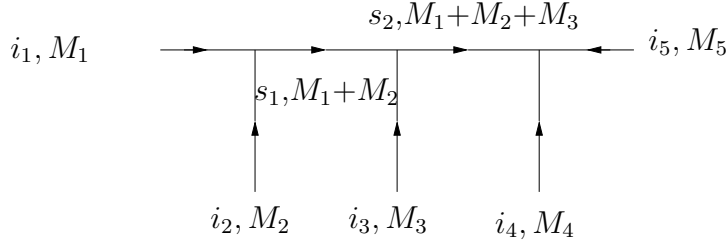}
\caption{Five-point coupling. No more independent Feynman-like diagram
for different insertion.}
\label{f:5pt}
\end{center} \end{figure}

As an illustrating example we show the result for the five-point
coupling.
We employ $s$-channel-like insertions, by naming
intermediate quantum numbers $s_i$ as in Fig. \ref{f:5pt}.
We have
\begin{equation} \begin{split}
 y_{i_1 i_2 i_3 i_4 i_5} =& \prod_{j=1}^5
 \jstheta{i_j/M_j}{0}(zM_i,\tau M_i) \\
 =& \sum_{s_1,s_2 }
 \jtheta{{M_2 s_1 - (M_1 +M_2)i_2 + M_2(M_1 +M_2)l_1 \over M_2
     (M_1+M_2)(M_1+2M_2) } \\ 0} (0,M_1 M_2 (M_1 +M_2)\tau)\\
 & \times
  \jtheta{{(M_1+ M_2) i_3 - M_3 s_1 + M_3(M_1 +M_2)l_2 \over M_3
 (M_1+M_2)(M_1+M_2+M_3) } \\ 0} (0,(
M_1+M_2)M_3(M_1+M_2+M_3))\\
 & \times
  \jtheta{{(M_1+M_2+M_3) i_4 - M_4 s_2  + M_4(M_1 +M_2+M_3)l_3 \over
      M_4 (M_1+M_2+M_3)(M_1+M_2+M_3+M_4) } \\ 0}(0,-(M_4+M_5)M_4 M_5
 \tau),
\end{split} \end{equation}
where
$$ s_1 \in \Z_{M_1+M_2}, \quad  s_2 \in \Z_{M_1 + M_2       + M_3}.
$$
{}From the regular patterns of increasing orders, we can
straightforwardly generalize the couplings to arbitrary order.

Now, taking into account full six internal dimensions, as in
three-coupling case (\ref{Yukawa}), we have various normalization
factors
besides the product of theta functions. Again, from the product
relation of
theta function (\ref{thetaprod}) we have
\begin{equation} \begin{split} \label{symfactor}
 s_L & g_{\rm YM}^{L-2} {\alpha'}^{(L-4+L_\chi/2)/2} \\
\times &\prod_{d=1}^3
\Bigg( {2 \Im \tau_d 
   \over A^2_d} \sum_{M_i^{(d)} > 0}|M_i^{(d)}| \Bigg)^{-\frac14}
\Bigg( {2 \Im \tau_d 
   \over A^2_d} \sum_{M_i^{(d)} < 0}|M_i^{(d)}| \Bigg)^{-\frac14}
\prod_{i=1}^L \left( {2 \Im \tau_d |M_i^{(d)}|
   \over A^2_d} \right)^{\frac14} .
\end{split} \end{equation}
Recall that $L_\chi$ is the number of fermions in the couplings 
(\ref{eq:L-coupling}). 
We have $g^{10}_L $ $=$ $s_L g_{\rm YM}^{L-2}$
${\alpha'}^{(L-4+L_\chi/2)/2}$ in (\ref{wavefnoverlap}), where 
symmetric factor $s_L$ comes from higher order expansions of
lower-level 
completion of Yang--Mills theory, having also an expansion parameter
$\alpha'$. 
In open string theory, it is 
the Dirac--Born--Infeld action, and it is unknown beyond the quartic
order in $\alpha' F$ \cite{Koerber:2002zb}.
The dependence of ten-dimensional gauge coupling $g_{\rm YM}$ and
Regge
slope $\alpha'$ can be easily accounted by order counting \cite{Po}.
Note that $g_{\rm YM}$ is dimensionful.
This factor (\ref{symfactor}) is non-holomorphic in the complex
structure $\tau$ and
complexified K\"ahler modulus $\alpha' J = B + i A/4 \pi^2$, where
$B_{z^d \bar z^d}$ is the antisymmetric tensor field component in
$d$-th
  two-torus. 
They are interpreted as originating from the K\"ahler potential
\cite{Cremades:2004wa,DiVecchia:2008tm}.
The product $\prod M_i^{1/4}$ is the leading order approximation of
Euler beta function and its multivariable generalization, which is the
property of dual amplitude.

As an example of full expressions, we show the four-point coupling 
among scalar fields, $Y_{ij\bar l \bar m}\phi^i \phi^j (\phi^l)^*
(\phi^m)^* $, where $\phi^i$ and $(\phi^l)^*$ 
($\phi^j$ and $(\phi^m)^*$) correspond to the magnetic flux 
$M_1^{(d)}$ ($M_2^{(d)}$).
For simplicity, we consider the case with vanishing Wilson lines 
and $\gcd (M_1,M_2)=1$.
The full coupling $Y_{ij\bar l \bar m}$ is obtained as 
\begin{equation}
Y_{ij\bar l \bar m} =  g_{\rm YM}^2\prod_{d=1}^3 \left({2 \Im \tau_d
\over
   A^2_d}{M_1^{(d)} M_2^{(d)} \over M_3^{(d)}}\right)^{1/2} 
\sum_{k \in \Z_{M^{(d)}_1+M^{(d)}_2}}  y^{(d)}_{ij\bar k} \
(y^{(d)})^*_{k
  \bar l \bar m} ,
\end{equation}
up to $s_L$, where 
\begin{equation}
y^{(d)}_{ij\bar k} = 
\jtheta{{M^{(d)}_2 k-M^{(d)}j+M^{(d)}_2 M^{(d)}r \over
      M^{(d)}_1M^{(d)}_2M^{(d)}} \\ 0}(0,\tau_d M^{(d)}_1
      M^{(d)}_2M^{(d)}).
\end{equation}
This scalar coupling with $s_L=1$ appears from ten-dimensional 
super Yang-Mills theory and satisfies the relation 
$Y_{ij\bar l \bar m} = Y_{ij \bar k} (Y)^*_{k \bar l \bar m}$ 
for the three-point coupling $Y_{ij \bar k}$ in 
eq.~(\ref{Yukawa}).

\subsection{Intersecting D-brane models}\label{sec:Intersecting}

Here we give comments on the relation between the results 
in the previous sections and higher order couplings 
in intersecting D-brane models, 
i.e. CFT-calculations.

There is well-known $T$-duality relation between magnetized and 
intersecting brane models.
In intersecting brane case, the wavefunctions are highly localized
around intersection points, whereas magnetized brane wavefunctions are
fuzzily delocalized over the entire space.

Under the `horizontal' duality with respect to real axis, 
$y_i \leftrightarrow 2 \pi \alpha' A_i$.
The parameter is changed as
\begin{equation}\label{T-dual}
 \tau \leftrightarrow J, \quad \zeta \leftrightarrow \nu .
\end{equation}
Still the translational offset $\nu$ is the Wilson line.
Thus, the magnetic flux gives the slope 
$A_{\bar z}^i = -\frac{i}{2} F_{z \bar z}^i z = \frac{\pi}{\Im
\tau}M_i$
and the corresponding quantum number is the
`relative angle,' for small angles,
\begin{equation}
 \pi \theta_{i} = \frac{M_i}{\Im J}.
\end{equation}
The selection rule due to the gauge invariance becomes
\begin{equation}
 M_1 + M_2 = M_3 \leftrightarrow \theta_1 + \theta_2 = \theta_3.
\end{equation}

In the intersecting brane case, as well as heterotic string case,
there have been CFT calculation of higher order
amplitude \cite{Atick:1987kd,AO,CK} using vertex operator insertion
\cite{Hamidi:1986vh,Cremades:2003qj,CP,BKM}.
There are vertex operators $V_i$ corresponding to massless modes.
We compute their $L$-point amplitude,
\begin{equation}\label{L-amp}
 \langle V_1 V_2 \dots V_L \rangle .
\end{equation}
We have operator product expansion (OPE),
\begin{equation}\label{ope}
V_i(z) V_j(0) \sim \sum_k \frac{c_{ijk}}{z^{h_{ijk}}}V_k(0),
\end{equation}
with $h_{ijk}= h(V_k) - h(V_i) - h(V_j)$,
where $h(V_l)$ is the conformal dimension of $V_l$.
This OPE corresponds to (\ref{wvprod}).
Furthermore, the coefficients $c_{ijk}$ correspond 
to the three-point couplings in four-dimensional effective 
field theory.
In Ref.~\cite{Cremades:2004wa}, it is shown that the above three-point 
coupling $c_{ijk}$ in intersecting D-brane models corresponds 
to the T-dual of the three-point couplings 
$Y_{ijk}$ in magnetized D-brane models.

Now, let us consider the 
$L$-point amplitude $\langle \prod_i V_i(z_i) \rangle $.
We use the OPE (\ref{ope}) to write the $L$-point 
amplitude in terms of $(L-1)$ point amplitudes.
Such a procedure is similar to one 
in the previous sections, where 
we write $L$-point couplings in terms of three-point 
couplings.

For example, the CFT calculations for the four-point 
couplings $c_{ijkl}$ in the intersecting D-brane models 
would lead 
\begin{equation}
c_{ijkl} \sim \sum_s c_{ij \bar s}c_{s kl},
\end{equation}
and
\begin{equation}
c_{ijkl} \sim \sum_t c_{ik\bar t}c_{t jl},
\end{equation}
depending on the order of OPE's, i.e. 
s-channel or t-channel.
Thus, the form of the four-point couplings 
as well as $L$-point couplings $(L>4)$ is almost 
the same as the results in the previous sections.
Note that in eq.(\ref{wvprod}), a product of two wavefunctions 
is decomposed in terms of only the 
lowest modes.
On the other hand, in RHS of Eq.~(\ref{ope}), higher modes 
as well as lowest modes may appear.
However, dominant contribution due to the
lowest modes are the same, because 
$c_{ijk}$ for the lowest modes $(i,j,k)$ 
corresponds exactly to $Y_{ijk}$ for the lowest modes.

Let us examine the correspondence of couplings 
between magnetized models and intersecting D-brane models 
by using concrete formulae.
In the intersecting D-brane models, 
the amplitude (\ref{L-amp}) is decomposed into the classical and the
quantum parts,
\begin{equation}\label{L-amp-2}
 \langle V_1 V_2 \dots V_L \rangle = {\cal Z}_{\rm qu} \cdot {\cal
 Z}_{\rm cl} =
{\cal Z}_{\rm qu} \cdot \sum_{\{X_{\rm cl}\}} \exp(-S_{\rm cl}),
\end{equation}
where $X_{\rm cl}$ is the solution to the classical equation of
motion.
The classical part is formally characterized as decomposable
part and physically gives instanton of worldsheet nature, via the
exchange of intermediate string. That gives
intuitive understanding via the `area rule', where 
the area corresponds to one, which intermediate string sweeps.

In the three-point amplitude, the summation of the classical action 
$\sum_{\{X_{\rm cl}\}} \exp(-S_{\rm cl})$  becomes the theta 
function~\cite{Cremades:2003qj}, where $S_{\rm cl}$ corresponds to 
the triangle area.
When we exchange $\tau$ and $J$ as (\ref{T-dual}) in the magnetized
models, 
the Yukawa coupling (\ref{3pt}) corresponds to the following expansion 
\begin{equation} \label{intersec3pt} \begin{split}
  y_{i j \bar k} & = \jtheta{{M_2 k - M_3 j + M_2 M_3 l \over M_1 M_2
      M_3} \\ 0} \left(0,iM_1 M_2 M_3 A/(4 \pi^2 \alpha') \right)
  \\
    & = \sum_{n \in \Z} \exp\left[-\frac{M_1 M_2 M_3 A}{4\pi \alpha'}
  \Big({M_2 k - M_3 j + M_2 M_3 l \over M_1 M_2 M_3}+n \Big)^2\right], 
\end{split} \end{equation}
by using the definition (\ref{jacobitheta}).
We have neglected the antisymmetric tensor component $B$. 
The exponent corresponds to the area (divided by $4 \pi \alpha'$) of 
possible formation of triangles and the one with $n=0$
corresponds to the minimal triangle.
Recall that the theta function part depends on only 
$\tau$ and $J$ in magnetized and intersecting D-brane models, 
respectively.

We have omitted the normalization factor, corresponding to the quantum
part ${\cal Z}_{\rm qu}$. It is obtained by comparing the coupling
(\ref{intersec3pt}) with (\ref{Yukawa}). We find the factor
\begin{equation}
 2^{-9/4} \pi^{-3} e^{\phi_4/2} \prod_{d=1}^3 \left(\Im \tau_d
 {M_1^{(d)}
    M_2^{(d)} \over M_3^{(d)}}  \right)^{1/4}, 
\end{equation}
in the magnetized brane side corresponds to 
\begin{equation}
 {\cal Z}_{\rm qu} = (2 \pi)^{-9/4} e^{\phi_4/2} \prod_{d=1}^3
 \left((\Im
 J_d)^2 {\theta_1^{(d)} \theta_2^{(d)} \over \theta_3^{(d)}}
 \right)^{1/4}, 
\end{equation}
in the intersecting brane side. We obtain the four dimensional dilaton
$\phi_4 = \phi_{10} - \ln |\Im \tau_1
\Im \tau_2 \Im \tau_3|$ from the ten dimensional one $\phi_{10}$, 
which is related with $g_{\rm YM}$ as  $g_{\rm YM} =
e^{\phi_{10}/2}\alpha^{\prime 3/2}$.
The vacuum expectation value of the dilaton gives gauge coupling $
e^{\langle \phi_4 \rangle /2} = g$.
In this case, the factor containing the angles is a leading order
approximation of the ratio of Gamma function
\begin{equation}
 { \Gamma(1-\theta_1) \Gamma(1-\theta_2) \Gamma(\theta_3) \over
 \Gamma(\theta_1) \Gamma(\theta_2) \Gamma(1-\theta_3)} 
\simeq {\theta_1 \theta_2 \over \theta_3},
\end{equation}
valid for small angles. 
Therefore, the three-point couplings coincide each other between 
magnetized and intersecting D-brane models.
That is the observation of \cite{Cremades:2004wa}.

\begin{figure}[t] \begin{center}
\includegraphics[height=2cm]{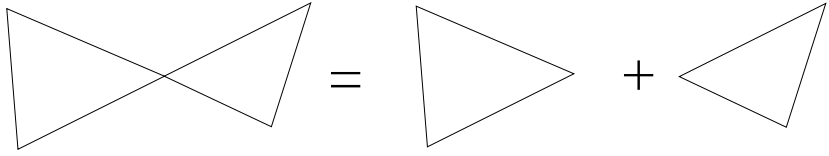}
\caption{Area of polygon, responsible for the classical part exponent,
  is decomposed in terms of those of three point functions. }
\label{f:areap}
\end{center} \end{figure}

Now, let us consider the four-point coupling 
of intersecting D-brane model corresponding to 
the left figure of Fig.~\ref{f:areap}.
The four-point amplitude is written as (\ref{L-amp-2}), 
where the classical action corresponds to 
the area of the left figure.
However, that can be decomposed into two triangles 
like the right figure, that is, 
the classical part can be decomposed into two parts, 
each of which corresponds to the classical part of 
three-point amplitude, i.e. 
\begin{equation}\label{decomp4-3}
\exp (-S_{\rm cl}^{(4)})=
\exp (-{S}_{\rm cl}^{(3)}) \exp (-{S'}_{\rm cl}^{(3)}),
\end{equation}
where $S_{\rm cl}^{(4)}$ corresponds to the area 
of the left figure of Fig.~\ref{f:areap} and 
${S}_{\rm cl}^{(3)}$ and ${S'}_{\rm cl}^{(3)}$ 
correspond to the triangle areas of the 
right figure.

On the other hand, our results in the previous sections show that 
the four-point
coupling in the magnetized model is also expanded as
(\ref{sdecomp-2}).
Each of theta functions in  (\ref{sdecomp-2}) corresponds to 
the classical parts of the three-point couplings 
in the intersecting D-brane models.
This relation corresponds to the 
above decomposition (\ref{decomp4-3}).
Thus, the theta function parts of the four-point couplings, 
i.e. the classical part, coincide each other between 
magnetized and intersecting D-brane models.  
That means that the holomorphic complex structure, $\tau$, 
dependence of the four-point couplings 
in the magnetized brane models is the same as 
the holomorphic K\"ahler moduli $J$ dependence in the intersecting 
D-brane models, since the theta function part in 
the magnetized (intersecting) D-brane models depends 
only on $\tau$ ($J$).
The other part in the magnetized brane models corresponds 
to normalization factors $N_M$.
When we take a proper normalization, these 
factors also coincide.

\subsection{Flavor symmetries}

We study order $L$ couplings including the three point 
couplings $L=3$ in four-dimensional effective theory, i.e., 
\begin{equation}
 Y_{i_1 \dots i_{L_\chi} i_{L_\chi+1}\cdots i_L} \chi^{i_1}(x) \cdots 
\chi^{i_{L_\chi}}(x) \phi^{i_{L_{\chi}+1}}(x) \dots \phi^{i_L}(x),
\end{equation}
with $L=L_\chi + L_\phi$, where $\chi$ and $\phi$ collectively 
represent four-dimensional components of fermions and bosons,
respectively.
In particular, the selection rule for allowed couplings 
is important.
The three-point couplings can appear from the dimensional 
reduction of ten-dimensional super-Yang--Mills theory 
and higher order coupling terms can be 
read off from the effective Lagrangian of 
the Dirac--Born--Infeld action with
supersymmetrization. 
The internal component of bosonic and 
fermionic wavefunctions is the same \cite{Cremades:2004wa}.
Thus, the couplings are determined by the wavefunction overlap in the
extra
dimensions,
\begin{equation}
 Y_{i_1 i_2 \dots i_L} = g_L^{10}  \int_{T^6} d^6z \ 
\prod_{d=1}^3 \psi^{i_1,M_1}_d(z) 
\psi^{i_2,M_2}_d(z) \dots \psi^{i_L,M_L}_d(z),
\end{equation}
where $g_L^{10}$ denotes the coupling in ten dimensions.
Here, as mentioned in the previous section, 
we concentrate on the two-dimensional $T^2$ part 
of the overlap integral of wavefunctions,
\begin{equation}
 y_{i_1 i_2 \dots i_L} = \int_{T^2} d^2z \ 
\psi^{i_1,M_1} (z) 
\psi^{i_2,M_2} (z) \dots \psi^{i_L,M_L}(z),
\end{equation}
where we have omitted the subscript $d$, again.

For example, we calculate the three-point couplings,
\begin{equation}
  y_{i_1i_2\bar i_3} = \int d^2 z \ 
\psi^{i_1,M_1}(z) \psi^{i_2,M_2}(z) \left( \psi^{i_3,M_3}(z) \right)^*
.
\end{equation}
For the moment, we consider the case with vanishing Wilson lines.
The gauge invariance requires that $M_1+M_2 =M_3$ and 
that the wavefunction $\left( \psi^{i_3,M_3}(z) \right)^*$ but not 
$\psi^{i_3,M_3}(z)$ appears in the allowed three-point couplings.
If these are not satisfied, 
there is not corresponding operators in the ten dimensions, 
i.e. $g^{10}_3=0$.
The results are obtained as~\cite{Cremades:2004wa}
\begin{equation}\label{yijk-1}
 y_{i_1 i_2\bar i_3} = \sum_{m \in Z_{M_3}}  \delta_{i_1+i_2+M_1m,i_3}
 ~\jtheta{{M_2 i_1 - M_1 i_2 + M_1 M_2 m \over M_1 M_2 M_3}
 \\ 0}(0,\tau M_1 M_2 M_3),
\end{equation}
where the numbers in the Kronecker delta is defined modulo $M_3$. 
Indeed, the Kronecker delta part leads to the selection rule 
for allowed couplings as 
\begin{equation} \label{deltaconstraint2}
 i_1+i_2-i_3 = M_3 l - M_1 m, \quad m \in { Z}_{M_3}, \ l \in {
   Z}_{M_1}.
\end{equation}
When $\gcd(M_1,M_2,M_3)=1$, every combination $(i_1,i_2,i_3)$ 
satisfies this constraint (\ref{deltaconstraint2}) 
because of Euclidean algorithm.
On the other hand, when $\gcd(M_1,M_2,M_3)=g$, 
the above constraint becomes
\begin{equation} \label{deltaconstraint-g}
 i_1+i_2-i_3 = 0  \qquad (~{\rm mod} \ g~).
\end{equation}
This implies that we can define $Z_g$ charges from 
$i_k$ 
for zero-modes and the allowed couplings are controlled by 
such $Z_g$ symmetry.
Indeed, each quantum number $i_k$ corresponds to 
quantized momentum defined with the $M_i$ modulo structure.
When $\gcd(M_1,M_2,M_3)=g$, the modulo structure 
becomes $Z_g$ and the conservation law of 
these discrete momenta corresponds to a
requirement due to the $Z_g$ invariance.

Let us consider higher order couplings.
In~\cite{Abe:2009dr}, it has been shown that 
higher order couplings can 
be decomposed as productions of three-point couplings.
For example, we consider the four-point coupling,
\begin{equation}
  y_{i_1i_2i_3\bar i_4} = \int d^2 z \ 
\psi^{i_1,M_1}(z) \psi^{i_2,M_2}(z) \psi^{i_3,M_3}(z) 
\left( \psi^{i_4,M_4}(z) \right)^* .
\end{equation}
This four-point coupling can be decomposed as 
\begin{equation}
y_{i_1i_2i_3\bar i_4} = \sum_{s \in Z_M} y_{i_1 i_2 \bar s} 
\ y_{s i_3 \bar i_4},
\end{equation}
where 
\begin{eqnarray}
  y_{i_1i_2\bar s} &=& \int d^2 z \ 
\psi^{i_1,M_1}(z) \psi^{i_2,M_2}(z) \left( \psi^{s,M}(z)
\right)^*, \nonumber \\
 y_{s i_3\bar i_4} &=& \int d^2 z \ 
\psi^{s,M}(z) \psi^{i_3,M_3}(z) \left( \psi^{i_4,M_4}(z)
\right)^*,
\end{eqnarray}
with $M=M_1+M_2=M_4-M_3$.
Here, $\psi^{s,M}(z)$ denotes the $s$-th zero-mode of 
Dirac equation with the relative magnetic flux $M$, 
and these modes correspond to intermediate states 
in the above decomposition.
Each of $y_{i_1 i_2 \bar s} $ and $y_{s i_3 \bar i_4}$ 
is obtained as eq.~(\ref{yijk-1}).
That is, the coupling selection rule is controlled by 
the $Z_g$ invariance (\ref{deltaconstraint2}), 
i.e. the conservation law of discrete momenta, 
and its modulo structure is determined by 
$\gcd(M_1,M_2,M_3,M_4)=g$.

Similarly, higher order couplings are decomposed 
as products of three-point couplings~\cite{Abe:2009dr}.
Therefore, the above analysis is generalized to 
generic order $L$ couplings.
That is, the coupling selection rule is 
given as the $Z_g$ invariance and its 
modulo structure is determined by 
$\gcd(M_1,\cdots ,M_L)=g$.

So far, we have considered the model with 
vanishing Wilson lines.
Non-vanishing Wilson lines do not affect 
the coupling selection rule due to the $Z_g$ 
invariance, but change 
values of couplings $y_{i_1 i_2 \bar i_3}$.
For example, when we introduce Wilson lines 
$\zeta_k$ for $\psi^{i_k,M_k}(z)$, 
the three-point coupling (\ref{yijk-1}) becomes  
\begin{eqnarray}\label{yijk-wl}
 y_{i_1 i_2\bar i_3} &=& \sum_{m \in \Z_{M_3}}
 \delta_{i_1+i_2+M_1m,i_3} e^{i\pi 
(\sum_{k=1}^3 M_k \zeta_k \Im \zeta_k)/\Im  \tau} \nonumber \\
& & \times 
 ~\jtheta{{M_2 i_1 - M_1 i_2 + M_1 M_2 m \over M_1 M_2 M_3}
 \\ 0}(M_2 M_3 (\zeta_2 - \zeta_3),\tau M_1 M_2 M_3),
\end{eqnarray}
where Wilson lines must satisfy $\zeta_3 M_3 = \zeta_1 M_1 + \zeta_2
M_2$.
Similarly, higher order couplings with non-vanishing 
Wilson lines can be obtained.

\subsection{Non-Abelian Wilson line}

In this section we calculate the Yukawa coupling
with non-Abelian Wilson lines.
Let us consider the following form of the magnetic fluxes,
\begin{equation} 
F = 
\begin{pmatrix}
\frac{m_a}{n_a} {\bf 1}_{N_a} & & \\
& \frac{m_b}{n_b} {\bf 1}_{N_b} & \\
& & \frac{m_c}{n_c} {\bf 1}_{N_c} 
\end{pmatrix},
\end{equation}
and non-Abelian Wilson lines similar to (\ref{eq:nA-WL-1}).
Then, there are three types of matter fields, 
$(N_a,\overline N_b)$, $(N_b,\overline N_c)$, 
$(N_c,\overline N_a)$ and their conjugates 
under $U(N_a) \times U(N_b) \times U(N_c)$, 
although they break to $U(P_a) \times U(P_b) \times U(P_c)$
by non-Abelian Wilson lines.
We consider the case with $\frac{m_a}{n_a}-\frac{m_b}{n_b}>0$, 
$\frac{m_b}{n_b}-\frac{m_c}{n_c}>0$
and  $\frac{m_a}{n_a}-\frac{m_c}{n_c}>0$.
Then, the three types of matter fields 
whose wavefunctions are denoted by $\Psi^{j,M_1}$, 
$\Psi^{k,M_2}$ and $(\Psi^{l,M_3})^*$, 
appear in the 
following off-diagonal elements,
\begin{equation} 
\begin{pmatrix}
{\rm{const}} & \Psi^{j,M_1} & \\
& {\rm{const}} &  \Psi^{k,M_2} \\
(\Psi^{l,M_3})^* & & {\rm{const}} 
\end{pmatrix}, 
\end{equation}
where $M_1= M_{ab}$, $M_2=M_{bc}$ and $M_3=M_{ac}$ for simplicity.
We use the same indices for $Q_{ab}$ and others, i.e. 
$Q_1= Q_{ab}$, $Q_2=Q_{bc}$ and $Q_3=Q_{ac}$. 
As already explained, in the background with fractional fluxes 
and non-Abelian Wilson lines, their fields are the matrix valued 
wavefunctions.
The Yukawa coupling can be calculated by computing the following
overlap 
integral of zero-modes in the $(y_4,y_5)$ compact space,
\begin{eqnarray}\label{eq:y-ijk-1}
y^{jkl}_{1,pqr} &=&
 \int_0^1 dy_4\int_0^1 dy_5 {\rm{Tr}}
[\Psi^{j,M_1}_{pq}\Psi^{k,M_2}_{qr}(\Psi^{l,M_3}_{pr})^* ]. 
\end{eqnarray}
The Yukawa coupling $Y^{ijk}$ in 4D effective theory is
obtained as 
their products on $(T^2)^n$, i.e. 
$Y^{ijk}_{1,pqr} = g_D\prod_{d=1}^{n/2} y^{ijk}_d$, where 
$y_{d,pqr}^{ijk}$ denotes the overall integral similar to
Eq.~(\ref{eq:y-ijk-1}) 
for the $d$-th torus $(T^2)$ and  
$g_D$ is the D-dimensional gauge coupling.
{}From this structure, one can see that the allowed couplings 
are restricted.
In order to see it,  we introduce the following parameters as
$k_1={\rm g.c.d.}(n_a,n_b)$, 
$k_2={\rm g.c.d.}(n_b,n_c)$, 
$k_3={\rm g.c.d.}(n_a,n_c)$
and $K={\rm g.c.d.}(k_1,k_2,k_3)={\rm g.c.d.}(n_a,n_b,n_c)$. 
Then the parameter of $K$ determines the allowed couplings of Yukawa
interactions.
If $K=1$, all of possible combinations $(p,q,r)$ appear 
in the above trace (\ref{eq:y-ijk-1}).
However, if 
$K \ne 1$, only restricted combinations of  $(p,q,r)$ appear 
in Eq.~(\ref{eq:y-ijk-1}), but not all combinations.
That is, the couplings are restricted by the $Z_K$ symmetry.
Indeed, allowed combinations of $(p,q,r)$ are 
controlled by the gauge invariance 
before the gauge symmetry breaking.
This $Z_K$ symmetry is unbroken symmetry in the 
original gauge symmetry.

Now, let us consider the following summation of wavefunction 
products, 
\begin{eqnarray}
I_{pqr}^{jkl} &=&
\Psi^j_{pq} \Psi^k_{qr} (\Psi^l_{pr})^* +\Psi^j_{p+1,q+1}
\Psi^k_{q+1,r+1} (\Psi^l_{p+1,r+1})^* 
\nonumber \\
&& 
+\cdots+ 
\Psi^j_{p+Q-1,q+Q-1} \Psi^k_{q+Q-1,r+Q-1} (\Psi^l_{p+Q-1,r+Q-1})^*,
\nonumber
\end{eqnarray}
where $Q={\rm l.c.m.}(Q_1,Q_2,Q_3)$.
One can represent $Q$ as $Q=Q_1 q_1=Q_2q_2=Q_3 q_3$.
To compute the integral it is useful to represent the wavefunctions
as follows
\begin{eqnarray}
\tilde{\Psi}^{j',M_1'}(y_4,y_5)_{pq}
&=&
C_{pq}^{j'} 
e^{-\pi \frac{M_1'}{Q}y_5^2}
\jtheta{\frac{j'}{M_1'} \\ 0}
\left( \frac{M_1'}{Q }z+ 
\left(\frac{m_a}{n_a}p-\frac{m_b}{n_b}q \right),\frac{M_1'}{Q} \tau
\right),
\nonumber \\
\tilde{\Psi}^{k',M_2'}(y_4,y_5)_{qr}
&=&
C_{qr}^{k'} 
e^{-\pi \frac{M_2'}{Q}y_5^2}
\jtheta{\frac{k'}{M_2'} \\ 0}
\left( \frac{M_2'}{Q} z+ 
\left(\frac{m_b}{n_b}q-\frac{m_c}{n_c}r \right),\frac{M_2'}{Q} \tau
\right),
\\
\tilde{\Psi}^{l',M_3'}(y_4,y_5)_{pr}
&=&
C_{pr}^{l'} 
e^{-\pi \frac{M_3'}{Q}y_5^2}
\jtheta{\frac{l'}{M_3'} \\ 0}
\left( \frac{M_3'}{Q} z+ 
\left(\frac{m_a}{n_a}p-\frac{m_c}{n_c}r \right),\frac{M_3'}{Q} \tau
\right),
\nonumber 
\end{eqnarray}
where $j'=q_1 j$ $k'=q_2 k$, $l'=q_3 l$ and $M_i'=q_i M_i$,
($i=1,2,3$).
Here the relation $M_1'+M_2'=M_3'$ holds. 
By using the production property of the theta function,
the product of $\Psi^{j,M_1}\Psi^{k,M_2}$ is represented by the sum of
the theta functions as
\begin{eqnarray}
\tilde{\Psi}^{j,M_1}_{pq}\tilde{\Psi}^{k,M_2}_{qr}
=
C_{pq}^{j'} C_{qr}^{k'} 
e^{\pi \frac{M_3'}{Q}y_5^2}
\sum_{m\in Z_{M_3'}} 
\jtheta{\frac{j'+k'+M_1'm}{M_3'}}
\left(\frac{M_3'}{Q} z+ 
\left(\frac{m_a}{n_a}p-\frac{m_c}{n_c}r \right),\frac{M_3'}{Q} \tau
\right) 
\nonumber \\
\times 
\jtheta{\frac{M_2'j'-M_1'k'+M_1'M_2'm}{M_1'M_2'M_3'}}
\left( \frac{m_a}{n_a}M_2'p-\frac{m_b}{n_b}M_2'q
      -\frac{m_b}{n_b}M_1'q+\frac{m_c}{n_c}M_1'r ,
\frac{M_1'M_2'M_3'}{Q} \tau \right).
\end{eqnarray}

Here one can use the properties of boundary conditions for non-Abelian
Wilson lines. Using the property of
$\Psi_{p,q}(y_4+1,y_5)=\Psi_{p+1,q+1}(y_4,y_5)$ 
the overlap integral reduces to the following integral 
\begin{eqnarray}
\int_0^1 dy_4 \int_0^1 dy_5 I^{ijk}_{pqr} =
\int_0^Q dy_4 \int_0^1 dy_5 \Psi^i_{pq} \Psi^j_{qr} (\Psi^k_{rp})^*
\end{eqnarray}
where $Q$ is again defined by $Q={\rm l.c.m.}(Q_1,Q_2,Q_3)$.
Therefore we can obtain the analytic form of Yukawa couplings
and similar flavor structures to the case with Abelian Wilson lines.   
By using the orthogonal condition for the matrix valued wavefunctions 
as
\begin{eqnarray}
\int_0^Q dy_4 \int_0^1 dy_5 \Psi^{j,M_1}_{pq}
(\Psi^{k,M_1}_{pq})^\dagger
=\delta_{j,k},
\end{eqnarray}
one can lead the following form of Yukawa couplings
\begin{eqnarray}
\int_0^Q dy_4 \int_0^1 dy_5 \Psi^i_{pq} \Psi^j_{qr} (\Psi^k_{rp})^*
= 
N_{M_1}N_{M_2}N_{M_3}^* C_{pq}^j C_{qr}^k (C_{pr}^l)^*
Q\sqrt{\frac{Q}{2M_3'}}
\sum_{m\in Z_{M_3'}}\delta_{j'+k'+M_1'm, ~l' (\rm{mod} {M_3'})}
\nonumber \\
\times
\jtheta{\frac{M_2'j'-M_1'k'+M_1'M_2'm}{M_1'M_2'(M_3')} \\ 0}
\left(  Q\left(
\frac{m_a}{n_a}\tilde{I}_{bc}p + \frac{m_b}{n_b}\tilde{I}_{ca}q 
+ \frac{m_c}{n_c}\tilde{I}_{ab}r  \right)  ,\frac{M_1'M_2'M_3'}{Q}
\tau \right).
\end{eqnarray}
Here, the Kronecker delta $\delta_{j'+k'+M_1'm, ~l' (\rm{mod}
{M_3'})}$
leads to the  coupling selection rule 
\begin{eqnarray}
j'+k'+M_1'm=l'\mod{M_3'},
\end{eqnarray}
where $m=0,1,...,M_3'-1$.
When $g={\rm g.c.d.}(M_1',M_2',M_3')={\rm g.c.d.}(M_1,M_2,M_3)$, 
the coupling selection rule is given by 
\begin{eqnarray}\label{eq:Zg}
j'+k'=l' \mod{g}.
\end{eqnarray}
That means that we can assign $Z_g$ charges to 
all of zero-modes.\footnote{See
  Refs.~\cite{Cremades:2003qj,Higaki:2005ie} 
for a similar selection rule in intersecting D-brane models.}

Here we study again the $Z_K$ symmetry, which we showed.
The total number of multiplicity of $\Psi_{ab}$ is nothing but
$|I_{ab}|$, 
and it is represented by two parameters of $k_{ab}$ and $M_{ab}$ as 
 $I_{ab}=k_{ab}M_{ab}$.
If $K={\rm g.c.d.}(k_{ab},k_{bc},k_{ca}) \ne 1$, 
they are divided to $K$ types of zero-modes and 
distinguished by labeling the component of each matrix.
We introduce such a kind of flavor indices as $\tilde{j}, \tilde{k}$
and $\tilde{l}$ for $ab$-, $bc$-, $ca$-sectors, respectively.
We define the relation between the flavor labeled by $\tilde{j}$ 
and the component of matrix $p,q$ as $\tilde{j}=p-q \mod{k_1}$.
Similarly the other sectors are also defined as
$\tilde{k}=q-r \mod{k_2}$  and $\tilde{l}=p-r \mod{k_3}$. 
Since the allowed couplings must be gauge invariance, 
there is the coupling selection rule for this kind of flavor indices, 
which is given by 
\begin{eqnarray}\label{eq:ZK}
\tilde{j}+\tilde{k}=\tilde{l} \mod{K}.
\end{eqnarray}
This is because the Yukawa couplings are 
restricted in the trace of the matrix.
Therefore we find two types of coupling selection rules, 
i.e. the $Z_g$ and $Z_K$ symmetries.

We can extend the computation of 3-point couplings to 
higher order couplings.
For example, we show the computation of 4-point
couplings.
We assume that $\tilde{I}_{ab}$, $\tilde{I}_{bc}$, $\tilde{I}_{cd}>0$
and $\tilde{I}_{da}<0$.
Four zero-mode wavefunctions are written as 
\begin{eqnarray}
\psi^{j,M_1}_{pq} &=&
C_{pq}e^{-\pi \frac{M_1'}{Q}y^2_5}
\jtheta{ j'/M_1' \\ 0 }\left( \frac{M_1'}{Q} z +(\frac{m_a}{n_a}p-
\frac{m_b}{n_b}q ), \frac{M_1'}{Q} \tau\right),
\nonumber \\
\psi^{k,M_2}_{qr} &=&
C_{qr}e^{-\pi \frac{M_2'}{Q}y^2_5}
\jtheta{ k'/M_2' \\ 0 }\left(\frac{ M_2'}{Q} z +(\frac{m_b}{n_b}q-
\frac{m_c}{n_c}r ), \frac{M_2'}{Q} \tau\right),
\nonumber \\
\psi^{l,M_3}_{rs} &=&
C_{rs}e^{-\pi \frac{M_3'}{Q}y^2_5}
\jtheta{ l'/M_3' \\ 0 }\left( \frac{M_3'}{Q} z +(\frac{m_c}{n_c}r-
\frac{m_d}{n_d}s ), \frac{M_3'}{Q} \tau\right),
\nonumber \\
\psi^{t,M_4}_{ps} &=&
C_{ps}e^{-\pi \frac{M_4'}{Q}y^2_5}
\jtheta{ t'/M_4' \\ 0 }\left( \frac{M_4'}{Q} z +(\frac{m_a}{n_a}p-
\frac{m_d}{n_b}s ), \frac{M_4'}{Q} \tau\right),
\nonumber 
\end{eqnarray}
where $Q$ is defined as $Q={\rm l.c.m.}(n_a,n_b,n_c,n_d)$.
First, the product of $\psi^{j,M_1}_{pq}$ and $\psi^{k,M_2}_{qr} $
becomes
\begin{eqnarray}
\psi^{j,M_1}_{pq} \psi^{k,M_2}_{qr} 
=
C_{pq}C_{qr}
e^{-\pi \frac{M'}{Q}y^2_5}
\sum_{m\in Z_{M'}} \jtheta{ \frac{j'+k'+M_1'm}{M'} \\ 0 }
\left( \frac{M'}{Q} z +(\frac{m_a}{n_a}p-\frac{m_c}{n_c}r ), 
\frac{M'}{Q} \tau\right) \nonumber \\
\ \ 
\times
\jtheta{\frac{M_2'j'-M_1'k'+M_1'M_2'm}{M_1'M_2'M'}}
\left(
M_2'(\frac{m_a}{n_a} p-\frac{m_b}{n_b} q)-
M_1'(\frac{m_b}{n_b}q-\frac{m_c}{n_c}r), \frac{M_1' M_2' M'}{Q} \tau
\right)
\end{eqnarray}
where $M'=M_1'+M_2'$.
Then we repeat this product for $\psi^{l,M_3}_{rs}$ and 
use the orthogonal condition for the $M_4'$ sector 
because $M_1'+M_2'+M_3'=M'+M_3'=M_4'$ hold by definition.
Finally we obtain the overlap integral for four
wavefunctions as
\begin{eqnarray}
&&
Y^{jklt}_{pqrs}
=
C^j_{pq}C^k_{qr}C^l_{rs}(C^t_{ps})^*
Q\sqrt{\frac{M_4'}{Q}}
\sum_{m\in Z_{M'}} \sum_{n\in Z_{M_4'}}
\delta_{j'+k'+M_1'm+l'+M'n, t'(\mod{M_4'})} \nonumber \\
&& 
\ \  \times
\jtheta{\frac{M_2'j'-M_1'k'+M_1'M_2'm}{M_1'M_2'M'}}
\left(
M_2'(\frac{m_a}{n_a} p-\frac{m_b}{n_b} q)-
M_1'(\frac{m_b}{n_b}q-\frac{m_c}{n_c}r), \frac{M_1' M_2' M'}{Q} \tau
\right)  \\
&& 
\ \ \times
\jtheta{\frac{M_3'(j'+k'+M_1'm)-M'l'+M'M_4'n}{M_3'M_4'M'}}
\left(
M_3'(\frac{m_a}{n_a} p-\frac{m_c}{n_c} r)-
M'(\frac{m_c}{n_c}r-\frac{m_d}{n_d} s), \frac{M' M_3' M_4'}{Q} \tau
\right) .      \nonumber
\end{eqnarray}
This result is just the product of two theta functions.
By solving the Kronecker delta, 
we obtain the sum of two theta functions like $\sum_m
y^{j'k'm}y^{l't'm'}$.
Therefore even including the non-Abelian Wilson lines we obtain 
results which are similar to Ref.~\cite{Abe:2009dr} 
for general four point couplings.

\subsection{Comments on soft supersymmetry breaking terms and
moduli stabilization}
In this section we discuss about the relation between moduli
parameters and low-energy supersymmetry breaking effects. If these
low-energy physics describe our world,  the supersymmetry must 
be broken softly. In the MSSM or its extension, 
supersymmetry breaking is parameterized by a set of soft supersymmetry
breaking terms. 
However the MSSM can not tell the microscopic origin of the 
soft supersymmetry breaking terms. They are generally free parameters 
and it needs some new physics mechanism for the supersymmetry breaking 
from the underlying theory such as string theory constructions.
A method to obtain the soft supersymmetry breaking terms of the
MSSM is to calculate the couplings of the matter sectors in the MSSM
and moduli fields. The spontaneous supersymmetry breaking can be 
induced by the non-vanishing $F$ and $D$ terms of some moduli fields. 
The super potential are already given in Eq.\ref{eq:W} and 
the F-term contribution of the tree-level scalar potential is given by 
\begin{align}
V_F(\mathcal{M},\mathcal{\bar{M}})
= e^G(G_M K^{MN}G_N-3),
\end{align}
then supersymmetry is broken if some of them have non-zero VEVs 
which are the SM gauge singlet scalar fields like as dilaton, 
or some geometric moduli as 
Kahler moduli and complex structure moduli.
We have seen that 
dimensional reduction scheme can give a key observation about 
such moduli field dependence about matter sectors as well as 
low-energy phenomena like chiral spectrums or generation number.
The soft supersymmetry breaking terms can be triggered by the 
spurious field methods \cite{Brignole:1997dp} as 
\begin{align}\label{eq:soft}
m_a &= \frac{1}{2\rm{Re}f_a}(F^M \partial_M f_a) \nonumber \\
m^2_{\alpha \beta} &= (m^2_{3/2}+ V_F ) 
-\bar{F}^{\bar{M}} F^N \partial_{\bar{M}}\partial_N \log{(K_{\alpha\beta})},
\\
A_{\alpha\beta\gamma} &= F^M (K_M+\partial_M \log{(Y_{\alpha\beta\gamma})}
-\partial \log{(K_{\alpha\beta} K_{\beta\gamma}K_{\gamma\alpha})} ) \nonumber 
\end{align}
where $m_a$, $m^2_{\alpha\beta}$ and $A_{\alpha\beta\gamma}$  are 
corresponding to the soft supersymmetry breaking terms for gaugino masses and scalar
masses and A-terms. 
Therefore these scenario enable to carry out the model independent
analysis for the low-energy physics.

We have some comment on the moduli stabilization and low-energy spectra.
First of all, these formulae for the soft terms are depending on the Yukawa
couplings. 
For instance as shown in section\ref{sec:Intersecting}, 
Yukawa coupling in magnetized D-brane side is represented by 
\begin{align}
Y_{IIB}=&
 2^{-9/4} \pi^{-3} e^{\phi_4/2} \prod_{d=1}^3 \left(\Im \tau_d
 {M_1^{(d)}  M_2^{(d)} \over M_3^{(d)}}  \right)^{1/4}
\nonumber \\
& \times \vartheta \left[ \begin{matrix}
    {M_2^{(d)}k - M_3^{(d)}j + M_2^{(d)} M_3^{(d)} l \over M_1^{(d)}
      M_2^{(d)} M_3^{(d)} } \\ 0 \end{matrix} 
  \right](0,\tau_d M_1^{(d)} M_2^{(d)} M_3^{(d)}). 
\end{align}
Therefore these supersymmetry breaking terms may affect on
the low-energy phenomena. 
The flavor dependent part only come from the theta functions. 
This structure has dependence of the parameters of complex structure
moduli, on the contrary, in the type IIA sides, flavor dependence is a
function of Kahler moduli.
As well known, 
there are some experimental constraint on the soft
supersymmetry breaking terms. The crucial constraint is the limitation 
of the flavor changing neutral currents (FCNC) which suggests the universal
squark mass for all generations. 
The simplest way to avoid the dangerous soft breaking terms is 
the scenario with dilaton moduli dominated scenario where 
it is assumed that the F-terms contributions of the moduli fields are
dominated by dilaton moduli $F^s$. 
Then the soft supersymmetry breaking terms are universal for all the
flavors and that is nicely acceptable for the experimental constraint. 
Furthermore from the expression of the Yukawa couplings in the scenario,
the flavor dependence of the physical Yukawa couplings can only  
appear as the parameters of the complex structure moduli.
Therefore the scenario with the dilaton and Kahler moduli dominant may
not affect the low-energy spectrum.
For generic case, we analyze carefully the soft supersymmetry breaking terms
mediated by those dilaton, Kahler and complex structure moduli by
using the formula in Eq.~(\ref{eq:soft}).

In order to specify the scenario to be selected 
it is necessary 
to study the moduli stabilization mechanism, 
because these F-term as $F^U$, $F^S$ and $F^K$ are usually
proportional to its vacuum expectation values.
There are several ways for stabilization mechanism in the string theory.
The dilaton and moduli stabilization mechanism using the three form
flux are very well studied in which these moduli are stabilized at the 
Planck scale VEVs of the background fluxes. 
The KKLT scenario can provide the novel way to stabilize the
overall Kahler moduli by non-perturbative super potential. 
The gauge flux can also stabilize some of Kahler moduli by F flatness
conditions like in Eq.\ref{eq:SUSY-condition-0}.
If we use the oblique flux for gauge, its generalized supersymmetry
conditions are obtained. This type of model constructions are 
explored in globally defined toroidal
compactifications
~\cite{Bianchi:2005yz,Antoniadis:2004pp}
with stabilized geometric moduli in a supersymmetric vacuum 
within a perturbative string description.
Combining the three form flux and magnetic flux
may stabilize all the geometric moduli. 
These scenario would give the moduli VEVs of the same
magnitude of the scale. They may occur some unwanted FCNC
process by induced soft supersymmetry breaking terms. 
However once we obtain the realistic patterns of Yukawa couplings, 
the characteristic patterns of the sparticle spectrum may be predicted.
Therefore analysis for the relations between low-energy spectrum and
moduli breaking parameters is important issue.

\newpage

\section{Non-Abelian flavor symmetries}

Here we study more presicely the flavor structures
by using the analysis on the coupling selection rule 
in the previous section.

\subsection{Generic case}\label{sec:generic}

First we study generic case with non-vanishing Wilson lines.
For simplicity, we restrict on the case with trivial 
torus background (integer flux). 
The case with fractional flux will be discussed later.
We consider the model with zero-modes $\psi^{i_k,M_k}$ 
for $k=1,\cdots, L$. 
We denote $\gcd (M_1,\cdots,M_L) =g$.
As studied in the previous section, 
these modes have $Z_g$ charges and their 
couplings are controlled by the $Z_g$ invariance.
For simplicity, suppose that $M_1=g$.
Then, there are $g$ zero-modes of  $\psi^{i_1,M_1}$.
The above $Z_g$ transformation acts on $\psi^{i_1,g}$ as 
$Z \psi^{i_1,g}$, where
\begin{eqnarray}
Z = \left(
\begin{array}{ccccc}
1 & & & & \\
  & \rho & & & \\
  & & \rho^2 & & \\
  & &   & \ddots & \\
  &  &  &    & \rho^{g-1} 
\end{array}
\right),
\end{eqnarray}
and $\rho = e^{2\pi i /g}$.

In addition to this $Z_g$ symmetry, 
the effective theory has another symmetry.
That is, the effective theory must be 
invariant under cyclic permutations 
\begin{equation}\label{eq:permutation}
\psi^{i_1,g} \rightarrow \psi^{i_1+n,g} ,
\end{equation}
with a universal integer $n$ for $i_1$.
That is nothing but a change of ordering and also 
has a geometrical meaning as a discrete shift 
of the origin, $z=0 \rightarrow z=-\frac{n}{g}$.
This symmetry also generates another $Z_g$ symmetry, 
which we denote by $Z_g^{(C)}$ and its generator is 
represented as 
\begin{eqnarray}\label{eq:C}
C = \left(
\begin{array}{cccccc}
0 & 1& 0 & 0 & \cdots & 0 \\
0  & 0 &1 & 0 & \cdots & 0\\
  &    &  & &\ddots & \\
1  &  0  & 0 &  & \cdots   & 0 
\end{array}
\right),
\end{eqnarray}
on $\psi^{i_1,g}$.
That is, the above permutation (\ref{eq:C}) 
is represented as $C^n \psi^{i_1,g}$.
These generators, $Z$ and $C$, do not commute each other,
i.e., 
\begin{equation}
CZ = \rho ZC .
\end{equation}
Then, the flavor symmetry corresponds to the closed algebra 
including $Z$ and $C$.
Diagonal matrices in this closed algebra are written as 
$Z^n(Z')^m$, 
where $Z'$ is the generator of another $Z'_g$ and 
written as 
\begin{eqnarray}
Z' = \left(
\begin{array}{ccc}
\rho & &  \\
    & \ddots & \\
    &    & \rho 
\end{array}
\right),
\end{eqnarray}
on $\psi^{i_1,g}$.
Hence, these would generate the non-abelian flavor symmetry 
$(Z_g \times Z'_g)\rtimes Z_g^{(C)}$, since 
$Z_g \times Z'_g$ is a normal subgroup.
These discrete flavor groups would include $g^3$ elements totally.

Let us study actions of $Z$ and $C$ on 
other zero-modes, $\psi^{i_k,M_k}$, with $M_k=gn_k$, 
where $n_k$ is an integer.
First, the generator $C$ acts as 
\begin{equation}
\psi^{i,gn_k} \rightarrow \psi^{i+n_k,gn_k},
\end{equation}
because the above discrete shift of the origin 
$z=0 \rightarrow z=-\frac{n}{g}$ can be written as 
$z=0 \rightarrow z=-\frac{nn_k}{gn_k}$
for these zero-modes.
Thus, the generator $C$ is represented as the same as 
(\ref{eq:C}) on the basis
\begin{eqnarray}\label{eq:g-plet}
\left(
\begin{array}{c}
\psi^{p,gn_k}  \\ 
\psi^{p+n_k,gn_k}  \\ 
\vdots  \\
\psi^{p+(g-1)n_k,gn_k} 
\end{array}
\right),
\end{eqnarray}
where $p$ is an integer.
Note that $\psi^{p+gn_k,gn_k}$ is identical to $\psi^{p,gn_k}$.
Furthermore, the generator $Z$ is represented on this basis 
(\ref{eq:g-plet}) as 
\begin{eqnarray}
Z = \rho^p \left(
\begin{array}{ccccc}
1 & & & & \\
  & \rho^{n_k} & & & \\
   & & \rho^{2n_k} & & \\
  &    & & \ddots & \\
  &    & &   & \rho^{(g-1)n_k} 
\end{array}
\right).
\end{eqnarray}
Thus, the zero-modes $\psi^{i_k,gn_k}$ include $n_k$ $g$-plet 
representations of the symmetry $(Z_g \times Z'_g) \rtimes Z_g^{(C)}$ 
and some of them may be reducible $g$-plet 
representations.
For example, when we consider 
the zero-modes corresponding to $n_k=g$, i.e. $M_k=g^2$, 
the generator $Z$ is represented as 
$\rho^p {\1}_g$ on the above $g$-plet (\ref{eq:g-plet}), 
where ${\1}_g$ is the $(g \times g)$ unit matrix.
In such a case, the generator $C$ can also be diagonalized.
Then, these zero-modes correspond to $g$ singlets of 
$(Z_g \times Z'_g) \rtimes Z_g^{(C)}$ 
including trivial and non-trivial singlets.

As illustrating examples, we consider the models with 
$g=2,3$ in the next subsections and study 
more concretely about non-abelian discrete flavor symmetries.

\subsubsection{$g=2$ case}

Here we consider the model with $g=2$, that is, 
all of relative magnetic fluxes $M_k$ are even.
Its flavor symmetry is given as the closed algebra of 
$Z_2$, $Z'_2$ and $Z_2^{(C)}$, and all of these elements are 
written as 
\begin{eqnarray}
\pm \left(
\begin{array}{cc}
1 & 0 \\
0 & 1 \\
\end{array}
\right), \quad 
\pm \left(
\begin{array}{cc}
0 & 1 \\
1 & 0 \\
\end{array}
\right), \quad 
\pm \left(
\begin{array}{cc}
0 & 1 \\  -1 & 0 \\
\end{array}
\right), \quad 
\pm \left(
\begin{array}{cc}
1 & 0 \\
0 & -1 \\
\end{array}
\right).
\end{eqnarray}
That is,  the flavor symmetry is $D_4$.
The zero-modes with the relative magnetic flux 
$M=2$,
\begin{eqnarray}
\left(
\begin{array}{c}
\psi^{0,2} \\
\psi^{1,2}
\end{array}
\right),
\end{eqnarray}
correspond to the doublet representation ${\bf 2}$ of $D_4$.
This result is the same as the non-abelian flavor symmetry 
appearing from heterotic orbifold models with $S^1/Z_2$, 
where twisted modes on two fixed points of $S^1/Z_2$ 
correspond to the $D_4$
doublet~\cite{Kobayashi:2004ya,Kobayashi:2006wq}.

Next, we consider the zero-modes corresponding to 
the relative magnetic flux $M=4$, $\psi^{i,4}$ ($0=0,1,2,3$).
As discussed in the previous subsection, 
in order to represent $C$, 
it may be convenient to decompose them into the $g$-plets 
(\ref{eq:g-plet})
\begin{eqnarray}
\left(
\begin{array}{c}
\psi^{0,4} \\
\psi^{2,4}
\end{array}
\right), \qquad 
\left(
\begin{array}{c}
\psi^{1,4} \\
\psi^{3,4}
\end{array}
\right).
\end{eqnarray}
However, they are reducible representations as follows.
Note that both $\psi^{0,4}$ and $\psi^{2,4}$ have 
even $Z_2$ charges, and that both $\psi^{1,4}$ and $\psi^{3,4}$ have 
odd $Z_2$ charges.
That is, the generator $Z$ is represented in the form $\pm {\bf 1}_2$,
where 
${\bf 1}_2$ is the $2 \times 2$ identity matrix.
Thus, the generator $C$ can be diagonalized and such a diagonalizing 
basis is obtained as
\begin{eqnarray}\label{eq:D4-singlets}
&&{\bf 1}_{++}:\ (\psi^{0,4} + \psi^{2,4}), \quad \
{\bf 1}_{+-}:\ (\psi^{0,4} - \psi^{2,4}), \quad \ \nonumber \\
&&{\bf 1}_{-+}:\ (\psi^{1,4} + \psi^{3,4}), \quad \
{\bf 1}_{--}:\ (\psi^{1,4} - \psi^{3,4}), \quad \
\end{eqnarray}
up to normalization factors.
Obviously, these correspond to four $D_4$ singlets, 
${\bf 1}_{++}$, ${\bf 1}_{+-}$, ${\bf 1}_{-+}$ 
and ${\bf 1}_{--}$.
The first subscript of two denotes $Z_2$ charges for $Z$ 
and the second one denotes $Z_2$ charges for $C$.
Hence, all of irreducible representations of $D_4$ appear from 
$\psi^{i,2}$ and $\psi^{i,4}$.
New representations can not appear in zero-modes 
$\psi^{i,M}$ with $M >4$.
For example,  we consider zero-modes corresponding to $M=6$, 
i.e. $\psi^{i,6}$.
They can be decomposed as 
\begin{equation}\label{eq:g2-6}
|\psi^{6}\rangle_1=
\left(
\begin{array}{c}
\psi^{0,6} \\ \psi^{3,6} 
\end{array} \right), \quad
|\psi^{6}\rangle_2=
\left(
\begin{array}{c}
\psi^{2,6} \\ \psi^{5,6} 
\end{array} \right), \quad
|\psi^{6}\rangle_3=
\left(
\begin{array}{c}
\psi^{4,6} \\ \psi^{1,6} 
\end{array} \right). \
\end{equation}
Each of $|\psi^{6}\rangle_i$ with $i=1,2,3$ is nothing but 
the $D_4$ doublet.
That is, we have three $D_4$ doublets in $\psi^{i,6}$.
The above representations appear repeatedly 
in $\psi^{i,M}$ with larger $M$.
These results are shown in Table~\ref{tab:g2}.

\begin{table}[t]
\begin{center}
\begin{tabular}{|c|c|} \hline
$M$ & Representation of $D_4$ \\ \hline \hline
2 & ${\bf 2}$ \\ 
4 & ${\bf 1}_{++}, \ {\bf 1}_{+-}, \ {\bf 1}_{-+}, \ {\bf 1}_{--}$ \\ 
6 & $3 \times {\bf 2}$\\  \hline
\end{tabular}
\end{center}
\caption{$D_4$ representations of zero-modes in the model with $g=2$.}
\label{tab:g2}
\end{table}

\subsubsection{$g=3$ case}

Here we consider the model with $g=3$, where 
all of relative magnetic fluxes are equal to $M_k=3n_k$.
Its flavor symmetry is given as $(Z_3 \times Z_3) \rtimes Z_3$, that
is, 
$\Delta(27)$~\cite{Branco:1983tn}.
This flavor symmetry is different from the flavor symmetry 
appearing from heterotic orbifold models with $T^2/Z_3$.
Later, we will explain what makes this difference.

The zero-modes corresponding to the relative magnetic flux $M=3$,
\begin{eqnarray}\label{eq:g3-3}
|\psi^{3}\rangle_1=\left(
\begin{array}{c}
\psi^{0,3} \\
\psi^{1,3}  \\
\psi^{2,3} 
\end{array}
\right),
\end{eqnarray}
correspond to the triplet representation ${\bf 3}$ of $\Delta(27)$.
Next, we consider the zero-modes corresponding to 
the relative magnetic flux $M=6$, i.e. $\psi^{i,6}$.
Again, it may be convenient to decompose 
them into the $g$-plets (\ref{eq:g-plet})
\begin{equation}\label{eq:g3-6}
|\psi^{6}\rangle_1=
\left(
\begin{array}{c}
\psi^{0,6} \\ \psi^{2,6} \\ \psi^{4,6} 
\end{array} \right), \quad
|\psi^{6}\rangle_2=
\left(
\begin{array}{c}
\psi^{3,6} \\ \psi^{5,6} \\ \psi^{1,6} 
\end{array} \right).
\end{equation} 
The generator $C$ is represented in the same way for 
$|\psi^{3}\rangle_1$ and $|\psi^{6}\rangle_i$ ($i=1,2$).
On the other hand, the representation of the generator $Z$ 
for $|\psi^{6}\rangle_i$ ($i=1,2$) is the complex conjugate 
to one for $|\psi^{3}\rangle_1$.
Thus, both $|\psi^{6}\rangle_i$  ($i=1,2$) correspond to 
$\bar {\bf 3}$ representations of $\Delta(27)$.

Moreover, let us consider the zero-modes with the relative 
magnetic flux $M=9$, i.e. $\psi^{i,9}$.
Then, we decompose 
them into the $g$-plets (\ref{eq:g-plet})
\begin{equation}
|\psi^{9}\rangle_1=
\left(
\begin{array}{c}
\psi^{0,9} \\ \psi^{3,9} \\ \psi^{6,9} 
\end{array} \right), \quad
|\psi^{9}\rangle_\omega=
\left(
\begin{array}{c}
\psi^{1,9} \\ \psi^{4,9} \\ \psi^{7,9} 
\end{array} \right), \quad
|\psi^{9}\rangle_{\omega^2}=
\left(
\begin{array}{c}
\psi^{2,9} \\ \psi^{5,9} \\ \psi^{8,9} 
\end{array} \right),
\end{equation} 
where $\omega = e^{2\pi i/3}$.
These (reducible) triplets $ |\psi^{9}\rangle_{\omega^n}$ have 
$Z_3$ charges, $n$ and are decomposed into 
nine singlets,
\begin{equation}\label{eq:g3-9singlets}
{\bf 1}_{\omega^n,\omega^m}:\psi^{n,9}+\omega^m\psi^{n+3m,9}
+\omega^{2m} \psi^{n+6m,9},
\end{equation}
up to normalization factors, where 
$n$ and $m$ are $Z_3$ charges for $Z$ and $C$, respectively.
In zero-modes with $M >9$, new representations do not appear, 
but the above representations appear repeatedly.
These results as well as zero-modes with $M >9$ 
are shown in Table~\ref{tab:g3}.
Similar analysis can be carried out in other models 
with $g > 3$.

\begin{table}[t]
\begin{center}
\begin{tabular}{|c|c|} \hline
$M$ & Representation of $\Delta(27)$ \\ \hline \hline
3 & ${\bf 3}$ \\ 
6 & $2 \times {\bar {\bf 3}}$  \\ 
9 & ${\bf 1}_{1}, \ {\bf 1}_{2}, \ {\bf 1}_{3}, \ {\bf 1}_{4}, 
\ {\bf 1}_{5}, \ {\bf 1}_{6}, \ {\bf 1}_{7}, \ {\bf 1}_{8}, 
\ {\bf 1}_{9}$ \\ 
12 & $4 \times {\bf 3}$\\  
15 & $5 \times {\bar {\bf 3}}$  \\ 
18 & $2 \times \{ {\bf 1}_{1}, \ {\bf 1}_{2}, \ {\bf 1}_{3}, \ {\bf
1}_{4}, 
\ {\bf 1}_{5}, \ {\bf 1}_{6}, \ {\bf 1}_{7}, \ {\bf 1}_{8}, 
\ {\bf 1}_{9} \}$ \\ \hline
\end{tabular}
\end{center}
\caption{$\Delta(27)$ representations of zero-modes in the model with
$g=3$.}
\label{tab:g3}
\end{table}

We comment on symmetries in subsectors.
Suppose that our model has zero-modes $\psi^{i_k,M_k}$ for 
$k = 1,\cdots, L$ with $\gcd(M_1,\cdots,M_L)=g$ and 
they are separated into two classes, 
$\psi^{i_l,M_l}$ $(l=1,\cdots,L_1)$ and 
$\psi^{i_m,M_m}$ $(m=L_1,\cdots,L)$, 
where $\gcd(M_1,\cdots,M_{L_1})=g_1$,  
$\gcd(M_{L_1},\cdots,M_{L})=g_2$ and 
$\gcd (g_1,g_2)=g$.
Coupling terms  including only the first class of 
fields $\psi^{i_l,M_l}$ $(l=1,\cdots,L_1)$  
in the four-dimensional effective theory have the 
symmetry $(Z_{g_1} \times Z_{g_1})\rtimes Z_{g_1}$, where 
$g_1$ would be larger than $g$.
However, such a symmetry is broken 
by terms including the second class of fields.
Thus, we would have a larger symmetry 
at least  at tree level for the subsectors.
Such larger symmetries in the subsectors would be interesting 
for model building.

\subsection{Cases without Continuous Wilson lines}

In the section \ref{sec:generic}, we have considered the 
models with non-vanishing Wilson lines.
Here, we study the models without Abelian Wilson lines.
In this case, flavor symmetries are enhanced.

When Wilson lines are vanishing, all of zero-modes
$\psi^{0,M_k}$ have the peak at the same point 
in the extra dimensions.
In the intersecting D-brane picture, 
this corresponds to the D-brane configuration, that 
all of D-branes intersect (at least) at a single point on $T^2$.
This model has the $Z_2$ rotation symmetry around such a point.
Here, we denote its generator as $P$.
In general, this acts as 
\begin{equation}
P: \ \psi^{i,M} \rightarrow \psi^{M-i,M}.
\end{equation}

As in the previous section, we consider the models with 
$g=2,3$ as illustrating models.

\subsubsection{$g=2$ case}

First, we consider the zero-modes with $M=2$, 
$\psi^{i,2}$, which correspond to the $D_4$ doublet.
For them, the generator $P$ acts as the identity.
That implies that the flavor symmetry is enhanced as 
$D_4 \times Z_2$ and $\psi^{i,2}$ correspond to 
${\bf 2}_+$, where the subscript denotes 
the $Z_2$ charge for $P$.\footnote{
Although this is just an enhancement by the factor $Z_2$, 
such an enhanced flavor symmetry $D_4 \times Z_2$ 
would be important to phenomenological model building.
See e.g.~\cite{Grimus}.}

We consider the zero-modes with $M=4$, 
$\psi^{i,4}$, which are decomposed as 
the four $D_4$ singlets, 
${\bf 1}_{++}$, ${\bf 1}_{+-}$, ${\bf 1}_{-+}$ 
and ${\bf 1}_{--}$
as (\ref{eq:D4-singlets}).
They have definite $Z_2$ charges for $P$ and 
are represented as 
\begin{eqnarray}\label{eq:D4-singlets-2}
&&{\bf 1}_{+++}:\ (\psi^{0,4} + \psi^{2,4}), \quad \
{\bf 1}_{+-+}:\ (\psi^{0,4} - \psi^{2,4}), \quad \ \nonumber \\
&&{\bf 1}_{-++}:\ (\psi^{1,4} + \psi^{3,4}), \quad \
{\bf 1}_{---}:\ (\psi^{1,4} - \psi^{3,4}), \quad \
\end{eqnarray}
where the third sign in the subscripts 
denotes $Z_2$ charges for $P$.

Now, let us consider the zero-modes with $M=6$, 
$\psi^{i,6}$, which are decomposed as 
three $D_4$ doublets (\ref{eq:g2-6}).
The doublet $|\psi^{6}\rangle_1$ has the even 
$Z_2$ charges for $P$.
However, other doublets $|\psi^{6}\rangle_2$ and 
$|\psi^{6}\rangle_3$ transform each other under $P$.
Thus, we take linear combinations of these two doublets as
\begin{eqnarray}
|\psi^{6}\rangle_\pm \equiv 
|\psi^{6}\rangle_2 \pm
|\psi^{6}\rangle_3 
=
\left(
\begin{array}{c}
\psi^{2,6} \\ \psi^{5,6} 
\end{array} \right) \pm
\left(
\begin{array}{c}
\psi^{4,6} \\ \psi^{1,6} 
\end{array} \right) ,
\end{eqnarray}
where $\pm$ also means $Z_2$ charge of $P$.
As a result, these zero-modes $\psi^{i,6}$ are 
decomposed as two ${\bf 2}_+$ and one ${\bf 2}_-$.

We can repeat these analysis for larger $M$.
For example, zero-modes with $M=8$, 
$\psi^{i,8}$, are decomposed as 
\begin{equation}
\{{\bf 1}_{+++},\ {\bf 1}_{+-+},\ 
           {\bf 1}_{+++},\ {\bf 1}_{+--},\
           {\bf 1}_{-++},\ {\bf 1}_{-+-},\ 
           {\bf 1}_{---},\ {\bf 1}_{--+}\} , 
\end{equation}
and zero-modes with $M=10$, 
$\psi^{i,10}$, are decomposed as 
three ${\bf 2}_+$ and two ${\bf 2}_-$.
These results are shown in Table~\ref{tab:g2-2}.

\begin{table}[t]
\begin{center}
\begin{tabular}{|c|c|} \hline
$M$ & Representation of $D_4\times Z_2$ \\ \hline \hline
2 & ${\bf 2}_+$ \\ 
4 & ${\bf 1}_{+++}, \ {\bf 1}_{+-+}, \ {\bf 1}_{-++}, \ {\bf 1}_{---}$
\\ 
6 & $2 \times {\bf 2}_+, \ {\bf 2}_-$\\  
8 & ${\bf 1}_{+++},\ {\bf 1}_{+-+},\ 
           {\bf 1}_{+++},\ {\bf 1}_{+--},\
           {\bf 1}_{-++},\ {\bf 1}_{-+-},\ 
           {\bf 1}_{---},\ {\bf 1}_{--+}$ \\ 
10 & $3 \times {\bf 2}_+, \ 2 \times {\bf 2}_-$\\  \hline
\end{tabular}
\end{center}
\caption{$D_4 \times Z_2$ representations of zero-modes in the model
with $g=2$.}
\label{tab:g2-2}
\end{table}

\subsubsection{$g=3$ case}

Here, we study the model with $g=3$.
First, we consider the zero-modes with $M=3$, $\psi^{i,3}$.
They correspond to a triplet of $\Delta(27)$ with non-vanishing 
Wilson lines.
At any rate, the generators, $Z$, $C$ and $P$, act on $\psi^{i,3}$ as 
\begin{eqnarray}\label{eq:delta-54}
Z=\left(
\begin{array}{ccc}
1 & 0 & 0 \\
0 & \omega & 0 \\
0 & 0 & \omega^2 
\end{array}
\right), \quad
C = \left(
\begin{array}{ccc}
0 & 1 & 0 \\
0 & 0 & 1 \\
1 & 0 & 0
\end{array}
\right), \quad
P = \left(
\begin{array}{ccc}
1 &  0 & 0 \\
0 & 0 & 1 \\
0 & 1 & 0
\end{array}
\right).
\end{eqnarray}
Their closed algebra is $\Delta(54)$.
Thus, the zero-modes $\psi^{i,3}$ correspond to 
the triplet of $\Delta(54)$, ${\bf 3}_1$.
This is the same as the flavor symmetry, which appears in
heterotic orbifold models with $T^2/Z_3$~\cite{Kobayashi:2006wq}.
Three fixed points on the orbifold $T^2/Z_3$ have the geometrical 
permutation symmetry $S_3$.
Such symmetry is enhanced in magnetized brane models, 
only when Wilson lines are vanishing.
Indeed, the closed algebra of generators $C$ and $P$ 
is $S_3$.

Similarly, we can consider the zero-modes with $M=6$, $\psi^{i,6}$.
We decompose them as (\ref{eq:g3-6}).
The generators, $C$ and $P$, act on 
$|\psi^{6}\rangle_i$ ($i=1,2$) in the same way as 
$\psi^{i,3}$, but the representation of the generator $Z$ 
for $|\psi^{6}\rangle_i$ ($i=1,2$) is the complex conjugate 
to one for $|\psi^{3}\rangle_1$.
Thus, both $|\psi^{6}\rangle_i$ correspond to 
$\bar {\bf 3}_1$ representations of $\Delta(54)$.
Recall that $|\psi^{6}\rangle_i$ are 
$\bar {\bf 3}$ representations of $\Delta(27)$.

Next, let us consider the zero-modes with $M=9$, $\psi^{i,9}$.
Recall that they correspond to nine singlets of $\Delta(27)$ as 
(\ref{eq:g3-9singlets}).
The following linear combination,
\begin{equation}
\psi^{0,9}+\psi^{3,9}+\psi^{6,9},
\end{equation}
is still a singlet under $\Delta(54)$, which is 
a trivial singlet ${\bf 1}_1$.
However, the others in linear combinations (\ref{eq:g3-9singlets}) 
transform each other under $P$.
Then, they correspond to four doublets of $\Delta(54)$,
\begin{eqnarray}
\begin{array}{cccc}
{\bf 2}_1 :&
\left(
\begin{array}{c}
\psi^{0,9}+\omega \psi^{3,9}+\omega^2 \psi^{6,9} \\
\psi^{0,9}+\omega^2 \psi^{3,9}+\omega \psi^{6,9}  
\end{array} 
\right), \quad &
{\bf 2}_2 :&
\left(
\begin{array}{c}
\psi^{1,9}+\psi^{4,9}+ \psi^{7,9} \\
\psi^{2,9}+\psi^{5,9}+ \psi^{8,9}  
\end{array} 
\right), \\
 & & & \\
{\bf 2}_3 :&
\left(
\begin{array}{c}
\psi^{1,9}+\omega \psi^{4,9}+\omega^2 \psi^{7,9} \\
\psi^{8,9}+\omega^2 \psi^{5,9}+\omega \psi^{2,9}  
\end{array} 
\right),  \quad &
{\bf 2}_4 :&
\left(
\begin{array}{c}
\psi^{1,9}+\omega^2 \psi^{4,9}+\omega \psi^{7,9} \\
\psi^{8,9}+\omega^2 \psi^{5,9}+\omega \psi^{2,9}  
\end{array} 
\right).
\end{array} 
\end{eqnarray}

Now, let us consider the zero-modes with $M=12$, $\psi^{i,12}$.
We decompose them into $g$-plets 
(\ref{eq:g-plet})
\begin{eqnarray}
 & & |\psi^{12} \rangle_1 =
\left(
\begin{array}{c}
\psi^{0,12} \\ \psi^{4,12} \\ \psi^{8,12} 
\end{array} \right), \qquad
|\psi^{12} \rangle_2 =
\left(
\begin{array}{c}
\psi^{6,12} \\ \psi^{10,12} \\ \psi^{2,12} 
\end{array} \right), \nonumber\\ 
 & & 
|\psi^{12} \rangle_3 =
\left(
\begin{array}{c}
\psi^{3,12} \\ \psi^{7,12} \\ 
\psi^{11,12}
\end{array} \right), \qquad
|\psi^{12} \rangle_4 =
\left(
\begin{array}{c}
\psi^{9,12} \\ \psi^{1,12} \\ 
\psi^{5,12} 
\end{array} \right). 
\end{eqnarray}
They correspond to four triplets of $\Delta(27)$.
Representations of the generators, $Z$, $C$ and $P$, 
on $|\psi^{12} \rangle_1$ and  $|\psi^{12} \rangle_2$ 
are the same as those on  $\psi^{i,3}$ like 
Eq.~(\ref{eq:delta-54}).
Thus, they correspond to ${\bf 3}_1$.
On the other hand, $|\psi^{12} \rangle_3$ and  $|\psi^{12} \rangle_4$ 
transform each other under $P$.
Hence, we take the following linear combinations,
\begin{eqnarray}
& &  |\psi^{12} \rangle_{\pm} =
\left(
\begin{array}{c}
\psi^{3,12}\pm \psi^{9,12} \\ \psi^{7,12}\pm \psi^{1,12} \\ 
\psi^{11,12}\pm \psi^{5,12} 
\end{array} \right).
\end{eqnarray}
Then, representations of $Z$, $C$ and $P$ on 
$|\psi^{12} \rangle_+$ are the same as (\ref{eq:delta-54}), 
and $|\psi^{12} \rangle_+$  corresponds to ${\bf 3}_1$.
On the other hand, representations of $Z$ and $C$ on 
$|\psi^{12} \rangle_-$ are the same as (\ref{eq:delta-54}), 
but the generator $P$ is represented on $|\psi^{12} \rangle_-$ as 
\begin{eqnarray}
P = \left(
\begin{array}{ccc}
 -1 &  0 & 0 \\
 0 & 0 & -1 \\
 0 & -1 & 0
\end{array}
\right).
\end{eqnarray}
That is, $|\psi^{12} \rangle_-$ corresponds to 
another triplet of $\Delta(54)$, i.e. ${\bf 3}_2$.
Furthermore, the zero-modes with $M=15$, $\psi^{i,15}$ 
correspond to 
\begin{eqnarray}
& & 3 \times \bar{{\bf 3}}_1,\quad 
           2 \times \bar{{\bf 3}}_2 ,
\end{eqnarray}
and the zero-modes with $M=18$, $\psi^{i,18}$ 
correspond to 
\begin{eqnarray}
& & 2 \times \{ {\bf 1}_1, \ {\bf 2}_1, \ {\bf 2}_2, \ {\bf 2}_3, \
{\bf 2}_4 \}  .
\end{eqnarray}
These results are shown in Table~\ref{tab:g3-2}.
Irreducible representations of $\Delta(54)$ are 
two triplets ${\bf 3}_1$, ${\bf 3}_2$, their conjugates 
${\bar {\bf 3}}_1$ ${\bar {\bf 3}}_2$, four doublets 
${\bf 2}_1$, ${\bf 2}_2$, ${\bf 2}_3$, ${\bf 2}_4$, 
trivial singlet ${\bf 1}$ and non-trivial singlet ${\bf 1}_2$.
All of them except the non-trivial singlet ${\bf 1}_2$ 
can appear in this model.

\begin{table}[t]
\begin{center}
\begin{tabular}{|c|c|} \hline
$M$ & Representation of $\Delta(54)$ \\ \hline \hline
3 & ${\bf 3}_1$ \\ 
6 & $2 \times {\bar {\bf 3}}_1$  \\ 
9 & ${\bf 1}_1, \ {\bf 2}_1, \ {\bf 2}_2, \ {\bf 2}_3, \ {\bf 2}_4$ \\ 
12 & $3 \times {\bf 3}_1, \ {\bf 3}_2$\\  
15 & $3 \times {\bar {\bf 3}}_1, \ 
2 \times {\bar {\bf 3}}_2$  \\ 
18 & $2 \times \{ {\bf 1}_1, \ {\bf 2}_1, \ {\bf 2}_2, \ {\bf 2}_3, \
{\bf 2}_4 \}$  \\
 \hline
\end{tabular}
\end{center}
\caption{$\Delta(54)$ representations of zero-modes in the model with
$g=3$.}
\label{tab:g3-2}
\end{table}

Similar analysis can be carried out in other models with $g>3$.
In generic case, the $Z$ and $P$ satisfy 
\begin{equation}
PZ = Z^{-1}P,
\end{equation} 
and the closed algebra of $C$ and $P$ is 
$D_g$.
Thus, the flavor symmetry, which is generated by 
$Z$, $C$ and $P$, would be written as $D_g \ltimes (Z_g \times Z_g)$.
Note that $S_3 \sim D_3$ and 
$\Delta(54)$ is $D_3 \ltimes (Z_3 \times Z_3)$.

\subsection{Cases with non-Abelian Wilson lines}

Here, we study the non-Abelian flavor symmetries, 
which can appear in our models.

\subsubsection{The case with $M_i \ne 1$ and $k_i=1$ }

First, we consider the models with $k_1=k_2=k_3=1$.
Then, the number of zero-modes are given by $|I_{ab}|=M_1$,
$|I_{bc}|=M_2$ and $|I_{ca}|=M_3$.
We consider the models with $g={\rm g.c.d.}(M_1,M_2,M_3)\ne 1$.
The Yukawa couplings do not depend on the matrix components  
$(p, q,r)$, and are reduced to the following form 
\begin{eqnarray}
\int_0^Q dy_4 \int_0^1 dy_5 \Psi^i_{pq} \Psi^j_{qr} \Psi^k_{rp}
= 
N_{M_1}N_{M_2}N_{M_3}^* 
Q\sqrt{\frac{Q}{2M_3'}}
\sum_{m\in Z_{M_3'}}\delta_{j'+k'+M_1'm, l' (\rm{mod} {M_3'})}
\nonumber \\
\times
\jtheta{\frac{M_2'j'-M_1'k'+M_1'M_2'm}{M_1'M_2'(M_3')} \\ 0}
\left(  0 ,M_1'M_2'M_3'/Q \tau \right),
\end{eqnarray}
where we have taken simply $p=q=r=0$ and 
the phase factor like $C_{pq}^j$ disappears.
This form is nothing but the case with integer fluxes and 
without non-Abelian Wilson lines.
In this types of Yukawa couplings,  
4D effective theory has another flavor symmetry called 
by the shift symmetry, which 
corresponds to the transformations of flavor indices as
\begin{eqnarray}\label{eq:permutation-g}
& & j' \to j'+M'_1/g, \nonumber \\ 
& & k' \to k'+M'_2/g, \\
& & l' \to l'+M'_3/g, \nonumber
\end{eqnarray}
simultaneously. 
Under this transformation, Yukawa couplings are invariant. 
This has also coupling selection rule as shown in the previous section
given by the $Z_g$ symmetry (\ref{eq:Zg}).
Then, they form the non-Abelian discrete flavor symmetries 
as the same as the 
case without non-Abelian Wilson lines.

For simplicity, suppose that $M'_1=g$.
Then, there are $g$ zero-modes of  $\Psi^{j',M'_1}$.
The selection rule (\ref{eq:Zg}) means that 4D effective theory is 
symmetric under the  $Z_g$ transformation, which 
acts on $\Psi^{j',g}$ as 
$Z \Psi^{j',g}$, where
\begin{eqnarray}
Z = \left(
\begin{array}{ccccc}
1 & & & & \\
  & \rho & & & \\
  & & \rho^2 & & \\
  & &   & \ddots & \\
  &  &  &    & \rho^{g-1} 
\end{array}
\right),
\end{eqnarray}
and $\rho = e^{2\pi i /g}$.
Furthermore, 
the effective theory has another symmetry (\ref{eq:permutation-g}).
That  can be written as cyclic permutations on  $\Psi^{j',g}$, 
\begin{equation}\label{eq:permutation}
\Psi^{j',g} \rightarrow \Psi^{j'+1,g} .
\end{equation}
That is nothing but a change of ordering and also 
has a geometrical meaning as a discrete shift 
of the origin, $z=0 \rightarrow z=-\frac{1}{g}$.
This symmetry also generates another $Z_g$ symmetry, 
which we denote by $Z_g^{(C)}$ and its generator is 
represented as 
\begin{eqnarray}\label{eq:C2}
C = \left(
\begin{array}{cccccc}
0 & 1& 0 & 0 & \cdots & 0 \\
0  & 0 &1 & 0 & \cdots & 0\\
  &    &  & &\ddots & \\
1  &  0  & 0 &  & \cdots   & 0 
\end{array}
\right),
\end{eqnarray}
on $\Psi^{j',g}$.
These generators, $Z$ and $C$, do not commute each other,
i.e., 
\begin{equation}
CZ = \rho ZC .
\end{equation}
Then, the flavor symmetry corresponds to the closed algebra 
including $Z$ and $C$.
Diagonal matrices in this closed algebra are written as 
$Z^n(Z')^m$, 
where $Z'$ is the generator of another $Z'_g$  
written as 
\begin{eqnarray}
Z' = \left(
\begin{array}{ccc}
\rho & &  \\
    & \ddots & \\
    &    & \rho 
\end{array}
\right),
\end{eqnarray}
on $\Psi^{j',g}$.
Hence, these would generate the non-Abelian flavor symmetry 
$(Z_g \times Z'_g)\rtimes Z_g^{(C)}$, since 
$Z_g \times Z'_g$ is a normal subgroup.
These discrete flavor groups would include $g^3$ elements totally.

For example, for $g=2$ and 3 these flavor symmetries are given as 
$Z_2 \rtimes Z_2 =D_4$ 
and $(Z_3 \times Z_3) \rtimes Z_3 = \Delta(27)$, respectively.
Then, the fields $\Psi^{j',g}$ correspond to 
${\bf 2}$ of $D_4$ and ${\bf 3}$ of $\Delta (27)$, as shown in 
Tables \ref{tab:g2} and \ref{tab:g3}, respectively.
When $M'/g$ is an integer larger than $1$, 
the $\Psi^{j',M'}$ fields correspond to other representations.
For smaller values of $M'/g$, the corresponding representations 
are shown in Tables \ref{tab:g2} and \ref{tab:g3}.

However we note that
their multiplets have several types of representation under 
this symmetry.
Because a $Z_g$ charge of fields labeled by $j$ is not 
$j$ but $j'=qj$.
Therefore even if they have same multiplicities ($M_1=M_2$),
their representations may be different from each other.

\subsubsection{The case with $M_i = 1$ and $k \ne 1$ }

Next, we consider the models with $M_i = 1$ and $k \ne 1$.
In this case, we also find similar flavor structures 
as well as the case without non-Abelian Wilson lines.  
Suppose all the components of zero-modes are given by 
$|I_{ab}|=k_1$, $|I_{bc}|=k_2$ and $|I_{ca}|=k_3$.
Then it is possible to take phase factors for each wavefunction 
$C^j_{pq}=1$.
We commonly use $K={\rm g.c.d.}(k_1,k_2,k_3)$.
The Yukawa couplings only depend on the indices $p,q$ and $r$
as a function $\theta_{pqr}$ given by 
\begin{eqnarray}
\theta_{pqr}&=&
Q\left( \frac{m_a}{n_a}\tilde{I}_{bc}p
+ \frac{m_b}{n_b}\tilde{I}_{ca}q
+ \frac{m_c}{n_c}\tilde{I}_{ab}r
\right) \nonumber \\
&=&
Q\left( \frac{m_a}{n_a}\tilde{I}_{bc}(\tilde{j}+n_1 k_1)
- \frac{m_c}{n_c}\tilde{I}_{ab}(\tilde{k}+n_2 k_2)
\right) ,
\end{eqnarray}
where we have used the relations $p-q=n_1 k_1+\tilde{j}$ and 
$l-r=n_2 k_2+\tilde{k}$ with $n_1, n_2 \in Z$.
We find that the Yukawa couplings are invariant under the following
transformation as
\begin{eqnarray}\label{eq:shift-2}
\tilde{j} &\to& \tilde{j} + \frac{m_c I_{ab}}{K}, \nonumber \\
\tilde{k} &\to& \tilde{k} + \frac{m_a I_{bc}}{K}, \\
\tilde{l} &\to& \tilde{l} + \frac{m_b I_{ac}}{K}. \nonumber 
\end{eqnarray}
It is obvious that this transformation is the permutation of flavor
index with order $K$.
Therefore we have two symmetries: one is 
the discrete $Z_K$ symmetry comes from the coupling selection rule and 
another is this shift symmetry.
By combining these two symmetries, it becomes the same non-Abelian
discrete flavor symmetry as the case without Non-Abelian
Wilson-lines. 
That is, these flavor symmetries are given as 
$Z_2 \rtimes Z_2 =D_4$ for $K=2$, 
$(Z_3 \times Z_3) \rtimes Z_3 = \Delta(27)$ for $K=3$ and 
$(Z_K \times Z_K) \rtimes Z_K $ for generic $K$.

We have two aspects of flavor structures which are characterized
by the parameters $M,K$.
In the latter case, the origin of flavor symmetry is  the gauge
symmetry.
The background breaks the continuous gauge symmetry, 
but discrete symmetry remains as the flavor symmetry.
In the former case, the flavor would not be directly originated from 
the gauge symmetry.
However, T-duals of both cases would correspond to 
similar intersecting $D$-brane models, where 
$n_a$ and $m_a$ have almost the same meaning, that is, 
winding numbers of $D$-branes for different directions.
Thus, these two pictures of flavor symmetries are 
related with each other by T-duality through the intersecting 
$D$-brane picture.

So far, we have considered the models with $M_i =1$ and $K \ne 1$ 
and found the flavor symmetry $(Z_K \times Z_K) \rtimes Z_K$.
Here we comment on generic case with  $M \ne 1$ and $K \ne 1$.
Even in such a case, the selection rules due to $Z_g$ and $Z_K$ 
symmetries hold exact.
However, the general formula of Yukawa couplings depend on both the
indices
$j$ and  $\tilde{j}$. 
Then, 4D effective Lagrangian is not always invariant under 
the above (independent) shift transformations 
(\ref{eq:permutation-g}) and (\ref{eq:shift-2}).

\subsubsection{Illustrating examples}

We show two illustrating examples.
We concentrate on  only the $T^2$ torus.
The first example is the model with $(I_1,I_2,I_3)=(2,4,2)$.
The background magnetic flux is taken as 
\begin{eqnarray}
F=
2\pi 
\begin{pmatrix}
\frac{1}{2} {\bf 1}_{N_a} & & \\ 
& \frac{3}{8} {\bf 1}_{N_b} & \\ 
& & \frac{1}{4} {\bf 1}_{N_c} \\ 
\end{pmatrix}.
\end{eqnarray}
Then the appearing chiral matters are denoted by 
\begin{eqnarray} 
\lambda=
\begin{pmatrix}
\rm{const} & L_{pq}^{j,M_1=1} & \\
& \rm{const} &  R_{qr}^{k,M_2=1} \\
H_{rp}^{l,M_3=1} & & \rm{const} 
\end{pmatrix}, 
\end{eqnarray}
where $p=$0, 1, $q=$0, 1, ..., 7 and $r=$0, 1, 2, 3.
The wavefunctions are represented by following theta functions as
\begin{eqnarray} 
L_{pq}^j(x,y)
&=&
N_{M_1}e^{-\pi/8 y^2}
\jtheta{0 \\ 0}(z/8+(1/2p-3/8q),\tau/8),  \nonumber \\
R_{qr}^k(x,y)
&=&
N_{M_2}e^{-\pi/8 y^2}
\jtheta{0 \\ 0}(z/8+(3/8q-1/4r),\tau/8),  \nonumber \\
H_{pr}^l(x,y)
&=&
N_{M_3}e^{-\pi/4 y^2}
\jtheta{0 \\ 0}(z/4+(1/2p-1/4r),2\tau/8)  ,
\end{eqnarray}
where we take $j=k=l=0$. 
The several parameters are also given by these fluxes.
We have $k_1=2$, $k_2=4$, $k_3=2$ and  
$K={\rm g.c.d.}(k_1,k_2,k_3)=2$. 
The gauge invariant 3-point couplings are divided to four types
of Yukawa couplings shown below 
\begin{eqnarray} 
\mathcal{L}
&=&
Y_{000}+Y_{010}+Y_{001}+Y_{100} , \nonumber \\
Y_{000} &=&
 L_{00}R_{00}H_{00}^\dagger+L_{11}R_{11}H_{11}^\dagger
+L_{02}R_{22}H_{02}^\dagger+L_{13}R_{33}H_{13}^\dagger 
\nonumber \\
&&
+L_{04}R_{40}H_{00}^\dagger+L_{15}R_{51}H_{11}^\dagger
+L_{06}R_{62}H_{02}^\dagger+L_{17}R_{73}H_{13}^\dagger ,
\nonumber \\
Y_{011} &=&
 L_{00}R_{01}H_{01}^\dagger+L_{11}R_{12}H_{12}^\dagger
+L_{02}R_{23}H_{03}^\dagger+L_{13}R_{30}H_{10}^\dagger
\nonumber \\
&&
+L_{04}R_{41}H_{01}^\dagger+L_{15}R_{52}H_{12}^\dagger
+L_{06}R_{63}H_{03}^\dagger+L_{17}R_{70}H_{10}^\dagger ,
\nonumber \\
Y_{101} &=&
 L_{10}R_{00}H_{10}^\dagger+L_{01}R_{11}H_{01}^\dagger
+L_{12}R_{22}H_{12}^\dagger+L_{03}R_{33}H_{03}^\dagger
\nonumber \\
&&
+L_{14}R_{40}H_{10}^\dagger+L_{05}R_{51}H_{01}^\dagger
+L_{16}R_{62}H_{12}^\dagger+L_{07}R_{73}H_{03}^\dagger ,
\nonumber \\
Y_{110} &=&
 L_{10}R_{01}H_{11}^\dagger+L_{01}R_{12}H_{01}^\dagger
+L_{12}R_{23}H_{13}^\dagger+L_{03}R_{30}H_{00}^\dagger
\nonumber \\
&&
+L_{14}R_{41}H_{11}^\dagger+L_{05}R_{52}H_{02}^\dagger
+L_{16}R_{63}H_{13}^\dagger+L_{07}R_{70}H_{00}^\dagger .
\nonumber 
\end{eqnarray}
As seen in these interaction terms, 
one finds that all the combinations $(p,q,r)$ are not allowed.
This is because it has $K={\rm g.c.d.}(2,4,2)=2$. 
Their fields $L,R,H$ are divided to two classes under the discrete
$Z_2$ charge.
For instance, for $R$ fields, 
the flavor index is defined by $\tilde{k}=q-r\mod{4}$.
We assign the $Z_2$ charges as  
\begin{eqnarray} 
Z_2 \ + &:& R^{\tilde{k}=0}, R^{\tilde{k}=2}, \nonumber \\
Z_2 \ - &:& R^{\tilde{k}=1}, R^{\tilde{k}=3},
\end{eqnarray}
and other fields are also assigned the $Z_2$ charges as
\begin{eqnarray} 
Z_2 \ + &:& L^{\tilde{k}=0}, H^{\tilde{k}=0}, \nonumber \\
Z_2 \ - &:& L^{\tilde{k}=1}, H^{\tilde{k}=1}.
\end{eqnarray}
That corresponds to the coupling selection rule as 
$\tilde{j}+\tilde{k}=\tilde{l} \mod{2}$.
The Yukawa couplings $Y_{jl}^{pqr}$ are calculated by the 
overlap integrals as follows
\begin{eqnarray} 
Y_{pqr}^{kl}
\propto 
\jtheta{0 \\ 0}(1/2p-3/4q+1/4r,\tau/4).
\end{eqnarray}
We also consider about the shift symmetry for this model, i.e.
\begin{eqnarray} 
\tilde{j} &\to& \tilde{j} + \frac{m_c I_{ab}}{K}=\tilde{j}+1 \mod{2},
\nonumber \\
\tilde{k} &\to& \tilde{k} + \frac{m_a I_{ab}}{K}=\tilde{k}+2 \mod{4},
\\
\tilde{l} &\to& \tilde{l} + \frac{m_b I_{ab}}{K}=\tilde{l}+1
\mod{2}. \nonumber
\end{eqnarray}
As shown in the previous section, the Yukawa couplings are 
invariant under this transformation.
These two operators make the $D_4=Z_2 \rtimes Z_2$ discrete flavor
symmetry. 
One can understand the representation for each field under $D_4$
symmetry.
As an analysis similar to the previous section, 
one can find that $L$ and $R$ correspond to doublets and $H$ fields
become 
four non-trivial singlets under $D_4$ symmetry.

As another example, we consider the model with 
$(I_1,I_2,I_3)=(3,3,3)$, which is not realized by only integer
fluxes. 
We choose fluxes as 
\begin{eqnarray}
F=
2\pi 
\begin{pmatrix}
3 {\bf 1}_{N_a} & & \\ 
& \frac{3}{2} {\bf 1}_{N_b} & \\ 
& & 0 {\bf 1}_{N_c} \\ 
\end{pmatrix}.
\end{eqnarray}
Then the appearing chiral matter fields are denoted as follows,
\begin{eqnarray} 
\lambda=
\begin{pmatrix}
\rm{const} & L_{0p}^{j,M_1=3} & \\
& \rm{const} &  R_{q0}^{k,M_2=3} \\
H_{00}^{l,M_3=3} & & \rm{const} 
\end{pmatrix}, 
\end{eqnarray}
where $p,q=0,1$.
This model has $k_i=1$ and $Q_1=2, Q_2=2, Q_1=1, Q=2$
$(j'=j,k'=k,l'=2l)$. 
The gauge invariant 3-point couplings are given as 
\begin{eqnarray} 
\mathcal{L}
&=&
{\rm{tr}}L_{pq}R_{qr}H_{pr}^\dagger \nonumber \\
&=&
L_{00}R_{00}H_{00}^\dagger+L_{01}R_{10}H_{00}^\dagger . 
\end{eqnarray}
The Yukawa couplings $Y_{jl}^{pqr}$ are calculated by overlap
integrals as follows
\begin{eqnarray} 
Y_{pqr}^{jkl}
&=&
\int_0^1 dy_4 \int_0^2 dy_5 L_{pq}(x,y)R_{qr}(x,y)H_{rp}(x,y)^*
\nonumber \\
&\propto& 
\sum_{m \in Z_6} 
\delta_{j'+k'+3m,l'}
\jtheta{\frac{3j'-3k'+9m}{54} \\ 0}(0,27\tau), 
\end{eqnarray}
where we take $p=q=r=0$.
From the structure of Kronecker delta,
one can read the selection rule as
\begin{eqnarray} 
&j'+k'+3m=l' \mod{6}& \nonumber \\ 
\to &j+k-2l=0 \mod{3}&.
\end{eqnarray}
Since $g$ is defined by $g={\rm g.c.d.}(M_1',M_2',M_3')=3$,
so this model has $\Delta(27) = (Z_3 \times Z_3) \rtimes Z_3$ flavor
symmetry.
Here we mention that the charge assignment is different from 
the case with Abelian Wilson line.
For the $H$ fields, their $Z_3$ charges are given as $l'=2l$, 
so they correspond to the multiplet of $\bar{\bf{3}}$ representations.
Other sectors of $L$ and $R$ correspond to $\bf{3}$
representations, and they can couple in the language of 
flavor symmetry. 
Therefore the extension to the non Abelian Wilson line case causes to 
have more various types of representations and flavor structures.

It is possible to introduce the constant gauge 
potential called the Abelian Wilson line.
We use the previous model with $(I_1,I_2,I_3)=(3,3,3)$.
We assume $N_a=4$, $N_b=4$, $N_c=2$.
The fractional fluxes can break the rank of gauge symmetry,
that is, the $U_b(4)$ gauge group breaks to $U_b(2)$ and 
the total gauge symmetry is $U(4)\times U(2)\times (2)$.
To break the gauge symmetry to the standard-model gauge group, 
Abelian Wilson line is introduced.
There are three types of gauge potential $A_a$, $A_b$ and $A_c$.
Their configurations are taken as follows 
\begin{eqnarray}
A_a=
\begin{pmatrix}
a_1 {\bf 1}_{3} & \\ 
& a_2 {\bf 1}_{1} \\ 
\end{pmatrix},\ \ 
A_b= b {\bf 1}_{2} ,\ \  
A_c=
\begin{pmatrix}
c_1 {\bf 1}_{1} & \\ 
& c_2 {\bf 1}_{1}  
\end{pmatrix}.
\end{eqnarray}
Then the (supersymmetric) standard model with three generations
is realized. 
Since the different Wilson line leads to different Yukawa couplings,
that would lead to various flavor structures.
For example, the above model leads to the $\Delta(27)$ flavor 
symmetry in generic values of Wilson lines as studied in 
the previous section.
However, the flavor symmetry is enhanced to 
the $\Delta(54)$ symmetry when Wilson lines vanish.
Thus by choosing the particular choice of Abelian Wilson lines,
we could realize that the flavor symmetry is large like $\Delta(54)$ 
in a subsector, 
e.g. in the lepton sector, but the other sector, e.g. the quark 
sector,  has the smaller flavor symmetry like $\Delta(27)$.\footnote{
Indeed, non-Abelian discrete flavor symmetries such as 
$D_4$, $\Delta(27)$ and $\Delta(54)$ would lead to 
phenomenologically interesting models  
\cite{Grimus,Branco:1983tn,Ishimori:2008uc}.}
This is the explicit example which can realize the {\it co-existence} 
of the different types of the flavor symmetries from the 
GUT type models~\cite{Abe:2009vi}.
Furthermore in the next section we will see that 
this mechanism plays an important role 
to obtain the realistic quark/lepton mass matrices and mixings.

\subsection{Phenomenological model construction}

Here the flavor structures we obtained are 
from their effective field theoretical constructions.
Therefore the effective three point couplings have common 
structures for each of four types of Yukawa couplings and 
does not depend on the gauge symmetry.
Among these structures, the specific example is U(8) Pati-Salam GUT
models where its matter sectors have three generations and 
up and down type Higgs sectors to couple through the Yukawa
interactions.
We assume the compactification with factorizable three $T^2$.
Then we also assume that all the flavors are generated at one torus
in order to obtain various types of Yukawa structures. 
We show such an example to generate three generations for quarks and
leptons and up and down type Higgs fields by introducing 
following magnetic flux and Wilson lines which break
the standard model gauge groups as 
\begin{equation} 
 F_{z \bar z} = {2 \pi i \over \Im \tau}  
\begin{pmatrix}
   m_1 {\bf 1}_4 & & \\
  & m_2 {\bf 1}_2  & \\ & & m_3 {\bf 1}_2 
\end{pmatrix}, 
\end{equation}
and  
\begin{equation} 
 A_{z \bar z} = 
\begin{pmatrix}
\begin{pmatrix}
 \xi_Y {\bf 1}_1 & \\
  & \xi_C {\bf 1}_3
\end{pmatrix}  & & \\ 
& \xi_L{\bf 1}_2 & \\
& & \begin{pmatrix}
\xi_u & \\
& \xi_d 
\end{pmatrix}
\end{pmatrix}, 
\end{equation}
where $\xi_Y$ and $\xi_C$ are the Wilson lines 
which should have different VEVs each other, otherwise 
it does not break $U(4) \to U(3)\times U(1)$.
Similarly $\xi_u$ and $\xi_d$ are also different VEVs and 
the remain gauge group goes to the $U(3)\times U(2)\times U(1)^3$.
Then  the bi-fundamental matter fields 
are affected by the difference of 
the Wilson line parameters stratched between each gauge sectors
like $A_\alpha - A_\beta$. These structures are also understood by
the Intersecting D-brane models in which the positions of the 
two stack of D-brane are parameterized by the two different open string 
moduli parameters. 
One can show the following wavefunction profiles 
\begin{eqnarray}
Q&:& \Theta^{j,3}(z+\xi_C-\xi_u)
\nonumber \\
L&:& \Theta^{j,3}(z+\xi_Y-\xi_d)
\nonumber 
\\
u&:& \Theta^{j,3}(z+\xi_C-\xi_L)
\nonumber \\
d&:& \Theta^{j,3}(z+\xi_C-\xi_d)
\\
e&:& \Theta^{j,3}(z+\xi_Y-\xi_d)
\nonumber \\
\nu&:& \Theta^{j,3}(z+\xi_Y-\xi_u), 
\nonumber  
\end{eqnarray}
and 
\begin{eqnarray}
H_u : \Theta^{j,3}(z+\xi_L-\xi_u), \quad \quad
H_d : \Theta^{j,3}(z+\xi_L-\xi_d).
\end{eqnarray}
From the above constructions, up type quark and 
Dirac neutrino $\nu$ couple to the same Higgs fields $H^u$ but 
their couplings have different Wilson lines.
Therefore even in this Pati-Salam models, 
it allows for quark and lepton to have different types of flavor
structures.
Then four dimensional Yukawa interactions are expressed by the flavor
indices and the 
parameters of Wilson line degrees of freedom as
\begin{align}
\mathcal{L}=
y^u_{jkl}(\xi_u^q) Q^j u^k H_u^l+ y^u_{jkl}(\xi_u^l) E^j \nu^k
H_u^l + y^d_{jkl}(\xi_d^q) Q^j d^k H_d^l + y^d_{jkl}(\xi_d^l) E^j e^k
H_d^l 
\end{align}
where Wilson lines $\xi_u^q$, $\xi_u^l$, $\xi_d^q$
and $\xi_d^l$ are generally different each other.
From the previous analysis this model has flavor structures of
$\Delta(27)$ discrete symmetry generally.
Three generations are corresponding to the multiplet of $\Delta(27)$, 
i.e. triplet ${\bf 3}$,
and 6 Higgs are two triplets.
We denote these multiplets as 
\begin{eqnarray}
L = 
\left(\begin{array}{c}
L_0 \\ L_1 \\ L_2
\end{array}\right), \ \ \ 
R = 
\left(\begin{array}{c}
R_0 \\ R_1 \\ R_2
\end{array}\right), \ \ \ 
H_a = 
\left(\begin{array}{c}
H_0 \\ H_2 \\ H_4
\end{array}\right), \ \ \ 
H_b = 
\left(\begin{array}{c}
H_3 \\ H_5 \\ H_1
\end{array}\right).
\end{eqnarray}

In order to get the quark/lepton masses and break the electro weak
gauge symmetry, Higgs VEVs are needed.
We consider about following typical breaking pattern of Higgs VEV 
\begin{eqnarray}
H \to  
\left(\begin{array}{c}
v \\ 0 \\ 0
\end{array}\right).
\end{eqnarray}
Allowed Yukawa coupling are calculated by overlap integrals,  
one can show the all possible patterns of Yukawa matrices as
\begin{eqnarray}
H_0 \left(\begin{array}{ccc}
y_a & 0 & 0 \\
0 & 0 & y_c \\
0 & y_e & 0
\end{array}\right), \ \ \ 
H_1 \left(\begin{array}{ccc}
0 & y_f & 0 \\
y_b & 0 & 0 \\
0 & 0 & y_d
\end{array}\right), \ \ \ 
H_2 \left(\begin{array}{ccc}
0 & 0 & y_e \\
0 & y_a & 0 \\
y_c & 0 & 0 
\end{array}\right), \nonumber \\
H_3 \left(\begin{array}{ccc}
y_d & 0 & 0 \\
0 & 0 & y_f \\
0 & y_b & 0
\end{array}\right), \ \ \ 
H_4 \left(\begin{array}{ccc}
0 & y_c & 0 \\
y_e & 0 & 0 \\
0 & 0 & y_a
\end{array}\right), \ \ \ 
H_5 \left(\begin{array}{ccc}
0 & 0 & y_b \\
0 & y_d & 0 \\
y_f & 0 & 0 
\end{array}\right). \nonumber 
\end{eqnarray}
There are six independent Yukawa couplings $y_a, y_b, \cdots, y_f$,
these numerical values depend on the complex structure moduli 
($\tau= \tau_1+ i\tau_2$) and Wilson line degrees of freedom in two
internal 
directions ($A_1+iA_2$).
In general, one can assign these coefficients as 
$|y_a| \ge |y_b|, |y_f| \ge |y_c|, |y_e| \ge |y_d|$.
Especially, taking $\xi=0$ or its equivalent configurations of
$\xi$  
they lead $y_b=y_f$ and $y_c=y_e$.
In that case, enhancement of symmetry occurs and approximate flavor
symmetry becomes $\Delta(54)$.

Here we provide semi realistic Yukawa patterns.
For up type quark sector we take $\tau=4i$ and $\xi=-4.2i$.
Here we note that the moduli parameter $\tau$ must be commonly taken
for down type quark sectors and lepton sectors. 
We take following up type Higgs VEVs as 
$(v_u^a)^T = (v_u \ 0 \ 0)$ and 
$(v_u^b)^T = (0 \ 0 \ 0)$.
Then Yukawa coefficients are 
$y_a=0.77$, $y_c=0.0033$, $y_e=9.5 \times 10^{-6}$ and 
quark mass ratio $m_c/m_t=0.0043$ and $m_u/m_c=0.0028$.
These values are roughly close to realistic ones.

Next we consider about down type quark masses.
Taking the down type Higgs VEVs as 
$v_0^d \ne0, v_1^d \ne 0$ and $\rm{others} = 0$ 
has following down type quark mass matrix 
\begin{eqnarray}
m_d = \left(
\begin{array}{ccc}
y_d v_1^d & y_c v_0^d & 0 \\
y_e v_0^d & 0 & y_f v_1^d \\
0 & y_b v_1^d &  y_a v_0^d
\end{array}\right).
\nonumber
\end{eqnarray}
To give the models explicitly, 
we provide some results by taking certain values of Wilson line.
For down type quark matrix, we take $A=0$ with $\tau=4i$,
then it has $m_c/m_b=0.0010$, $m_d/m_c=0.053$,
we assume that the Higgs VEVs are $v^d_0=v^d$ and
$v^d_1=\frac{v_d}{4}$. 
It have also small quark mixing as
\begin{eqnarray}
V_{CKM} = \left(
\begin{array}{ccc}
0.97 & 0.23 & -0.0070 \\
0.23 & -0.97 & 0.029 \\
6.9 \times 10^{-6} & 0.030 & 1.0 \\
\end{array}\right).
\nonumber
\end{eqnarray}
The details of the numerical values are shown below
\begin{eqnarray}
y_a = 1, \quad y_b= 0.12, && 
y_c = 0.00023 \quad y_d= 1.3 \times 10^{-8} \nonumber \\
y_f = y_b,  &&  y_c=y_e.  
\nonumber
\end{eqnarray}

For charged lepton sectors, 
we can take other Wilson line, 
it gives $m_\mu/m_\tau$ =0.049, $m_e/m_\mu=1.8\times 10^{-5}$
with  $A=-40i$ and commonly used $\tau=4i$. 
It gives following charged lepton mass matrix as 
\begin{eqnarray}
y_a = 3.4 \times 10^{-7}, &y_b= 0.30 \times 10^{-3}, & 
y_c = 0.39 \nonumber \\
y_d= 0.79,  &y_e = 0.024,&  y_f=1.2 \times 10^{-5}.  
\nonumber
\end{eqnarray}
This matrix gives very small mixing.

For the right handed neutrino masses, it is necessary to 
make use of seesaw mechanism.
In this model it is forbidden to have Majorana neutrino mass terms at 
tree level. Therefore it may be generated via higher dimensional
operators and need additional vector like matter fields. 
It is obvious that our set up is not defined globally, we must 
introduce other gauge sectors beyond U(8) gauge symmetry.
Even in such case, it is possible to calculate the higher order
couplings in principle and these couplings obey the selection rule 
from the overlap integrals of localized wavefunction. 
We assume that the effective Majorana mass
terms can also have structures similar to three point couplings,
$M_{ij}(N_\nu)_i (N_\nu)_j $ and $y_{ijk} (H^u)_k E_i (N_\nu)_j $.
Therefore we take the following form of the Majorana mass matrix
$M_{ij}$ as 
\begin{eqnarray}
M_{ij} = 
\left(
\begin{array}{ccc}
M & M' & M' \\
M' & M & M' \\
M' & M' & M 
\end{array}\right). 
\end{eqnarray}
These structures respect the $\Delta(54)$ 
discrete flavor symmetries.
Moreover one can construct the Dirac mass matrices simply
by assuming the specific vacuum alignments.
Here again we must take following up type Higgs VEVs as 
$(v_u^a)^T = (v_u \ 0 \ 0)$ and 
$(v_u^b)^T = (0 \ 0 \ 0)$.
It leads following Dirac mass matrices
\begin{eqnarray}
M_D = v_u \left(
\begin{array}{ccc}
y_a & 0 & 0 \\
0 & 0 & y_c \\
0 & y_e & 0 
\end{array}\right).
\nonumber
\end{eqnarray}
As shown before, these coupling $y_a,y_c,y_e$ have generally deferent
values. However taking some special combinations of Wilson lines like 
an enhancement point to $\Delta(54)$,
one can obtain $y_a=y_c \ne y_e$.
Then Dirac mass matrices has 
\begin{eqnarray}
M_D = v_u \left(
\begin{array}{ccc}
a & 0 & 0 \\
0 & 0 & b \\
0 & b & 0 
\end{array}\right).
\nonumber
\end{eqnarray}

Here we use the formula of the light neutrino mass formula within
seesaw mechanism as $m_D=- M_D^T M_R^{-1} M_D$.
That has following neutrino mass matrices as
\begin{eqnarray}
m_D = - \left(
\begin{array}{ccc}
x & y & y \\
y & z & w \\
y & w & z 
\end{array}\right),
\nonumber
\end{eqnarray}
where coefficients $x,y,z,w$ are 
\begin{eqnarray}
x &=& a^2 (M+M')D, \quad \quad \  y=-abM'D \nonumber \\
z &=& b^2 (M+M')D, \quad \quad \  w=-b^2M'D \nonumber \\
D &=& \frac{v_u^2}{M^2 +MM'-2M'^2}.
\nonumber
\end{eqnarray}
As well known, this structure of matrix can be diagonalized by 
a unitary matrix $U$ as 
\begin{eqnarray}
U = \left(
\begin{array}{ccc}
\cos{\theta} & \sin{\theta} & 0 \\
-\sin{\theta}\sqrt{2} & \cos{\theta}/\sqrt{2} & 1/\sqrt{2} \\
-\sin{\theta}\sqrt{2} & \cos{\theta}/\sqrt{2} & -1/\sqrt{2} 
\end{array} \right).
\nonumber
\end{eqnarray}
For realistic neutrino mixing angles it requires 
the constraint as $x+y-z-w=0$ which means 
$M'/M=-(a+b)/a$ and  this gives rise to tri-bimaximal neutrino mixing.
One can calculate the tree-level analysis of the coefficient of 
$a, b$ which depend on the moduli parameters. 
For the case $a > b$, naively one can expect that 
$M' \sim -M$ and for the case of $b > a$ it means 
that the $M'/M  \sim b/a$.
Since we would expect that the Majorana masses $M,M'$ also
depend on the same moduli parameters as $a, b$, 
these structures are naturally understood.
In fact, if Dirac neutrino mass matrix are taken as 
$a=1.0$ and $b=0.00023$ with assuming $M'= -1.00023 M \sim -M$ as a
consequence of tri-bi maximal neutrino mixing. 
By combing two results from up and down type lepton mass matrix,
we obtain the following mixing matrix
\begin{eqnarray}
V_{MNS} = \left(
\begin{array}{ccc}
0.81 & 0.59 & -0.012 \\
-0.41 &  0.55 & -0.72 \\ 
-0.42 & 0.59 & 0.69 
\end{array}\right).
\nonumber
\end{eqnarray}
As shown here, we can obtain the semi-realistic values of 
not only lepton mixing but also quark mass mixing 
by shifting the Wilson line parameters.
Actually these mass hierarchies in particular lowest mode
(e.g. up/down
quark or electron ) are less than realistic ones 
but these results are tree-level analysis,
so the higher order couplings can give small deviations 
which may have the large contributions to the small Yukawa sectors.
Moreover we can analyze it including the full parameter spaces of
moduli fields.
Then we would expect that these Yukawa structures have
fully realistic structures of quark/lepton mass hierarchies and
mixings. 
We have shown that it is useful to obtain the realistic flavor
structures for the {\it co-existence} of the different types of flavor
symmetries like $\Delta(54)$ and $\Delta(27)$.
Actually they are related as the breaking of the larger flavor
symmetries.
Recently, many interesting discrete symmetries and its subgroups 
are discussed in~\cite{Ishimori:2010au} and 
this model is an example for such a scenario. 
Therefore it is also interesting to study the other different types of
the flavor symmetries for quark and lepton as a bottom up approach.

\newpage

\section{Magnetized orbifold models}

In this section, we study orbifold models with non-vanishing 
magnetic fluxes, in particular N=1 super Yang-Mills theory 
on such a background.
Orbifolding the extra dimensions is another way to 
derive chiral theories \cite{Dixon}.
We will show that four-dimensional effective field theories 
on magnetized orbifolds 
have a rich structure and they lead to interesting aspects, 
which do not appear in magnetized torus models.
In particular, it will be found that a new type of 
flavor structures can appear.
We also show semi-realistic models on magnetized orbifolds.
Furthermore we study more about 
these backgrounds such as consistency conditions, zero-mode profiles
and phenomenological aspects of 4D effective theory.

Effects of Wilson lines on the torus with magnetic fluxes 
are gauge symmetry breaking and shift of wavefunction profiles.
For the same magnetic flux, the numbers of chiral zero-modes 
between the torus compactification and orbifold compactification 
are different from each other and zero-modes profiles are 
different~\cite{Abe:2008fi,Abe:2008sx}.
Adjoint matter fields remain massless on the torus with 
magnetic fluxes, those are projected out on the
orbifold~\footnote{Within 
the framework of intersecting D-brane models, analogous results have
been obtained by considering 
D6-branes wrapping rigid 3-cycles~\cite{Blumenhagen:2005tn}.}.
These differences lead to 
phenomenologically interesting aspects~\cite{Abe:2008sx}.
In the latter of this section 
we study more about Wilson line backgrounds such as consistency conditions,
zero-mode profiles and phenomenological aspects of 4D effective
theory~\cite{Abe:2009uz}.

\subsection{$U(1)$ gauge theory on magnetized orbifold $T^2/Z_2$}

Now, let us study $U(1)$ gauge theory on the orbifold $T^2/Z_2$ 
with the coordinates $(y_4,y_5)$, 
which are transformed as 
\begin{equation}
y_4 \rightarrow -y_4, \qquad y_5 \rightarrow -y_5,
\end{equation}
under the $Z_2$ orbifold twist.
Then, we introduce the same magnetic flux $F_{45}=2 \pi M$ as 
one in section 2.2 and use the same gauge as (\ref{eq:gauge}).
Note that this magnetic flux is invariant under the $Z_2$ orbifold
twist and consistent with fractional flux with non-Abelian Wilson
line. 
In the followings we focus on the integer flux case and 
it is straightforward to extend to the case with non-Abelian Wilson
line which is discussed later.

We study the spinor field $\psi(y)$ on the above background.
The spinor field $\psi(y)$ with the $U(1)$ charge $q=\pm 1$ 
satisfies the same equation as one on $T^2$, i.e. (\ref{eq:Dirac-T2}).
Then, we require $\psi(y)$ transform under the $Z_2$ twist as 
\begin{equation}
\psi(-y_4,-y_5) = (-i)\tilde \Gamma^4 \tilde \Gamma^5
P\psi(-y_4,-y_5),
\end{equation}
where $P$ depends on the charge $q$ like $P=(-1)^{q+n}$ 
with $n=$ integer and it should satisfy $P^2 =1$.
Suppose that $qM >0$.
Then, there are $M$ independent zero-modes for 
$\psi$ when we do not take into account the $Z_2$ projection.
However, some of them are projected out 
by the above $Z_2$ boundary condition.
For example, for $(-i)\tilde \Gamma^4 \tilde \Gamma^5P=1$, 
only even functions remain, while 
only odd functions remain for $(-i)\tilde \Gamma^4 \tilde
\Gamma^5P=-1$.
Note that 
\begin{equation}
\Theta^j(-y_4,-y_5)=\Theta^{M-j}(y_4,y_5),
\end{equation}
where $\Theta^{M}(y_4,y_5)=\Theta^{0}(y_4,y_5)$.
That is, even and odd functions are given by 
\begin{eqnarray}
\Theta^j_{\rm even}&=& \frac{1}{\sqrt 2}(\Theta^j + \Theta^{M-j}), \\
\Theta^j_{\rm odd}&=& \frac{1}{\sqrt 2}(\Theta^j - \Theta^{M-j}), 
\end{eqnarray}
respectively.
Hence, for $M=2k$ with $k=$ integer and $k>0$, 
the number of zero-modes $\psi_+$ for $P=1$ 
and $P=-1$ are equal to $k+1$ and $k-1$, respectively.
On the other hand, for $M=2k+1$ with $k=$ integer and $k\geq 0$, 
the number of zero-modes $\psi_+$ for $P=1$ 
and $P=-1$ are equal to $k+1$ and $k$, respectively.
It is interesting that odd functions can correspond to zero-modes 
in magnetized orbifold models.
On the orbifold with vanishing magnetic flux $M=0$, 
odd modes correspond to not zero-modes, but massive modes.
However, odd modes, which would correspond to massive modes 
for $M=0$, mix to lead to zero-modes in the case with $M\neq 0$.
It would be convenient to write these results explicitly 
for later discussions.
Table \ref{even-odd-zero-modes} shows the numbers of zero-modes with even 
and odd wavefunctions for $M \leq 10$.
Note that the degree of continuous Wilson line,
which we have on the torus,  
is ruled out on the orbifold.

\begin{table}[htb]
\begin{center}
\begin{tabular}{|c|ccccccccccc|}\hline
$M$ & 0 & 1 & 2 & 3 & 4& 5 & 6 & 7 & 8 & 9& 10 \\ \hline 
even & 1 & 1 & 2 & 2 & 3 & 3 & 4 & 4 & 5 & 5 & 6 \\ \hline
odd & 0 & 0 & 0& 1 & 1 & 2 & 2 & 3 & 3 & 4 & 4 \\ \hline
\end{tabular}
\end{center}
\caption{The numbers of zero-modes with even and odd wavefunctions.
\label{even-odd-zero-modes}}
\end{table}

\subsection{$U(N)$ gauge theory on magnetized orbifold $T^2/Z_2$}

Now, let us study $U(N)$ gauge theory on the orbifold $T^2/Z_2$.
We consider the same magnetic flux as (\ref{eq:F45-UN}), 
which breaks the gauge group 
$U(N) \rightarrow \prod_{a=1}^n U(N_a)$.
Furthermore, we associate the $Z_2$ twist with the $Z_2$ 
action in the gauge space as 
\begin{equation}
A_\mu(x,-y) = P A_\mu(x,y)P^{-1}, \qquad A_m(x,y) = -P A_m
(x,y)P^{-1}.
\end{equation}
In general, the $Z_2$ boundary condition breaks the gauge group 
$\prod_{a=1}^n U(N_a)$ further.
For simplicity, here we restrict ourselves to 
the $Z_2$ action, which remains the gauge group $\prod_{a=1}^n U(N_a)$ 
unbroken.
Thus, the $Z_2$ action is trivial for the unbroken gauge group, 
but it is not trivial for spinor fields as well as scalar fields.

Here, let us study spinor fields.
We focus on the $U(N_a) \times U(N_b)$ block (\ref{eq:F-ab-block}) 
and use the same gauge as (\ref{eq:gauge}), 
i.e. $A_4 = F_{45}y_5$ and $A_5 = 0$.
We consider the spinor fields, $\lambda^{aa}_\pm$, 
$\lambda^{ab}_\pm$, $\lambda^{ba}_\pm$ and $\lambda^{bb}_\pm$, 
where $\pm$ denotes the chirality in the extra dimension like 
(\ref{eq:two-spinor}).
Their $Z_2$ boundary conditions are given by 
\begin{equation}
\lambda_{\pm}(x,-y) = \pm P \lambda_{\pm}(x,y) P^{-1},
\end{equation}
for $\lambda^{aa}_\pm$, 
$\lambda^{ab}_\pm$, $\lambda^{ba}_\pm$ and $\lambda^{bb}_\pm$.
First, we study the gaugino fields,  $\lambda^{aa}_\pm$ 
and $\lambda^{bb}_\pm$ for the unbroken gauge group.
Since the $Z_2$ action $P$ is trivial for the unbroken gauge 
indices, the above   $Z_2$ boundary conditions reduce to 
$\lambda^{aa}_{\pm}(x,-y)=\pm \lambda^{aa}_{\pm}(x,y)$ and 
$\lambda^{bb}_{\pm}(x,-y)=\pm \lambda^{bb}_{\pm}(x,y)$.
In addition, the magnetic flux does not appear in their 
zero-mode equations.
Thus, $\lambda^{aa}_{+}(x,y)$ as well as $\lambda^{bb}_{+}(x,y)$
has a zero-mode, but $\lambda^{aa}_{-}(x,y)$ and 
$\lambda^{bb}_{-}(x,y)$ are projected out by the $Z_2$ orbifold 
projection as the usual $Z_2$ orbifold without the magnetic flux.

Next, let us study the bi-fundamental matter fields 
$\lambda^{ab}_\pm$ and $\lambda^{ba}_\pm$.
The magnetic flux $M_a - M_b$ appears in their zero-mode 
equations.
Without the $Z_2$ projection, there are $|M_a - M_b|$ 
zero-modes.
For example, when $M_a - M_b>0$, $\lambda^{ab}_+$ as well as 
$\lambda^{ba}_-$ has  $(M_a - M_b)$ zero-modes with 
the wavefunctions $\Theta^j$ for $j=0, \cdots, (M_a - M_b -1)$.
When we consider the $Z_2$ projection, either even or odd modes 
remain.
For example, when we consider the projection $P$ such that 
$\lambda^{ab}_+(x,-y) =\lambda^{ab}_+(x,y)$,  
only zero-modes corresponding to $\Theta^j_{\rm even}$ remain 
and the number of zero-modes is equal to $(M_a - M_b)/2 + 1$ for 
$(M_a - M_b)=$ even and $(M_a - M_b+1)/2 $ for $(M_a - M_b)=$ odd.
On the other hand, when we consider the projection $P$ such that 
$\lambda^{ab}_+(x,-y) =-\lambda^{ab}_+(x,y)$,  
only zero-modes corresponding to $\Theta^j_{\rm odd}$ remain 
and the number of zero-modes is equal to $(M_a - M_b)/2 - 1$ for 
$(M_a - M_b)=$ even and $(M_a - M_b - 1)/2 $ for $(M_a - M_b)=$ odd.
The same holds true for $\lambda^{ba}_-$.
Furthermore, when $M_a - M_b <0$, the situation is the same 
except replacing  $(M_a - M_b)$, $\lambda^{ab}_+$ and
$\lambda^{ba}_-$ by $|M_a - M_b|$, $\lambda^{ab}_-$ and
$\lambda^{ba}_+$, respectively.

The 3-point couplings among modes corresponding to 
the wavefunctions, $\Theta^i_{\rm even,odd}$, 
$\Theta^j_{\rm even,odd}$ and $\Theta^k_{\rm even,odd}$ are 
given by the overlap integral like (\ref{eq:Yukawa}).
Note that 
\begin{equation}\label{eq:vanish-overlap}
\int dy~\Theta^i_{\rm even}(y) \cdot \Theta^j_{\rm even}(y) 
\cdot \Theta^k_{\rm odd}(y) = 
\int dy~\Theta^i_{\rm odd}(y) \cdot \Theta^j_{\rm odd}(y) 
\cdot \Theta^k_{\rm odd}(y) = 0,
\end{equation}
while $\int dy~\Theta^i_{\rm even}(y) \cdot \Theta^j_{\rm odd}(y) 
\cdot \Theta^k_{\rm odd}(y) $ and 
$\int dy~\Theta^i_{\rm even}(y) \cdot \Theta^j_{\rm even}(y) 
\cdot \Theta^k_{\rm even}(y) $ are nonvanishing.

\subsection{$U(N)$ gauge theory on magnetized orbifolds 
$T^6/Z_2$ and $T^6/(Z_2\times Z'_2)$}

Here, we can extend the previous analysis on 
the two-dimensional orbifold $T^2/Z_2$ to 
the $U(N)$ gauge theory on the six-dimensional 
orbifolds $T^6/Z_2$ and $T^6/(Z_2\times Z'_2)$.
We consider two types of six-dimensional orbifolds, 
$T^6/Z_2$ and $T^6/(Z_2\times Z'_2)$.
For the orbifold $T^6/Z_2$, the $Z_2$ twist 
acts on the six-dimensional coordinates $y_m$ ($m=4,\cdots, 9$) as 
\begin{equation}
y_m \rightarrow -y_m~~({\rm for~~}m=4,5,6,7), \qquad 
y_n \rightarrow y_n~~({\rm for~~}n=8,9).
\end{equation}
In addition to this $Z_2$ action, we introduce another 
independent $Z'_2$ action,
\begin{equation}
y_m \rightarrow -y_m~~({\rm for~~}m=4,5,8,9), \qquad 
y_n \rightarrow y_n~~({\rm for~~}n=6,7),
\end{equation}
for the orbifold $T^6/(Z_2\times Z'_2)$.
If magnetic flux is vanishing, we realize 
four-dimensional N=2 and N=1 supersymmetric gauge 
theories for the orbifolds,  $T^6/Z_2$ and 
$T^6/(Z_2\times Z'_2)$, respectively.

Now, let us introduce 
the same magnetic flux as (\ref{eq:6D-flux}).
The gauge group $U(N)$ is broken as 
$U(N) \rightarrow \prod_{a=1}^n U(N_a)$ with $N=\sum_a N_a$.
This magnetic flux is invariant under 
both $Z_2$ and $Z'_2$ actions.
Furthermore, we associate the $Z_2$ and $Z'_2$ twists with 
the $Z_2$ and $Z'_2$ actions in the gauge space as 
\begin{eqnarray}
A_\mu(x,-y_m,y_n)   &=& PA_\mu(x,-y_m,y_n)P^{-1}, \nonumber \\
A_m(x,-y_m,y_n)   &=& -PA_m(x,-y_m,y_n)P^{-1}, \\
A_n(x,-y_m,y_n)   &=& PA_n(x,-y_m,y_n)P^{-1}, \nonumber
\end{eqnarray}
for $m=4,5,6,7$ and $n=8,9$, and 
\begin{eqnarray}
A_\mu(x,-y_m,y_n)   &=& P'A_\mu(x,-y_m,y_n){P'}^{-1}, \nonumber \\
A_m(x,-y_m,y_n)   &=& -P'A_m(x,-y_m,y_n){P'}^{-1}, \\
A_n(x,-y_m,y_n)   &=& P'A_n(x,-y_m,y_n){P'}^{-1}, \nonumber
\end{eqnarray}
for $m=4,5,8,9$ and $n=6,7$.
In general, these $Z_2$ boundary conditions break the gauge group 
$\prod_{a=1}^n U(N_a)$ further.
For simplicity, here we restrict to the $Z_2$ and $Z_2'$ 
projections, which remain the gauge group $\prod_{a=1}^n U(N_a)$ 
unbroken.
That is, both the $Z_2$ and $Z_2'$ actions are trivial 
for the unbroken gauge group.

Now, we study spinor fields.
We focus on the $U(N_a) \times U(N_b)$ block as
(\ref{eq:6D-F-ab-block}) and use the same gauge as
(\ref{eq:6D-gauge}).
We consider the spinor fields $\lambda^{aa}_{s_1,s_2,s_3}$, 
$\lambda^{ab}_{s_1,s_2,s_3}$, $\lambda^{ba}_{s_1,s_2,s_3}$ 
and $\lambda^{bb}_{s_1,s_2,s_3}$, where $s_i$ 
denotes the chirality corresponding to the $i$-th $T^2$.
Their $Z_2$ boundary conditions are given by 
\begin{equation}
\lambda_{s_1,s_2,s_3}(x,-y_m,y_n) =
s_1s_2P\lambda_{s_1,s_2,s_3}(x,y_m,y_n) P^{-1},
 \end{equation}
with $m=4,5,6,7$ and $n=8,9$ for $\lambda^{aa}_{s_1,s_2,s_3}$, 
$\lambda^{ab}_{s_1,s_2,s_3}$, $\lambda^{ba}_{s_1,s_2,s_3}$ 
and $\lambda^{bb}_{s_1,s_2,s_3}$.
Similarly, the $Z'_2$ boundary conditions are given by 
\begin{equation}
\lambda_{s_1,s_2,s_3}(x,-y_m,y_n) =
s_1s_3P'\lambda_{s_1,s_2,s_3}(x,y_m,y_n) P'^{-1},
 \end{equation}
with $m=4,5,8,9$ and $n=6,7$.

First, we study the gaugino fields $\lambda^{aa}_{s_1,s_2,s_3}$ 
and $\lambda^{bb}_{s_1,s_2,s_3}$ for the unbroken gauge group.
Their zero-mode equations have no effect due to 
magnetic fluxes, but only the $Z_2$ and $Z'_2$ orbifold twists 
play a role.
Since the $Z_2$ and $Z'_2$ twists, $P$ and $P'$, are trivial 
for the unbroken gauge sector, the boundary conditions are 
given by 
\begin{equation}
\lambda^{aa(bb)}_{s_1,s_2,s_3}(x,-y_m,y_n) =
s_1s_2\lambda^{aa(bb)}_{s_1,s_2,s_3}(x,y_m,y_n) \qquad{\rm for~~} Z_2,
 \end{equation}
with $m=4,5,6,7$ and $n=8,9$, and 
\begin{equation}
\lambda^{aa(bb)}_{s_1,s_2,s_3}(x,-y_m,y_n) =
s_1s_3\lambda^{aa(bb)}_{s_1,s_2,s_3}(x,y_m,y_n) \qquad{\rm for~~}
Z'_2,
 \end{equation}
with $m=4,5,8,9$ and $n=6,7$.
Hence, zero-modes of $\lambda^{aa(bb)}_{+,+,\pm}$ and 
$\lambda^{aa(bb)}_{-,-,\pm}$ survive on $T^6/Z_2$, 
that is, two kinds of gaugino fields with a fixed 
four-dimensional chirality.
Furthermore, on $T^6/(Z_2 \times Z'_2)$, 
zero-modes of $\lambda^{aa(bb)}_{+,+,+}$ and 
$\lambda^{aa(bb)}_{-,-,-}$ survive, that is, a single sort of 
gaugino fields with a fixed four-dimensional chirality.

Next, let us study the bi-fundamental matter fields, 
$\lambda^{ab}_{s_1,s_2,s_3}$ and $\lambda^{ba}_{s_1,s_2,s_3}$.
Without the $Z_2$ projection, they have zero-modes, 
whose number is $I_{ab}=I^1_{ab}I^{2}_{ab}I^3_{ab}$ and 
wavefunctions are given by 
$\Theta^{j_1}(y_4,y_5) \Theta^{j_2}(y_6,y_7) \Theta^{j_3}(y_8,y_9)$ 
($j_i=0,\cdots,(I^i_{ab}-1)$).
We assume that $I^i_{ab} > 0$ for $i=1,2,3$.
Then, the zero-modes correspond to $\lambda^{ab}_{+,+,+}$.
On $T^6/Z_2$, some of them are projected out.
Suppose that the $Z_2$ boundary condition is given by 
\begin{equation}
\lambda^{ab}_{s_1,s_2,s_3}(x,-y_m,y_n) =
s_1s_2\lambda^{ab}_{s_1,s_2,s_3}(x,y_m,y_n),
 \end{equation}
with $m=4,5,6,7$ and $n=8,9$.
Then, surviving zero-modes correspond to \\
$\Theta^{j_1}_{\rm even}(y_4,y_5) \Theta^{j_2}_{\rm even}(y_6,y_7) 
\Theta^{j_3}(y_8,y_9)$ and 
$\Theta^{j_1}_{\rm odd}(y_4,y_5) \Theta^{j_2}_{\rm odd}(y_6,y_7) 
\Theta^{j_3}(y_8,y_9)$.
Further modes are projected out on $T^6/(Z_2 \times Z'_2)$.
Suppose that the $Z'_2$ boundary condition is given by 
\begin{equation}
\lambda^{ab}_{s_1,s_2,s_3}(x,-y_m,y_n) =
s_1s_3\lambda^{ab}_{s_1,s_2,s_3}(x,y_m,y_n),
 \end{equation}
with $m=4,5,8,9$ and $n=6,7$.
Then, the surviving modes through the $Z_2 \times Z'_2$ projection 
correspond to 
$\Theta^{j_1}_{\rm even}(y_4,y_5) \Theta^{j_2}_{\rm even}(y_6,y_7) 
\Theta^{j_3}_{\rm even}(y_8,y_9)$ and 
$\Theta^{j_1}_{\rm odd}(y_4,y_5) \Theta^{j_2}_{\rm odd}(y_6,y_7) 
\Theta^{j_3}_{\rm odd}(y_8,y_9)$.
Similarly, we can analyze surviving zero-modes through 
the $Z_2 \times Z'_2$ projection in the models with 
different signs of $I^i_{ab}$ and different $Z_2 \times Z'_2$ 
projections.
It would be convenient to introduce the notation, 
$I^i_{ab({\rm even})}$ and $I^i_{ab({\rm odd})}$, 
such that  $I^i_{ab({\rm even})}$ and $I^i_{ab({\rm odd})}$ 
denote the number of even and odd functions, $\Theta^j_{\rm even}$ 
and $\Theta^j_{\rm odd}$, respectively, among $|I^i_{ab}|$ functions 
$\Theta^j$ for the $i$-th $T^2$.
Note that $I^i_{ab({\rm even})}, I^i_{ab({\rm odd})} \geq 0$ 
in the above definition, while $I^i_{ab}$ can be negative.

\subsubsection{Discrete flavor symmetry for orbifold models}

We have found that several non-abelian discrete flavor symmetries 
like $D_4$, $\Delta(27)$ and $\Delta(54)$ can appear.
However, these exact symmetries may be rather large 
to explain realistic mass matrices of quarks and leptons.
Their breaking would be preferable.
Such symmetry breaking can happen within the framework 
of four-dimensional effective field theory, 
that is, 
scalar fields with non-trivial representations 
are assumed to develop their vacuum expectation values.
On the other hand, a certain type of symmetry breaking 
can happen on the orbifold background, which is 
called magnetized orbifold models~\cite{Abe:2008fi,Abe:2008sx}.
Here, we discuss the flavor structure in 
magnetized orbifold models.

The orbifold $T^2/Z_2$ is constructed by 
dividing $T^2$ by the $Z_2$ projection $z \rightarrow -z$.
Furthermore, on such an orbifold, we require periodic 
or anti-periodic boundary condition for matter fields 
as well as gauge fields,
\begin{eqnarray}\label{eq:orbifold-bc}
& & 
\psi (-z) = \pm  \ \psi (z).
\end{eqnarray}
Since such boundary conditions are consistent in models 
with vanishing Wilson lines, we consider 
the case without Wilson lines.
Indeed, zero-mode wavefunctions in models without 
Wilson lines satisfy the following relation,
\begin{eqnarray}
& & 
\psi^{j,M}(-z) = \psi^{M-j,M}(z).
\end{eqnarray}
Thus, even and odd zero-modes are obtained as 
their linear combinations, 
\begin{eqnarray}\label{eq:even-odd-zero-modes}
& & 
\psi^j_{\pm}(z)  = \psi^{j,M}(z) \pm \psi^{M-j,M}(z),
\end{eqnarray}
up to a normalization factor.
Which modes among even and odd modes are selected 
depends on how to embed the $Z_2$ orbifold projection 
into the gauge space, 
that is, model dependent.
At any rate, either even or odd zero-modes are 
projected out for each kind of matter fields\footnote{Within 
the framework of intersecting D-brane
  models, analogous results have been obtained by considering 
D6-branes wrapping rigid 3-cycles~\cite{Blumenhagen:2005tn}.}.
Note that the $Z_2$ orbifold parity of $\psi^j_{\pm}(z) $ 
is the same as the $Z_2$ charge of $P$.
Thus, through the orbifold projection 
zero-modes with either even or odd $Z_2$ charge of $P$ 
survive for each kind of matter fields.

Let us consider examples.
First we study the model with $g=2$.
This model has the non-abelian flavor symmetry 
$D_4 \times Z_2$.
The zero-modes with $M=2$, $\psi^{i,2}$, correspond 
to ${\bf 2}_+$ of $D_4 \times Z_2$.
When we require the periodic boundary condition, 
they survive.
On the other hand, they are projected out 
for the anti-periodic boundary condition.
Similarly, the zero-modes with 
$M=4$, $\psi^{i,4}$, correspond 
to ${\bf 1}_{+++}$, ${\bf 1}_{+-+}$, ${\bf 1}_{-++}$ and 
${\bf 1}_{---}$, where the third subscript denotes 
the  $Z_2$ charge of $P$.
Thus, the zero-modes corresponding to 
 ${\bf 1}_{+++}$, ${\bf 1}_{+-+}$ and ${\bf 1}_{-++}$
survive for the periodic boundary condition, 
while only ${\bf 1}_{---}$ survives 
for the anti-periodic boundary condition.
Similarly, we can identify which modes can survive 
through the $Z_2$ orbifold projection.
The number of matter fields are reduced through 
the $Z_2$ orbifold projection.
However, four-dimensional effective field theory 
after orbifolding has the flavor symmetry $D_4 \times Z_2$.
The reason why the flavor symmetry $D_4 \times Z_2$ 
remains unbroken is that the flavor symmetry 
is the direct product between $D_4$ and $Z_2$.

Next, let us consider the model with $g=3$.
This model has the flavor symmetry $\Delta(54)$.
The zero-modes with $M=3$, $\psi^{i,3}$, correspond 
to ${\bf 3}_1$ of $\Delta(54)$.
However, the eigenstates of $Z_2$ are 
$\psi^{0,3}$ and $\psi^{1,3} \pm \psi^{2,3}$.
Hence, when we project out $Z_2$ even or odd modes, 
the triplet structure is broken, that is, 
the flavor symmetry $\Delta(54)$ is completely 
broken.
However, such symmetry breaking is non-trivial, 
because the original theory has the $\Delta(54)$ symmetry 
and we project out certain modes from such a theory.\footnote{
This type of flavor symmetry breaking has been proposed in
not magnetized brane models, but orbifold 
models~\cite{Haba:2006dz,Kobayashi:2008ih,Seidl:2008yf}.}

Orbifold models with larger $g$, $g > 3$ have a similar 
structure on flavor symmetries.
The original theory before orbifolding has 
a large non-abelian flavor symmetry.
By orbifolding, certain matter fields are projected out 
and the flavor symmetry is broken although some symmetries 
like abelian discrete symmetries remain unbroken.
However, there remains a footprint of the larger flavor symmetry
in four-dimensional effective theory, 
that is, coupling terms are constrained.

As an illustrating example, let us consider explicitly 
the model with three zero-modes, which have 
relative magnetic fluxes, $(M_1,M_2,M_3)=(4,4,8)$, 
that is, $g=4$.
The generators, $Z$, $C$ and $P$, are represented 
on the zero-modes with $M_1=4$ as 
\begin{equation}
Z=
\left(
\begin{array}{cccc}
1 & & & \\  & i & & \\ & & -1 & \\ & & & -i 
\end{array}
\right), \ \ 
C=
\left(
\begin{array}{cccc}
& 1 & & \\  & & 1 & \\ & & & 1  \\ 1 & & &  
\end{array}
\right), \ \ 
P=
\left(
\begin{array}{cccc}
1 & & & \\  &  & & 1 \\ & & 1 & \\ & 1 & &  
\end{array}
\right).
\end{equation}
Obviously, we find $[P,Z] \ne 0$ and $[C,P] \ne 0$.
Thus, eigenstates of $P$ are not eigenstates for 
$Z$ or $C$.
Since eigenstates with $P=1$ or $P=-1$ are 
projected out by orbifolding, the flavor symmetry 
is broken.
However, one can find that $[P,Z^2]=[P,C^2]=0$.
The symmetry generated by $Z^2$, $C^2$ and $P$ 
remains unbroken after orbifolding.
Thus, the flavor symmetry is reduced to 
$Z_2 \times Z_2 \times Z_2$.
The first two $Z_2$ factors are originally subgroups of 
$Z_4 \ltimes (Z_4 \times Z_4)$ generated by $Z$ and $C$ algebra and 
they are abelian groups.

For concreteness, let us consider the following 
$Z_2$ boundary conditions,
\begin{eqnarray}
& & \psi^{i_1,M_1}(-z)=\psi^{i_1,M_1}(z), \quad 
\psi^{i_2,M_2}(-z)=\psi^{i_2,M_2}(z),  \quad 
\psi^{i_3,M_3}(-z)=\psi^{i_3,M_3}(z) ,
\end{eqnarray}
for three types of zero-modes.
Then, we assign the first and second modes with left-handed 
and right-handed fermions, $L_i$ and $R_j$, 
while the third is assigned with Higgs fields $H_k$.
There are three $Z_2$ even modes for $M_1 = M_2 =4$, 
that is, the three generation model~\cite{Abe:2008fi,Abe:2008sx}, 
while there are five $Z_2$ even modes for $M_3=8$.
Their wavefunctions are shown in Table~\ref{tab:orbifold-model}.

\begin{table}[t]
\begin{center}
\begin{tabular}{|c|c|c|c|}\hline 
$i,j,k$ & $L_i $ & $R_j $ & 
$H_k $ 
\\ \hline \hline
0 & 
$\psi^{0,4}$ & $\psi^{0,4}$ & $\psi^{0,8}$ \\  
1 & 
$\frac{1}{\sqrt{2}}\left(\psi^{1,4}+\psi^{3,4}\right)$ & 
$\frac{1}{\sqrt{2}}\left(\psi^{1,4}+\psi^{3,4}\right)$ & 
$\frac{1}{\sqrt{2}}\left(\psi^{1,8}+\psi^{7,8}\right)$ \\  
2 & 
$\psi^{2,4}$ & 
$\psi^{2,4}$ & 
$\frac{1}{\sqrt{2}}\left(\psi^{2,8}+\psi^{6,8}\right)$ \\  
3 & - & - & 
$\frac{1}{\sqrt{2}}\left(\psi^{3,8}+\psi^{5,8}\right)$ \\  
4 & - & - & 
$\psi^{4,8}$ \\  \hline
\end{tabular}
\end{center}
\caption{Wavefunctions in the orbifold model.}
\label{tab:orbifold-model}
\end{table}

After orbifold projection, Yukawa couplings $Y_{ijk} L_i R_j H_k$ 
in this model are given by ~\cite{Abe:2008sx}
\begin{eqnarray}\label{eq:448}
Y_{ijk}H_k &=& 
\left( \begin{array}{ccc}
y_a H_0 + y_e H_4 & y_f H_3 + y_b H_1 & y_c H_2 \\
y_f H_3 + y_b H_1 & \frac{1}{\sqrt{2}}(y_a+y_e)H_2 
+ y_c(H_0+H_4) & y_b H_3+y_d H_1 \\
y_c H_2 & y_b H_3+y_d H_1 & y_e H_0 + y_a H_4 
\end{array} \right) .
\end{eqnarray}
Here, Yukawa coupling strengths, $y_a, y_b, \cdots, y_f$, 
are written as functions of moduli and they are, 
in general, different from each other.

We can take the basis of $L_i, R_j, H_k$ as 
eigenstates of $Z^2$ and $C^2$.
Such a basis is shown in Table~\ref{tab:z2c2}. 
Thus, if this effective theory has 
only $Z_4 \times Z_2 \times Z_2$ symmetry, 
the following couplings would be allowed,  
\begin{eqnarray}
Y_{ijk}H_k &=& 
\left( \begin{array}{ccc}
y_1 H_0+ y_2 H_2 +y_3 H_4   & 
y_4 H_1 + y_5 H_3 & 
y_6 H_0 + y_7 H_2 + y_8 H_4  \\
y_4' H_1 + y_5' H^3 &
y_9 (H_0+H_4) + y_{10} H_2 &
y_5' H_1 + y_4' H_3 \\ 
y_8 H_0 + y_7 H_2 +y_6 H_4 &
y_5  H_1 + y_4 H_3 &
y_3 H_0 + y_2 H_2 + y_1 H_4
\end{array} \right),  
\end{eqnarray}
where coupling strengths like $y_1$,$y_2$, etc. 
are independent parameters.
For example, the $Z_4 \times Z_2 \times Z_2$ 
symmetry allows non-vanishing couplings 
of $y_2$, $y_6$ and $y_8$.
However, these couplings are forbidden 
by the symmetry $Z_4 \ltimes (Z_4 \times Z_4) $ 
and such couplings do not appear in Eq.~(\ref{eq:448}).
Thus, Yukawa couplings derived from orbifolding 
are constrained more compared with 
the model, which has only the $Z_4 \times Z_2 \times Z_2$ 
flavor symmetry.

\begin{table}[t]
\begin{center}
\begin{tabular}{|c|cc|c|cc|c|cc|} \hline
$L_i$  & $Z^2$ & $C^2$ & $R_j $& $Z^2$ & $C^2$ 
& $H_k$ & $Z^2$ & $C^2$ \\ \hline \hline
$\frac{1}{\sqrt{2}} (L^0+L^2)$ &  1 &  1 &
$\frac{1}{\sqrt{2}} (R^0+R^2)$ &  1 &  1 &
$\frac{1}{\sqrt{2}} (H^0+H^4)$ &  1 &  1 \\
$\frac{1}{\sqrt{2}} (L^0-L^2)$ &  1 & --1 &
$\frac{1}{\sqrt{2}} (R^0-R^2)$ &  1 & --1 &
$\frac{1}{\sqrt{2}} (H^0-H^4)$ &  1 & --1 \\
$L_1$                          & --1 &  1 &
$R_1$                          & --1 &  1 &
$\frac{1}{\sqrt{2}} (H^1+H^3)$ & --1 &  1 \\
--                            &  -- &  -- & 
--                            &  -- &  -- & 
$\frac{1}{\sqrt{2}} (H^1-H^3)$ & --1 & --1 \\
--                            &  -- &  -- & 
--                            &  -- &  -- & 
$H_2$                          &  1 &  1 \\ \hline
\end{tabular}
\end{center}
\caption{Eigenstates of $Z^2$ and $C^2$}
\label{tab:z2c2} 
\end{table}

Similarly, other orbifold models have more constraints 
at least at tree level 
compared with unbroken symmetry as a footprint 
of larger flavor symmetries before orbifolding.
Such a structure would be useful for 
phenomenological applications.

Finally let us consider generic situation of unbroken flavor symmetry. 
Here we use the properties of the algebra for the discrete symmetries.
All the elements $h$ are represented by 
$h=\omega^t Z^r C^s$, $(r,s,t=0, 1, \cdots, g-1)$.
The remain generator with respect to unbroken symmetry should commute
with generator $P$. Using the following properties
\begin{align}
Z^g = C^g = \omega^g = 1, \ \ CZ=\omega ZC, \nonumber \\
PC= C^{-1}P, \ \ PZ=Z^{-1}P
\end{align} 
elements satisfying the conditions 
($2r=0 \mod{g}$ and $2s=0 \mod{g}$) only remain as unbroken
symmetry. Obviously the case with $g=\rm{odd}$ has trivial discrete
symmetries as $Z_g$. Cases with $g=\rm{even}$ are divided two
possibilities as $g=2m$ or $g=2m+2$, ($m \in \mathcal{Z}$).
For the former case all the elements are commutable,  remain symmetry
is $Z_g \times Z_2 \times Z_2$.
On the other hand, in the latter case one find 
two elements $C^{g/2}$ and $Z^{g/2}$ have
$C^{g/2}Z^{g/2}=-Z^{g/2}C^{g/2}$. Therefore the remain symmetry is 
non-Abelian discrete symmetry $Z_2 \ltimes (Z_g\times Z_2)$.

\subsection{Three generation magnetized orbifold
models}\label{sec:three-gene} 

In this section, 
we consider the $U(N_a)\times U(N_b)\times U(N_c)$ models,
which lead to three 
families of bi-fundamental matter fields, 
$(N_a,\bar N_b)$ and $(\bar N_a,N_c)$.
Such a gauge group is derived by starting with the $U(N)$ group 
and introducing the following form of the magnetic flux,
\begin{eqnarray}
F_{45} &=& 2 \pi \left(
\begin{array}{ccc}
M^{(1)}_a {\bf 1}_{N_a\times N_a} & & 0  \\
 & M^{(1)}_b {\bf 1}_{N_b\times N_b} & \\
0 & & M^{(1)}_c {\bf 1}_{N_c\times N_c}
\end{array}\right),  
\nonumber \\
F_{67} &=& 2 \pi\left(
\begin{array}{ccc}
M^{(2)}_a {\bf 1}_{N_a\times N_a} & & 0 \\
 & M^{(2)}_b {\bf 1}_{N_b\times N_b} & \\
0 & & M^{(2)}_c {\bf 1}_{N_c\times N_c}
\end{array}\right),  
\nonumber \\
F_{89} &=& 2 \pi \left(
\begin{array}{ccc}
M^{(3)}_a {\bf 1}_{N_a\times N_a} & & 0 \\
 & M^{(3)}_b {\bf 1}_{N_b\times N_b} & \\
0 & & M^{(3)}_c {\bf 1}_{N_c\times N_c}
\end{array}\right), 
\nonumber
\end{eqnarray}
where $N=N_a+N_b+N_c$.
For $N_a=4, N_b=2$ and $N_c=2$, we can realize 
the Pati-Salam gauge group up to $U(1)$ factors, 
some of which may be anomalous and become massive 
by the Green-Schwarz mechanism.
Then, the bi-fundamental matter fields, 
$(N_a,\bar N_b)$ and $(\bar N_a,N_c)$ correspond 
to left-handed and right-handed matter fields.
In addition, the bi-fundamental matter fields 
$(N_b,\bar N_c)$ correspond to higgsino fields.
We assume that supersymmetry is preserved 
at least locally at the $a-b$ sector, $b-c$ sector 
and $c-a$ sector.\footnote{See for the supersymmetric conditions 
e.g. Ref.~\cite{Cremades:2004wa,Troost:1999xn}.}
Then, the number of Higgs scalar fields are the same 
as the number of higgsino fields.
There are no tachyonic modes at the tree level.
Indeed, in intersecting D-brane models 
it would be one of convenient ways towards  
realistic models to derive the Pati-Salam model at some stage 
and to break the gauge group to 
the group $SU(3) \times SU(2)_L \times U(1)$. 
(See e.g. Ref.~\cite{Cvetic:2001tj,Blumenhagen:2002gw} and references 
therein.)\footnote{
See for the Pati-Salam model in heterotic orbifold 
models e.g. Ref.~\cite{Kobayashi:2004ud}, 
where $SU(4)\times SU(2)_L \times SU(2)_R$ 
is broken to the standard gauge group by vacuum expectation values 
of scalar fields, $(4,1,2)$ and $(\bar 4,1,2)$, 
while in the intersecting D-brane models 
$SU(4)\times SU(2)_L \times SU(2)_R$ is broken by splitting 
D-branes, that is, vacuum expectation values of adjoint scalar 
fields.}
At the end of this section, we give a comment on 
breaking of $SU(4) \times SU(2)_L \times SU(2)_R$
to $SU(3) \times SU(2)_L \times U(1)$.

In both cases with and without orbifolding, 
the total number of chiral matter fields is a product 
of the numbers of zero-modes corresponding to the 
$i$-th $T^2$ for $i=1,2,3$.
That is, the three generations are realized in the 
models, where the $i$-th $T^2$ has three 
zero-modes while each of the other tori has 
a single zero-mode.
Thus, there are two types of flavor structures.
That is, in one type the three zero-modes 
corresponding to both left-handed matter fields 
$(N_a,\bar N_b)$ and right-handed matter fields 
$(\bar N_a, N_c)$ appear in the same $i$-th $T^2$, 
while each of the other tori has a single zero-mode 
for $(N_a,\bar N_b)$ as well as $(\bar N_a,N_c)$.
In the other type, three zero-modes of 
 $(N_a,\bar N_b)$ and $(\bar N_a,N_c)$ are 
originated from different tori.
The Yukawa coupling for 4D effective field theory 
is evaluated by the following overlap integral 
of zero-mode wavefunctions \cite{Green:1987mn}
\begin{eqnarray}
Y_{ij} &=& \int d^6y \psi_{Li}(y) \psi_{Rj}(y)  \phi_H(y), 
\nonumber
\end{eqnarray}
where $\psi_L(y)$, $\psi_R(y)$ and $\phi_H(y)$ 
denote zero-mode wave-functions of 
the left-handed, right-handed matter fields 
and Higgs field, respectively.
Note that the integral corresponding to each torus is 
factorized in the Yukawa coupling.
In the second type of flavor structure, 
one obtains the following 
form of Yukawa matrices,
\begin{eqnarray}
Y_{ij} &=& a_i b_j,
\nonumber
\end{eqnarray}
at the tree-level, because the flavor structure 
of left-handed and right-handed matter fields are 
originated from different tori.
This matrix, $Y_{ij}$, has rank one 
and that is not phenomenologically interesting, 
unless certain corrections appear.
Hence, we concentrate on the first type 
of the flavor structure.
In the first type, the flavor structure is 
originated from the single torus, where 
both three zero-modes of $(N_a,\bar N_b)$ 
and $(\bar N_a,N_c)$ appear.
We assign this torus with the first torus. 
On the other hand, 
the other tori, the second and third tori,  
do not lead to flavor-dependent aspects.
That is, Yukawa matrices are obtained 
as the following form,
\begin{eqnarray}
Y_{ij} &=& a^{(2)}a^{(3)}a^{(1)}_{ij},
\nonumber
\end{eqnarray}
where the structure of $a^{(1)}_{ij}$ is determined 
by only the first torus corresponding to three 
zero-modes $(N_a,\bar N_b)$ and $(\bar N_a,N_c)$ 
while the other tori contribute to overall factors 
$a^{(2)}$ and $a^{(3)}$.
Thus, we concentrate on the single torus, 
where both of three zero-modes $(N_a,\bar N_b)$ 
and $(\bar N_a,N_c)$ appear, i.e. the first torus.

\begin{table}[t]
\begin{center}
\begin{tabular}{c|ccc} 
    &  $\lambda^{ab}$ 
    &  $\lambda^{ca}$ 
    &  $\lambda^{bc}$ 
\\ \hline 
I   &  even & even & even  \\
II  &  even & odd  & odd   \\
II' &  odd  & even & odd   \\
III &  odd  & odd  & even  \\
\end{tabular}
\end{center}
\caption{Possible patterns of wavefunctions 
with non-vanishing Yukawa couplings 
for the first torus.}
\label{class}
\end{table}

Zero-mode wavefunctions are classified into even and odd modes 
under the $Z_2$ twist.
Only even or odd modes remain through 
the orbifold projection.
Furthermore, the 4D Yukawa couplings are 
non-vanishing for combinations among 
(even, even, even) wavefunctions and 
(even, odd, odd) wavefunctions, while Yukawa couplings 
vanish for combinations among 
(even, even, odd) wavefunctions and 
(odd, odd, odd) wavefunctions.
Thus, we study only the former case
with non-vanishing Yukawa couplings, 
that is, the combinations among 
(even, even, even) wavefunctions and 
(even, odd, odd) wavefunctions.
Hence, we are interested in four types of 
combinations of wavefunctions for the first torus, 
as shown in Table \ref{class}.
The II' type of combinations is obtained by 
exchanging the left and right-handed matter fields 
in the II type.
Thus, we study explicitly the three types, I, II and III.

We can realize three even zero-modes when $|I^{(1)}_{ab}|=4,5$,
as shown in Table \ref{even-odd-zero-modes}.
On the other hand, three odd zero-modes can appear 
when $|I^{(1)}_{ab}|=7,8$.
Furthermore, the consistency condition on magnetic fluxes 
requires 
\begin{eqnarray}
|I_{bc}^{(1)}| &=& |I_{ab}^{(1)}|\pm |I_{ca}^{(1)}|.
\nonumber
\end{eqnarray}
Thus, the number of Higgs and higgsino fields are 
constrained.
Table \ref{higgs} shows all of possible magnetic fluxes 
for the three types, I, II and III.
The fourth and fifth columns of the table show 
possible sizes of magnetic fluxes for $|I_{bc}^{(1)}|$ 
and the number of zero-modes corresponding to 
the Higgs fields.
As a result, flavor structures of our models 
with Yukawa couplings are classified into 20 classes.
However, the model with 
$(|I_{ab}^{(1)}|,|I_{ca}^{(1)}|,|I_{bc}^{(1)}|) = (5,7,2)$ 
has no zero-modes for the Higgs fields.
Thus, we do not consider this case, 
but we will study the other 19 classes 
in Table \ref{higgs}.
Therefore, we study possible flavor structures explicitly 
by deriving the coupling selection rule 
and evaluating values of Yukawa couplings in 
these 19 classes.
That is the purpose of the next section.

\begin{table}[t]
\begin{center}
\begin{tabular}{c|ccc|c} 
    &  $|I_{ab}^{(1)}|$ 
    &  $|I_{ca}^{(1)}|$ 
    &  $|I_{bc}^{(1)}|$ 
    &  the numbers of \\ 
    &  &  &  & Higgs zero-modes \\ \hline 
I   &  4 & 4 & 8  & 5  \\
    &  4 & 4 & 0  & 1  \\
    &  4 & 5 & 9  & 5  \\
    &  4 & 5 & 1  & 5  \\
    &  5 & 5 & 10 & 6  \\
    &  5 & 5 & 0  & 1  \\ \hline
II  &  4 & 7 & 11 & 5  \\
    &  4 & 7 & 3  & 1  \\
    &  4 & 8 & 12 & 5  \\
    &  4 & 8 & 4  & 1  \\
    &  5 & 7 & 12 & 5  \\
    &  5 & 7 & 2  & 0  \\
    &  5 & 8 & 13 & 6  \\
    &  5 & 8 & 3  & 1  \\ \hline
III &  7 & 7 & 14 & 8  \\
    &  7 & 7 & 0  & 1  \\
    &  7 & 8 & 15 & 8  \\
    &  7 & 8 & 1  & 1  \\
    &  8 & 8 & 16 & 9  \\
    &  8 & 8 & 0  & 1  \\
\end{tabular}
\end{center}
\caption{The number of Higgs fields of $(T^2)^1$ 
with non-vanishing Yukawa couplings.}
\label{higgs}
\end{table}

Before explicit study on flavor structures of 
19 classes in the next section, 
we give a comment on breaking of $SU(4)\times SU(2)_L \times SU(2)_R$.
At any rate, we need the $SU(3) \times SU(2)_L \times U(1)$ 
gauge group at low-energy.
When the magnetic flux and orbifold projections lead to 
the $SU(4)\times SU(2)_L \times SU(2)_R$ gauge group from 
$U(8)$ as we have discussed so far, 
we need further breaking of $SU(4)\times SU(2)_L \times SU(2)_R$ 
to $SU(3) \times SU(2)_L \times U(1)$.
Such breaking can be realized by assuming 
non-vanishing vacuum expectation values (VEVs) 
of Higgs fields like adjoint scalar fields for 
$SU(4)$ and $SU(2)_R$ and/or 
bi-fundamental scalar fields like $(4,1,2)$ and 
$(\bar 4, 1,2)$ on fixed points.
Note that our models have degree of freedom to add any modes at 
the fixed points from the viewpoint of point particle field theory.
The above breaking may affect the structure of Yukawa matrices 
as higher dimensional operators.
However, we will show results on Yukawa matrices 
without such corrections.

Alternatively, magnetic fluxes and/or 
orbifold projections break $U(8)$ into 
$U(3)\times U(1)_1 \times U(2)_L \times U(1)_2 \times U(1)_3$.
The gauge group $U(3)\times U(1)_1$ would correspond to $U(4)$ and 
$U(1)_2 \times U(1)_3$ would correspond to $U(2)_R$. 
We assume that all the bi-fundamental matter fields under 
$U(3) \times U(1)_1$, i.e. extra colored modes, are projected out.
The bi-fundamental matter fields for 
$U(3) \times U(1)_2$ and $U(3) \times U(1)_3$ 
correspond to up and down sectors of right-handed quarks,
respectively.
Similarly, up and down sectors of Higgs fields 
and right-handed charged leptons and neutrinos 
are obtained.
In this case, the classification of this section and 
patterns of Yukawa matrices, which will be studied 
in the next section and Appendix, are available for 
up-sector and down-sector quarks as well as the 
lepton sector.
However, the up sector and down sector can 
correspond to different classes of Table \ref{higgs}.
On the other hand, the up sector and 
down sector correspond to 
the same class in Table \ref{higgs}, when 
the  $SU(4)\times SU(2)_L \times SU(2)_R$ 
is broken by VEVs of Higgs fields 
on fixed points as discussed above.

\subsubsection{Yukawa couplings in three generation models}

Following \cite{Cremades:2004wa,DiVecchia:2008tm}, 
first we show computation of Yukawa interactions 
on the torus with the magnetic flux.
Omitting the gauge structure and spinor structure,
the Yukawa coupling among left, right-handed matter fields 
and Higgs field corresponding to three zero-mode 
wavefunctions, $\Theta^{i,M_1}(z)$,  $\Theta^{j,M_2}(z)$ 
and $(\Theta^{k,M_3}(z))^*$, is written by
\begin{eqnarray}
Y_{ijk} &=& c \int dz d\bar{z}
\Theta^{i,M_1}(z) \Theta^{j,M_2}(z) (\Theta^{k,M_3}(z))^*,
\label{eq:yukawa}
\end{eqnarray}
where $z=x_4+\tau y_5$, 
$M_1 \equiv I^{(1)}_{ab}$, 
$M_2 \equiv I^{(1)}_{ca}$, 
$M_3 \equiv I^{(1)}_{cb}$ 
and $c$ is a flavor-independent contribution due to the other tori.
Note that $M_1+M_2=M_3$. 
Because of the gauge invariance, not the wavefunction 
$\Theta^{k,M_3}(z)$, but $(\Theta^{k,M_3}(z))^*$ 
appears in the Yukawa coupling~\cite{Cremades:2004wa}. 

By using the formula of the $\vartheta$ function, 
\begin{eqnarray}
\lefteqn{
\vartheta
\left[\begin{array}{c} r/N_1 \\ 0 \end{array} \right]
\left(z_1,N_1\tau \right)
\,\times\, 
\vartheta
\left[\begin{array}{c} s/N_2 \\ 0 \end{array} \right]
\left(z_2,N_2\tau \right)
}
\nonumber \\ &=& 
\sum_{m\in \mathcal{Z}_{N_1+N_2}}
\vartheta
\left[\begin{array}{c} \frac{r+s+N_1m}{N_1+N_2} \\ 0 \end{array}
\right]
\left(z_1+z_2,\tau(N_1+N_2) \right) 
\nonumber \\ && \qquad\qquad \,\times\,
\vartheta
\left[\begin{array}{c} \frac{N_2r-N_1s+N_1N_2m}{N_1N_2(N_1+N_2)} \\ 0 
\end{array} \right]
\left(z_1N_2-z_2N_1,\tau N_1N_2(N_1+N_2) \right),
\nonumber
\end{eqnarray}
we can decompose $\Theta^{i,M_1}(z)\Theta^{j,M_2}(z)$ as 
\begin{eqnarray}
\Theta^{i,M_1}(z)\,\Theta^{j,M_2}(z) 
&=&\sum_{m\in \mathcal{Z}_{M_3}}\Theta^{i+j+M_1m,M_3}(z) 
\,\times\, 
\vartheta
\left[\begin{array}{c} 
\frac{M_2i-M_1j+M_1M_2m}{M_1M_2M_3} \\ 0
\end{array} \right]
\left(0,\tau M_1M_2M_3 \right). 
\nonumber
\end{eqnarray}
Wavefunctions satisfy the orthogonal condition
\begin{eqnarray}
\int dzd\bar{z}\,\Theta^{i,M}\,(\Theta^{j,M})^* &=& \delta_{ij}.
\nonumber
\end{eqnarray}
Then, the integral of three wavefunctions is represented by 
\begin{eqnarray}
Y_{ijk} 
&=& c \int
dzd\bar{z}\,\Theta^{i,M_1}\,\Theta^{j,M_2}(\Theta^{k,M_3})^* 
\nonumber \\ &=& 
c \sum_{m=0}^{|M_3|-1} \vartheta 
\left[\begin{array}{c} 
\frac{M_2i-M_1j+M_1M_2m}{M_1M_2M_3} \\ 0
\end{array} \right]
\left(0,\tau M_1M_2M_3 \right) 
\times \delta_{i+j+M_1m,\,k+M_3 \ell},
\nonumber
\end{eqnarray}
where $\ell =$ integer.

Thus, we have the selection rule for allowed Yukawa
couplings as 
\begin{eqnarray}
i+j=k,
\nonumber
\end{eqnarray}
where $i, j$ and $k$ are defined up to mod 
$M_1, M_2$ and $M_3$, respectively.\footnote{
See for the selection rule in intersecting D-brane models, e.g. 
Ref.~\cite{Cremades:2003qj,Higaki:2005ie}.}
In addition, the Yukawa coupling $Y_{ijk}$, in particular 
its flavor-dependent part, is written by the $\vartheta$ function.
When $g.c.d.(M_1,M_3)=1$, a single $\vartheta$ function appears 
in $Y_{ijk}$.
When $g.c.d.(M_1,M_3)=g \neq 1$, $g$ terms appear in $Y_{ijk}$ 
as 
\begin{eqnarray}
 Y_{ijk} = c \sum_{n=1}^g \vartheta \left[ \begin{array}{c}
    {M_2k - M_3j + M_2 M_3 \ell_0 \over M_1 M_2 M_3 } + {n \over g} \\
    0 
\end{array}
  \right](0,\tau M_1 M_2 M_3),
\nonumber
\end{eqnarray}
where $\ell_0$ is an integer corresponding 
to a particular solution of $M_3 l_0 = M_1 m_0 + i + j - k $ 
with integer $m_0$.

Zero-mode wavefunctions on the orbifold with the magnetic flux are 
obtained as even or odd linear combinations of 
wavefunctions on the torus with the magnetic 
flux (\ref{eq:even-odd-zero-modes}).
Thus, it is straightforward to extend the above computations of
Yukawa couplings on the torus to Yukawa couplings 
on the orbifold.
As a result, Yukawa couplings on the orbifold are obtained 
as proper linear combinations of Yukawa couplings 
on the torus, i.e. linear combinations of $\vartheta$ functions.
Here we introduce the following short notation for 
the Yukawa coupling,
\begin{eqnarray}
\eta_{N} &=& \vartheta
\left[\begin{array}{c} 
\frac{N}{M} \\ 0
\end{array} \right] \left(0,\tau M \right), 
\label{eq:notation-y}
\end{eqnarray}
where 
\begin{eqnarray}
M &=& M_1M_2M_3. \qquad 
\nonumber
\end{eqnarray}
Since the value of $M$ is unique in one model, 
we omit the value of $M$ as well as $\tau$ 
for a compact presentation of long equations.

Four models in Table \ref{higgs} has $|I^{(1)}_{bc}|=0$, 
where the Higgs zero-mode corresponds to the even 
function, that is, the constant profile.
We can repeat the above calculation for this case,  
that is, the case where, one of wavefunctions in (\ref{eq:yukawa}),
e.g. 
$\Theta^{i,M_1}(z)$ is constant.
As a result, the Yukawa matrix is proportional to 
the $(3 \times 3)$ unit matrix, $Y_{jk} = c' \delta_{jk}$.
That is not realistic.
Thus, we will not consider such models.

At any rate, we can apply the above selection rule and $\eta_{N}$ 
for 20 classes of models, which have been classified 
in section \ref{sec:three-gene}, 
in order to analyze explicitly all of possible 
patterns of Yukawa matrices.
In the next subsection, we show one example of 
Yukawa matrix among 20 classes of models.
In Appendix C, we show all of possible Yukawa 
matrices for 15 classes of models in Table \ref{higgs} 
except models with $I^{(1)}_{bc}=0$ and the model without zero-modes 
for the Higgs fields. 

\begin{table}[t]
\begin{center}
\begin{tabular}{c|c|c|c}
& $L_i (\lambda^{ab})$ & $R_j(\lambda^{ca})$ & $H_k(\lambda^{bc})$ 
\\ \hline
0 & 
$\frac{1}{\sqrt{2}}\left(\Theta^{1,7}-\Theta^{6,7}\right)$ & 
$\frac{1}{\sqrt{2}}\left(\Theta^{1,7}-\Theta^{6,7}\right)$ & 
$\Theta^{0,14}$ \\  
1 & 
$\frac{1}{\sqrt{2}}\left(\Theta^{2,7}-\Theta^{5,7}\right)$ & 
$\frac{1}{\sqrt{2}}\left(\Theta^{2,7}-\Theta^{5,7}\right)$ & 
$\frac{1}{\sqrt{2}}\left(\Theta^{1,14}+\Theta^{13,14}\right)$ \\  
2 & 
$\frac{1}{\sqrt{2}}\left(\Theta^{3,7}-\Theta^{4,7}\right)$ & 
$\frac{1}{\sqrt{2}}\left(\Theta^{3,7}-\Theta^{4,7}\right)$ & 
$\frac{1}{\sqrt{2}}\left(\Theta^{2,14}+\Theta^{12,14}\right)$ \\  
3 & - & - & 
$\frac{1}{\sqrt{2}}\left(\Theta^{3,14}+\Theta^{11,14}\right)$ \\  
4 & - & - & 
$\frac{1}{\sqrt{2}}\left(\Theta^{4,14}+\Theta^{10,14}\right)$ \\  
5 & - & - & 
$\frac{1}{\sqrt{2}}\left(\Theta^{5,14}+\Theta^{9,14}\right)$ \\  
6 & - & - & 
$\frac{1}{\sqrt{2}}\left(\Theta^{6,14}+\Theta^{8,14}\right)$ \\  
7 & - & - & 
$\Theta^{7,14}$ \\  
\end{tabular}
\end{center}
\caption{Zero-mode wavefunctions in the 7-7-14 model.}
\label{7-7-14model}
\end{table}

\subsubsection{An illustrating example: 7-7-14 model}
\label{ssec:7-7-14}

Let us study the model with 
$(|I^{(1)}_{ab}|,|I^{(1)}_{ca}|,|I^{(1)}_{bc}|)=(7,7,14)$.
Following Table \ref{higgs}, we consider the combination of zero-mode 
wavefunctions, where zero-modes of 
left and right-handed matter fields and Higgs fields correspond to 
odd, odd and even wavefunctions, respectively. 
Their wavefunctions are shown in Table~\ref{7-7-14model}.
Hereafter, for concreteness, 
we denote left and right-handed matter fields 
and Higgs fields by $L_i$, $R_j$ and $H_k$, respectively.
This model has eight zero-modes for Higgs fields.

Then, their Yukawa couplings $Y_{ijk}L_iR_jH_k$ are 
written by 
\begin{eqnarray}
Y_{ijk}H_k &=& 
  y_{ij}^0 H_0 + y_{ij}^1 H_1 + y_{ij}^2 H_2 
+ y_{ij}^3 H_3 + y_{ij}^4 H_4 + y_{ij}^5 H_5
+ y_{ij}^6 H_6 + y_{ij}^7 H_7, 
\nonumber
\end{eqnarray}
where 
\begin{eqnarray}
y_{ij}^0 &=& 
\left( \begin{array}{ccc}
-y_c & 0 & 0 \\
0 & -y_e & 0 \\
0 & 0 & -y_g 
\end{array} \right), \quad
y_{ij}^1 \ = \ 
\left( \begin{array}{ccc}
0 & -\frac{1}{\sqrt{2}}y_d & 0 \\
-\frac{1}{\sqrt{2}}y_d & 0 & -\frac{1}{\sqrt{2}}y_f \\
0 & -\frac{1}{\sqrt{2}}y_f & \frac{1}{\sqrt{2}}y_h  
\end{array} \right), 
\nonumber \\
y_{ij}^2 &=& 
\left( \begin{array}{ccc}
\frac{1}{\sqrt{2}}y_a & 0 &  -\frac{1}{\sqrt{2}}y_e  \\
0 & 0 & \frac{1}{\sqrt{2}}y_g \\ 
-\frac{1}{\sqrt{2}}y_e & \frac{1}{\sqrt{2}}y_g & 0   
\end{array} \right), \quad 
y_{ij}^3 \ = \ 
\left( \begin{array}{ccc}
0 & \frac{1}{\sqrt{2}}y_b  &  \frac{1}{\sqrt{2}}y_f \\
\frac{1}{\sqrt{2}}y_b & \frac{1}{\sqrt{2}}y_h & 0 \\ 
\frac{1}{\sqrt{2}}y_f & 0 & 0 
\end{array} \right), 
\nonumber \\
y_{ij}^4 &=& 
\left( \begin{array}{ccc}
0 & \frac{1}{\sqrt{2}}y_g  &  \frac{1}{\sqrt{2}}y_c \\
\frac{1}{\sqrt{2}}y_g & \frac{1}{\sqrt{2}}y_a & 0 \\ 
\frac{1}{\sqrt{2}}y_c & 0 & 0 
\end{array} \right), \quad 
y_{ij}^5 \ = \ 
\left( \begin{array}{ccc}
\frac{1}{\sqrt{2}}y_h & 0 &  -\frac{1}{\sqrt{2}}y_d  \\
0 & 0 & \frac{1}{\sqrt{2}}y_b \\ 
-\frac{1}{\sqrt{2}}y_d & \frac{1}{\sqrt{2}}y_b & 0   
\end{array} \right), 
\nonumber \\
y_{ij}^6 &=& 
\left( \begin{array}{ccc}
0 & -\frac{1}{\sqrt{2}}y_e & 0 \\
-\frac{1}{\sqrt{2}}y_e & 0 & -\frac{1}{\sqrt{2}}y_c \\
0 & -\frac{1}{\sqrt{2}}y_c & \frac{1}{\sqrt{2}}y_a  
\end{array} \right), \quad 
y_{ij}^7 \ = \ 
\left( \begin{array}{ccc}
-y_f & 0 & 0 \\
0 & -y_d & 0 \\
0 & 0 & -y_b 
\end{array} \right), 
\label{eq:yij}
\end{eqnarray}
and 
\begin{eqnarray}
y_a &=& 
\eta_0 +2\eta_{98} +2\eta_{196} +2\eta_{294}, 
\nonumber\\
y_b &=& 
\eta_{7} + \eta_{91} + \eta_{105} + \eta_{189} 
+ \eta_{203} + \eta_{287} +\eta_{301}, 
\nonumber \\ 
y_c &=& 
\eta_{14} + \eta_{84} + \eta_{112} + \eta_{182} 
+ \eta_{210} + \eta_{280} +\eta_{308},  
\nonumber \\
y_d &=& 
\eta_{21} + \eta_{77} + \eta_{119} + \eta_{175} 
+ \eta_{217} + \eta_{273} +\eta_{315}, 
\nonumber \\ 
y_e &=& 
\eta_{28} + \eta_{70} + \eta_{126} + \eta_{168} 
+ \eta_{224} + \eta_{266} +\eta_{322},  
\nonumber \\
y_f &=& 
\eta_{35} + \eta_{63} + \eta_{133} + \eta_{161} 
+ \eta_{231} + \eta_{259} +\eta_{329}, 
\nonumber \\ 
y_g &=& 
\eta_{42} + \eta_{56} + \eta_{140} + \eta_{154} 
+ \eta_{238} + \eta_{252} +\eta_{336},  
\nonumber \\
y_h &=& 
2\eta_{49} +2\eta_{147} +2\eta_{245} + \eta_{343}. 
\nonumber
\end{eqnarray}
Here we have used the short notation $\eta_N$ defined in
Eq.~(\ref{eq:notation-y}) 
with the omitted value $M=M_1M_2M_3=686$.

\subsubsection{Numerical examples in 7-7-14 model}
\label{ssec:numerical}

\begin{figure}[t]
\begin{center}
\begin{minipage}{0.45\linewidth}
\begin{center}
\epsfig{file=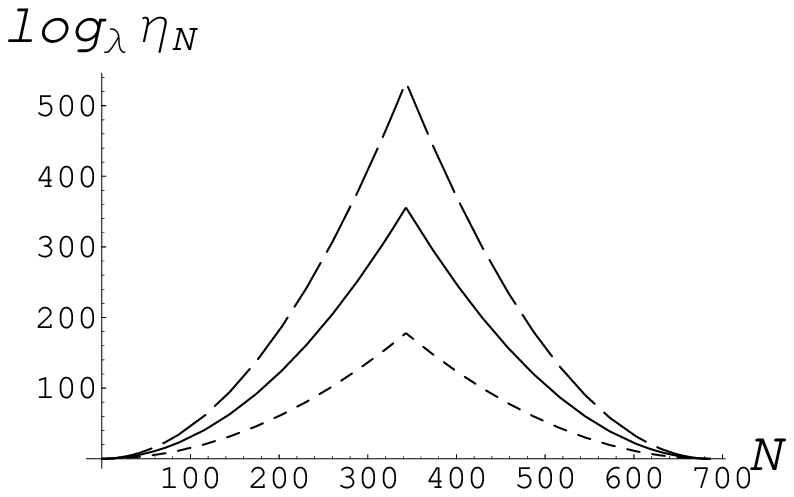,width=\linewidth}
\end{center}
\end{minipage}
\begin{minipage}{0.45\linewidth}
\begin{center}
\epsfig{file=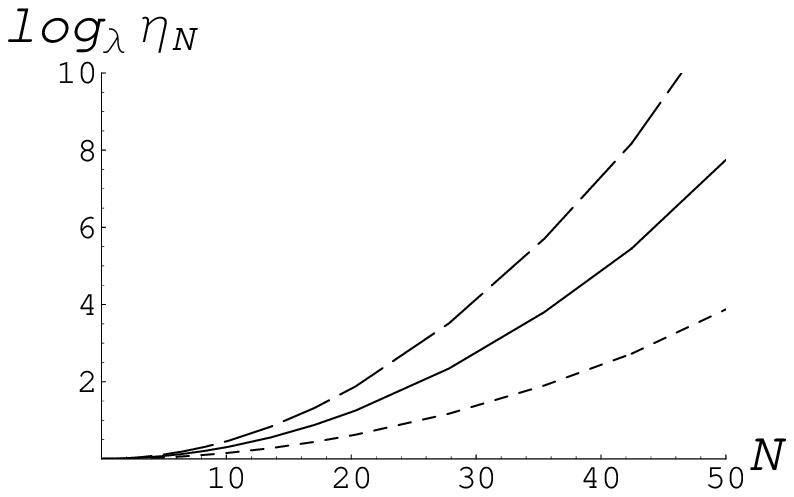,width=\linewidth}
\end{center}
\end{minipage}
\end{center}
\caption{The $N$-dependence of $\log_\lambda \eta_N$ 
in the 7-7-14 model ($M=686$), where $\lambda=0.22$ 
is chosen to the Cabibbo angle. 
The solid, dashed and dotted curves correspond to 
$\tau=i$, $1.5i$ and $0.5i$, respectively. 
Note that $\eta_N$ has a periodicity 
$\eta_{N+nM}=\eta_N$ with an integer $n$.}
\label{fig:eta}
\end{figure}

Here, we give examples of numerical studies by using the 7-7-14 model, 
which is discussed in the previous subsection.
For such studies, the numerical values of $\eta_N$ defined in 
Eq.~(\ref{eq:notation-y}) are useful. 
The $N$-dependence of $\eta_N$ is shown in Fig.~\ref{fig:eta}. 

We assume that both the up-sector and the down-sector of quarks 
as well as their Higgs fields have the Yukawa matrix, 
which is led in the 7-7-14 model.
Such situation is realized in the case that 
we start with the $U(8)$ gauge group and break  
it to $U(4) \times U(2)_L \times U(2)_R$ by the magnetic flux, 
and then the Pati-Salam gauge group is 
broken to the Standard gauge group 
by assuming VEVs of Higgs fields on fixed points.
Alternatively, we break the $U(8)$ gauge group 
to $U(3)\times U(1)_1 \times U(2)_L \times U(1)_2 \times U(1)_3$ 
by magnetic fluxes and orbifold projections 
as discussed in section 3.
Then, both the up-sector and down-sector 
of quarks can correspond to the Yukawa matrix 
led in the 7-7-14 model, although the 
up-sector and down-sector can generically 
correspond to different patterns of Yukawa matrices.
In both cases, VEVs of the up-sector and down-sector 
Higgs fields are independent.

First, we consider the case that 
VEVs of $H_d^6$, $H_d^7$ and $H_u^0$ 
are non-vanishing and the other VEVs vanish.
In this case, the relevant Yukawa couplings are 
\begin{eqnarray}
Y^u_{ijk}H_k &=& 
\left( \begin{array}{ccc}
-y_c & &  \\
& -y_e & \\
 & & -y_g 
\end{array} \right) H_u^0, 
\nonumber \\
Y^d_{ijk}H_k &=& 
\left( \begin{array}{ccc}
-y_f H_d^7& -\frac{1}{\sqrt{2}}y_e H_d^6& 0 \\
-\frac{1}{\sqrt{2}}y_e H_d^6& -y_d H_d^7& -\frac{1}{\sqrt{2}}y_c H_d^6
-\\
0 & -\frac{1}{\sqrt{2}}y_c H_d^6& \frac{1}{\sqrt{2}}y_aH_d^6 -y_b
-H_d^7
\end{array} \right). 
\nonumber
\end{eqnarray}
Let us assume $\langle H_d^6\rangle =-\langle H_d^7 \rangle $ for
their VEVs.
Then, quark mass ratios are obtained from these matrices as
\begin{eqnarray}
(m_u, m_c, m_t)/m_t &\sim& 
(7.6 \times 10^{-4},\, 6.8 \times 10^{-2},\, 1.0), 
\nonumber \\
(m_d, m_s, m_b)/m_b &\sim& 
(7.5 \times 10^{-4},\, 5.1\times 10^{-2},\, 1.0), 
\nonumber
\end{eqnarray}
for $\tau =i$.
Furthermore, the mixing angles are obtained as 
\begin{eqnarray}
|V_{CKM}| &\sim& 
\left( \begin{array}{ccc}
0.97 & 0.24 & 0.0025 \\
0.24 & 0.95 & 0.20 \\
0.046 & 0.19 & 0.98 
\end{array} \right). 
\nonumber
\end{eqnarray}
Similarly, for $\tau = 1.5 i$, 
quark mass ratios are obtained as 
\begin{eqnarray}
(m_u, m_c, m_t)/m_t &\sim& 
(2.1 \times 10^{-5},\, 1.8 \times 10^{-2},\, 1.0), 
\nonumber \\
(m_b,m_s,m_d)/m_b &\sim& 
(1.4 \times 10^{-4},\, 1.7 \times 10^{-2},\, 1.0), 
\nonumber
\end{eqnarray}
and the mixing angles are obtained as 
\begin{eqnarray}
|V_{CKM}| &\sim& 
\left( \begin{array}{ccc}
0.99 & 0.13 & 0.00029 \\
0.13 & 0.98 & 0.13 \\
0.017 & 0.13 & 0.99 
\end{array} \right). 
\nonumber
\end{eqnarray}

Let us consider another type of VEVs.
We assume that VEVs of $H_u^0$, $H_u^2$, $H_d^1$ and 
$H_d^7$ are non-vanishing and the other VEVs vanish.
Furthermore, we consider the case with 
$\langle H_u^0 \rangle = -\langle H_u^2\rangle $ 
and $\langle H_d^1 \rangle = \langle H_d^7 \rangle /3$. 
In this case, the mass ratios are given by 
\begin{eqnarray}
(m_u, m_c, m_t)/m_t &\sim& 
(2.9 \times 10^{-5},\, 2.5 \times 10^{-2},\, 1.0), 
\nonumber \\
(m_d, m_s, m_b)/m_b &\sim& 
(4.4 \times 10^{-3},\, 0.18,\, 1.0), 
\nonumber
\end{eqnarray}
for $\tau =i$, and the mixing angles are given by 
\begin{eqnarray}
|V_{CKM}| &\sim& 
\left( \begin{array}{ccc}
0.98 & 0.22 & 0.018 \\
0.22 & 0.98 & 0.0014 \\
0.017 & 0.0052  & 1.0
\end{array} \right). 
\nonumber
\end{eqnarray}
Similarly, for $\tau =1.5i$ the mass ratios 
and the mixing angles are given by
\begin{eqnarray}
(m_u, m_c, m_t)/m_t &\sim& 
(5.6\times 10^{-6},\, 4.7 \times 10^{-3},\, 1.0), 
\nonumber \\
(m_d, m_s, m_b)/m_b &\sim& 
(3.3 \times 10^{-3},\, 7.1 \times 10^{-2},\, 1.0), 
\nonumber 
\end{eqnarray}
\begin{eqnarray}
|V_{CKM}| &\sim& 
\left( \begin{array}{ccc}
0.98 & 0.22  & 0.0034 \\
0.22 & 0.98 & 0.000081 \\
0.0033 & 0.00081 & 1.0 
\end{array} \right). 
\nonumber
\end{eqnarray}
Thus, these values can realize experimental values 
of quark masses and mixing angles at a certain level 
by using a few parameters, i.e. $\tau$ and 
a couple of VEVs of Higgs fields.
If we consider more non-vanishing VEVs of Higgs fields, 
we could obtain more realistic values.
For example, we assume that VEVs of $H_u^0$, $H_u^1$, $H_u^2$, 
$H_d^1$ and $H_d^7$ are non-vanishing and they satisfy 
$-\langle H_u^0 \rangle = \langle H_u^1\rangle = \langle H_u^2\rangle
$ 
and $\langle H_d^1 \rangle = - \langle H_d^7 \rangle /2$
while the other VEVs vanish. 
For $\tau = 1.5 i$, we realize the mass ratios, 
$m_u/m_t \sim 2.7 \times 10^{-5}$, $m_c/m_t \sim 3.5 \times 10^{-3}$, 
 $m_d/m_b \sim 7.3 \times 10^{-3}$ and  $m_s/m_b \sim 7.5 \times
10^{-2}$, 
and mixing angles, $V_{us} \sim 0.2$, $V_{cb} \sim 0.03$ and 
$V_{ub} \sim 0.006$. 
When we consider more non-vanishing VEVs of Higgs fields, 
it is possible to derive completely realistic values.
Similarly, we can study other classes of models 
and they have a rich flavor structure.

\subsection{Orbifold background with non-Abelian Wilson line}

Since we have obtained the explicit wavefunctions on the 
torus with non-Abelian Wilson lines, 
we can easily extend above analysis to the case with non-Abelian
Wilson line. 
we study the $T^2/Z_2$ orbifold,
which is constructed by dividing $T^2$ by the $Z_2$ projection $z \to
-z$.
Furthermore, we require the field projection of periodic or
anti-periodic boundary conditions with consistent of $Z_2$ orbifold,
\begin{eqnarray}
 \Psi(-y_4,-y_5)=P \Psi(y_4,y_5), 
\end{eqnarray}
where $P$ is $+1$ or $-1$.
One can show that the matter wavefunctions satisfy the following
property
\begin{eqnarray}\label{eq:z2-property} 
\Psi_{pq}^j(-y_4,-y_5)=\Psi_{-p,-q}^{-j}(y_4,y_5).
\end{eqnarray}
Obviously, in the case with Abelian Wilson, this result reduces to 
$\Psi(-y_4,-y_5)^j=\Psi(y_4,y_5)^j$.
For the case with $k=1$, this relation holds, 
because every sector of $(p,q)$ is related by the boundary
conditions, so the labels $(p,q)$ have no meaning.
However $k \ne 1$ case, they have $k \times M$ independent zero-modes 
and we symbolically denote them by $\Psi^{j,\tilde j}$ 
($j=0,1,...,M-1$ and $\tilde j=0,1,...,k-1$).
For example, in the case with $n_a=n_b=3$, we may 
use the following notations
\begin{eqnarray} 
\Psi_{00}^j,\  \Psi_{11}^j,\  \Psi_{22}^j \to  \Psi^{j,\tilde j=0},
\nonumber \\
\Psi_{01}^j,\  \Psi_{12}^j,\  \Psi_{20}^j \to  \Psi^{j,\tilde j=1},
\\
\Psi_{02}^j,\  \Psi_{10}^j,\  \Psi_{21}^j \to  \Psi^{j,\tilde
j=2}. \nonumber 
\end{eqnarray}
where $\tilde j= p-q$ mod $K$.
Then, the above property (\ref{eq:z2-property}) can be written as
\begin{eqnarray} 
\Psi^{j,\tilde j}(-y_4,-y_5)=\Psi^{-j,-\tilde j}(y_4,y_5).
\end{eqnarray}
Then the even and odd wave-functions are easily obtained.
For the case with $M=3$, there are $3\times 3$ independent fields 
and they are divided into the following even and odd wavefunctions
\begin{eqnarray} 
{\rm{even}} &:& \Psi^{0,0},\  
\Psi^{1,0} +\Psi^{2, 0},\  
\Psi^{0, 1}+\Psi^{0, 2},\ 
\Psi^{1,1}+\Psi^{2,2},
\Psi^{2,1}+\Psi^{1,2}, \nonumber \\ 
{\rm{odd}}  &:& \Psi^{1,0}-\Psi^{2,0},\
\Psi^{1,1}-\Psi^{2,2},\  
\Psi^{2,1}-\Psi^{1,2}.
\end{eqnarray}
Note that these represent the wavefunctions
e.g. $\Psi_{12}^1+\Psi_{21}^2$ by $\Psi^{1,1}+\Psi^{2,2}$. 
As examples, the zero-mode numbers of even and odd wavefunctions 
for smaller values of $k$ and $M$ 
are shown in Table \ref{tab:orbifold}.

\begin{table}[httbp]
\begin{center}
$k=1$
\begin{tabular}{c|cccccc} 
   $M$ & 1 & 2 & 3 & 4 & 5 & 6 \\ \hline 
even & 1 & 2 & 2 & 3 & 3 & 4 \\
odd  & 0 & 0 & 1 & 1 & 2 & 2 
\end{tabular} \ \ \ 
$k=2$
\begin{tabular}{c|cccccc} 
   $M$ & 1 & 2 & 3 & 4 & 5 & 6 \\ \hline 
even & 2 & 4 & 4 & 6 & 6 & 8 \\
odd  & 0 & 0 & 2 & 2 & 4 & 4 
\end{tabular}
\\ 
$k=3$
\begin{tabular}{c|cccccc} 
   $M$ & 1 & 2 & 3 & 4 & 5 & 6  \\ \hline 
even & 2 & 4 & 5 & 7 & 8 & 10 \\
odd  & 1 & 2 & 4 & 5 & 7 & 8 
\end{tabular} \ \ \ 
$k=4$
\begin{tabular}{c|cccccc} 
   $M$ & 1 & 2 & 3 &  4 &  5 &  6 \\ \hline 
even & 3 & 6 & 7 & 10 & 11 & 14 \\
odd  & 1 & 2 & 5 &  6 &  9 & 10 
\end{tabular}
\end{center}
\caption{The numbers of even and odd zero-modes}
\label{tab:orbifold}
\end{table}

Yukawa couplings as well as higher order couplings 
can be computed on the orbifold background by 
overlap integrals of wavefunctions in a way 
similar to the torus models.

\subsection{Further direction to orbifold}
Here, we study orbifold models with magnetic fluxes and 
Wilson lines.
The $T^2/Z_2$ orbifold is constructed by identifying 
$z \sim -z$ on $T^2$.
We also embed the $Z_2$ twist into the gauge space as $P$.
Note that under the $Z_2$ twist, magnetic flux background is 
invariant.
That is, we have no constraint on magnetic fluxes due to 
orbifolding.  
Furthermore, zero-mode wavefunctions satisfy
\begin{eqnarray}
\Theta^{j,M}(-z) = \Theta^{M-j,M}(z).
\end{eqnarray}
Note that $\Theta^{0,M}(z) = \Theta^{M,M}(z)$.
Hence, the $Z_2$ eigenstates are written as \cite{Abe:2008fi}
\begin{eqnarray}\label{eq:wv-orbi}
\Theta^{j,M}_\pm(z) = \frac{1}{\sqrt 2} 
\left(\Theta^{j,M}(z) \pm \Theta^{M-j,M}(z) \right),
\end{eqnarray}
for $j\neq 0, M/2, M$.
The wavefunctions $\Theta^{j,M}(z)$ for $j=0, M/2$ are 
the $Z_2$ eigenstates with the $Z_2$ even parity.
Either of $\Theta^{j,M}_+(z)$ and $\Theta^{j,M}_-(z)$ 
is projected out by the orbifold projection.
Odd wavefunctions can also correspond to 
massless modes in the magnetic flux background, 
although on the orbifold without magnetic flux odd modes 
always correspond to massive modes, but not 
massless modes.
Before orbifolding, the number of zero-modes is equal to 
the magnetic flux $M$.
For example, we have to choose $M=3$ in order to realize 
the three families.
On the other hand, the number of zero-modes on the orbifold also 
depends on the boundary conditions under the $Z_2$ twist, 
even or odd functions.
For $M=$ even, 
the number of zero-modes with even (odd) functions 
are equal to $M/2+1$ $(M/2-1)$.
For $M=$ odd, 
the number of zero-modes with even and odd functions 
are equal to $(M+1)/2$ and $(M-1)/2$, respectively.

Now, let us introduce Wilson lines~\cite{Ibanez:1987xa}
with some different types of gauge groups. 
For example, we consider $U(1)_a \times SU(2)$ theory.
Then we introduce magnetic flux in $U(1)_a$ like
Eq.~(\ref{eq:magne-a}).
In addition, we embed the $Z_2$ twist $P$ into the $SU(2)$ 
gauge space.
For example, we consider the $SU(2)$ doublet 
\begin{eqnarray}
\left(
\begin{array}{c}
\lambda_{1/2} \\ \lambda_{-1/2} 
\end{array}
\right),
\end{eqnarray}
with the $U(1)_a$ charge $q_a$.
We embed the $Z_2$ twist $P$ in the gauge space as 
\begin{eqnarray}\label{eq:P-su2}
P = \left( 
\begin{array}{cc}
0 & 1 \\
1 & 0 
\end{array}
\right),
\end{eqnarray}
for the doublet.
Obviously, we can diagonalize $P$ as 
$P'={\rm diag} (1,-1)$, 
if there is no Wilson line 
along the other $SU(2)$ directions.
However, we introduce a Wilson line along the 
Cartan direction of $SU(2)$, i.e, the following 
direction
\begin{eqnarray}
 \left( 
\begin{array}{cc}
1 & 0 \\
0 & -1 
\end{array}
\right),
\end{eqnarray}
in the $P$ basis.
Thus, we use the above basis for $P$.
For the $SU(2)$ gauge sector, there is no effect due to 
the magnetic flux.
Then, the situation is the same as one on 
the orbifold without magnetic flux.
The $SU(2)$ gauge group is broken completely, that is, 
all of $SU(2)$ vector multiplets become massive.

Before orbifolding, the $SU(2)$ is not broken and 
both $\lambda_{1/2}$ and $\lambda_{-1/2}$
have $M=q_am_a$ independent zero-modes, which we denote by 
$\Theta^{j,M}_{1/2}(z)$ and $\Theta^{j,M}_{-1/2}(z)$, respectively.
Here, we have put the indices, $1/2$ and $-1/2$ in order to 
make it clear that they correspond to $\lambda_{1/2}$ and
$\lambda_{-1/2}$, respectively.
However, the form of wavefunctions are the same, i.e. 
$\Theta^{j,M}_{1/2}(z) = \Theta^{j,M}_{-1/2}(z)$.
When we impose the orbifold boundary conditions with 
the above $P$ in (\ref{eq:P-su2}), 
the zero-modes on the orbifold without Wilson lines are written as 
\begin{eqnarray}\label{eq:Z2-state}
\frac{1}{\sqrt 2} \left(\Theta^{j,M}_{1/2}(z) 
+ \Theta^{M-j,M}_{-1/2}(z) \right),
\end{eqnarray}
for $j=0,\cdots, M-1$.
Note that there are $M$ independent zero-modes.
It may be useful to explain remaining zero-modes in the basis 
for $P'$.
Before orbifolding, both $\lambda'_{1/2}$ and $\lambda'_{-1/2}$
have $M=q_am_a$ independent zero-modes in the basis for $P'$.
Then by orbifolding with $P'$, even modes corresponding 
to $\Theta^{j,M}_+ (z)$ remain for $\lambda'_{1/2}$, 
while  $\lambda'_{-1/2}$ has only odd modes $\Theta^{j,M}_- (z)$.
Their total number is equal to $M$.

Then, we introduce the Wilson lines along the 
Cartan direction in the basis for $P$.
The corresponding zero-mode wavefunctions are shifted as 
\begin{eqnarray}\label{eq:wf-cWL}
\frac{1}{\sqrt 2} \left(\Theta^{j,M}_{1/2}(z+C^b/2M) 
+ \Theta^{M-j,M}_{-1/2}(z-C^b/2M) \right),
\end{eqnarray}
for $j=0,\cdots, M-1$, where $C^b$ is a continuous parameter.
Note that $\lambda_{1/2}$ and $\lambda_{-1/2}$
have opposite charges under the $SU(2)$ Cartan element.
Then, their wavefunctions are shifted to opposite directions   
by the same Wilson lines $C^b$
as $\Theta^{j,M}_{1/2}(z+C^b/2M) $ and
$\Theta^{j,M}_{-1/2}(z-C^b/2M)$.
We can also consider the $Z_2$ twist $P$ in the doublet such that 
the following wavefunction 
\begin{eqnarray}\label{eq:wf-cWL-}
\frac{1}{\sqrt 2} \left(\Theta^{j,M}_{1/2}(z+C^b/2M) -
 \Theta^{M-j,M}_{-1/2}(z-C^b/2M) \right),
\end{eqnarray}
remains.

The above aspect would be important to applications for 
particle phenomenology.
We compute Yukawa couplings among 
two $SU(2)$ doublet fields and a singlet field.
We assume that two $SU(2)$ doublet fields have $U(1)_a$ charges 
$q^1_a$ and $q^2_a$, while the singlet field has the $U(1)_a$ 
charge $q^3_a$.
We introduce the magnetic flux $m_a$ in $U(1)_a$ and 
the same $SU(2)$ Wilson line as the above. 
Then, the zero-mode wavefunctions of 
two $SU(2)$ doublets and the singlet can be obtained 
on the orbifold as 
\begin{eqnarray}\label{eq:3wf-cWL}
\left(
\begin{array}{c}
\lambda^1_{1/2} \\ \lambda^1_{-1/2} 
\end{array}
\right) &:& \frac{1}{\sqrt 2} \left(\Theta^{i,M_1}_{1/2}(z+C^b/2M_1) 
+ \Theta^{M_1-i,M_1}_{-1/2}(z-C^b/2M_i) \right), \nonumber \\
\left(
\begin{array}{c}
\lambda^2_{1/2} \\ \lambda^2_{-1/2} 
\end{array}
\right) &:& \frac{1}{\sqrt 2} \left(\Theta^{j,M_2}_{1/2}(z+C^b/2M_2) 
+ \Theta^{M_2-j,M_2}_{-1/2}(z-C^b/2M_j) \right), \\
\lambda^3_0 ~~~~~~~&:& \frac{1}{\sqrt 2} \left(\Theta^{k,M_3}_{0}(z) 
+ \Theta^{M_3-k,M_3}_{0}(z) \right)^*,  \nonumber
\end{eqnarray}
where $M_i=q^i_am_a$.
Note that the Wilson line $C^b$ has no effect on the wavefunctions 
of the $SU(2)$ singlet field $\lambda^3_0$ because 
$\lambda^3_0$ has no $SU(2)$ charges.
Here we have taken the same orbifold projection $P$ as 
Eq.~(\ref{eq:P-su2}), but we can study other orbifold 
projections.
Then, their Yukawa couplings are obtained by the following 
overlap integral,
\begin{eqnarray}\label{eq:yukawa-orbi}
& & \frac{1}{2\sqrt 2}\int d^2z \{
\Theta^{i,M_1}_{1/2}(z+C^b/2M_1)\Theta^{j,M_2}_{-1/2}(z-C^b/2M
_2) 
\left( \Theta^{k,M_3}_0(z)+\Theta^{M_3-k,M_3}_{0}(z) \right)^*
\nonumber \\
& & +
\Theta^{M_1-i,M_1}_{-1/2}(z-C^b/2M_1)\Theta^{M_2-j,M_2}_{1/2}(z+C^b/2M_2) 
\left( \Theta^{k,M_3}_0(z)+\Theta^{M_3-k,M_3}_{0}(z) \right)^* \}.
\end{eqnarray}
This integral is computed as 
\begin{eqnarray}\label{eq:yukawa-orbi-2}
& & 
\sum_{m \in \Z_{M_3}}  \delta_{i+j+M_1m,k}  
 \times \jtheta{{M_2 i - M_1 j + M_1 M_2 m \over M_1 M_2 M_3}
 \\ 0}(C^b(M_1+M_2)/2),\tau M_1 M_2 M_3), \nonumber   \\
& & + \sum_{m \in \Z_{M_3}}  \delta_{i+j+M_1m,-k}  
 \times \jtheta{{M_2 i - M_1 j + M_1 M_2 m \over M_1 M_2 M_3}
 \\ 0}(C^b(M_1+M_2)/2),\tau M_1 M_2 M_3), \nonumber   \\
& & + \sum_{m \in \Z_{M_3}}  \delta_{-i-j+M_1m,k}  
 \times \jtheta{{-M_2 i + M_1 j + M_1 M_2 m \over M_1 M_2 M_3}
 \\ 0}(-C^b(M_1+M_2)/2),\tau M_1 M_2 M_3),   \\
& & +  \sum_{m \in \Z_{M_3}}  \delta_{-i-j+M_1m,-k}  
 \times \jtheta{{-M_2 i + M_1 j + M_1 M_2 m \over M_1 M_2 M_3}
 \\ 0}(-C^b(M_1+M_2)/2),\tau M_1 M_2 M_3), \nonumber 
\end{eqnarray}
up to the normalization factor $N_1N_2/(2 \sqrt 2 N_3)$,
where the Kronecker delta $\delta_{i+j+M_1m,k}$ 
in the first term means $i+j+M_1m =k$ 
mod $M_3$ and others have the same meaning.
Obviously, the result depends non-trivially on 
the Wilson line $C^b$.
Thus, the Wilson lines have important effects on 
the Yukawa couplings.

For comparison, we study another dimensional representations, 
e.g. a triplet
\begin{eqnarray}
\left(
\begin{array}{c}
\lambda_{1} \\ \lambda_{0} \\\lambda_{-1} 
\end{array}
\right),
\end{eqnarray}
with the $U(1)_a$ charge $q_a$.
Suppose that we embed the $Z_2$ twist $P$ in the three dimensional 
gauge space as 
\begin{eqnarray}\label{eq:P-su2-3}
P = \left( 
\begin{array}{ccc}
0 & 0 & 1 \\
0 & 1 & 0 \\
1 & 0 & 0
\end{array}
\right),
\end{eqnarray}
for the triplet.
Then, zero-modes on the orbifold are written as 
\begin{eqnarray}
 & &  \Theta^{j,M}_{1}(z) 
+ \Theta^{M-j,M}_{-1}(z) , \nonumber \\
 & &\Theta^{j,M}_{0}(z) 
+ \Theta^{M-j,M}_{0}(z) ,
\end{eqnarray}
up to the normalization factor $1/\sqrt{2}$.
The former corresponds to $\lambda_{1}$ and $\lambda_{-1}$ and 
there are $M$ zero-modes.
The latter corresponds to $\lambda_0$ and there are 
$(M/2+1)$ zero-modes and $(M+1)/2$ zero-modes when $M$ is even and
odd, 
respectively.
When we introduce the continuous Wilson lines along the Cartan 
direction, the wavefunctions of these zero-modes shift as 
\begin{eqnarray}
 & &  \Theta^{j,M}_{1}(z+C_b/M) 
+ \Theta^{M-j,M}_{-1}(z-C_b/M) , \nonumber \\
 & &\Theta^{j,M}_{0}(z) 
+ \Theta^{M-j,M}_{0}(z) ,
\end{eqnarray}
up to the normalization factor $1/\sqrt{2}$.

Similarly to the above,  
here let us compute the Yukawa couplings among two $SU(2)$ doublets 
corresponding to Eq.~(\ref{eq:3wf-cWL}) and 
the triplet $(\lambda^3_1,\lambda^3_0,\lambda^3_{-1})$.
In particular, we compute the couplings including 
$\lambda^3_1$ and $\lambda^3_{-1}$, whose zero-mode wavefunctions are 
obtained by
\begin{eqnarray}
 & & \frac{1}{2\sqrt 2} \left( \Theta^{k,M_3}_{1}(z+C_b/M_3) 
+ \Theta^{M_3-k,M_3}_{-1}(z-C_b/M_3) \right)^*, 
\end{eqnarray}
with $M_3=q^3_a m_a$ after orbifolding.
Their Yukawa couplings are obtained by the following 
overlap integral,
\begin{eqnarray}\label{eq:yukawa-orbi-3}
& & \frac{1}{2\sqrt 2}\int d^2z \{
\Theta^{i,M_1}_{1/2}(z+C^b/2M_1)\Theta^{j,M_2}_{1/2}(z +C^b/2M_2) 
\left( \Theta^{M_3-k,M_3}_{-1}(z-C_b/M_3) \right)^*
\nonumber \\
& & + \Theta^{M_1-i,M_1}_{-1/2}(z-C^b/2M_1)
\Theta^{M_2-j,M_2}_{-1/2}(z-C^b/2M_2) 
\left( \Theta^{k,M_3}_1(z+C_b/M_3)\right)^* \}.
\end{eqnarray}
This integral is computed as 
\begin{eqnarray}\label{eq:yukawa-orbi-4}
& & 
\sum_{m \in \Z_{M_3}}  \delta_{i+j+M_1m,k}  
 \times \jtheta{{M_2 i - M_1 j + M_1 M_2 m \over M_1 M_2 M_3}
 \\ 0}(C^b(M_2 -M_1)/2),\tau M_1 M_2 M_3), \nonumber   \\
& & +  \sum_{m \in \Z_{M_3}}  \delta_{-i-j+M_1m,-k}  
 \times \jtheta{{-M_2 i + M_1 j + M_1 M_2 m \over M_1 M_2 M_3}
 \\ 0}(C^b(M_1 -M_2)/2),\tau M_1 M_2 M_3),  
\end{eqnarray}
up to the normalization factor $N_1N_2/(2 \sqrt 2 N_3)$.
This result is different from Eq.~(\ref{eq:yukawa-orbi-2}), 
in particular from the viewpoint of 
Wilson line dependence.
Thus, the Wilson lines have phenomenologically important effects, 
depending on the directions of Wilson lines and the 
representations of matter fields.

We can extend the above analysis to larger gauge groups.
Here, we show a rather simple example.
We consider $U(1)_a \times SU(3)$ theory with 
the magnetic flux in $U(1)_a$ like Eq.~(\ref{eq:magne-a}).
Then, we consider the $SU(3)$ triplet, 
\begin{eqnarray}
\left(
\begin{array}{c}
\lambda_{0} \\ \lambda_{1/2} \\ \lambda_{-1/2} 
\end{array}
\right),
\end{eqnarray}
with the $U(1)_a$ charge $q_a$, where 
the subscripts $(0,1/2,-1/2)$ denote the $U(1)_b$ charge 
along one of $SU(3)$ Cartan directions.
Now, we embed the $Z_2$ twist $P$ in the gauge space as 
\begin{eqnarray}\label{eq:P-su3}
P = \left( 
\begin{array}{ccc}
1 & 0 & 0 \\
0 & 0 & 1 \\
0 & 1 & 0
\end{array}
\right),
\end{eqnarray}
for the triplet.
In addition, we introduce the Wilson line $C^b$ along 
the $U(1)_b$ direction.
The gauge group is broken as $SU(3) \rightarrow U(1)$.\footnote{
This remaining $U(1)$ symmetry might be anomalous.
If so, the remaining $U(1)$ would also be broken by 
the Green-Schwarz mechanism.}
There are $M$ zero-modes for linear combinations of 
$\lambda_{1/2}$ and $\lambda_{-1/2}$ with the wavefunctions,
\begin{eqnarray} 
\Theta^{j,M}_{1/2}(z+C^b/2M) +
 \Theta^{M-j,M}_{-1/2}(z-C^b/2M),
\end{eqnarray}
 up to the normalization factor.
Also, the zero-modes for $\lambda_{0}$ are written as 
\begin{eqnarray} 
\Theta^{j,M}_{0}(z) +
 \Theta^{M-j,M}_{0}(z),
\end{eqnarray}
up to the normalization factor.
The number of zero-modes is equal to 
$(M/2+1)$ and $(M+1)/2$ when $M$ is even and odd, respectively.
Thus, the situation is almost the same as the 
above $SU(2)$ case with the triplet.
Although the above example is rather simple, 
we can consider various types of breaking for larger groups.
For example, when the gauge group includes 
two or more $SU(2)$ subgroups, 
we could embed the $Z_2$ twist in two of $SU(2)$'s and 
introduce independent Wilson lines along their Cartan directions.
Similarly, we can investigate such models and other types of  
various embedding of $P$ and Wilson lines.

In section 2.2, we have considered 10D theory on $T^6$.
Also, we can consider the $T^6/Z_2$ orbifold, where the $Z_2$ 
twist acts e.g. 
\begin{eqnarray}
Z_2 \ : \ z_1 \rightarrow -z_1, \qquad 
z_2 \rightarrow -z_2, \qquad 
z_3 \rightarrow z_3.
\end{eqnarray}
For $T^2_1$ and $T^2_2$, we can introduce the type 
of Wilson lines, which we have considered in this section, 
while for $T^2_3$ we can introduce the type of Wilson lines, 
which are considered in the previous section.
Then, we have a richer structure of models on 
the $T^6/Z_2$ orbifold.
Furthermore, we could consider another independent $Z'_2$ twist 
as 
\begin{eqnarray}
Z_2 \ : \ z_1 \rightarrow -z_1, \qquad 
z_2 \rightarrow z_2, \qquad 
z_3 \rightarrow -z_3,
\end{eqnarray}
on the $T^6/(Z_2 \times Z'_2)$ orbifold.
In this case, we can consider another independent 
embedding $P'$ of $Z'_2$ twist on the gauge space.
Using these two $Z_2$ twist embedding and Wilson lines, we 
could construct various types of models.
For example, when the gauge group includes two or 
more $SU(2)$ subgroups, we could embed $P$ on one of $SU(2)$ 
and $P'$ on other $SU(2)$ and introduce independent 
Wilson lines along their Cartan directions.
Other various types of model building would be possible.
Thus, it would be interesting to study such model building 
elsewhere.


\newpage

\section{Anomalies for field theory and string theory}

In the previous section, we have seen some phenomenological 
interesting features of non-Abelian discrete symmetries 
which can appear dynamically as flavor symmetries.
In general, symmetries at the tree-level can be 
broken by quantum effects, i.e. anomalies.
Anomalies of continuous symmetries, in particular 
gauge symmetries, have been studied well.
Here we study about anomalies of non-Abelian discrete symmetries.
For our purpose, the path integral approach is convenient.
Thus, we use Fujikawa's method \cite{Fujikawa:1979ay,Fujikawa:1980eg} 
to derive anomalies of discrete symmetries.

\subsection{General formula for anomalies}

Here we consider a gauge theory with a (non-Abelian) gauge group $G_g$ 
and a set of fermions 
$\Psi = [\psi^{(1)}, \cdots, \psi^{(M)}]$.
Then, we assume that their Lagrangian is invariant under 
the following chiral transformation,
\begin{eqnarray}\label{eq:U-trans}
\Psi(x) \rightarrow U \Psi(x),
\end{eqnarray}
with $U=exp(i\alpha P_L)$ and $\alpha = \alpha^A T_A$, 
where $T_A$ denote the generators of the transformation and 
$P_L$ is the left-chiral projector.
It is not always necessary for above transformation to be a gauge
transformation. 
The fermions $\Psi(x)$ are the (irreducible) $M$-plet 
representation $\boldsymbol{R}^M$.
For the moment, we suppose that $\Psi(x)$ correspond to a
(non-trivial) 
singlet under the flavor symmetry while they correspond to 
the $\boldsymbol{R}^M$ 
representation under the gauge group $G_g$.
Since the generator $T_A$ as well as $\alpha$ is represented on 
$\boldsymbol{R}^M$ as a $(M \times M)$ matrix, 
we use the notation, $T_A(\boldsymbol{R}^M)$ 
and $\alpha(\boldsymbol{R}^M)=\alpha^AT_A(\boldsymbol{R}^M)$.

The anomaly appears in Fujikawa's method from 
the transformation of the path integral measure as the 
Jacobian,
\begin{equation}
\mathcal{D}\Psi \mathcal{D}\bar \Psi \rightarrow J(\alpha)
\mathcal{D}\Psi \mathcal{D}\bar \Psi,\ \ 
J(\alpha)=
\exp \left( i \int d^4x \mathcal{A}(x;\alpha)  \right).
\end{equation}
The anomaly function  $\mathcal{A}$ decomposes into a gauge part 
and a gravitational  part
\cite{AlvarezGaume:1983ig,AlvarezGaume:1984dr,Fujikawa:1986hk}
\begin{equation}\label{eq:AnomalyFunctionA}
 \mathcal{A}~=~
  \mathcal{A}_\mathrm{gauge}+\mathcal{A}_\mathrm{grav}
\;.
\end{equation}
The gauge part is given by 
\begin{equation}
 \mathcal{A}_\mathrm{gauge}(x;\alpha)~=~\frac{1}{32\,\pi^2}
 \mathrm{Tr}\left[\alpha(\boldsymbol{R}^M)\,F^{\mu\nu}(x)\,
\widetilde{{F}}_{\mu\nu}(x)\right]\;,
\end{equation}
where ${F}^{\mu\nu}$ denotes the field strength of 
the gauge fields, ${F}_{\mu\nu}=[D_\mu,D_\nu]$, and 
$\widetilde{{F}}_{\mu\nu}$ denotes its dual, 
$\widetilde{{F}}^{\mu\nu}=
\varepsilon^{\mu\nu\rho\sigma}{F}_{\rho\sigma}$.
The trace `Tr' runs over all internal indices. 
When the transformation corresponds to a continuous symmetry, 
this anomaly can be calculated by the triangle diagram with 
external lines of two gauge bosons and one current corresponding 
to the symmetry for Eq.~\ref{eq:U-trans}.

Similarly, the gravitation part is obtained as 
\begin{equation}
 \mathcal{A}_\mathrm{grav}
 ~=~
 -\mathcal{A}_\mathrm{grav}^\mathrm{Weyl\:fermion} 
   \mathrm{tr} \left[\alpha(\boldsymbol{R}^{(M)})\right]\;, 
\label{eq:Agrav}
\end{equation}
where 
`$\mathrm{tr}$' is the trace for the matrix $(M \times M)$
$T_A(\boldsymbol{R}^M)$. 
The contribution  of a single Weyl fermion to the
gravitational anomaly is given by 
\begin{equation}
 \mathcal{A}_\mathrm{grav}^\mathrm{Weyl\:fermion}  ~=~
 \frac{1}{384 \pi^2}\,\frac{1}{2}\,
 \varepsilon^{\mu\nu\rho\sigma}\, R_{\mu\nu}{}^{\lambda\gamma}\,
 R_{\rho\sigma\lambda\gamma}\;.
\end{equation}
When other sets of $M_i$-plet fermions $\Psi_{M_i}$ 
are included in a theory, 
the total gauge and gravity anomalies are obtained as their
summations, 
$\sum_{\Psi_{M_i}} \mathcal{A}_\mathrm{gauge}$ and 
$\sum_{\Psi_{M_i}} \mathcal{A}_\mathrm{grav}$.

For the evaluation of these anomalies, it is useful to recall the
index theorems~\cite{AlvarezGaume:1983ig,AlvarezGaume:1984dr}, which
imply
\begin{subequations}\label{eq:indexTheorems}
\begin{eqnarray}
 \int d^4 x\, \frac{1}{32\pi^2}\,\varepsilon^{\mu\nu\rho\sigma}\,
 	F_{\mu\nu}^a\,F_{\rho\sigma}^b\,
	\mathrm{tr}\left[\mathsf{t}_a\,\mathsf{t}_b\right]
 & \in & \mathbbm{Z}
 \;, \label{eq:index1}\\*
 \frac{1}{2}\,\int d^4 x\, \frac{1}{384 \pi^2}\,\frac{1}{2}\,
 \varepsilon^{\mu\nu\rho\sigma}\, R_{\mu\nu}{}^{\lambda\gamma}\,
 R_{\rho\sigma\lambda\gamma}
 & \in & \mathbbm{Z}
 \;,\label{eq:index2}
\end{eqnarray}
\end{subequations}
where $\mathsf{t}_a$ are generators of $G_g$ 
in the fundamental representation.
We use the convention that 
$\mathrm{tr}[\mathsf{t}_a\,\mathsf{t}_b]=\frac{1}{2}\delta_{ab}$.
The factor $\frac12$ in eq.~(\ref{eq:index2}) follows from Rohlin's
theorem~\cite{Rohlin:1959}, as
discussed in~\cite{Csaki:1997aw}.
Of course, these indices are independent of each other.
The path integral includes all possible configurations 
corresponding to different index numbers.

First of all, we study anomalies of the continuous $U(1)$ symmetry.
We consider a theory with a (non-Abelian) gauge symmetry $G_g$ as well
as the continuous $U(1)$ symmetry, which may be gauged.
This theory include fermions with $U(1)$ charges, $q^{(f)}$ and 
representations $\boldsymbol{R}^{(f)}$.
Those anomalies vanish if and only if the Jacobian is trivial, 
i.e. $J(\alpha)=1$ for an arbitrary value of $\alpha$. 
Using the index theorems, one can find that 
the anomaly-free conditions require
\begin{equation}\label{eq:A[U(1)-G-G]}
 A_\mathrm{U(1)-G_g-G_g}~\equiv~
  \sum_{\boldsymbol{R}^{(f)}} q^{(f)}\,T_2(\boldsymbol{R}^{(f)})~=~0,
\end{equation}
for the mixed $U(1)- G_g-G_g$ anomaly, and 
\begin{equation}
 A_\mathrm{U(1)-\mathrm{grav}-\mathrm{grav}}
 ~\equiv~ 
 \sum_{f} q^{(f)}~=~0,
\end{equation}
the $U(1)$--gravity--gravity anomaly.
Here, $T_2(\boldsymbol{R}^{(f)})$ is the Dynkin index of 
the $\boldsymbol{R}^f$ representation, i.e.
\begin{equation}
\mathrm{tr}\left[\mathsf{t}_a\left(\boldsymbol{R}^{(f)}\right)\,
 	\mathsf{t}_b\left(\boldsymbol{R}^{(f)}\right)\right]
      ~=~\delta_{ab}
T_2(\boldsymbol{R}^{(f)})
 \;.
\end{equation}

\subsection{Discrete flavor symmetry anomalies}

Next, let us study anomalies of the abelian discrete symmetry, 
i.e. the $Z_N$ symmetry.
For the $Z_N$ symmetry, we write $\alpha =2 \pi Q_N/N$, 
where $Q_N$ is the $Z_N$ charge operator and its eigenvalues are 
integers.
Here we denote $Z_N$ charges of fermions as $q^{(f)}_N$.
Then we can evaluate the $Z_N - G_g-G_g$ and $Z_N$-gravity-gravity 
anomalies as the above $U(1)$ anomalies.
However, the important difference is that $\alpha$ takes a discrete 
value.
Then, the anomaly-free conditions, i.e., $J(\alpha)=1$ for a discrete 
transformation, require 
\begin{eqnarray}\label{eq:A_Z_N-G-G}
 A_{Z_{N}-G_g-G_g} & = & \frac{1}{N}\sum_{\boldsymbol{R}^{(f)}}
 q^{(f)_N} \,
 \big(2\,T_2(\boldsymbol{R}^{(f)})\big) \in \mathbbm{Z}
 \;,
\end{eqnarray}
for the $Z_N - G_g-G_g$  anomaly, and 
\begin{eqnarray}\label{eq:A_Z_N-grav-grav}
 A_{Z_{N}-\mathrm{grav}-\mathrm{grav}} &= & 
 \frac{2}{N}
 \sum_{f} q^{(f)}_N \,
 \dim\boldsymbol{R}^{(f)} \in \mathbbm{Z} \; ,
\end{eqnarray}
for the $Z_N$-gravity-gravity anomaly.
These anomaly-free conditions reduce to 
\begin{subequations}\label{eq:ZNconditions}
\begin{eqnarray}
& &  \sum_{\boldsymbol{R}^{(f)}} q^{(f)}_N \,
 T_2(\boldsymbol{R}^{(f)}) ~=~ 0\mod
 N/2\;,\label{eq:condition-gauge}\\
& & 
 \sum_{f}
 q^{(f)}_N \,\dim\boldsymbol{R}^{(f)} ~=~0\mod
 N/2\;.\label{eq:condition-grav}
\end{eqnarray}
\end{subequations}
Note  that the $Z_2$ symmetry is always free from 
the $Z_2$-gravity-gravity anomaly.

Finally, we study anomalies of non-Abelian discrete symmetries $G$.
A discrete group $G$ consists of the finite number of elements,
$g_i$. Hence, the non-Abelian discrete symmetry is anomaly-free 
if and only if the Jacobian is vanishing for the transformation 
corresponding to each element $g_i$.
Furthermore, recall that $(g_i)^{N_i} =1$. 
That is, each element $g_i$ in the non-Abelian discrete group 
generates a $Z_{N_i}$ symmetry.
Thus, the analysis on non-Abelian discrete anomalies reduces to 
one on Abelian discrete anomalies.
One can take the field basis such that $g_i$ is represented in 
a diagonal form.
In such a basis, each field has a definite $Z_{N_i}$ charge, 
$q^{(f)}_{N_i}$.
The anomaly-free conditions for the $g_i$ transformation 
are written as 
\begin{subequations}\label{eq:ZNi-conditions}
\begin{eqnarray}
& &  \sum_{\boldsymbol{R}^{(f)}} q^{(f)}_{N_i} \,
 T_2(\boldsymbol{R}^{(f)}) ~=~ 0\mod
 N_i/2\;,\label{eq:condition-gauge-i}\\
& & 
 \sum_{f}
 q^{(f)}_{N_i} \,\dim\boldsymbol{R}^{(f)} ~=~0\mod N_i/2\;.
\label{eq:condition-grav-i}
\end{eqnarray}
\end{subequations}
If these conditions are satisfied for all of $g_i \in G$, 
there are no anomalies of the full non-Abelian symmetry $G$.
Otherwise, the non-Abelian symmetry is broken completely 
or partially to its subgroup by quantum effects.

In principle, we can investigate anomalies of non-Abelian 
discrete symmetries $G$ following the above procedure.
However, we give a practically simpler way to analyze those anomalies.
Here, we consider again the transformation similar to
\eqref{eq:U-trans} 
for a set of fermions 
$\Psi = [\psi^{(1)}, \cdots, \psi^{(Md_\alpha)}]$, 
which correspond to the $\boldsymbol{R}^M$ irreducible representation 
of the gauge group $G_g$ 
and the ${\bf r}^\alpha$ irreducible representation of 
the non-Abelian discrete symmetry $G$ with the dimension $d_\alpha$.
Let $U$ correspond to one of group elements $g_i \in G$, 
which is represented by the matrix $D_\alpha(g_i)$ on ${\bf
r}^\alpha$.
Then, the Jacobian is proportional to its determinant, $\det D(g_i)$. 
Thus, the representations with $\det D_\alpha(g_i) = 1$ 
do not contribute to anomalies.
Therefore, the non-trivial Jacobian, i.e. anomalies 
are originated from representations with $\det D_\alpha(g_i) \neq 1$.
Note that $\det D_\alpha(g_i) = \det D_\alpha(gg_ig^{-1})$ 
for $g \in G$, that is, 
the determinant is constant in a conjugacy class.
Thus, it would be useful to calculate the determinants 
of elements on each irreducible representation.
Such a determinant for the  conjugacy class $C_i$ can be written by 
\begin{eqnarray}
\det(C_i)_\alpha= e^{2\pi i q_{\hat N_i}^\alpha/\hat N_i},
\end{eqnarray}  
on the 
irreducible representation ${\bf r}^\alpha$.
Note that $\hat N_i$ is a divisor of $N_i$, where 
$N_i$ is the order of $g_i$ in the conjugacy class $C_i$, i.e.
$g^{N_i}=e$, such that $q_{\hat N_i}^\alpha$ are normalized 
to be integers for all of the 
irreducible representation ${\bf r}^\alpha$.
We consider the $Z_{\hat N_i}$ symmetries and 
their anomalies.
Then, we obtain the anomaly-free conditions similar to 
\eqref{eq:ZNi-conditions}.
That is, the anomaly-free conditions for the 
conjugacy classes $C_i$ are written as 
\begin{subequations}\label{eq:ZhatNi-conditions}
\begin{eqnarray}
& &  \sum_{\boldsymbol{r}^{(\alpha)},\boldsymbol{R}^{(f)}} 
q^{\alpha(f)}_{\hat N_i} \,
 T_2(\boldsymbol{R}^{(f)}) 
~=~ 0\mod \hat N_i/2\;,\label{eq:condition-gauge-hati}\\
& & 
 \sum_{\alpha. f}
 q^{\alpha(f)}_{\hat N_i} \,\dim\boldsymbol{R}^{(f)} ~=~0\mod \hat
 N_i/2\; ,
\label{eq:condition-grav-hati}
\end{eqnarray}
\end{subequations}
for the theory including fermions with the $\boldsymbol{R}^{(f)}$ 
representations of the gauge group $G_g$ and the ${\bf r}^{\alpha(f)}$
representations of the flavor group $G$, which correspond to 
the $Z_{\hat N_i}$ charges, $q^{\alpha (f)}_{\hat N_i}$.
Note that the fermion fields with the $d_\alpha$-dimensional 
representation ${\bf r}^\alpha$ contribute on these anomalies, 
$q^{\alpha(f)}_{\hat N_i} \,
 T_2(\boldsymbol{R}^{(f)})$ and 
$q^{\alpha(f)}_{\hat N_i} \,\dim\boldsymbol{R}^{(f)}$, but not 
$d_\alpha q^{\alpha(f)}_{\hat N_i} \,
 T_2(\boldsymbol{R}^{(f)})$ and 
$d_\alpha q^{\alpha(f)}_{\hat N_i} \,\dim\boldsymbol{R}^{(f)}$.
If these conditions are satisfied for all of conjugacy classes of 
$G$, the full non-Abelian symmetry $G$ is free from anomalies.
Otherwise, the non-Abelian symmetry is broken by 
quantum effects.
As we see as follows, in concrete examples, the above 
anomaly-free conditions often lead to the same conditions 
between different conjugacy classes. 
Note, when $\hat N_i=2$, the symmetry is always free from 
the mixed gravitational anomalies.  
We study explicitly more for concrete groups.

\vskip .5cm
{$\bullet$ \bf  ${\bf D_{4}}$ }

\vskip .2cm
We study anomalies of $D_4$. 
As shown in appendix \ref{sec:D4}, 
the $D_4$ group has the four 
singlets, ${\bf 1}_{\pm \pm}$ and one doublet ${\bf 2}$.
All of the $D_4$ elements can be written as 
products of two elements, $Z$ and $C$.
Their determinants on ${\bf 2}$ are obtained as 
$\det (Z) =-1$ and $\det (C) = -1$.
Similarly, we can obtain determinants of $Z$ and $C$ on 
four singlets, ${\bf 1}_{\pm \pm}$.
Indeed, four singlets are classified by values of 
$\det (Z)$ and $\det (C)$, that is, 
$\det (Z) =1$ for  ${\bf 1}_{+\pm}$, 
$\det (Z) =-1$ for  ${\bf 1}_{-\pm}$,
$\det (C) =1$ for  ${\bf 1}_{\pm+}$ and 
$\det (C) =-1$ for  ${\bf 1}_{\pm-}$.
Those determinants are summarized in Table
\ref{tab:D4}.
That implies that two $Z_2$ symmetries can be anomalous.
One $Z_2$ corresponds to $Z$ and the other $Z_2'$ corresponds to 
$C$.
Under these $Z_2 \times Z_2'$ symmetry, each representation 
has the following behavior,
\begin{eqnarray}
Z_2 ~{\rm even} &:& {\bf 1}_{+\pm},\\                   
Z_2 ~{\rm odd} &:& {\bf 1}_{-\pm}, \quad {\bf 2},
\end{eqnarray}
\begin{eqnarray}
Z'_2 ~{\rm even} &:& {\bf 1}_{\pm +},\\                   
Z'_2 ~{\rm odd} &:& {\bf 1}_{\pm -}, \quad {\bf 2}.
\end{eqnarray}
Then, the anomaly-free conditions are written as 
\begin{eqnarray}
\sum_{{\bf 1}_{- \pm}}\sum_{\boldsymbol{R}^{(f)}}
T_2(\boldsymbol{R}^{(f)})
+ \sum_{{\bf 2}}\sum_{\boldsymbol{R}^{(f)}} \,
 T_2(\boldsymbol{R}^{(f)}) 
~=~ 0\mod 1\;  ,
\end{eqnarray}
for the $Z_2-G_g-G_g$ anomaly and 
\begin{eqnarray}
\sum_{{\bf 1}_{\pm -}}\sum_{\boldsymbol{R}^{(f)}}
T_2(\boldsymbol{R}^{(f)})
+ \sum_{{\bf 2}}\sum_{\boldsymbol{R}^{(f)}} \,
 T_2(\boldsymbol{R}^{(f)}) 
~=~ 0\mod 1\;  ,
\end{eqnarray}
for the $Z_2'-G_g-G_g$ anomaly.

\begin{table}[t]
\begin{center}
\begin{tabular}{|c|c|c|c|c|c|}
\hline
     &${\bf 1}_{++}$&${\bf 1}_{+-}$& ${\bf 1}_{-+}$& ${\bf 1}_{--}$
      &${\bf 2}$ \\ \hline
$\det(Z)$&  $1$   &   $1$     &   $-1$    &  $-1$  & $-1$ \\ \hline
$\det(C)$&  $1$   &   $-1$     &   $1$    &  $-1$ & $-1$\\ 
\hline
\end{tabular}
\end{center}
\caption{Determinants on $D_4$ representations}
\label{tab:D4}
\end{table}

\vskip .5cm
{$\bullet$ \bf  ${\bf \Delta (27)}$ }

\vskip .2cm
Similarly, we can study anomalies of $\Delta(27)$.
As shown in section \ref{sec:delta27}, 
the $\Delta(27)$ group has 
nine singlets, ${\bf 1}_{r s}$ 
and two triplets, ${\bf 3}$ and $\bar{{\bf 3}}$.
All elements $\Delta(27)$  can be written 
by products of $Z$, $C$.
On all of triplet representations, 
their determinants are obtained as 
$\det(Z)=\det (C)=1$.
Only anomaly coefficients come from the nine singlets fields.
These results are shown in Table \ref{tab:delta27}.
That implies that two independent $Z_3$ symmetries can be 
anomalous.
One corresponds to $Z$ and the other corresponds to 
$C$.
For the $Z_3$ symmetry corresponding to $Z$, 
each representation has the following $Z_3$ charge $q_3$,
\begin{eqnarray}
q_3=0 &:& {\bf 1}_{0 s}, \quad {\bf 3}, \bar{{\bf 3}} \\                   
q_3=1 &:& {\bf 1}_{1 s}, \\
q_3=2 &:& {\bf 1}_{2 s} ,
\end{eqnarray}
while for $Z'_3$ symmetry corresponding to $C$, 
each representation has the following $Z_3$ charge $q'_3$,
\begin{eqnarray}
q'_3=0 &:& {\bf 1}_{r 0 }, \quad {\bf 3}, \bar{{\bf 3}} \\             
q'_3=1 &:& {\bf 1}_{r 1 }, \\
q'_3=2 &:& {\bf 1}_{r 2 } .
\end{eqnarray}
Then, the anomaly-free conditions are written as 
\begin{eqnarray}
\sum_{{\bf 1}_{1 s}}\sum_{\boldsymbol{R}^{(f)}}
T_2(\boldsymbol{R}^{(f)})
+ 2\sum_{{\bf 1}_{2 s}}\sum_{\boldsymbol{R}^{(f)}} \,
 T_2(\boldsymbol{R}^{(f)}) 
~=~ 0\mod 3/2\;  ,
\end{eqnarray}
for the $Z_3-G_g-G_g$ anomaly and 
\begin{eqnarray}
\sum_{{\bf 1}_{1 s}}\sum_{\boldsymbol{R}^{(f)}}  \dim
\boldsymbol{R}^{(f)}
+ 2\sum_{{\bf 1}_{2 s}}\sum_{\boldsymbol{R}^{(f)}} \,
 \dim \boldsymbol{R}^{(f)} 
~=~ 0\mod 3/2\;  ,
\end{eqnarray}
for the $Z_3$-gravity-gravity anomaly.
Similarly, for the $Z_3'$ symmetry, 
the anomaly-free conditions are written as 
\begin{eqnarray}
\sum_{{\bf 1}_{r 1}}\sum_{\boldsymbol{R}^{(f)}}
T_2(\boldsymbol{R}^{(f)})
+ 2\sum_{{\bf 1}_{r 2 }}\sum_{\boldsymbol{R}^{(f)}} \,
 T_2(\boldsymbol{R}^{(f)}) 
~=~ 0\mod 3/2\;  ,
\end{eqnarray}
for the $Z'_3-G_g-G_g$ anomaly and 
\begin{eqnarray}
\sum_{{\bf 1}_{r 1 }}\sum_{\boldsymbol{R}^{(f)}}  \dim
\boldsymbol{R}^{(f)}
+ 2\sum_{{\bf 1}_{r 2 }}\sum_{\boldsymbol{R}^{(f)}} \,
 \dim \boldsymbol{R}^{(f)} 
~=~ 0\mod 3/2\;  ,
\end{eqnarray}
for the $Z'_3$-gravity-gravity anomaly.

\begin{table}[t]
\begin{center}
\begin{tabular}{|c|c|c|c|}
\hline
     &${\bf 1}_{rs}$ & ${\bf 3}$ & $\bar{{\bf 3}}$ \\ \hline
$\det(Z)$&  $\omega^r$    &   $1$  & $1$  \\ \hline
$\det(C)$&  $\omega^s$    &   $1$  & $1$      \\  
\hline
\end{tabular}
\end{center}
\caption{Determinants on $\Delta(27)$ }
\label{tab:delta27}
\end{table}

\vskip .5cm
{$\bullet$ \bf  ${\bf \Delta (54)}$ }

\vskip .2cm
Finally, we also show anomalies of $\Delta(54)$.
As shown in section \ref{sec:delta54}, 
the $\Delta(54)$ group has 
two singlets, ${\bf 1}_{1,2}$ 
and four doublets, ${\bf 2}_{1,2,3,4}$ and 
four triplets ${\bf 3}_{1,2}$ and $\bar{{\bf 3}}_{1,2}$.
All elements $\Delta(54)$  can be written 
by products of $Z$, $C$ and $P$.
First of all, we obtain $\det(Z)=\det(C)=1$.
This implies that the only anomaly is arising the symmetries of 
$P$ which is $Z_2$ symmetry.
The determinants for each representation of $P$ are summarized  
in Table \ref{tab:delta54}.
Then, the anomaly-free condition is given by 
\begin{eqnarray}
\sum_{{\bf 1}_{2}}\sum_{\boldsymbol{R}^{(f)}}
T_2(\boldsymbol{R}^{(f)})
+ \sum_{{\bf 2}_{1, 2, 3, 4}}\sum_{\boldsymbol{R}^{(f)}} \,
 T_2(\boldsymbol{R}^{(f)}) 
+ \sum_{{\bf 3}_{1},\bar{{\bf 3}} }\sum_{\boldsymbol{R}^{(f)}} \,
 T_2(\boldsymbol{R}^{(f)}) 
~=~ 0\mod 1\;  ,
\end{eqnarray}
for the $Z_2-G_g-G_g$ anomaly.

\begin{table}[t]
\begin{center}
\begin{tabular}{|c|c|c|c|c|c|c|c|c|c|c|c|}
\hline
  & ${\bf 1}_1$ & ${\bf 1}_2$ & ${\bf 2}_1$ & ${\bf 2}_2$ &
${\bf 2}_3$ & ${\bf 2}_4$ &
${\bf 3}_1$ & $\bar{{\bf 3}}_1$ &
${\bf 3}_2$ & $\bar{{\bf 3}}_2$ \\ \hline
$\det(P)$&  $1$ & $-1$ & $-1$ & $-1$ & $-1$ & $-1$ & $-1$ & $-1$ & $1$  & $1$  \\ \hline
\end{tabular}
\end{center}
\caption{Determinants on $\Delta(54)$ }
\label{tab:delta54}
\end{table}

\vskip.5cm
Similarly, we can analyze on anomalies for other 
non-Abelian discrete symmetries.

\subsection{Other anomalies in string models}
In this section, we also introduce other interesting anomalies for 
discrete symmetries. 
Here  the discrete symmetries
we discuss are discrete R-symmetries for heterotic orbifold
models. 
It is widely assumed that
superstring theory leads to anomaly-free effective theories.
In fact the anomalous $U(1)$ symmetries
are restored by the GS 
mechanism \cite{Green:1984sg,Witten:1984dg,Ibanez:1998qp}.
For this mechanism to work,
the mixed anomalies between the anomalous $U(1)$ and other continuous 
gauge symmetries have to satisfy a certain set of conditions,
the GS conditions,
at the field theory level.
In particular, in heterotic string theory the mixed anomalies between 
the anomalous $U(1)$ symmetries and other continuous gauge 
symmetries must be universal for different gauge 
groups up to their Kac-Moody 
levels \cite{Schellekens:1986xh,Kobayashi:1996pb}.
Therefore stringy-originated discrete symmetries are strongly constrained
 due to stringy consistency, and 
 it is phenomenologically  and theoretically
 important to study  anomalies of discrete symmetries,
 as it is pointed out in \cite{Ibanez:1991hv}
and the example of T-duality shows.
We shall investigate the mixed anomalies between the discrete
R-symmetries and the continuous gauge symmetries in concrete orbifold models. 
We will also study relations between the discrete $R$-anomalies, 
one-loop beta-function coefficients (scale anomalies).

In orbifold models, the 6D compact space is chosen to be   
6D orbifold.
A 6D orbifold is a division of 6D torus $T^6$ by a twist $\theta$, 
while the torus $T^6$ is obtained as $R^6/\Lambda^6$, 
where $\Lambda^6$ is 6D lattice.
Eigenvalues of the twist $\theta$ are denoted as 
$e^{2\pi i v_1}, e^{2\pi i v_2}$ 
and $e^{2\pi i v_3}$ in the complex basis $Z_i$ ($i=1,2,3$).

It is convenient to bosonize right-moving fermionic strings.
Here we write such bosonized fields by $H^t$ ($t=1,\cdots,5$).
Their momenta $p_t$ are quantized and span the SO(10) weight lattice.
Space-time bosons correspond to SO(10) vector momenta, 
and space-time fermions correspond to SO(10) spinor momenta.
The 6D compact part, i.e. the SO(6) part, $p_i$ ($i=1,2,3$) 
is relevant to our study.
All of $\Z_N$ orbifold models have three untwisted 
sectors, $U_1$, $U_2$ and $U_3$, and their massless bosonic modes 
have the following SO(6) momenta,
\begin{equation}
U_1:(1,0,0), \qquad U_2:(0,1,0), \qquad U_3:(0,0,1).
\label{H-momenta-U}
\end{equation}
On the other hand, the twisted sector $T_k$ has 
shifted $SO(6)$ momenta, $r_i=p_i+kv_i$.
Table \ref{tab:H-momenta-1} and Table \ref{tab:H-momenta-2} show
explicitly $H$-momenta $r_i$ of massless bosonic states.
That implies their $SO(6)$ $H$-momenta are obtained as 
\begin{equation}
r_i = |kv_i|-{\rm Int}[|kv_i|],
\label{H-momentum:Zn}
\end{equation}
where ${\rm Int}[a]$ denotes an integer part of fractional number $a$.
This relation is not available for the untwisted sectors, 
and $r_i$ is obtained as Eq.~(\ref{H-momenta-U}).

\begin{table}[t]
\begin{center}
\small
\begin{tabular}{|c|c|c|c|c|c|}
\hline
 & $\Z_3$ & $\Z_4$ & $\Z_6$-I & $\Z_6$-II & $\Z_7$  \\
$v_i$ & $(1,1,-2)/3$ & $(1,1,-2)/4$ & $(1,1,-2)/6$ & $(1,2,-3)/6$ &
$(1,2,-3)/7$  \\ \hline \hline
$T_1$ & $(1,1,1)/3$ & $(1,1,2)/4$ & $(1,1,4)/6$ & $(1,2,3)/6$  &
$(1,2,4)/7$ \\
$T_2$ & --- & $(2,2,0)/4$ & $(2,2,2)/6$ & $(2,4,0)/6$ & $(2,4,1)/7$ \\
$T_3$ & --- & --- & $(3,3,0)/6$ &  $(3,0,3)/6$ &  --- \\
$T_4$ & --- & --- & --- & $(4,2,0)/6$ & $(4,1,2)/7$ \\ \hline
\end{tabular}
\end{center}
\caption{$H$-momenta for $\Z_3$, $\Z_4$, $\Z_6$-I, $\Z_6$-II and
$\Z_7$ orbifolds} \label{tab:H-momenta-1}
\end{table}

\begin{table}[t]
\begin{center}
\small
\begin{tabular}{|c|c|c|c|c|}
\hline
 &  $\Z_8$-I & $\Z_8$-II & $\Z_{12}$-I & $\Z_{12}$-II  \\
$v_i$ &  $(1,2,-3)/8$ & $(1,3,-4)/8$ & $(1,4,-5)/12 $ &
$(1,5,-6)/12$ \\  \hline \hline
$T_1$ & $(1,2,5)/8$ & $(1,3,4)/8$ & $(1,4,7)/12$ & $(1,5,6)/12$ \\
$T_2$ & $(2,4,2)/8$ & $(2,6,0)/8$ & $(2,8,2)/12$ & $(2,10,0)/12$ \\
$T_3$ & ---         & $(3,1,4)/8$ & $(3,0,9)/12$ & $(3,3,6)/12$ \\
$T_4$ & $(4,0,4)/8$ & $(4,4,0)/8$ & $(4,4,4)/12$ & $(4,8,0)/12$ \\
$T_5$ & $(5,2,1)/8$ & ---         & ---          & $(5,1,6)/12$ \\
$T_6$ & ---         & ---         & $(6,0,6)/12$ & $(6,6,0)/12$ \\
$T_7$ & ---         & ---         & $(7,4,1)/12$ & ---          \\
$T_8$ & ---         & ---         & ---          & ---          \\
$T_9$ & ---         & ---         & $(9,0,3)/12$ & ---          \\
$T_{10}$ & ---      & ---         & ---          & $(10,2,0)/12$ \\
\hline
\end{tabular}
\end{center}
\caption{$H$-momenta for $\Z_8$-I, $\Z_8$-II, $\Z_{12}$-I and
$\Z_{12}$-II orbifolds} \label{tab:H-momenta-2}
\end{table}

The gauge sector can also be broken and 
gauge groups smaller than $E_8 \times E_8$ are obtained.
Matter fields have some representations under such 
unbroken gauge symmetries.

Massless modes for 4D space-time bosons correspond to 
the following vertex operator \cite{Friedan:1985ge,Hamidi:1986vh},
\begin{equation}
V_{-1} = e^{-\phi}\prod_{i=1}^3(\partial Z_i)^{{\cal N}_i} (\partial
\bar
Z_i)^{\bar {\cal N}_i}e^{ir_tH^t}e^{iP^IX^I}e^{ikX}
\sigma_k,
\end{equation}
in the $(-1)$-picture, where $\phi$ is the bosonized ghost, 
$kX$ corresponds to the 4D part and $P^IX^I$ corresponds 
to the gauge part.
Oscillators of the left-mover are denoted by 
$\partial Z_i$ and $\partial \bar Z_i$, and 
${\cal N}_i$ and $\bar {\cal N}_i$ are oscillator numbers, which are
included 
in these massless modes.
In addition, $\sigma_k$ denotes the twist field for 
the $T_k$ sector.
Similarly, we can write the vertex operator for 4D space-time 
massless fermions as 
\begin{equation}
V_{-\frac12} = e^{-\frac12 \phi}\prod_{i=1}^3(\partial Z_i)^{N_i}
(\partial \bar Z_i)^{\bar N_i}e^{ir_t^{(f)}H_t}e^{iP^IX^I}e^{ikX}
\sigma_{k},
\end{equation}
in the $(-1/2)$-picture.
The $H$-momenta for space-time fermion and boson, $r_i^{(f)}$ and
$r_i$ in the same supersymmetric multiplet are
related each other as 
\begin{equation}
r_i = r_i^{(f)} + (1,1,1)/2.
\end{equation}

We need vertex operators $V_0$ with the 0-picture when we compute 
generic n-point couplings.
We can obtain such vertex operators $V_0$ by operating 
the picture changing operator, $Q$, on $V_{-1}$, 
\cite{Friedan:1985ge},
\begin{equation}
Q=e^\phi (e^{-2 \pi i r^v_i H_i}\bar \partial Z_i + e^{2 \pi i r^v_i
H_i}\bar \partial \bar Z_i), \label{p-change}
\end{equation}
where $r^v_1=(1,0,0)$, $r^v_2=(0,1,0)$ and $r^v_3=(0,0,1)$.

Next we briefly review on $\Z_N \times \Z_M$ orbifold 
models \cite{Font:1988mk}.
In $\Z_N \times \Z_M$ orbifold models, we introduce two independent 
twists $\theta$ and $\omega$, whose twists are represented by 
$e^{2\pi i v^1_i}$ and $e^{2\pi i v^2_i}$, respectively 
in the complex basis.
Two twists are chosen such that 
each of them breaks 4D N=4 SUSY to 4D N=2 SUSY and 
their combination preserves only N=1 SUSY.
Thus, eigenvalues $v^1_i$ and $v^2_i$ are chosen as 
\begin{equation}
v^1_i=(v^1,-v^1,0), \qquad v^2_i=(0,v^2,-v^2),
\end{equation}
where $v^1,v^2 \neq {\rm integer}$.
In general, $\Z_N \times \Z_M$ orbifold models 
have three untwisted sectors, $U_1$, $U_2$ and $U_3$, 
and their massless bosonic modes have the same $SO(6)$ $H$-momenta
$r_i$ as 
Eq.~(\ref{H-momenta-U}).
In addition, there are $\theta^k \omega^\ell$-twisted sectors, 
and their $SO(6)$ $H$-momenta are obtained as 
\begin{equation}
r_i = |kv^1_i|+|\ell v^2_i| - {\rm Int}[|kv^1_i|+|\ell v^2_i|].
\label{H-momentum:ZnZm}
\end{equation}
Vertex operators are also constructed in a similar way.
Recently, non-factorizable $\Z_N \times \Z_M$ orbifold 
models have been studied \cite{Faraggi:2006bs}.
The above aspects are the same for such non-factorizable models.

\subsubsection{Discrete R-symmetries}

Here we define R-charges.
We consider n-point couplings including two fermions.
Such couplings are computed by the following 
n-point correlation function of vertex operators,
\begin{equation}
\langle V_{-1}V_{-1/2}V_{-1/2}V_0\cdots V_0 \rangle .
\end{equation}
They must have the total ghost charge $-2$, because the background 
has the ghost number 2.
When this n-point correlation function does not vanish, 
its corresponding n-point coupling in effective theory is 
allowed.
That is, selection rules for allowed n-point correlation functions 
in string theory correspond to symmetries in effective theory.

The vertex operator consists of several parts, the 
4D part $e^{kX}$, the gauge part $e^{iPX}$, 
the 6D twist field $\sigma_k$, the 6D left-moving oscillators 
$\partial Z_i$ and the bosonized fermion $e^{irH}$.
Each part has its own selection rule for allowed couplings.
For the 4D part and the gauge part, 
the total 4D momentum $\sum k$ and the total momentum 
of the gauge part $\sum P$ should be conserved.
The latter is nothing but the requirement of gauge invariance.
The selection rule for 6D twist fields $\sigma_k$ is controlled by 
the space group selection rule 
\cite{Hamidi:1986vh,Kobayashi:1991rp}.

Similarly, the total $H$-momenta can be conserved 
\begin{equation}
\sum r_i =1.
\end{equation}
Here we take a summation over the $H$-momenta for 
scalar components, using the fact that 
the $H$-momentum of fermion component differs by $-1/2$.
Another important symmetry is the twist symmetry of
oscillators.
We consider the following twist of oscillators,
\begin{eqnarray}
& & \partial Z_i \rightarrow e^{2 \pi i v_i}\partial Z_i, \qquad
\partial \bar Z_i \rightarrow e^{-2 \pi i v_i}\partial \bar Z_i, 
\nonumber \\
& & \bar \partial Z_i \rightarrow e^{2 \pi i v_i}\bar \partial Z_i,
\qquad \bar \partial \bar Z_i \rightarrow e^{-2 \pi i v_i}\bar
\partial \bar Z_i.
\end{eqnarray}
Allowed couplings may be invariant under the above $Z_N$ twist.

Indeed, for 3-point couplings corresponding to 
$\langle V_{-1}V_{-1/2}V_{-1/2}\rangle$, we can require 
$H$-momentum conservation and $Z_N$ twist invariance of oscillators 
independently.
However, we have to compute generic n-point couplings 
through picture changing, and the picture changing operator $Q$ 
includes non-vanishing $H$-momenta and right-moving oscillators 
$\bar \partial
Z_i$ and
 $\bar \partial \bar Z_i$. 
Consequently, the definition of the H-momentum  
of each vertex operator depends on the choice of
the picture and so its physical meaning remains somewhat obscure.
We therefore use a picture independent quantity as 
follows,
\begin{equation}
R_i \equiv r_i + {\cal N}_i - \bar {\cal N}_i,
\end{equation}
which can be interpreted as an R-charge  \cite{Kobayashi:2004ud}.
This R-symmetry is a discrete 
surviving symmetry of the continuous $SU(3)~(\subset SU(4))$
R-symmetry 
under orbifolding. 
Here we do not
distinguish oscillator numbers for the left-movers and right-movers,
because they have the same phase under $Z_N$ twist. Indeed,
physical states with $-1$ picture have vanishing oscillator number
for the right-movers, while the oscillator number for the
left-movers can be non-vanishing. Thus, hereafter   ${\cal N}_i$ and
$\bar {\cal N}_i$ denote the oscillator number for the left-movers,
because we study the physical states with $-1$ picture from now. For
simplicity, we use the notation $\Delta {\cal N}_i = {\cal N}_i -
\bar {\cal N}_i$. 
Now, we can write the selection rule due to $R$-symmetry as 
\begin{equation}
\sum R_i = 1 \quad {\rm mod} \quad N_i,
\end{equation}
where $N_i$ is the minimum integer satisfying $N_i = 1/\hat v_i$, 
where $\hat v_i= v_i + m$ with any integer $m$. 
For example, for $Z_6$-II orbifold, we have $v_i=(1,2,-3)/6$, and
$N_i=(6,3,2)$.
Thus, these are discrete symmetries.
Note that the above summation is taken over scalar components.

Discrete R symmetry itself is defined as the following 
transformation,
\begin{equation}
\vert R_i \rangle \rightarrow e^{2\pi i v_i R_i} \vert R_i \rangle,
\label{eq:R-trans}
\end{equation}
for states with discrete $R$-charges, which are defined 
mod $N_i$.
For later convenience, we show discrete $R$-charges for 
fermions in Table~\ref{tab:R}.
As shown there, gaugino fields always have 
$R$-charge $(1/2,1/2,1/2)$.

\begin{table}[t]
\begin{center}
\small
\begin{tabular}{|c|c|}  \hline
 & $R_i$   \\ \hline
gaugino & $(1/2,1/2,1/2)$ \\
$U_1$   & $(1/2,-1/2,-1/2)$ \\
$U_2$   & $(-1/2,1/2,-1/2)$  \\
$U_3$ & $(-1/2,-1/2,1/2)$  \\
$T_k$ & $kv_i - {\rm Int}[kv_i]-1/2+\Delta {\cal N}_i$ \\ \hline
\end{tabular}
\end{center}
\caption{Discrete $R$-charges of fermions in 
$\Z_N$ orbifold models} \label{tab:R}
\end{table}

\subsubsection{Discrete R-anomalies}

Let us study anomalies of discrete R-symmetry.
Under the R-transformation like Eq.~(\ref{eq:R-trans}),
the anomaly coefficients $A^{R_i}_{G_a}$ are obtained as 
\begin{equation}
A_{G_a}^{R_i}=\sum R_i T({\bf R}_{G_a}),
\end{equation}
where $T({\bf R}_{G_a})$ is the Dynkin index for ${\bf R}_{G_a}$
representation under $G_a$.

By use of our discrete $R$ charge, the anomaly coefficients are
written as 
\begin{equation}
A_{G_a}^{R_i}= \frac{1}{2} C_2(G_a) + 
\sum_{\rm matter}  (r^{}_i-\frac{1}{2} +\Delta {\cal N}_i) T({\bf
R}_{G_a}),
\end{equation}
where $C_2(G_a)$ is quadratic Casimir.
Note that $r_i$ denotes the SO(6) shifted momentum for 
bosonic states.
The first term in the right hand side is a contribution from 
gaugino fields and the other is the contribution from matter fields.

If these anomalies are canceled by the 
Green-Schwarz mechanism, these mixed anomalies must 
satisfy the following condition,
\begin{equation}
\frac{A_{G_a}^{R_{i}}}{k_a}= \frac{A_{G_b}^{R_i}}{k_b},
\label{GS-R}
\end{equation}
for different gauge groups, $G_a$ and $G_b$, where 
$k_a$ and $k_b$ are Kac-Moody levels.
In the simple orbifold construction, we have the Kac-Moody level 
$k_a=1$ for non-abelian gauge groups.
Note again that anomalies are defined modulo
$N_iT({\bf R}^{(f)}_{G_a})$.
The above GS condition has its meaning mod
$N_iT({\bf R}^{(f)}_{G_a})/k_a$.

As illustrating examples, let us study explicitly one $Z_3$ model 
and one $Z_4$ model.
Their gauge groups and massless spectra are shown 
in Table~\ref{tab:Z3} and Table~\ref{tab:Z4}.\footnote{
See for explicit massless spectra Ref.~\cite{Katsuki:1989cs}, 
where a typographical error is included in the $U_3$ sector 
of the $Z_4$ orbifold model.
It is corrected in Table~\ref{tab:Z4}.}
First, we study R-anomalies in the $Z_3$ orbifold model.
Since $v_i=(1,1,-2)/3$, we have $N_i=3$.
For both $E_6$, mixed R-anomalies are computed as 
\begin{equation}
A^{R_{i}}_{E_6}= \frac{3}{2}+9n^i_{E_6},
\end{equation}
where $n^i_{E_6}$ is integer.
The second term in the right hand side appears because 
anomalies are defined modulo $N_iT(27)$ with 
$N_i=3$ and $T(27)=3$ for $E_6$.
Similarly, mixed R-anomalies for $SU(3)$ are computed as 
\begin{equation}
A^{R_i}_{SU(3)}=-12 +\frac{3}{2}n^i_{SU(3)},
\end{equation}
where $n^i_{SU(3)}$ is integer.
The second term in the right hand side appears through 
$N_iT(3)$ with $N_i=3$ and $T(3)=1/2$ for $SU(3)$.
Thus, in this model, mixed R-anomalies satisfy 
\begin{equation}
A^{R_i}_{E_6}=A^{R_i}_{SU(3)} \qquad ({\rm mod}~~3/2).
\end{equation}

\begin{table}[t]
\begin{center}
\small
\begin{tabular}{|c|c|}  \hline
gauge group & $E_6 \times SU(3) \times E_6 \times SU(3)$ \\ \hline
\hline
sector & massless spectrum   \\ \hline
$U_1$   & (27,3;1,1)+ (1,1;27,3)\\
$U_2$   & (27,3;1,1)+ (1,1;27,3)  \\
$U_3$ & (27,3;1,1)+ (1,1;27,3)  \\  
$T_1$ & $27(1,\bar 3;1,\bar3)$ \\ \hline
\end{tabular}
\end{center}
\caption{Massless spectrum in a 
$\Z_3$ orbifold model} \label{tab:Z3}
\end{table}

\begin{table}[t]
\begin{center}
\small
\begin{tabular}{|c|c|}  \hline
gauge group & $SO(10)  \times SU(4) \times SO(12) \times SU(2)
\times U(1)$ \\ \hline \hline
sector & massless spectrum   \\ \hline
$U_1$   & $(16_c,4;1,1)+ (1,1;32_c,1)+(1,1;12_v,2)$ \\
$U_2$   & $(16_c,4;1,1)+  (1,1;32_c,1)+(1,1;12_v,2)$           \\
$U_3$ & $(10_v,6;1,1)+ (1,1;32_c,2) +2(1,1,;1,1)$  \\
$T_1$ & $16(1,4;1,2)$ \\ 
$T_2$ & $16(10_v,1;1,1)+16(1,6;1,1)$ \\ \hline
\end{tabular}
\end{center}
\caption{Massless spectrum in a 
$\Z_4$ orbifold model} \label{tab:Z4}
\end{table}

Next, we study R-anomalies in the $Z_4$ orbifold model 
with the gauge group 
$SO(10)\times SU(4) \times SO(12) \times SU(2) \times U(1)$.
Since the $Z_4$ orbifold has $v_i=(1,1,-2)/4$, 
we have $N_i=(4,4,2)$.
Mixed anomalies between $R_{1,2}$ and $SO(10)$ are 
computed as 
\begin{equation}
A^{R_{1,2}}_{SO(10)} = 1 + 4 n^{1,2}_{SO(10)},
\end{equation}
with integer $n^{1,2}_{SO(10)}$, 
where the second term appears through $N_iT({\bf R}_a)$ 
with $N_i=4$ and $T(10)=1$ for $SO(10)$.
Similarly, mixed anomalies between $R_3$ and $SO(10)$ 
is computed as 
\begin{equation}
A^{R_{3}}_{SO(10)} = -9 + 2 n^{3}_{SO(10)},
\end{equation}
with integer $n^{3}_{SO(10)}$.
Furthermore, mixed R-anomalies for other non-abelian groups 
are obtained as 
\begin{eqnarray}
 & & A^{R_{1,2}}_{SU(4)} = -7 + 2 n^{1,2}_{SU(4)},  \qquad 
A^{R_{3}}_{SU(4)} = -9 + n^{3}_{SU(4)},  \nonumber \\
 & & A^{R_{1,2}}_{SO(12)} = 1 + 4 n^{1,2}_{SO(12)},  \qquad 
A^{R_{3}}_{SO(12)} = 3 + 2n^{3}_{SO(12)},   \\
& & A^{R_{1,2}}_{SU(2)} = -15 + 2 n^{1,2}_{SU(2)},  \qquad 
A^{R_{3}}_{SU(2)} = 3 + n^{3}_{SU(2)},  
\nonumber 
\end{eqnarray}
with integer $n^{i}_{G_a}$, where the second terms 
appear through $N_iT({\bf R}_a)$ with $N_i=(4,4,2)$, and 
$T(12)=1$ for $SO(12)$, $T(4)=1/2$ for $SU(4)$ and 
$T(2)=1/2$ for $SU(2)$.
These anomalies satisfy the GS condition,
\begin{eqnarray}
 & & A^{R_{1,2}}_{SO(10)}=A^{R_{1,2}}_{SU(4)} =  A^{R_{1,2}}_{SO(12)}
 = 
A^{R_{1,2}}_{SU(2)} \qquad ({\rm mod}~~2), \nonumber \\
 & & A^{R_{3}}_{SO(10)}=A^{R_{3}}_{SU(4)} =  A^{R_{3}}_{SO(12)} = 
A^{R_{3}}_{SU(2)} \qquad ({\rm mod}~~1).
\end{eqnarray}

The GS condition is satisfied in the above models without Wilson
lines. However, it is not satisfied in explicit models with Wilson
lines for naively defined R-charges~\cite{Araki:2008ek}.
Anomalies for discrete shifts are important.

\subsubsection{Relation with beta-function}

Here we study the relation between discrete R anomalies and 
one-loop beta-functions.
We find 
\begin{equation}
\sum_{i=1,2,3}r_i=1,
\end{equation}
{}from Eqs.~(\ref{H-momentum:Zn}) and (\ref{H-momentum:ZnZm}) 
as well as Table~\ref{tab:H-momenta-1} and 
Table~\ref{tab:H-momenta-2}.
By using this, we can write the sum of R-anomalies as 
\begin{eqnarray}
A^R_{G_a} &=& \sum_{i=1,2,3}A^{R_i}_{G_a} \nonumber \\
  &=& \frac{3}{2}C_2(G_a) + \sum_{\rm matter}  T({\bf
  R}_{G_a})(-\frac{1}{2}+ 
\sum_i\Delta {\cal N}_i).
\end{eqnarray}
Thus, when $\sum_i\Delta {\cal N}_i=0$, the total anomaly 
$A^R_{G_a}$ is proportional to the one-loop beta-function 
coefficient, i.e. the scale anomaly, $b_{G_a}$,
\begin{equation}
b_{G_a} = 3 C_2(G_a) - \sum_{\rm matter} T({\bf R}_{G_a}).
\end{equation}
When we use the definition of R charge $\tilde R_i = 2 R_i$, 
we would have $A^{\tilde R}_{G_a} = b_{G_a}$.
It is not accidental that 
$A^R_{G_a}$ is proportional to $b_{G_a}$
\cite{jones,piguet1}.
The sum of the R-charges $\sum_{i=1,2,3}R_i$
  of a supermultiplet is
nothing but the R-charge (up to an overall normalization)
associated with the R-current
which is a bosonic component of the supercurrent \cite{ferrara}, 
when the R-charge is universal for all of matter fields, 
i.e. $\sum_i\Delta {\cal N}_i=0$. 
Using the supertrace identity \cite{piguet2} it is
in fact possible to show \cite{piguet1} that
$A^R_{G_a}$ is proportional to $b_{G_a}$ to all orders in perturbation 
theory.

In explicit models, non-abelian groups except $SU(2)$ 
have few massless matter fields with non-vanishing oscillator 
numbers, while massless matter fields with oscillators  
can appear as singlets as well as $SU(2)$ doublets.
Thus, in explicit models the total R-anomaly $A^R_{G_a}$ 
is related with the one-loop beta-function coefficient $b_{G_a}$,
\begin{equation}
2A^R_{G_a} = b_{G_a},
\label{anomR-b}
\end{equation}
modulo $N_iT({\bf R}_a)$ for most of non-abelian groups.
Since the total R-anomalies satisfy the GS condition, 
$A^R_{G_a}=A^R_{G_b}$, the above relation between 
$A^R_{G_a}$ and $b_{G_a}$ leads to 
\begin{equation}
b_{G_a} = b_{G_b},
\label{GS-b}
\end{equation}
modulo $2N_iT({\bf R}_a)$.

For example, the explicit $Z_3$ orbifold model 
and $Z_4$ orbifold model in Table~\ref{tab:Z3} and Table~\ref{tab:Z4} 
have only non-oscillated massless modes except singlets.
The $Z_3$ orbifold model has the following total R-anomalies and 
one-loop beta-function coefficient,
\begin{eqnarray}
 & & A^R_{E_6}=\frac{9}{2}+9n_{E_6}, \qquad 
 b_{E_6} = 9, \nonumber \\
 & & A^R_{SU(3)}=-36 + \frac{3}{2}n_{SU(3)}, \qquad 
b_{SU(3)}=-72.
\end{eqnarray}
Hence, this model satisfy $2A^R_{G_a}=b_{G_a}$ and its 
one-loop beta-function coefficients satisfy
\begin{equation}
b_{E_6}=b_{SU(3)} \qquad ({\rm mod}~~3).
\end{equation}
Similarly, the $Z_4$ orbifold model in Table~\ref{tab:Z4}
has the total R-anomalies and
one-loop beta-function coefficients as,
\begin{eqnarray}
 & & A^R_{SO(10)}= -7 + 2 n_{SO(10)}, \qquad  b_{SO(10)}= -14
 \nonumber \\
 & & A^R_{SU(4)}= -23 + n_{SU(4)}, \qquad b_{SU(4)} = -46   
\nonumber \\
 & & A^R_{SO(12)}=5 + 2n_{SO(10)}, \qquad b_{SO(12)}= 10    \\
 & & A^R_{SU(2)}=-27 + n_{SU(2)}, \qquad b_{SU(2)}= -54.
\nonumber
\end{eqnarray}
Thus, this model also satisfies $2A^R_{G_a}=b_{G_a}$
and its one-loop beta-function coefficients satisfy 
\begin{equation}
b_{SO(10)}=b_{SU(4)}=b_{SO(12)}=b_{SU(2)} \qquad ({\rm mod}~~2).
\end{equation}

\subsubsection{Relation with T-duality anomaly}

Here we study the relation between R-anomalies and T-duality
anomalies.
The relation between R-symmetries and T-duality has also been 
studied in Ref.~\cite{Ibanez:1992uh}.
The T-duality anomalies are obtained 
as \cite{Derendinger:1991hq,Ibanez:1992hc}
\begin{equation}
A^{T_i}_{G_a} = -C_2({G_a}) +\sum_{\rm matter} T({\bf R}_{G_a}) 
(1+2n_i),
\end{equation}
where $n_i$ is the modular weight of matter fields for 
the $i$-th torus.
The modular weight is related with $r_i$ as 
\begin{eqnarray}
n_i &=& -1 {\rm~~for~~} r_i=1,\nonumber \\
&=& 0 {\rm~~for~~} r_i=0,\\
&=&  r_i-1 - \Delta {\cal N}_i {\rm~~for~~} r_i \neq 0,1.
\nonumber
\end{eqnarray}
Note that $n_i = -r_i$ for $r_i=0,1/2,1$.
Thus, in the model, which includes only matter fields with 
 $r_i=0,1/2,1$, the T-duality anomalies and R-anomalies are 
proportional to each other, 
\begin{equation}
A^{T_i}_{G_a} = -2A^{R_i}_{G_a}.
\label{relation:T-R}
\end{equation}
In generic model, such relation is violated, but 
T-duality anomalies and R-anomalies are still related with each other 
as 
\begin{equation}
A^{T_i}_{G_a} = -2A^{R_i}_{G_a} -2 \sum_{r_i \neq 0,1/2,1}
(2  r_i -1).
\end{equation}

T-duality should also satisfy the GS condition,
\begin{equation}
\frac{A^{T_i}_{G_a}}{k_a} =\frac{A^{T_i}_{G_b}}{k_b},
\end{equation}
for the $i$-th torus, which does not include the N=2 subsector.
Thus, the requirement that T-duality anomalies and R-anomalies 
should satisfy the GS condition, leads to a similar condition for 
\begin{equation}
\Delta_a^i= 2 \sum_{r^b_i \neq 0,1/2,1}
(2  r^b_i -1).
\end{equation}

For the $i$-th torus, which includes 
N=2 subsector, T-duality anomalies can be canceled by 
the GS mechanism and T-dependent threshold correction
\cite{Dixon:1990pc}.
Thus, for such torus, the T-duality anomalies has no 
constrain from the GS condition.
However, even for such torus, R-anomaly should satisfy 
the GS condition.

For example, the $Z_4$ orbifold model in Table~\ref{tab:Z4} 
has the following T-duality anomalies,
\begin{eqnarray}
& & A^{T_{1,2}}_{SO(10)}=-2, \qquad A^{T_3}_{SO(10)}=18, \nonumber \\
& & A^{T_{1,2}}_{SU(4)}=-2, \qquad A^{T_3}_{SU(4)}=18, \nonumber \\
& & A^{T_{1,2}}_{SO(12)}=-2, \qquad A^{T_3}_{SO(12)}=-6, \\
& & A^{T_{1,2}}_{SU(2)}=-2, \qquad A^{T_3}_{SU(2)}=-6.
\nonumber
\end{eqnarray}
They satisfy the GS condition,
\begin{equation}
A^{T_{1,2}}_{SO(10)}=A^{T_{1,2}}_{SU(4)}=A^{T_{1,2}}_{SO(12)}=
A^{T_{1,2}}_{SU(2)}.
\end{equation}
On the other hand, for the third torus, T-duality anomalies
$A^{T_3}_{G_a}$ 
do not satisfy the GS condition, that is, anomalies $A^{T_3}_{G_a}$ 
are not universal, because there is the N=2 subsector 
and one-loop gauge kinetic functions depend on the $T_3$ moduli 
with non-universal coefficients \cite{Dixon:1990pc}.
However, they satisfy 
\begin{eqnarray}
& & A^{T_3}_{SO(10)}=-2A^{R_3}_{SO(10)}, \qquad 
A^{T_3}_{SU(4)}=-2A^{R_3}_{SU(4)}, \nonumber \\
& & A^{T_3}_{SO(12)}=-2A^{R_3}_{SO(12)}, \qquad 
A^{T_3}_{SU(2)}=-2A^{R_3}_{SU(2)}, 
\end{eqnarray}
because this model has only massless modes with $r_3=0,1/2,1$.
Indeed, all of $Z_4$ orbifold models include only 
massless modes with $r_3=0,1/2,1$.
Furthermore, all of $Z_N$ orbifold models with $v_i=1/2$ 
have only massless modes with $r_i=0,1/2,1$.
Thus, the above relation (\ref{relation:T-R}) 
holds true in such $Z_N$ orbifold models.
That is also true for $R_1$-anomalies in $Z_2 \times Z_M$ 
orbifold models with $v_1=(1/2,-1/2,0)$ and $v_2=(0,v_2,-v_2)$.

Such relation between T-duality anomalies and 
R-anomalies (\ref{relation:T-R}) would be important, 
because the GS condition on R-anomalies leads to a certain 
condition on the T-duality anomalies even including the 
N=2 subsector.
For example, in the above $Z_4$ orbifold model, 
the following condition is required 
\begin{equation}
A^{T_3}_{SO(10)}=A^{T_3}_{SU(4)}=
A^{T_3}_{SO(12)}=A^{T_3}_{SU(2)} \qquad ({\rm mod}~~2).
\end{equation}

\subsubsection{Symmetry breaking of
the discrete R-symmetries}

\begin{itemize}
\item Non-perturbative breaking
\end{itemize}

If the discrete R-symmetries are anomalous, they
  are broken by non-perturbative effects  at  low-energy.
This is because,  for
the GS mechanism to take place,
  the axionic part of the dilaton $S$ should transform
non-linearly under the anomalous symmetry.
This means that a term like $e^{-aS}$
with a constant $a$ has a
definite charge $R_i^S$ under the anomalous symmetry.

Non-perturbative effects can therefore induce
terms like $e^{-aS}\Phi^1 \cdots \Phi^n$
with matter fields $\Phi^a$, where the total
charge satisfies the condition for allowed couplings, 
i.e.  $R^S_i+\sum_a R^a_{i}=1$ (mod $N_i$).
This implies that  below the scale of the 
vacuum expectation value (VEV) of $S$,
  such non-invariant terms can
  appear in a low-energy effective Lagrangian.
  The canonical dimension of the non-invariant operator
$e^{-aS}\Phi^1 \cdots \Phi^n$  that can be generated by the
non-perturbative effects depends of course on the R charge $R^S$.
If the smallest dimension is lager than four, they will be
suppressed
by certain powers of the string scale.
However,  the operator can produce
non-invariant mass terms like $m \Phi \Phi'$,
because some of the chiral superfields may acquire VEVs.
One should worry about such cases.
Needless to say that small higher dimensional  terms  would be useful
in phenomenological applications such as explaining
fermion masses.

In the case that
the smallest dimension is smaller than three,
the anomalous discrete R symmetry
has less power to constrain the low-energy theory.

\begin{itemize}
\item Spontaneous breaking
\end{itemize}

In the discussion above, we have considered
R-symmetry breaking by non-perturbative effects when
R-symmetries are anomalous.
Here we comment on another type of symmetry breaking;
they can be broken spontaneously by the VEVs of scalar fields in the
form
$U(1)\times R \rightarrow R'$.
That is, we consider a spontaneous symmetry breaking, where
some scalar fields with non-vanishing $U(1)$ and $R$ charges
develop their VEVs and they break
$U(1)$ and $R$ symmetries in such a way that  an unbroken $R'$
symmetry
remains intact.
(Its order is denoted by $N'$ below.)
Even in such symmetry breaking, we can obtain
the GS condition for the unbroken $R'$ from the GS condition for the
$U(1)$
and R-anomalies.
Suppose that we have the GS condition for the $U(1)$ symmetry as
\begin{equation}
Tr Q T({\bf R}_{G_a})/k_a= Tr Q T({\bf R}_{G_b})/k_b,
\end{equation}
where $Q$ is the $U(1)$ charge.
Since the unbroken $R'$ charge is a linear combination of $R_i$ and
$Q$,
 the mixed anomalies for $R'$ should also satisfy the
GS condition,
\begin{equation}
Tr R' T({\bf R}_{G_a})/k_a= Tr R' T({\bf R}_{G_b})/k_b.
\end{equation}
Here the anomaly coefficients $Tr R' T({\bf R}_{G_a})$ are
defined modulo $N'T({\bf R}^{(f)}_{G_a})$.

Through the symmetry breaking $U(1)\times R \rightarrow R'$,
some matter fields may  gain mass terms like
\begin{equation}
W\sim m \Phi \bar \Phi.
\end{equation}
Such a pair of the matter fields $\Phi$ and $\bar \Phi$ should form
 a vector-like representation of $G_{a}$ and have
opposite $R'$ charges of the unbroken $R'$ symmetry.
The heavy modes of this type have therefore no contribution to the 
mixed anomalies between the gauge symmetry $G_a$ and
the unbroken $R'$ symmetry.
This implies that  the above GS condition for the unbroken $R'$
 remains
unchanged even after the spontaneous symmetry breaking.
The symmetry breaking
$U(1)\times R \rightarrow R'$ also allows
 Majorana mass terms like
\begin{equation}
W\sim m \Phi \Phi.
\end{equation}
This type of Majorana mass terms can appear for an even
 order $N'$ of the $R'$ symmetry if the $R'$
charge of $\Phi$ is $N'/2$ and $\Phi$ is
in a real representation  ${\bf R}_{G_a}$
of the unbroken gauge group $G_a$.
The field $\Phi$ contributes to the anomaly coefficient as
$\frac{N'}{2}T({\bf R}_{G_a})$.
That however may change only the modulo-structure of the anomaly
coefficients.
For $SU(N)$ gauge group, this contribution is
obtained as $\frac{N'}{2}\times
({\rm integer})$.
Thus, the modulo-structure does not change, that is,
the anomaly coefficients  $Tr R' T({\bf R}_{G_a})$ are defined
modulo $N'/2$.
However, for other gauge groups, the modulo-structure
of the anomaly coefficients may change.

\begin{itemize}
\item Gravity-induced supersymmetry and Gauge symmetry breaking
\end{itemize}

The most important difference of the discrete R-symmetries
compared with T-duality in phenomenological applications
comes from the fact that (for the heterotic orbifold
string models) the moduli and  dilaton superfields have vanishing 
R-charges.
The VEVs of their bosonic components do not therefore violate
the discrete R-symmetries in  the  perturbation theory.
(We have discussed above the non-perturbative effects due to the VEV of 
the dilaton, which may be small in a wide class of  models.)
However,
the F-components of the moduli and  dilaton superfields
have non-zero R-charges. Therefore, since the VEVs of these
F-components generate soft-supersymmetry breaking (SSB) terms
at low-energy, the SSB terms do not have to respect the discrete 
R-symmetries. \footnote{
Whether
  the non-perturbative effects due to the VEV of the dilaton
  do play an important roll in the SSB sector depends on the
  R charge of the dilaton, and one has to check it explicitly
  for a given model.}
Fortunately,  in the visible sector,
the scale of the R-symmetry breaking must be of the same
order as that of  supersymmetry breaking. 
 If the order of the discrete R-symmetry is even, 
the VEVs of these F-components break the discrete R-symmetry
down to its subgroup $Z_2$, an R-parity.
That is  an interesting observation because it may be an origin
of the R-parity of the MSSM.

Gauge symmetry breaking can be achieved by VEVs of chiral
supermultiplets in a non-trivial representation of the gauge group
or by non-trivial Wilson lines. Clearly, if the chiral supermultiplets
have  vanishing R-charges and only their scalar
components acquire VEVs,  the discrete  R-symmetries remain
unbroken. Similarly, the Wilson lines do not break the discrete
R-symmetries
because gauge fields have no R charge.
As a consequence, the discrete R-symmetries have a good chance
to be intact at low-energy if the non-perturbative effects are small.

\subsubsection{Constraints on low-energy beta-functions}
Only anomaly-free discrete R-symmetries
remain as intact symmetries in a low-energy effective theory.
Obviously, the model with anomaly-free
discrete R-symmetries corresponds to $A^{R_i}_{G_a}=0$
(mod $N_iT({\bf R}^{(f)}_{G_a}))$.
Consider  for instance $SU(N)$ gauge groups for which
$T({\bf R}^{(f)}_{G_a})=1/2$ is usually satisfied.
Then in models, which have
no oscillator mode in a non-trivial representations of $SU(N)$,
  the relation between R-anomalies
and beta-function coefficients lead to
\begin{equation}
b_a = 2 A^{}_{G_a}=0,
\end{equation}
mod $N_i$ for any gauge group $G_a$.
For example, the $Z_3$ orbifold model
with anomaly-free R-symmetries leads to
$b_a=3n_a$ with integer $n_a$, while
the $Z_4$ orbifold model with anomaly-free R-symmetries
leads to $b_a=2n_a$.
Similarly,
$b_a=1$ would be possible in $Z_6$-II orbifold models
because $N_i=(6,3,2)$ as one can see from Table 1.

Even for anomalous discrete R-symmetries,
the GS condition for R-anomalies and the relation between
beta-function coefficients (\ref{GS-R}), (\ref{anomR-b}), 
(\ref{GS-b}) would have phenomenological
implications. As discussed at the beginning in this section,
the non-perturbative effects  can generate
operators like $e^{-aS}\Phi^1 \cdots \Phi^n$.
If its canonical dimension is larger than four, its contribution
to low-energy beta-functions may be assumed to be small.
\footnote{If the operator produces
non-invariant mass terms like $M \Phi \Phi'$ with $M$
larger than the low-energy scale,
the low-energy spectrum may change. Then
the power of the discrete R-symmetries decreases.}

As for the MSSM
we find $b_3=-3$ and $b_2=1$ for $SU(3)$ and $SU(2)$, respectively.
That is, we have $b_2 - b_3=4$, implying
the MSSM can not be realized, e.g. in $Z_3$ orbifold models,
because $Z_3$ orbifold models require
$b_a - b_b=0$ mod $3$ if the effects of the symmetry breaking
of the discrete R-symmetries can be neglected.
Similarly, the model with $b_2 - b_3=4$ can not be
obtained in the $Z_6$-I, $Z_7$ or $Z_{12}$-I orbifold models.

Finally, we comment on the symmetry breaking effects by quantum
effect. 
When a discrete (flavor) symmetry is anomalous, 
breaking terms can appear in Lagrangian, e.g. 
by instanton effects, such as 
$\frac{1}{M^n}\Lambda^m \Phi_1 \cdots \Phi_k$, 
where $\Lambda$ is a dynamical scale and $M$ is a 
typical (cut-off) scale.
Within the framework of string theory
discrete anomalies as well as anomalies of continuous 
gauge symmetries can be canceled by the GS mechanism 
unless discrete symmetries are accidental.
In the GS mechanism, dilaton and moduli fields, 
i.e. the so-called GS fields $\Phi_{GS}$, transform 
non-linearly under anomalous transformation.
The anomaly cancellation due to the GS mechanism imposes 
certain relations among anomalies.
(See e.g. Ref.~\cite{Araki:2008ek}.) 
Stringy non-perturbative effects as well as field-theoretical effects 
induce terms in Lagrangian such as 
$\frac{1}{M^n}e^{-a\Phi_{GS}} \Phi_1 \cdots \Phi_k$.
The GS fields $\Phi_{GS}$, i.e. dilaton/moduli fields 
are expected to develop non-vanishing vacuum expectation values 
and above terms correspond to breaking terms of discrete symmetries.

\newpage

\section{Conclusion}
Here we conclude by summarizing the results of this thesis
and considering the future prospects.

In this thesis we have studied ten dimensional N=1 super 
Yang-Mills theory on various types of compactifications.
These results can be also applied in lower dimensions as $D=6,8$.  
In the theory we considered field theoretical approach is possible to
obtain the chiral fermion coupled under non-abelian gauge symmetries
and also calculate matter spectrums, Yukawa couplings and other
couplings related to the low-energy physics.   
It is quite interesting for phenomenology to survey a successful
string compactifications.
Although the torus compactifications with magnetic flux is one of simple
background configurations, one can calculate explicitly the form of
wavefunctions and Yukawa couplings. 
We have extended these analysis to other more complicated
compactifications like orbifold background, toron background with
non-Abelian Wilson line. 
We have seen in such constructions there are many interesting features
for low-energy physics and the set up of these studies may apply in 
general Calabi-Yau compactifications in principle.
It enables us to survey more widely range of the theory. 

In section 3, we studied low-energy effective action namely
superpotential and Kahler potential. Yukawa couplings itself are 
important to link the SM and high energy UV completion underlying
theory. Following the analysis of the three wavefunction overlap, we 
have obtained generic n-point couplings.
We have found that higher order couplings are written 
as products of three-point couplings.
This behavior is the same as higher order 
amplitudes of CFT, that is, higher order amplitudes 
are decomposed as products of three-point 
amplitudes in intersecting D-brane models.
Our results on higher order couplings 
would be useful in phenomenological applications.
Numerical analysis on higher order couplings is also possible.

In section 4, We have shown the non-abelian flavor symmetries 
can appear dynamically in the couplings.
Because these are constrained by coupling selection rule as well as
heterotic orbifold models and they are easily understood geometrically.
We have found that $D_4$, $\Delta(27)$ and other 
$Z_g \ltimes (Z_g \times Z_g)$ 
flavor symmetries can appear from magnetized brane models 
with non-vanishing Wilson lines.
Matter fields with several representations of 
these discrete flavor symmetries can appear.
When we consider vanishing Wilson lines, 
these flavor symmetries are enhanced like 
$D_4 \times Z_2$, $\Delta(54)$, etc.
They propovided a realization of the {\it co-existence} 
of the different types of the flavor symmetries in GUT type models.
These results are interesting for model building 
of realistic quark/lepton mass matrices and mixings.

For the purposes to survey the low-energy effective theory, it is
important to study other background.
Using the field theoretical approaches one can study the widely range
of the background. In section 5, we studied 
the orbifolding with flux background which is one of the explicit
examples of the non-trivial background. 
Even in a simple construction, i.e. $T^2/Z_2$
orbifold, it has a rich structure.
Odd modes can have zero-modes 
and couplings are controlled by 
the orbifold periodicity of wavefunctions.
We have also discussed the flavor symmetry breaking 
on the orbifold background.

It is important to study anomalies of non-abelian 
flavor symmetries.
If string theory leads to anomaly-free effective low-energy 
theories including discrete symmetries, 
anomalies of discrete symmetries must be canceled by 
the Green-Schwarz mechanism.
In section 6 we study those discrete anomalies within the framework of
heterotic orbifold models in~\cite{Araki:2008ek}, and it was shown
that discrete anomalies can be canceled by 
the Green-Schwarz mechanism.
We found the important relations of discrete R-anomalies 
with U(1) anomalies and others.
Furthermore we have studied the possible anomaly of discrete flavor
symmetries come from several types of string models e.g. heterotic
orbifold models and D-brane models.
In addition, there are many constraints from the stringy consistency
conditions. 
The most important consistency condition for string theory model with
D-branes is the RR charge cancellation condition. 
This condition arises as a consequence of Gauss law 
constraint of the internal space.
Since the globally defined string model must satisfy above conditions,
several constraints on D-brane configurations are obtained. 
As a result of this constraint, it allows us to know all the spectrums including 
chiral and non-chiral multiplets and remaining gauge symmetry. 
Therefore it is quite important to investigate globally 
the string compactification models. 

To distinguish string vacuum it is rather important to study the moduli
parameters. Since $\mathcal{N}=1$ supersymmetry must be broken in a
certain scale we have discussed the soft supersymmetry breaking terms.  
Even in the type IIB theory the soft supersymmetry
breaking terms are dominated u moduli contributions, soft
supersymmetry terms still have constraint from this kind of symmetries.
Actually it was found explicitly that certain models which have
discrete flavor symmetries prohibit dangerous FCNC~\cite{Grimus}. 

It is also interesting to study the flavor structure in other 
background. 
There are already many kinds of explicit construction of
wavefunctions, for example, sphere background, warped
compactification. 
It is possible to survey more the flavor structure in such a
background.  
It is also important for the phenomenological view point.
It may give some hints to derive the realistic pattern of Yukawa
couplings and help to construct realistic vacua.
In addition, we can discuss the phenomenological aspects of the flavor
sector. 
Once if we have a mechanism to break low-energy supersymmetry, 
the relevant soft supersymmetry breaking terms would be also related
to flavor structures. 
Thus we can analyze the low-energy spectrum including the super
particle for future collider experiments.
  
Another application of the moduli field is the inflation.
It is a challenging issue to realize a successful scenario of inflation
within the framework of high energy underlying theory. Some of moduli
fields have naturally flat directions due to the supersymmetry and
they could be naturally candidates of the scalar fields responsible for 
inflation, inflaton.   
There are many studies for the natural inflation potential in particle 
physics of string theory.
The scale of inflation might be the same magnitude of the scale of 
the low-energy supersymmetry breaking and such models could be
implemented to the connection between underlying theory and cosmology
or phenomenology.  
Indeed cosmological observation is predicted in a certain model of
the moduli potential.  
It is quite interesting to investigate the moduli stabilization
mechanism and low-energy supersymmetry breaking, 
in which we can also discuss about flavor phenomenology.
All the above topics are left for near future. 

\subsection*{Acknowledgment}
I would like to thank my supervisor Tatsuo Kobayashi for fruitful
discussions, collaboration and leading me during my doctor course.
I would like to be grateful to H. Abe, H. Aoyama, 
K.-S. Choi, T. Eguchi, M. Fukuma, S. Hashimoto, H. Hata,
K. Izawa,  E. Itou,  T. Kaneko, H. Kawai, T. kugo, M. Kurachi,
H. Matsufuru, J. Noaki, Y. Omura, T. Onogi, S. Sasakura, R. Takahashi,
M. Tanimoto, T. Uematsu, K. Sugiyama, H. Kunitomo, S. Terashima,
K. Yoshioka, K. Yoshida, N. Yamada and T. Yamazaki for fruitful
discussions and study meeting.  
The author is supported in part by the Grant-in-Aid for 
Scientific Research No.~21$\cdot$897 from the 
Ministry of Education, Culture, Sports, Science and Technology of
Japan.

\appendix

\section{Dimensional reduction and the low-energy effective
action}\label{dimensional} 

Here we construct the effective
four dimensional super Yang-Mills theory.
We start with ten dimensional $\mathcal{N}=1$ super Yand-Mills theory
which is the low-energy limit of the DBI action,  
\begin{align}
S= 
\frac{1}{g^2}\int d^{10}x {\rm{Tr}}
\left(
-\frac{1}{4}F_{MN}F^{MN}+\frac{i}{2}\bar{\lambda}\Gamma^M D_M \lambda
\right),
\end{align}
where $g^2$ can be related to the string theory as 
$g^2=4\pi e^{\phi_{10}} (2\pi\sqrt{\alpha'})^6$.
We take the gauge groups  $U(N)$ and the generators are divided 
into two parts, the Cartan parts $U_a$ and off diagonal elements
$e_{ab}$
\begin{align}
(U_a)_{ij}= \delta_{ai} \delta_{aj}, \ \ \
(e_{ab})_{ij}=\delta_{ai} \delta_{bj}.
\end{align}
The gauge fields are consist of 
\begin{align}
A_M= B_M +W_M = B_M^a U_a + W_M^{ab}e_{ab}
\end{align}
and gaugino fields are also expanded in the same way.
We also expand the gauge fields as background configurations as
\begin{align}
B_i^a    &= \langle B_i^a(y^i) \rangle+b_i^a(x,y) \nonumber \\
W_i^{ab} &= \Phi_i^{ab}(x,y). 
\end{align}


In the following we will not rewrite the entire action in terms of
the fields
introduced above, but we will only write the relevant terms, namely
the quadratic
terms involving the scalar and fermion fields and the trilinear terms
involving a scalar and two fermions: we will derive the K{\"{a}}hler
metrics from the former
and  the Yukawa couplings from the latter. We will also restrict our
considerations to  toroidal compactifications.

The quadratic terms for the
fields $ \Phi_M^{ab}(x^{\mu} , y^{i})$ are followings 
\begin{eqnarray}
\mathcal{L}_{2}^{(\Phi)} =
-\frac{1}{2g^2} 
{\rm{tr}}\left[ 
D_\mu \Phi_{i} D^\mu \Phi^i +  \tilde{D}_i \Phi_j \tilde{D}^i  \Phi^j   
-  \tilde{D}_i \Phi_j \tilde{D}^j  \Phi^i   
-i  G_{ij} [\Phi^i, \Phi^j]
\right]  
\label{quacsca1}
\end{eqnarray}
where
\begin{eqnarray}
& {D}_\mu \Phi_{j}= \partial_{\mu} \Phi_{j}-i [B_{\mu}+W_\mu,
\Phi_{j}] 
,
 \tilde{D}_i \Phi_{j}= \partial_i \Phi_{j} - i 
 [ \langle B_i \rangle, \Phi_j],  &  \\
&G_{ij}  \equiv \partial_i \langle B \rangle_j 
- \partial_j \langle B \rangle_i &
\label{F-F}
\end{eqnarray}
where $G_{ij}$  is the field strength obtained from
the background field $B$.
By using the properties of Lie algebra $U$ and $e$ 
one can express the above quadratic terms as 
\begin{align}
\mathcal{L}_{2}^{(\Phi)} =&
-\frac{1}{2g^2}\left[
(D_\mu \Phi_i)^{ab} (D^\mu \Phi^i)^{ba}
+(\tilde{D}_i \Phi_j)^{ab} (\tilde{D}^i \Phi^j)^{ba}
-(\tilde{D}_i \Phi_j)^{ab} (\tilde{D}^j \Phi^i)^{ba}
\right]
\nonumber \\
&+
\frac{i}{2g^2}\Phi^{i,ab}(G_{ij}^a-G_{ij}^b)\Phi^{j,ba}
\end{align}
where $ (D_\mu \Phi_i)^{ab}=\partial_\mu
  \Phi_i-i(B_\mu^a-B_\mu^b)\Phi_i^{ab}
-i (W_\mu^{ac}\Phi_i^{cb}-\Phi_i^{ac}W_\mu^{cb}  )$.
Taking integration by part for $(\tilde{D}_i \Phi_j)^{ab} (\tilde{D}^i
  \Phi^j)^{ba}$,
we obtain 
\begin{align}
(\tilde{D}_i \Phi_j)^{ab} (\tilde{D}^i \Phi^j)^{ba}
= - \Phi_j^{ab} (\tilde{D}_i \tilde{D}^i) \Phi^{j,ba},
\end{align}
and similarly we obtain 
\begin{align}
(\tilde{D}_i \Phi_j)^{ab} (\tilde{D}^j \Phi^i)^{ba}
&= - \Phi_j^{ab} (\tilde{D}_i \tilde{D}^j \Phi^i)^{ba} \nonumber \\
&= - \Phi_j^{ab} ([\tilde{D}_i, \tilde{D}^j]+
\tilde{D}^j\tilde{D}_j ) \Phi^{i,ba}  \nonumber \\
&=- \Phi_j^{ab} ([\tilde{D}_i, \tilde{D}^j])\Phi^{i,ba},
\end{align}
where we use the gauge fixing condition $\tilde{D}_i \Phi^i=0$.
The commutator $[\tilde{D}_i,\tilde{D}_j]$ is given by
$[\tilde{D}_i,\tilde{D}_j]=-i (G_{ij}^a-G_{ij}^b)$.
Then we combine these results to rewrite the Lagrangian, 
\begin{align}
\mathcal{L}_{2}^{(\Phi)} =&
\frac{1}{2g^2}
\left[
\Phi_j^{ab}(D_\mu)^2 \Phi^{j,ba}
+\Phi_j^{ab}(\tilde{D}_i \tilde{D}^i) \Phi^{j,ba}
+2i \Phi_j^{ab}(G^{j\ ,a}_{\ i}-G^{j\ ,b}_{\ i})\Phi^{j,ba}
\right].
\end{align}
Thus, we obtain the equation of motion for $\Phi_j^{ab}$ as followings  
\begin{align}
\tilde{D}^2\Phi_j^{ab}+2i(F^{i\ ,a}_{\ j}-F^{i\ ,b}_{\ j})
\Phi_i^{ab}=-m^2 \Phi_j^{ab}, 
\end{align}
where $-m^2$ means the eigenvalue for the operator defined
in left hand side.
Therefore zero-mode wavefunctions are corresponding to the 
solution with vanishing $m^2$.
We use the usual Kaluza-Klein expansions for the field $\Phi$ as 
\begin{eqnarray}
\Phi^{ab}_{i} (x,y) = \sum_n \varphi^{ab}_{n,i} (x^{\mu}  )
\otimes
\phi^{ab}_{n} (y^{i} )
~~;~~\Psi^{ab} (x,y) =  \sum_n \psi_n^{ab}(x^{\mu}) \otimes
\eta_n^{ab}(y^i)\,\,.
\label{exps63}
\end{eqnarray}

The spectrum of the Kaluza-Klein states and their
wavefunctions along the compact directions are
obtained by  solving the eigenvalue equations for the six-dimensional
Laplace and Dirac operators:
\begin{eqnarray}
-{\tilde{D}}_{k}  {\tilde{D}}^{k}  \phi^{ab}_{n}  =
 m_{n}^{2} \phi^{ab}_{n}
 ~~,~~
i \gamma_{(6)}^i \tilde{D}_i \eta^{ab}_n =
\lambda_n\,\eta^{ab}_n
\label{eq88}
\end{eqnarray}
with the correct periodicity conditions along the compactified
directions.

Inserting Eq. (\ref{exps63}) and the first equation in (\ref{eq88})
in Eq. (\ref{quacsca1}) and
using the coordinates $z$ and ${\bar{z}}$ 
for describing the torus $T^2$, one gets scalar mass terms for six dimensions.
We see that there are
 two towers  of Kaluza-Klein states
for each torus, with masses given by:
\begin{eqnarray}
m_{n}^{2} = 
\frac{1}{(2 \pi R)^2} \left[\sum_{s=1}^{3}
\frac{2 \pi |M^{s}_{ab}| }{\mathcal{A}^{(s)}}
\left( 2N_s +1 \right) \pm
\frac{4 \pi I^r_{ab}}{ \mathcal{A}^{(r)}} \right]
\label{KKmass}
\end{eqnarray}
where $N_s$ is an integer given by the oscillator number operator. The
presence of
the oscillator number is a consequence of the fact that
the Laplace operator can be written in terms of the creation and
annihilation operators
of an harmonic oscillator.
One can have a massless state only if the following
condition is satisfied for $I_r>0$ or $I_r<0$
\begin{eqnarray}
\sum_{s=1}^{3} \frac{2 \pi |I_{s}| }{{\mathcal{A}}^{(s)}}   -
\frac{4 \pi |I_r| }{{\mathcal{A}}^{(r)}}  =0 \Longrightarrow
\frac{1}{2} \sum_{s=1}^{3} \frac{ |I_{s}|}{{\mathcal{A}}^{(s)}}   -
\frac{ |I_r|}{{\mathcal{A}}^{(r)}}  =0 \,\,.
\label{zerom}
\end{eqnarray}
In this case one  keeps ${\cal{N}}=1$ supersymmetry because there is a
massless
scalar that is in the same  chiral multiplet as a fermion that
we will study later. If one of the $I_r$'s is vanishing and the other
two are equal, then
we have an additional massless excitation corresponding to  an
extended
${\cal{N}}=2$ supersymmetry.

The SUSY conditions given in Eq. (\ref{zerom}) show that only one
of the two scalars is massless. In particular,  by choosing in such
equation $r=1$ and
$I_{1}>0$, we see that $\varphi_{1,-}$ is the massless scalar.
The corresponding  internal wave-function has been determined in
Ref.~\cite{Cremades:2004wa} and is the product of three eigenfunctions.
Instead,  by taking $r=1$ and $I_{1}<0$ we have that $\varphi_{1,+}$
becomes the massless mode.
It is useful to notice that
$(\phi_{r,+}^{ab;n_r})^\dag=\phi_{r,-}^{ba;n_r}$, and
furthermore, the reality of the scalar action implies:
\begin{eqnarray}
\phi^{ba }_{0} = ( \phi_{0}^{ab})^{*}.
\label{comco8}
\end{eqnarray}
In conclusion, by performing the Kaluza-Klein reduction of the
low-energy
world-volume action of a stack of D9 branes on
$R^{3,1} \times T^2 \times T^2 \times T^2$, we have found two  towers
of Kaluza-Klein states for each of  the scalar fields $\varphi_{r,
\pm}$
for $r=1,2,3$. In general,  only the lowest state of
one
of the two towers and for a particular value of  $r$ (say $r=1$ if
Eq. (\ref{zerom})
is satisfied for $r=1$) is
massless,  depending on the sign of $I_{1}$.
We have now all the elements for computing the
K{\"{a}}hler metric of the scalars $\varphi_{\pm}$.

\section{Models}

Here we give two examples of models, whose family numbers 
of bulk modes differ from three.
That is, one model has two bulk families and 
the other has eighteen bulk families.
We start with the ten-dimensional $U(N)$ super Yang-Mills theory 
on the background $R^{3,1}\times T^6/(Z_2 \times Z'_2)$.
We consider the trivial orbifold projections 
$P=P'=1$.

In the first model, we introduce the following magnetic flux,
\begin{eqnarray}\label{eq:model-3}
F_{45} &=& \left(
\begin{array}{ccc}
0 \times {\bf 1}_{N_a \times N_a} & & 0  \\
 & -2 \times {\bf 1}_{N_b \times N_b} & \\
0 & & 2 \times {\bf 1}_{N_c\times N_c}
\end{array}\right),  \nonumber \\
F_{67} &=& \left(
\begin{array}{ccc}  
0 \times  {\bf 1}_{N_a \times N_a} & & 0 \\
 & -1 \times {\bf 1}_{N_b \times N_b} & \\
0 & & 1 \times  {\bf 1}_{N_c\times N_c}
\end{array}\right),  \\
F_{89} &=& \left(
\begin{array}{ccc}  
  0 \times {\bf 1}_{N_a \times N_a} & & 0 \\
 & -1 \times {\bf 1}_{N_b \times N_b} & \\
0 & & 1 \times {\bf 1}_{N_c\times N_c}
\end{array}\right). \nonumber 
\end{eqnarray}
This magnetic flux satisfies the condition (\ref{eq:SUSY-condition}) 
and breaks the gauge group 
$U(N) \rightarrow U(N_a)\times U(N_b) \times U(N_c)$, 
although the orbifold projections are trivial $P=P'=1$.
Then, we can analyze the zero-modes as section 3.4.
The result is shown in Table 4.
This model has two bulk families, when we consider $\lambda^{ab}$ 
and $\lambda^{ca}$ as left-handed and right-handed matter fields.
This flavor number is not realistic.
However, in orbifold models 
it is possible to assume that one family appears on one of fixed
points.

\begin{table}[htb]
\begin{center}
\begin{tabular}{|c|c|c|c|c|} \hline
     & $I^i_{ef}$ & chirality & wavefunction & the total number  \\
     &            &           &               & of zero-modes \\
     \hline 
$\lambda^{ab}$ & $(2,1,1)$ & $\lambda^{ab}_{+,+,+}$ & 
$\Theta^{j_1}_{\rm even}\Theta^{j_2}_{\rm even}
\Theta^{j_3}_{\rm even}$ & 2 \\ 
$\lambda^{ca}$ & $(2,1,1)$ & $\lambda^{ca}_{+,+,+}$ & 
$\Theta^{j_1}_{\rm even}\Theta^{j_2}_{\rm even}
\Theta^{j_3}_{\rm even}$ & 2 \\ 
$\lambda^{cb}$ & $(4,2,2)$ & $\lambda^{cb}_{+,+,+}$ & 
$\Theta^{j_1}_{\rm even}\Theta^{j_2}_{\rm even}
\Theta^{j_3}_{\rm even}$ & 12 \\ 
\hline \end{tabular}
\end{center}
\caption{Two-family model from the bulk.
\label{Model-3}}
\end{table}

We give another example.
We use the same orbifold projections, i.e. $P=P'=1$.
We introduce the following magnetic flux,
\begin{eqnarray}\label{eq:model-4}
F_{45} &=& \left(
\begin{array}{ccc}
0 \times {\bf 1}_{N_a \times N_a} & & 0  \\
 & -6 \times {\bf 1}_{N_b \times N_b} & \\
0 & & 6 \times {\bf 1}_{N_c\times N_c}
\end{array}\right),  \nonumber \\
F_{67} &=& \left(
\begin{array}{ccc}  
0 \times  {\bf 1}_{N_a \times N_a} & & 0 \\
 & -3 \times {\bf 1}_{N_b \times N_b} & \\
0 & & 3 \times  {\bf 1}_{N_c\times N_c}
\end{array}\right),  \\
F_{89} &=& \left(
\begin{array}{ccc}  
  0 \times {\bf 1}_{N_a \times N_a} & & 0 \\
 & -3 \times {\bf 1}_{N_b \times N_b} & \\
0 & & 3 \times {\bf 1}_{N_c\times N_c}
\end{array}\right). \nonumber 
\end{eqnarray}
We study the spinor fields $\lambda^{ab}$, in whose Dirac 
equations the difference of magnetic fluxes 
$I^i_{ab}=(6,3,3)$ appears.
Their zero-modes correspond to $\lambda^{ab}_{+,+,+}$, 
which transform 
$\lambda^{ab}_{+,+,+}(x,y_m,y_n) \rightarrow
\lambda^{ab}_{+,+,+}(x,-y_m,y_n)$
for both $Z_2$ and $Z'_2$ actions.
These boundary conditions are satisfied with 
the wavefunctions  
$\Theta^{j_1}_{\rm even}(y_4,y_5) \Theta^{j_2}_{\rm even}(y_6,y_7) 
\Theta^{j_3}_{\rm even}(y_8,y_9) $ and 
$\Theta^{j_1}_{\rm odd}(y_4,y_5) \Theta^{j_2}_{\rm odd}(y_6,y_7) 
\Theta^{j_3}_{\rm odd}(y_8,y_9) $.
The number of zero-modes corresponding to 
the former wavefunctions is given by the product of 
$I^1_{ab({\rm even})}=4$, $I^2_{ab({\rm even})}=2$ and 
$I^3_{ab({\rm even})}=2$, while the zero-mode number 
corresponding to the latter is given by the product of 
$I^1_{ab({\rm odd})}=2$, $I^2_{ab({\rm odd})}=1$ and 
$I^3_{ab({\rm odd})}=1$.
Thus, the total number of $\lambda^{ab}$ zero-modes is 
equal to 18$(=16+2)$.
Similarly, we can analyze zero-modes for $\lambda^{bc}$ 
and $\lambda^{ca}$.
The result is shown in Table 5.
For these zero-modes, only two forms of wavefunctions 
are allowed, that is, one is 
$\Theta^{j_1}_{\rm even}(y_4,y_5) \Theta^{j_2}_{\rm even}(y_6,y_7) 
\Theta^{j_3}_{\rm even}(y_8,y_9) $ and the other is 
$\Theta^{j_1}_{\rm odd}(y_4,y_5) \Theta^{j_2}_{\rm odd}(y_6,y_7) 
\Theta^{j_3}_{\rm odd}(y_8,y_9) $.
The numbers of zero-modes corresponding to the former 
and latter are shown in the third and fourth columns.
Six-dimensional chirality of all zero-modes 
correspond to $\lambda_{+,+,+}$ and they are 
omitted in the table.

This model has 18 families.
It seems that this family number is too large.
However, we can reduce the light family number 
if we assume anti-families of $(\bar N_a,N_b)$ and 
$(N_a,\bar N_c)$ matter fields on fixed points 
and  their mass terms with the above families of matter fields.
Such mass terms are possible for zero-modes 
corresponding to $\Theta^{j_1}_{\rm even}(y_4,y_5) 
\Theta^{j_2}_{\rm even}(y_6,y_7) \Theta^{j_3}_{\rm even}(y_8,y_9) $.
Thus, when we assume $n$ anti-families, the number of 
light families reduces to $(18-n)$.
This type of models has an interesting aspect, that is, 
some families of matter fields correspond to 
$\Theta^{j_1}_{\rm even}(y_4,y_5) 
\Theta^{j_2}_{\rm even}(y_6,y_7) \Theta^{j_3}_{\rm even}(y_8,y_9) $ 
and other families of matter fields correspond to 
$\Theta^{j_1}_{\rm odd}(y_4,y_5) \Theta^{j_2}_{\rm odd}(y_6,y_7) 
\Theta^{j_3}_{\rm odd}(y_8,y_9) $.
In general, other combinations of wavefunctions can appear 
in zero-modes of matter fields.
Such asymmetry appears in this type of models.
Thus, their flavor structure is rich.

\begin{table}[htb]
\begin{center}
\begin{tabular}{|c|c|c|c|c|} \hline
     & $I^i_{ef}$ & No. of zero-modes & No. of
     zero-modes & the total number  \\
     &            &  $\Theta^{j_1}_{\rm even}\Theta^{j_2}_{\rm even}
\Theta^{j_3}_{\rm even}$  &   $\Theta^{j_1}_{\rm odd}\Theta^{j_2}_{\rm
     odd}
\Theta^{j_3}_{\rm odd}$         & of zero-modes \\ \hline 
$\lambda^{ab}$ & $(6,3,3)$ & 16 & 2 & 18 \\ 
$\lambda^{ca}$ & $(6,3,3)$ & 16 & 2 & 18 \\
$\lambda^{cb}$ & $(12,6,6)$ & 112 & 20 & 132 \\ 
\hline \end{tabular}
\end{center}
\caption{Eighteen-family model from the bulk.
\label{Model-4}}
\end{table}

\section{Possible patterns of Yukawa matrices}\label{sec:possibel-Yukawa}

In this appendix, we show explicitly all of possible Yukawa 
matrices for 15 classes of models in Table \ref{higgs} 
except the models with $I^{(1)}_{bc}=0$ and the model without
zero-modes 
for the Higgs fields.

\subsection{(Even-Even-Even) wavefunctions}

Here, we study the patterns of Yukawa matrices 
in the models, where zero-modes of left, right-handed 
matter fields and Higgs fields correspond to 
even, even and even functions, respectively.

\subsubsection{4-4-8 model}

Let us study the model with 
$(|I^{(1)}_{ab}|,|I^{(1)}_{ca}|,|I^{(1)}_{bc}|)=(4,4,8)$.
The following table shows zero-mode wavefunctions of left,
right-handed 
matter fields and Higgs fields.
\begin{center}
\begin{tabular}{c|c|c|c}
& $L_i (\lambda^{ab})$ & $R_j (\lambda^{ca})$ & $H_k (\lambda^{bc})$ 
\\ \hline
0 & 
$\Theta^{0,4}$ & $\Theta^{0,4}$ & $\Theta^{0,8}$ \\  
1 & 
$\frac{1}{\sqrt{2}}\left(\Theta^{1,4}+\Theta^{3,4}\right)$ & 
$\frac{1}{\sqrt{2}}\left(\Theta^{1,4}+\Theta^{3,4}\right)$ & 
$\frac{1}{\sqrt{2}}\left(\Theta^{1,8}+\Theta^{7,8}\right)$ \\  
2 & 
$\Theta^{2,4}$ & 
$\Theta^{2,4}$ & 
$\frac{1}{\sqrt{2}}\left(\Theta^{2,8}+\Theta^{6,8}\right)$ \\  
3 & - & - & 
$\frac{1}{\sqrt{2}}\left(\Theta^{3,8}+\Theta^{5,8}\right)$ \\  
4 & - & - & 
$\Theta^{4,8}$ \\  
\end{tabular}
\end{center}
This model has five zero-modes for the Higgs fields.
Yukawa couplings $Y_{ijk} L_i R_j H_k$ are given by 
\begin{eqnarray}
Y_{ijk}H_k &=& 
\left( \begin{array}{ccc}
y_a H_0 + y_e H_4 & y_f H_3 + y_b H_1 & y_c H_2 \\
y_f H_3 + y_b H_1 & \frac{1}{\sqrt{2}}(y_a+y_e)H_2 
+ y_c(H_0+H_4) & y_b H_3+y_d H_1 \\
y_c H_2 & y_b H_3+y_d H_1 & y_e H_0 + y_a H_4 
\end{array} \right), 
\nonumber
\end{eqnarray}
where
\begin{eqnarray}
\begin{array}{rclcrcl}
y_a &=& \eta_0 + 2\eta_{32}+ \eta_{64}, & & 
y_b &=& \eta_{4} + \eta_{28} + \eta_{36}+ \eta_{60}, 
\nonumber \\
y_c &=& \eta_{8} + \eta_{24} + \eta_{40}+ \eta_{56}, & & 
y_d &=& \eta_{12} + \eta_{20} + \eta_{44} + \eta_{52}, 
\nonumber \\
y_e &=& 2\eta_{16} + 2\eta_{48}, & & & & 
\end{array}
\nonumber
\end{eqnarray}
in the short notation $\eta_N$ defined in Eq.~(\ref{eq:notation-y})
with 
$M=M_1M_2M_3=128$.

\subsubsection{4-5-9 model}

Here we show the model with 
$(|I^{(1)}_{ab}|,|I^{(1)}_{ca}|,|I^{(1)}_{bc}|)=(4,5,9)$.
The following table shows zero-mode wavefunctions of left,
right-handed 
matter fields and Higgs fields.
\begin{center}
\begin{tabular}{c|c|c|c}
& $L_i (\lambda^{ab})$ & $R_j (\lambda^{ca})$ & $H_k (\lambda^{bc})$ 
\\ \hline
0 & 
$\Theta^{0,4}$ & $\Theta^{0,5}$ & $\Theta^{0,9}$ \\  
1 & 
$\frac{1}{\sqrt{2}}\left(\Theta^{1,4}+\Theta^{3,4}\right)$ & 
$\frac{1}{\sqrt{2}}\left(\Theta^{1,5}+\Theta^{4,5}\right)$ & 
$\frac{1}{\sqrt{2}}\left(\Theta^{1,9}+\Theta^{8,9}\right)$ \\  
2 & 
$\Theta^{2,4}$ & 
$\frac{1}{\sqrt{2}}\left(\Theta^{2,5}+\Theta^{3,5}\right)$ & 
$\frac{1}{\sqrt{2}}\left(\Theta^{2,9}+\Theta^{7,9}\right)$ \\  
3 & - & - & 
$\frac{1}{\sqrt{2}}\left(\Theta^{3,9}+\Theta^{6,9}\right)$ \\  
4 & - & - & 
$\frac{1}{\sqrt{2}}\left(\Theta^{4,9}+\Theta^{5,9}\right)$ \\  
\end{tabular}
\end{center}
This model has five zero-modes for Higgs fields.
Yukawa couplings $Y_{ijk} L_i R_j H_k$ are given by 
\begin{eqnarray}
Y_{ijk}H_k &=& 
y_{ij}^0 H_0 + y_{ij}^1 H_1 + y_{ij}^2 H_2 
+ y_{ij}^3 H_3 + y_{ij}^4 H_4, 
\nonumber
\end{eqnarray}
where
\begin{eqnarray}
y_{ij}^0 &=& 
\left( \begin{array}{ccc}
\eta_0 & \sqrt{2}\eta_{36} & \sqrt{2}\eta_{72} \\
\sqrt{2}\eta_{45} & \eta_{9}+\eta_{81} & \eta_{27}+\eta_{63} \\
\eta_{90} & \sqrt{2}\eta_{54} & \sqrt{2}\eta_{18}
\end{array} \right), 
\nonumber \\ 
y_{ij}^1 &=& 
\left( \begin{array}{ccc}
\frac{1}{\sqrt{2}}(\eta_{20}+\eta_{40}) & \eta_{4}+\eta_{76} &
\eta_{32}+\eta_{68} \\
\eta_{5}+\eta_{85} &
\frac{1}{\sqrt{2}}(\eta_{31}+\eta_{41}+\eta_{49}+\eta_{59}) 
& \frac{1}{\sqrt{2}}(\eta_{13}+\eta_{23}+\eta_{67}+\eta_{77}) \\
\sqrt{2}\eta_{50} & \eta_{44}+\eta_{64} & \eta_{22}+\eta_{58} 
\end{array} \right), 
\nonumber \\
y_{ij}^2 &=& 
\left( \begin{array}{ccc}
\frac{1}{\sqrt{2}}(\eta_{20}+\eta_{40}) & \eta_{44}+\eta_{64} &
\eta_{8}+\eta_{28} \\
\eta_{35}+\eta_{55} &
\frac{1}{\sqrt{2}}(\eta_{1}+\eta_{19}+\eta_{71}+\eta_{89}) 
& \frac{1}{\sqrt{2}}(\eta_{17}+\eta_{37}+\eta_{53}+\eta_{73}) \\
\sqrt{2}\eta_{10} & \eta_{26}+\eta_{46} & \eta_{62}+\eta_{82}
\end{array} \right), 
\nonumber \\ 
y_{ij}^3 &=& 
\left( \begin{array}{ccc}
\frac{1}{\sqrt{2}}(\eta_{60}+\eta_{80}) & \eta_{24}+\eta_{84} &
\eta_{12}+\eta_{48} \\
\eta_{15}+\eta_{75} &
\frac{1}{\sqrt{2}}(\eta_{21}+\eta_{39}+\eta_{51}+\eta_{69}) 
& \frac{1}{\sqrt{2}}(\eta_{3}+\eta_{33}+\eta_{57}+\eta_{87}) \\
\sqrt{2}\eta_{30} & \eta_{6}+\eta_{26} & \eta_{42}+\eta_{78}
\end{array} \right), 
\nonumber \\
y_{ij}^4 &=& 
\left( \begin{array}{ccc}
\frac{1}{\sqrt{2}}(\eta_{60}+\eta_{80}) & \eta_{16}+\eta_{56} &
\eta_{52}+\eta_{88} \\
\eta_{25}+\eta_{65} &
\frac{1}{\sqrt{2}}(\eta_{11}+\eta_{29}+\eta_{61}+\eta_{79}) 
& \frac{1}{\sqrt{2}}(\eta_{7}+\eta_{43}+\eta_{47}+\eta_{83}) \\
\sqrt{2}\eta_{70} & \eta_{34}+\eta_{74} & \eta_{2}+\eta_{38} 
\end{array} \right), 
\nonumber 
\end{eqnarray}
in the short notation $\eta_N$ defined in Eq.~(\ref{eq:notation-y})
with 
$M=M_1M_2M_3=180$.

\subsubsection{4-5-1 model}

Here we show the model with 
$(|I^{(1)}_{ab}|,|I^{(1)}_{ca}|,|I^{(1)}_{bc}|)=(4,5,1)$.
The following table shows zero-mode wavefunctions of left,
right-handed 
matter fields and Higgs field.
\begin{center}
\begin{tabular}{c|c|c|c}
& $L_i (\lambda^{ab})$ & $R_j (\lambda^{ca})$ & $H_k (\lambda^{bc})$ 
\\ \hline
0 & $\Theta^{0,4}$ & $\Theta^{0,5}$ & $\Theta^{0,1}$ \\  
1 & $\frac{1}{\sqrt{2}}\left(\Theta^{1,4}+\Theta^{3,4}\right)$ & 
$\frac{1}{\sqrt{2}}\left(\Theta^{1,5}+\Theta^{4,5}\right)$ &  \\  
2 & $\Theta^{2,4}$ & 
$\frac{1}{\sqrt{2}}\left(\Theta^{2,5}+\Theta^{3,5}\right)$ &  \\  
\end{tabular}
\end{center}
This model has a single zero-modes for the Higgs field.
Yukawa couplings $Y_{ijk} L_i R_j H_k$ are given 
\begin{eqnarray}
Y_{ijk}H_k &=& 
\left( \begin{array}{ccc}
\eta_0 & \sqrt{2}\eta_4 & \sqrt{2}\eta_8 \\
\sqrt{2}\eta_5 & (\eta_1+\eta_9) & (\eta_3+\eta_7) \\
\eta_{10} & \sqrt{2}\eta_6 & \sqrt{2}\eta_2 
\end{array} \right)H_0.
\nonumber
\end{eqnarray}
Here we have used the short notation $\eta_N$ defined in
Eq.~(\ref{eq:notation-y}) 
with the omitted value $M=M_1M_2M_3=20$.

\subsubsection{5-5-10 model}

Here we show the model with 
$(|I^{(1)}_{ab}|,|I^{(1)}_{ca}|,|I^{(1)}_{bc}|)=(5,5,10)$.
The following table shows zero-mode wavefunctions of left,
right-handed 
matter fields and Higgs fields.
\begin{center}
\begin{tabular}{c|c|c|c}
& $L_i (\lambda^{ab})$ & $R_j (\lambda^{ca})$ & $H_k (\lambda^{bc})$ 
\\ \hline
0 & 
$\Theta^{0,5}$ & $\Theta^{0,5}$ & $\Theta^{0,10}$ \\  
1 & 
$\frac{1}{\sqrt{2}}\left(\Theta^{1,5}+\Theta^{4,5}\right)$ & 
$\frac{1}{\sqrt{2}}\left(\Theta^{1,5}+\Theta^{4,5}\right)$ & 
$\frac{1}{\sqrt{2}}\left(\Theta^{1,10}+\Theta^{9,10}\right)$ \\  
2 & 
$\frac{1}{\sqrt{2}}\left(\Theta^{2,5}+\Theta^{3,5}\right)$ & 
$\frac{1}{\sqrt{2}}\left(\Theta^{2,5}+\Theta^{3,5}\right)$ & 
$\frac{1}{\sqrt{2}}\left(\Theta^{2,10}+\Theta^{8,10}\right)$ \\  
3 & - & - & 
$\frac{1}{\sqrt{2}}\left(\Theta^{3,10}+\Theta^{7,10}\right)$ \\  
4 & - & - & 
$\frac{1}{\sqrt{2}}\left(\Theta^{4,10}+\Theta^{6,10}\right)$ \\  
5 & - & - & 
$\Theta^{5,10}$ \\  
\end{tabular}
\end{center}
This model has six zero-modes for Higgs fields.
Yukawa couplings $Y_{ijk} L_i R_j H_k$ are obtained as
\begin{eqnarray}
\lefteqn{Y_{ijk}H_k} 
\nonumber \\ &=& 
\left( \begin{array}{ccc}
y_a H_0 + y_e H_5 & y_b H_1 + y_e H_4 & y_c H_2 + y_d H_3 \\
y_b H_1 + y_e H_4 & y_c H_0 + \frac{1}{\sqrt{2}}(y_a H_2+y_f H_3) +
y_d H_5 
& \frac{1}{\sqrt{2}}(y_d H_1+y_e H_2 + y_b H_3 + y_c H_4) \\
y_c H_2 + y_d H_3 & \frac{1}{\sqrt{2}}(y_d H_1+y_e H_2 + y_b H_3 + y_c
H_4)
& y_b H_0 + \frac{1}{\sqrt{2}}(y_f H_1 + y_a H_4) + y_a H_5 
\end{array} \right). 
\nonumber
\end{eqnarray}
\begin{eqnarray}
\begin{array}{rclcrcl}
y_a &=& \eta_0 + 2\eta_{50}+ 2\eta_{100}, & & 
y_b &=& \eta_{5} + \eta_{45} + \eta_{55} + \eta_{95} + \eta_{105}, 
\nonumber \\
y_c &=& \eta_{10} + \eta_{40} + \eta_{60} + \eta_{90} + \eta_{110}, &
& 
y_d &=& \eta_{15} + \eta_{35} + \eta_{65} + \eta_{85} + \eta_{115}, 
\nonumber \\ 
y_e &=& \eta_{20} + \eta_{30} + \eta_{70} + \eta_{80} + \eta_{120}, &
& 
y_f &=& 2\eta_{25} + 2\eta_{75} + \eta_{125}, 
\end{array}
\nonumber 
\end{eqnarray}
in the short notation $\eta_N$ defined in Eq.~(\ref{eq:notation-y})
with 
$M=M_1M_2M_3=250$.

\subsection{(Even-Odd-Odd) wavefunctions}

Here, we study the patterns of Yukawa matrices 
in the models, where zero-modes of left, right-handed 
matter fields and Higgs fields correspond to 
even, odd and odd functions, respectively.

\subsubsection{4-7-11 model}

Here we show the model with 
$(|I^{(1)}_{ab}|,|I^{(1)}_{ca}|,|I^{(1)}_{bc}|)=(4,7,11)$.
The following table shows zero-mode wavefunctions of left,
right-handed 
matter fields and Higgs fields.
\begin{center}
\begin{tabular}{c|c|c|c}
& $L_i (\lambda^{ab})$ & $R_j (\lambda^{ca})$ & $H_k (\lambda^{bc})$ 
\\ \hline
0 & 
$\Theta^{0,4}$ & $\frac{1}{\sqrt{2}}(\Theta^{1,7}-\Theta^{6,7})$ & 
$\frac{1}{\sqrt{2}}(\Theta^{1,11}-\Theta^{10,11})$ \\  
1 & 
$\frac{1}{\sqrt{2}}\left(\Theta^{1,4}+\Theta^{3,4}\right)$ & 
$\frac{1}{\sqrt{2}}\left(\Theta^{2,7}-\Theta^{5,7}\right)$ & 
$\frac{1}{\sqrt{2}}\left(\Theta^{2,11}-\Theta^{9,11}\right)$ \\  
2 & 
$\Theta^{2,4}$ & 
$\frac{1}{\sqrt{2}}\left(\Theta^{3,7}-\Theta^{4,7}\right)$ & 
$\frac{1}{\sqrt{2}}\left(\Theta^{3,11}-\Theta^{8,11}\right)$ \\  
3 & - & - & 
$\frac{1}{\sqrt{2}}\left(\Theta^{4,11}-\Theta^{7,11}\right)$ \\  
4 & - & - & 
$\frac{1}{\sqrt{2}}\left(\Theta^{5,11}-\Theta^{6,11}\right)$ \\  
\end{tabular}
\end{center}
This model has five zero-modes for the Higgs fields.
Yukawa couplings $Y_{ijk} L_i R_j H_k$ are given by
\begin{eqnarray}
Y_{ij}^kH_k=
  y_{ij}^0 H_0 + y_{ij}^1 H_1 + y_{ij}^2 H_2 
+ y_{ij}^3 H_3 + y_{ij}^4 H_4, 
\nonumber
\end{eqnarray}
where 
\begin{eqnarray}
y_{ij}^0 &=& \frac{1}{\sqrt{2}} 
\left( \begin{array}{ccc}
\sqrt{2}(\eta_4-\eta_{136}) & 
\sqrt{2}(\eta_{92}-\eta_{48}) & 
\sqrt{2}(\eta_{128}-\eta_{40}) \\
\eta_{81}-\eta_{59}-\eta_{95}+\eta_{73} & 
\eta_{139}-\eta_{29}-\eta_{125}+\eta_{15} & 
\eta_{51}-\eta_{117}-\eta_{37}+\eta_{103} \\ 
\sqrt{2}(\eta_{150}-\eta_{18}) & 
\sqrt{2}(\eta_{62}-\eta_{16}) & 
\sqrt{2}(\eta_{26}-\eta_{114}) 
\end{array} \right), 
\nonumber \\ 
y_{ij}^1 &=& \frac{1}{\sqrt{2}} 
\left( \begin{array}{ccc}
\sqrt{2}(\eta_{80}-\eta_{52}) & 
\sqrt{2}(\eta_{8}-\eta_{36}) & 
\sqrt{2}(\eta_{96}-\eta_{124}) \\
\eta_{3}-\eta_{25}-\eta_{129}+\eta_{151} & 
\eta_{85}-\eta_{113}-\eta_{41}+\eta_{69} & 
\eta_{135}-\eta_{107}-\eta_{47}+\eta_{19} \\ 
\sqrt{2}(\eta_{74}-\eta_{102}) & 
\sqrt{2}(\eta_{146}-\eta_{118}) & 
\sqrt{2}(\eta_{58}-\eta_{30}) 
\end{array} \right), 
\nonumber \\
y_{ij}^2 &=& \frac{1}{\sqrt{2}} 
\left( \begin{array}{ccc}
\sqrt{2}(\eta_{144}-\eta_{32}) & 
\sqrt{2}(\eta_{76}-\eta_{120}) & 
\sqrt{2}(\eta_{12}-\eta_{100}) \\
\eta_{87}-\eta_{109}-\eta_{45}+\eta_{67} & 
\eta_{1}-\eta_{111}-\eta_{43}+\eta_{153} & 
\eta_{89}-\eta_{23}-\eta_{131}+\eta_{65} \\ 
\sqrt{2}(\eta_{10}-\eta_{122}) & 
\sqrt{2}(\eta_{78}-\eta_{34}) & 
\sqrt{2}(\eta_{142}-\eta_{54}) 
\end{array} \right), 
\nonumber \\
y_{ij}^3 &=& \frac{1}{\sqrt{2}} 
\left( \begin{array}{ccc}
\sqrt{2}(\eta_{148}-\eta_{104}) & 
\sqrt{2}(\eta_{148}-\eta_{104}) & 
\sqrt{2}(\eta_{72}-\eta_{16}) \\
\eta_{171}-\eta_{115}-\eta_{39}+\eta_{17} & 
\eta_{83}-\eta_{27}-\eta_{127}+\eta_{71} & 
\eta_{5}-\eta_{61}-\eta_{13}+\eta_{149} \\ 
\sqrt{2}(\eta_{94}-\eta_{38}) & 
\sqrt{2}(\eta_{6}-\eta_{50}) & 
\sqrt{2}(\eta_{82}-\eta_{138}) 
\end{array} \right), 
\nonumber \\
y_{ij}^4 &=& \frac{1}{\sqrt{2}} 
\left( \begin{array}{ccc}
\sqrt{2}(\eta_{24}-\eta_{108}) & 
\sqrt{2}(\eta_{64}-\eta_{20}) & 
\sqrt{2}(\eta_{152}-\eta_{68}) \\
\eta_{53}-\eta_{31}-\eta_{123}+\eta_{101} & 
\eta_{141}-\eta_{57}-\eta_{7}+\eta_{13} & 
\eta_{79}-\eta_{145}-\eta_{9}+\eta_{75} \\ 
\sqrt{2}(\eta_{130}-\eta_{46}) & 
\sqrt{2}(\eta_{90}-\eta_{134}) & 
\sqrt{2}(\eta_{2}-\eta_{86}) 
\end{array} \right), 
\nonumber 
\end{eqnarray}
in the short notation $\eta_N$ defined in Eq.~(\ref{eq:notation-y})
with 
$M=M_1M_2M_3=308$.

\subsubsection{4-7-3 model}

Here we show the model with 
$(|I^{(1)}_{ab}|,|I^{(1)}_{ca}|,|I^{(1)}_{bc}|)=(4,7,3)$.
The following table shows zero-mode wavefunctions of left,
right-handed 
matter fields and Higgs fields.
\begin{center}
\begin{tabular}{c|c|c|c}
& $L_i (\lambda^{ab})$ & $R_j (\lambda^{ca})$ & $H_k (\lambda^{bc})$ 
\\ \hline
0 & 
$\Theta^{0,4}$ & $\frac{1}{\sqrt{2}}(\Theta^{1,7}-\Theta^{6,7})$ 
& $\frac{1}{\sqrt{2}}(\Theta^{1,3}-\Theta^{2,3})$ \\  
1 & 
$\frac{1}{\sqrt{2}}\left(\Theta^{1,4}+\Theta^{3,4}\right)$ & 
$\frac{1}{\sqrt{2}}\left(\Theta^{2,5}-\Theta^{5,7}\right)$ & 
- \\  
2 & 
$\Theta^{2,4}$ & 
$\frac{1}{\sqrt{2}}\left(\Theta^{3,5}-\Theta^{4,5}\right)$ & 
- \\  
\end{tabular}
\end{center}
This model has a single zero-modes for Higgs fields.
Yukawa couplings $Y_{ijk} L_i R_j H_k$ are obtained as 
\begin{eqnarray}
Y_{ij}^k H_k &=& \frac{1}{\sqrt{2}} H_0 
\left( \begin{array}{ccc}
\sqrt{2}(\eta_4-\eta_{32}) & 
\sqrt{2}(\eta_{20}-\eta_8) & 
\sqrt{2}(\eta_{40}-\eta_{16}) \\
\eta_{17} +\eta_{25}-\eta_{11}-\eta_{31} & 
\eta_{1} +\eta_{41}-\eta_{13}-\eta_{29} & 
\eta_{19} +\eta_{23}-\eta_{5}-\eta_{37} \\
\sqrt{2}(\eta_{38}-\eta_{10}) & 
\sqrt{2}(\eta_{22}-\eta_{34}) & 
\sqrt{2}(\eta_{2}-\eta_{26}) 
\end{array} \right), 
\nonumber   
\end{eqnarray}
in the short notation $\eta_N$ defined in Eq.~(\ref{eq:notation-y})
with 
$M=M_1M_2M_3=84$.

\subsubsection{4-8-12 model}

Here we show the model with 
$(|I^{(1)}_{ab}|,|I^{(1)}_{ca}|,|I^{(1)}_{bc}|)=(4,8,12)$.
The following table shows zero-mode wavefunctions of left,
right-handed 
matter fields and Higgs fields.
\begin{center}
\begin{tabular}{c|c|c|c}
& $L_i (\lambda^{ab})$ & $R_j (\lambda^{ca})$ & $H_k (\lambda^{bc})$
\\ \hline
0 & 
$\Theta^{0,4}$ & $\frac{1}{\sqrt{2}}(\Theta^{1,8}-\Theta^{7,8})$ & 
$\frac{1}{\sqrt{2}}(\Theta^{1,12}-\Theta^{11,12})$ \\  
1 & 
$\frac{1}{\sqrt{2}}\left(\Theta^{1,4}+\Theta^{3,4}\right)$ & 
$\frac{1}{\sqrt{2}}\left(\Theta^{2,8}-\Theta^{6,8}\right)$ & 
$\frac{1}{\sqrt{2}}\left(\Theta^{2,12}-\Theta^{10,12}\right)$ \\  
2 & 
$\Theta^{2,4}$ & 
$\frac{1}{\sqrt{2}}\left(\Theta^{3,8}-\Theta^{5,8}\right)$ & 
$\frac{1}{\sqrt{2}}\left(\Theta^{3,12}-\Theta^{9,12}\right)$ \\  
3 & - & - & 
$\frac{1}{\sqrt{2}}\left(\Theta^{4,12}-\Theta^{8,12}\right)$ \\  
4 & - & - & 
$\frac{1}{\sqrt{2}}\left(\Theta^{5,12}-\Theta^{7,12}\right)$ \\  
\end{tabular}
\end{center}
This model has five zero-modes for the Higgs fields.
Yukawa couplings $Y_{ijk} L_i R_j H_k$ are given by 
\begin{eqnarray}
Y_{ij}^k H_k &=& 
  y_{ij}^0 H_0 + y_{ij}^1 H_1 + y_{ij}^2 H_2 
+ y_{ij}^3 H_3 + y_{ij}^4 H_4, 
\nonumber
\end{eqnarray}
where 
\begin{eqnarray}
y_{ij}^0 &=& 
\left( \begin{array}{ccc}
y_b & 0 & -y_{l} \\
0 & \frac{1}{\sqrt{2}}(y_{e}-y_{i}) & 0 \\ 
-y_{f} & 0 & y_{h} 
\end{array} \right), \quad
y_{ij}^1 \ = \ 
\left( \begin{array}{ccc}
0 & y_c-y_{k} & 0 \\
\frac{1}{\sqrt{2}}(y_{b}-y_{h}) & 0 & \frac{1}{\sqrt{2}}(y_{f}-y_{l})
\\ 
0 & 0 & 0
\end{array} \right), 
\nonumber \\
y_{ij}^2 &=& 
\left( \begin{array}{ccc}
-y_{j} & 0 & y_{d} \\
0 & \frac{1}{\sqrt{2}}(y_{a}-y_{m}) & 0 \\ 
y_{d} & 0 & -y_{j} 
\end{array} \right), \quad 
y_{ij}^3 \ = \ 
\left( \begin{array}{ccc}
0 & 0 & 0 \\
\frac{1}{\sqrt{2}}(y_{f}-y_{l}) & 0 & \frac{1}{\sqrt{2}}(y_{b}-y_{h})
\\ 
0 & y_c-y_{k} & 0 
\end{array} \right), 
\nonumber \\
y_{ij}^4 &=& 
\left( \begin{array}{ccc}
y_h & 0 & -y_{f} \\
0 & \frac{1}{\sqrt{2}}(y_{e}-y_{i}) & 0 \\ 
-y_{l} & 0 & y_{b} 
\end{array} \right), 
\nonumber 
\end{eqnarray}
and 
\begin{eqnarray}
\begin{array}{rclcrcl}
y_a &=& \eta_0 + \eta_{96} + \eta_{192}  + \eta_{96}, & & 
y_b &=& \eta_{4} + \eta_{100} + \eta_{188} + \eta_{92}, 
\nonumber \\ 
y_c &=& \eta_{8} + \eta_{104} + \eta_{184} + \eta_{88}, & & 
y_d &=& \eta_{12} + \eta_{108} + \eta_{180} + \eta_{84}, 
\nonumber \\ 
y_e &=& \eta_{16} + \eta_{112} + \eta_{176} + \eta_{80}, & & 
y_f &=& \eta_{20} + \eta_{116} + \eta_{172} + \eta_{76}, 
\nonumber \\ 
y_g &=& \eta_{24} + \eta_{120} + \eta_{168} + \eta_{72}, & & 
y_h &=& \eta_{28} + \eta_{124} + \eta_{164} + \eta_{68}, 
\nonumber \\ 
y_i &=& \eta_{32} + \eta_{128} + \eta_{160} + \eta_{64}, & & 
y_j &=& \eta_{36} + \eta_{132} + \eta_{156} + \eta_{60}, 
\nonumber \\ 
y_k &=& \eta_{40} + \eta_{136} + \eta_{152} + \eta_{56}, & & 
y_l &=& \eta_{44} + \eta_{140} + \eta_{148} + \eta_{52}, 
\nonumber \\ 
y_m &=& \eta_{48} + \eta_{144} + \eta_{144} + \eta_{48}, & & & & 
\end{array}
\nonumber
\end{eqnarray}
in the short notation $\eta_N$ defined in Eq.~(\ref{eq:notation-y})
with 
$M=M_1M_2M_3=384$.

\subsubsection{4-8-4 model}

Here we show the model with 
$(|I^{(1)}_{ab}|,|I^{(1)}_{ca}|,|I^{(1)}_{bc}|)=(4,8,4)$.
The following table shows zero-mode wavefunctions of left,
right-handed 
matter fields and Higgs fields.
\begin{center}
\begin{tabular}{c|c|c|c}
& $L_i (\lambda^{ab})$ & $R_j (\lambda^{ca})$ & $H_k (\lambda^{bc})$ 
\\ \hline
0 & 
$\Theta^{0,4}$ & $\frac{1}{\sqrt{2}}(\Theta^{1,8}-\Theta^{7,8})$ 
& $\frac{1}{\sqrt{2}}(\Theta^{1,4}-\Theta^{3,4})$ \\  
1 & 
$\frac{1}{\sqrt{2}}\left(\Theta^{1,4}+\Theta^{3,4}\right)$ & 
$\frac{1}{\sqrt{2}}\left(\Theta^{2,7}-\Theta^{6,7}\right)$ & 
- \\  
2 & 
$\Theta^{2,4}$ & 
$\frac{1}{\sqrt{2}}\left(\Theta^{3,7}-\Theta^{5,7}\right)$ & 
- \\  
\end{tabular}
\end{center}
This model has a single zero-modes for Higgs fields.
Yukawa couplings $Y_{ijk} L_i R_j H_k$ are obtained as
\begin{eqnarray}
Y_{ij}^k H_k &=& H_0
\left( \begin{array}{ccc}
y_b  & 0 & -y_c \\
0    & \frac{1}{\sqrt{2}}(y_a - y_d) & 0 \\
-y_c & 0 & y_b 
\end{array} \right), 
\nonumber   
\end{eqnarray}
where
\begin{eqnarray}
\begin{array}{rclcrcl}
y_a &=& \eta_0 + 2\eta_{32} + \eta_{64}, & & 
y_b &=& \eta_{4} + \eta_{28} + \eta_{36} + \eta_{60}, 
\nonumber \\ 
y_c &=& \eta_{12} + \eta_{20} + \eta_{44} + \eta_{52}, & & 
y_d &=& 2\eta_{16} + 2\eta_{48}, 
\end{array}
\nonumber  
\end{eqnarray}
in the short notation $\eta_N$ defined in Eq.~(\ref{eq:notation-y})
with 
$M=M_1M_2M_3=128$.

\subsubsection{5-7-12 model}

Here we show the model with 
$(|I^{(1)}_{ab}|,|I^{(1)}_{ca}|,|I^{(1)}_{bc}|)=(5,7,12)$.
The following table shows zero-mode wavefunctions of left,
right-handed 
matter fields and Higgs fields.
\begin{center}
\begin{tabular}{c|c|c|c}
& $L_i (\lambda^{ab})$ & $R_j (\lambda^{ca})$ & $H_k (\lambda^{bc})$ 
\\ \hline
0 & 
$\Theta^{0,5}$ & $\frac{1}{\sqrt{2}}(\Theta^{1,7}-\Theta^{6,7})$ & 
$\frac{1}{\sqrt{2}}(\Theta^{1,12}-\Theta^{11,12})$ \\  
1 & 
$\frac{1}{\sqrt{2}}\left(\Theta^{1,5}+\Theta^{4,5}\right)$ & 
$\frac{1}{\sqrt{2}}\left(\Theta^{2,7}-\Theta^{5,7}\right)$ & 
$\frac{1}{\sqrt{2}}\left(\Theta^{2,12}-\Theta^{10,12}\right)$ \\  
2 & 
$\frac{1}{\sqrt{2}}\left(\Theta^{2,5}+\Theta^{3,5}\right)$ &
$\frac{1}{\sqrt{2}}\left(\Theta^{3,7}-\Theta^{4,7}\right)$ & 
$\frac{1}{\sqrt{2}}\left(\Theta^{3,12}-\Theta^{9,12}\right)$ \\  
3 & - & - & 
$\frac{1}{\sqrt{2}}\left(\Theta^{4,12}-\Theta^{8,12}\right)$ \\  
4 & - & - & 
$\frac{1}{\sqrt{2}}\left(\Theta^{5,12}-\Theta^{7,12}\right)$ \\  
\end{tabular}
\end{center}
This model has five zero-modes for the Higgs fields.
Yukawa coupling $Y_{ijk}L_iR_j H_k$ are given by 
\begin{eqnarray}
Y_{ijk}H_k &=& 
  y_{ij}^0 H_0 + y_{ij}^1 H_1 + y_{ij}^2 H_2 
+ y_{ij}^3 H_3 + y_{ij}^4 H_4, 
\nonumber
\end{eqnarray}
where
\begin{eqnarray}
y_{ij}^0 &=& \frac{1}{\sqrt{2}} 
\left( \begin{array}{ccc}
\sqrt{2}(\eta_5-\eta_{65}) & 
\sqrt{2}(\eta_{185}-\eta_{115}) & 
\sqrt{2}(\eta_{55}+\eta_{125}) \\
\eta_{173}-\eta_{103}-\eta_{187}+\eta_{163} & 
\eta_{67}-\eta_{137}-\eta_{53}+\eta_{17} & 
\eta_{113}-\eta_{43}-\eta_{127}+\eta_{197} \\ 
\eta_{79}-\eta_{149}-\eta_{19}+\eta_{89} & 
\eta_{101}-\eta_{31}-\eta_{199}+\eta_{151} & 
\eta_{139}-\eta_{209}-\eta_{41}+\eta_{29} 
\end{array} \right), 
\nonumber \\ 
y_{ij}^1 &=& \frac{1}{\sqrt{2}} 
\left( \begin{array}{ccc}
\sqrt{2}(\eta_{170}-\eta_{110}) & 
\sqrt{2}(\eta_{10}-\eta_{130}) & 
\sqrt{2}(\eta_{190}+\eta_{50}) \\
\eta_{2}-\eta_{142}-\eta_{58}+\eta_{82} & 
\eta_{178}-\eta_{38}-\eta_{122}+\eta_{158} & 
\eta_{62}-\eta_{202}-\eta_{118}+\eta_{22} \\ 
\eta_{166}-\eta_{26}-\eta_{194}+\eta_{94} & 
\eta_{74}-\eta_{206}-\eta_{46}+\eta_{94} & 
\eta_{106}-\eta_{34}-\eta_{134}+\eta_{146} 
\end{array} \right), 
\nonumber \\
y_{ij}^2 &=& \frac{1}{\sqrt{2}} 
\left( \begin{array}{ccc}
\sqrt{2}(\eta_{75}-\eta_{135}) & 
\sqrt{2}(\eta_{165}-\eta_{45}) & 
\sqrt{2}(\eta_{15}-\eta_{195}) \\
\eta_{177}-\eta_{33}-\eta_{117}+\eta_{93} & 
\eta_{3}-\eta_{207}-\eta_{123}+\eta_{87} & 
\eta_{183}-\eta_{27}-\eta_{57}+\eta_{153} \\ 
\eta_{9}-\eta_{201}-\eta_{51}+\eta_{81} & 
\eta_{171}-\eta_{39}-\eta_{129}+\eta_{81} & 
\eta_{69}-\eta_{141}-\eta_{111}+\eta_{99} 
\end{array} \right), 
\nonumber \\ 
y_{ij}^3 &=& \frac{1}{\sqrt{2}} 
\left( \begin{array}{ccc}
\sqrt{2}(\eta_{100}-\eta_{140}) & 
\sqrt{2}(\eta_{80}-\eta_{200}) & 
\sqrt{2}(\eta_{160}-\eta_{20}) \\
\eta_{68}-\eta_{208}-\eta_{128}+\eta_{152} & 
\eta_{172}-\eta_{32}-\eta_{52}+\eta_{88} & 
\eta_{8}-\eta_{148}-\eta_{188}+\eta_{92} \\ 
\eta_{184}-\eta_{44}-\eta_{124}+\eta_{164} & 
\eta_{4}-\eta_{136}-\eta_{116}+\eta_{164} & 
\eta_{176}-\eta_{104}-\eta_{64}+\eta_{76} 
\end{array} \right), 
\nonumber \\
y_{ij}^4 &=& \frac{1}{\sqrt{2}} 
\left( \begin{array}{ccc}
\sqrt{2}(\eta_{145}-\eta_{205}) & 
\sqrt{2}(\eta_{95}-\eta_{25}) & 
\sqrt{2}(\eta_{85}-\eta_{155}) \\
\eta_{107}-\eta_{37}-\eta_{47}+\eta_{23} & 
\eta_{73}-\eta_{143}-\eta_{193}+\eta_{157} & 
\eta_{167}-\eta_{97}-\eta_{13}+\eta_{83} \\ 
\eta_{61}-\eta_{131}-\eta_{121}+\eta_{11} & 
\eta_{179}-\eta_{109}-\eta_{59}+\eta_{11} & 
\eta_{1}-\eta_{71}-\eta_{181}+\eta_{169} 
\end{array} \right), 
\nonumber 
\end{eqnarray}
in the short notation $\eta_N$ defined in Eq.~(\ref{eq:notation-y})
with 
$M=M_1M_2M_3=420$.

\subsubsection{5-8-13 model}

Here we show the model with 
$(|I^{(1)}_{ab}|,|I^{(1)}_{ca}|,|I^{(1)}_{bc}|)=(5,8,13)$.
The following table shows zero-mode wavefunctions of left,
right-handed 
matter fields and Higgs fields.
\begin{center}
\begin{tabular}{c|c|c|c}
& $L_i (\lambda^{ab})$ & $R_j (\lambda^{ca})$ & $H_k (\lambda^{bc})$ 
\\ \hline
0 & 
$\Theta^{0,5}$ & $\frac{1}{\sqrt{2}}(\Theta^{1,8}-\Theta^{7,8})$ & 
$\frac{1}{\sqrt{2}}(\Theta^{1,13}-\Theta^{12,13})$ \\  
1 & 
$\frac{1}{\sqrt{2}}\left(\Theta^{1,5}+\Theta^{4,5}\right)$ & 
$\frac{1}{\sqrt{2}}\left(\Theta^{2,8}-\Theta^{6,8}\right)$ & 
$\frac{1}{\sqrt{2}}\left(\Theta^{2,13}-\Theta^{11,13}\right)$ \\  
2 & 
$\frac{1}{\sqrt{2}}\left(\Theta^{2,5}+\Theta^{3,5}\right)$ &
$\frac{1}{\sqrt{2}}\left(\Theta^{3,8}-\Theta^{5,8}\right)$ & 
$\frac{1}{\sqrt{2}}\left(\Theta^{3,13}-\Theta^{10,13}\right)$ \\  
3 & - & - & 
$\frac{1}{\sqrt{2}}\left(\Theta^{4,13}-\Theta^{9,13}\right)$ \\  
4 & - & - & 
$\frac{1}{\sqrt{2}}\left(\Theta^{5,13}-\Theta^{8,13}\right)$ \\  
5 & - & - & 
$\frac{1}{\sqrt{2}}\left(\Theta^{6,13}-\Theta^{7,13}\right)$ \\  
\end{tabular}
\end{center}
This model has six zero-modes for the Higgs fields.
Yukawa couplings $Y_{ijk} L_i R_j H_k$ are given by 
\begin{eqnarray}
Y_{ij}^k H_k &=& 
  y_{ij}^0 H_0 + y_{ij}^1 H_1 + y_{ij}^2 H_2 
+ y_{ij}^3 H_3 + y_{ij}^4 H_4 + y_{ij}^5 H_5, 
\nonumber
\end{eqnarray}
where 
\begin{eqnarray}
y_{ij}^0 &=& \frac{1}{\sqrt{2}} 
\left( \begin{array}{ccc}
\sqrt{2}(\eta_5-\eta_{125}) & 
\sqrt{2}(\eta_{190}-\eta_{70}) & 
\sqrt{2}(\eta_{135}-\eta_{255}) \\
\eta_{203}+\eta_{213}-\eta_{83}-\eta_{187} & 
\eta_{122}-\eta_{138}+\eta_{18}-\eta_{242} & 
\eta_{73}-\eta_{57}+\eta_{177}-\eta_{47} \\ 
\eta_{109}-\eta_{21}+\eta_{99}-\eta_{229} & 
\eta_{86}-\eta_{174}+\eta_{226}-\eta_{34} & 
\eta_{239}-\eta_{151}+\eta_{31}-\eta_{161} 
\end{array} \right), 
\nonumber \\ 
y_{ij}^1 &=& \frac{1}{\sqrt{2}} 
\left( \begin{array}{ccc}
\sqrt{2}(\eta_{205}-\eta_{75}) & 
\sqrt{2}(\eta_{10}-\eta_{250}) & 
\sqrt{2}(\eta_{185}-\eta_{55}) \\
\eta_{3}+\eta_{237}-\eta_{133}-\eta_{107} & 
\eta_{198}-\eta_{62}+\eta_{132}-\eta_{152} & 
\eta_{127}-\eta_{257}+\eta_{23}-\eta_{153} \\ 
\eta_{211}-\eta_{179}+\eta_{101}-\eta_{29} & 
\eta_{114}-\eta_{146}+\eta_{94}-\eta_{166} & 
\eta_{81}-\eta_{49}+\eta_{231}-\eta_{159} 
\end{array} \right), 
\nonumber \\
y_{ij}^2 &=& \frac{1}{\sqrt{2}} 
\left( \begin{array}{ccc}
\sqrt{2}(\eta_{115}-\eta_{245}) & 
\sqrt{2}(\eta_{210}-\eta_{50}) & 
\sqrt{2}(\eta_{15}-\eta_{145}) \\
\eta_{197}+\eta_{137}-\eta_{67}-\eta_{93} & 
\eta_{2}-\eta_{258}+\eta_{102}-\eta_{158} & 
\eta_{193}-\eta_{63}+\eta_{223}-\eta_{167} \\ 
\eta_{11}-\eta_{141}+\eta_{219}-\eta_{171} & 
\eta_{206}-\eta_{54}+\eta_{106}-\eta_{154} & 
\eta_{119}-\eta_{249}+\eta_{89}-\eta_{41} 
\end{array} \right), 
\nonumber \\ 
y_{ij}^3 &=& \frac{1}{\sqrt{2}} 
\left( \begin{array}{ccc}
\sqrt{2}(\eta_{85}-\eta_{45}) & 
\sqrt{2}(\eta_{110}-\eta_{150}) & 
\sqrt{2}(\eta_{135}-\eta_{255}) \\
\eta_{123}+\eta_{163}-\eta_{253}-\eta_{227} & 
\eta_{202}-\eta_{58}+\eta_{98}-\eta_{162} & 
\eta_{7}-\eta_{137}+\eta_{97}-\eta_{33} \\ 
\eta_{189}-\eta_{59}+\eta_{19}-\eta_{149} & 
\eta_{6}-\eta_{254}+\eta_{214}-\eta_{46} & 
\eta_{201}-\eta_{71}+\eta_{111}-\eta_{241} 
\end{array} \right), 
\nonumber \\
y_{ij}^4 &=& \frac{1}{\sqrt{2}} 
\left( \begin{array}{ccc}
\sqrt{2}(\eta_{235}-\eta_{155}) & 
\sqrt{2}(\eta_{90}-\eta_{170}) & 
\sqrt{2}(\eta_{105}-\eta_{25}) \\
\eta_{77}+\eta_{157}-\eta_{53}-\eta_{27} & 
\eta_{118}-\eta_{142}+\eta_{222}-\eta_{38} & 
\eta_{207}-\eta_{183}+\eta_{103}-\eta_{233} \\ 
\eta_{131}-\eta_{259}+\eta_{181}-\eta_{51} & 
\eta_{194}-\eta_{66}+\eta_{14}-\eta_{246} & 
\eta_{1}-\eta_{129}+\eta_{209}-\eta_{79} 
\end{array} \right), 
\nonumber \\
y_{ij}^5 &=& \frac{1}{\sqrt{2}} 
\left( \begin{array}{ccc}
\sqrt{2}(\eta_{35}-\eta_{165}) & 
\sqrt{2}(\eta_{230}-\eta_{30}) & 
\sqrt{2}(\eta_{95}-\eta_{225}) \\
\eta_{243}+\eta_{43}-\eta_{147}-\eta_{173} & 
\eta_{82}-\eta_{178}+\eta_{22}-\eta_{238} & 
\eta_{113}-\eta_{17}+\eta_{217}-\eta_{87} \\ 
\eta_{69}-\eta_{61}+\eta_{139}-\eta_{251} & 
\eta_{126}-\eta_{134}+\eta_{186}-\eta_{74} & 
\eta_{199}-\eta_{191}+\eta_{9}-\eta_{121} 
\end{array} \right), 
\nonumber 
\end{eqnarray}
in the short notation $\eta_N$ defined in Eq.~(\ref{eq:notation-y})
with 
$M=M_1M_2M_3=520$.

\subsubsection{5-8-3 model}

Here we show the model with 
$(|I^{(1)}_{ab}|,|I^{(1)}_{ca}|,|I^{(1)}_{bc}|)=(5,8,3)$.
The following table shows zero-mode wavefunctions of left,
right-handed 
matter fields and Higgs fields.
\begin{center}
\begin{tabular}{c|c|c|c}
& $L_i (\lambda^{ab})$ & $R_j (\lambda^{ca})$ & $H_k (\lambda^{bc})$ 
\\ \hline
0 & 
$\Theta^{0,5}$ & $\frac{1}{\sqrt{2}}(\Theta^{1,8}-\Theta^{7,8})$ 
& $\frac{1}{\sqrt{2}}(\Theta^{1,3}-\Theta^{2,3})$ \\  
1 & 
$\frac{1}{\sqrt{2}}\left(\Theta^{1,5}+\Theta^{4,5}\right)$ & 
$\frac{1}{\sqrt{2}}\left(\Theta^{2,8}-\Theta^{6,8}\right)$ & 
- \\  
2 & 
$\frac{1}{\sqrt{2}}(\Theta^{2,5}+\Theta^{3,5}) $ & 
$\frac{1}{\sqrt{2}}\left(\Theta^{3,8}-\Theta^{5,8}\right)$ & 
- \\  
\end{tabular}
\end{center}
This model has a single zero-mode for the Higgs field.
Yukawa couplings $Y_{ijk} L_i R_j H_k$ are given by 
\begin{eqnarray}
Y_{ij}^k H_k &=& \frac{1}{\sqrt{2}} 
\left( \begin{array}{ccc}
\sqrt{2}(\eta_5-\eta_{35}) & 
\sqrt{2}(\eta_{50}-\eta_{10}) & 
\sqrt{2}(\eta_{25}-\eta_{55}) \\
\eta_{43} -\eta_{37}-\eta_{13}+\eta_{53} & 
\eta_{2} -\eta_{38}-\eta_{58}+\eta_{22}  &
\eta_{47} -\eta_{7}-\eta_{17}+\eta_{23} \\
\eta_{29} -\eta_{11}-\eta_{59}+\eta_{19} &
\eta_{46} -\eta_{34}-\eta_{14}+\eta_{26} &
\eta_{1} -\eta_{41}-\eta_{31}+\eta_{49} 
\end{array} \right), 
\nonumber   
\end{eqnarray}
in the short notation $\eta_N$ defined in Eq.~(\ref{eq:notation-y})
with 
$M=M_1M_2M_3=120$.

\subsection{(Odd-Odd-Even) wavefunctions}

Here, we study the patterns of Yukawa matrices 
in the models, where zero-modes of left, right-handed 
matter fields and Higgs fields correspond to 
odd, odd and even functions, respectively.

\subsubsection{7-7-14 model}

Here we show the model with 
$(|I^{(1)}_{ab}|,|I^{(1)}_{ca}|,|I^{(1)}_{bc}|)=(7,7,14)$. 
This model is studied in the subsections~\ref{ssec:7-7-14} 
and \ref{ssec:numerical} in detail. 
The zero-mode wavefunctions of left, right-handed 
matter fields and Higgs fields are shown in Table~\ref{7-7-14model}. 

This model has eight zero-modes for the Higgs fields.
Yukawa couplings $Y_{ijk}L_j R_jH_k$ are obtained as 
\begin{eqnarray}
Y_{ijk}H_k &=& 
  y_{ij}^0 H_0 + y_{ij}^1 H_1 + y_{ij}^2 H_2 
+ y_{ij}^3 H_3 + y_{ij}^4 H_4 + y_{ij}^5 H_5
+ y_{ij}^6 H_6 + y_{ij}^7 H_7, 
\nonumber
\end{eqnarray}
where $y_{ij}^k$ is shown in Eq.~(\ref{eq:yij}) 
with $M=M_1M_2M_3=686$.

\subsubsection{7-8-15 model}

Here we show the model with 
$(|I^{(1)}_{ab}|,|I^{(1)}_{ca}|,|I^{(1)}_{bc}|)=(7,8,15)$.
The following table shows zero-mode wavefunctions of left,
right-handed 
matter fields and Higgs fields.
\begin{center}
\begin{tabular}{c|c|c|c}
& $L_i (\lambda^{ab})$ & $R_j (\lambda^{ca})$ & $H_k (\lambda^{bc})$ 
\\ \hline
0 & 
$\frac{1}{\sqrt{2}}\left(\Theta^{1,7}-\Theta^{6,7}\right)$ & 
$\frac{1}{\sqrt{2}}\left(\Theta^{1,8}-\Theta^{7,8}\right)$ & 
$\Theta^{0,15}$ \\  
1 & 
$\frac{1}{\sqrt{2}}\left(\Theta^{2,7}-\Theta^{5,7}\right)$ & 
$\frac{1}{\sqrt{2}}\left(\Theta^{2,8}-\Theta^{6,8}\right)$ & 
$\frac{1}{\sqrt{2}}\left(\Theta^{1,15}+\Theta^{14,15}\right)$ \\  
2 & 
$\frac{1}{\sqrt{2}}\left(\Theta^{3,7}-\Theta^{4,7}\right)$ & 
$\frac{1}{\sqrt{2}}\left(\Theta^{3,7}-\Theta^{5,8}\right)$ & 
$\frac{1}{\sqrt{2}}\left(\Theta^{2,15}+\Theta^{13,15}\right)$ \\  
3 & - & - & 
$\frac{1}{\sqrt{2}}\left(\Theta^{3,15}+\Theta^{12,15}\right)$ \\  
4 & - & - & 
$\frac{1}{\sqrt{2}}\left(\Theta^{4,15}+\Theta^{11,15}\right)$ \\  
5 & - & - & 
$\frac{1}{\sqrt{2}}\left(\Theta^{5,15}+\Theta^{10,15}\right)$ \\  
6 & - & - & 
$\frac{1}{\sqrt{2}}\left(\Theta^{6,15}+\Theta^{9,15}\right)$ \\  
7 & - & - & 
$\frac{1}{\sqrt{2}}\left(\Theta^{7,15}+\Theta^{8,15}\right)$ \\  
\end{tabular}
\end{center}
This model has eight zero-modes for the Higgs fields.
Yukawa couplings $Y_{ijk} L_i R_j H_k$ are given by
\begin{eqnarray}
Y_{ij}^k H_k &=& 
  y_{ij}^0 H_0 + y_{ij}^1 H_1 + y_{ij}^2 H_2 
+ y_{ij}^3 H_3 + y_{ij}^4 H_4 + y_{ij}^5 H_5
+ y_{ij}^6 H_6 + y_{ij}^7 H_7, 
\nonumber
\end{eqnarray}
where 
\begin{eqnarray}
y_{ij}^0 &=& 
\left( \begin{array}{ccc}
\eta_{225}-\eta_{15} & 
\eta_{330}-\eta_{90} &
\eta_{405}-\eta_{195} \\
\eta_{345}-\eta_{135} & 
\eta_{390}-\eta_{30} &
\eta_{285}-\eta_{75} \\
\eta_{375}-\eta_{255} & 
\eta_{270}-\eta_{150} &
\eta_{165}-\eta_{45} \\
\end{array} \right), 
\nonumber \\
y_{ij}^1 &=& \frac{1}{\sqrt{2}} 
\left( \begin{array}{ccc}
\eta_{113}-\eta_{97}-\eta_{127}+\eta_{337} & 
\eta_{218}-\eta_{202}-\eta_{22}+\eta_{398} & 
\eta_{323}-\eta_{307}-\eta_{83}+\eta_{293} \\
\eta_{233}-\eta_{23}-\eta_{247}+\eta_{383} &
\eta_{338}-\eta_{82}-\eta_{142}+\eta_{278} &
\eta_{397}-\eta_{187}-\eta_{37}+\eta_{173} \\
\eta_{353}-\eta_{143}-\eta_{367}+\eta_{263} &
\eta_{382}-\eta_{38}-\eta_{262}+\eta_{158} &
\eta_{277}-\eta_{67}-\eta_{157}+\eta_{53} \\
\end{array} \right), 
\nonumber \\
y_{ij}^2 &=& \frac{1}{\sqrt{2}} 
\left( \begin{array}{ccc}
\eta_{1}-\eta_{209}-\eta_{239}+\eta_{391} &
\eta_{106}-\eta_{314}-\eta_{134}+\eta_{286} &
\eta_{211}-\eta_{419}-\eta_{29}+\eta_{181} \\
\eta_{121}-\eta_{89}-\eta_{359}+\eta_{271} &
\eta_{226}-\eta_{194}-\eta_{254}+\eta_{166} &
\eta_{331}-\eta_{299}-\eta_{149}+\eta_{61} \\
\eta_{241}-\eta_{31}-\eta_{361}+\eta_{151} &
\eta_{346}-\eta_{74}-\eta_{374}+\eta_{46} &
\eta_{389}-\eta_{179}-\eta_{269}+\eta_{59} \\
\end{array} \right), 
\nonumber \\
y_{ij}^3 &=& \frac{1}{\sqrt{2}} 
\left( \begin{array}{ccc}
\eta_{111}-\eta_{321}-\eta_{351}+\eta_{279} &
\eta_{6}-\eta_{414}-\eta_{246}+\eta_{174} &
\eta_{99}-\eta_{309}-\eta_{141}+\eta_{69} \\
\eta_{9}-\eta_{201}-\eta_{369}+\eta_{159} &
\eta_{114}-\eta_{306}-\eta_{366}+\eta_{54} &
\eta_{219}-\eta_{411}-\eta_{261}+\eta_{51} \\
\eta_{129}-\eta_{81}-\eta_{249}+\eta_{39} &
\eta_{234}-\eta_{186}-\eta_{354}+\eta_{66} &
\eta_{339}-\eta_{291}-\eta_{381}+\eta_{171} \\
\end{array} \right), 
\nonumber \\
y_{ij}^4 &=& \frac{1}{\sqrt{2}} 
\left( \begin{array}{ccc}
\eta_{223}-\eta_{407}-\eta_{377}+\eta_{167} &
\eta_{118}-\eta_{302}-\eta_{358}+\eta_{62} &
\eta_{13}-\eta_{197}-\eta_{253}+\eta_{43} \\
\eta_{103}-\eta_{313}-\eta_{257}+\eta_{47} &
\eta_{2}-\eta_{418}-\eta_{362}+\eta_{58} &
\eta_{107}-\eta_{317}-\eta_{373}+\eta_{163} \\
\eta_{17}-\eta_{193}-\eta_{137}+\eta_{73} &
\eta_{122}-\eta_{298}-\eta_{242}+\eta_{178} &
\eta_{227}-\eta_{403}-\eta_{347}+\eta_{283} \\
\end{array} \right), 
\nonumber \\
y_{ij}^5 &=& \frac{1}{\sqrt{2}} 
\left( \begin{array}{ccc}
\eta_{335}-\eta_{295}-\eta_{265}+\eta_{55} &
\eta_{230}-\eta_{190}-\eta_{370}+\eta_{50} &
\eta_{125}-\eta_{85}-\eta_{365}+\eta_{155} \\
\eta_{215}-\eta_{415}-\eta_{145}+\eta_{65} &
\eta_{110}-\eta_{310}-\eta_{250}+\eta_{170} &
\eta_{5}-\eta_{205}-\eta_{355}+\eta_{275} \\
\eta_{95}-\eta_{305}-\eta_{25}+\eta_{185} &
\eta_{10}-\eta_{410}-\eta_{130}+\eta_{290} &
\eta_{115}-\eta_{325}-\eta_{235}+\eta_{395} \\
\end{array} \right), 
\nonumber \\
y_{ij}^6 &=& \frac{1}{\sqrt{2}} 
\left( \begin{array}{ccc}
\eta_{393}-\eta_{183}-\eta_{153}+\eta_{57} &
\eta_{342}-\eta_{78}-\eta_{258}+\eta_{162} &
\eta_{237}-\eta_{27}-\eta_{363}+\eta_{267} \\
\eta_{327}-\eta_{303}-\eta_{33}+\eta_{177} &
\eta_{222}-\eta_{198}-\eta_{138}+\eta_{282} &
\eta_{117}-\eta_{93}-\eta_{243}+\eta_{387} \\
\eta_{207}-\eta_{417}-\eta_{87}+\eta_{297} &
\eta_{102}-\eta_{318}-\eta_{18}+\eta_{402} &
\eta_{3}-\eta_{213}-\eta_{123}+\eta_{333} \\
\end{array} \right), 
\nonumber \\
y_{ij}^7 &=& \frac{1}{\sqrt{2}} 
\left( \begin{array}{ccc}
\eta_{281}-\eta_{71}-\eta_{41}+\eta_{169} &
\eta_{386}-\eta_{34}-\eta_{146}+\eta_{274} &
\eta_{349}-\eta_{139}-\eta_{251}+\eta_{379} \\
\eta_{401}-\eta_{191}-\eta_{79}+\eta_{289} &
\eta_{334}-\eta_{86}-\eta_{26}+\eta_{394} &
\eta_{229}-\eta_{19}-\eta_{131}+\eta_{341} \\
\eta_{319}-\eta_{311}-\eta_{199}+\eta_{409} &
\eta_{214}-\eta_{206}-\eta_{94}+\eta_{326} &
\eta_{109}-\eta_{101}-\eta_{11}+\eta_{221} \\
\end{array} \right), 
\nonumber
\end{eqnarray}
in the short notation $\eta_N$ defined in Eq.~(\ref{eq:notation-y})
with 
$M=M_1M_2M_3=840$.

\subsubsection{7-8-1 model}

Here we show the model with 
$(|I^{(1)}_{ab}|,|I^{(1)}_{ca}|,|I^{(1)}_{bc}|)=(7,8,1)$.
The following table shows zero-mode wavefunctions of left,
right-handed 
matter fields and Higgs fields.
\begin{center}
\begin{tabular}{c|c|c|c}
& $L_i (\lambda^{ab})$ & $R_j (\lambda^{ca})$ & $H_k (\lambda^{bc})$ 
\\ \hline
0 & 
$\frac{1}{\sqrt{2}}(\Theta^{1,7}-\Theta^{6,7})$ & 
$\frac{1}{\sqrt{2}}(\Theta^{1,8}-\Theta^{7,8})$ 
& $\Theta^{0,1}$ \\  
1 & 
$\frac{1}{\sqrt{2}}\left(\Theta^{2,7}-\Theta^{5,7}\right)$ & 
$\frac{1}{\sqrt{2}}\left(\Theta^{2,8}-\Theta^{6,8}\right)$ & 
- \\  
2 & 
$\frac{1}{\sqrt{2}}\left(\Theta^{3,7}-\Theta^{4,7}\right)$ & 
$\frac{1}{\sqrt{2}}\left(\Theta^{3,8}-\Theta^{5,8}\right)$ & 
- \\  
\end{tabular}
\end{center}
This model has a single zero-mode for the Higgs field.
Yukawa couplings $Y_{ijk} L_i R_j H_k$ are given by
\begin{eqnarray}
Y_{ij}^k H_k &=& \frac{1}{\sqrt{2}} H_0 
\left( \begin{array}{ccc}
\sqrt{2}(\eta_5-\eta_{35}) & 
\sqrt{2}(\eta_{50}-\eta_{10}) & 
\sqrt{2}(\eta_{25}-\eta_{55}) \\
\eta_{43} -\eta_{37}-\eta_{13}+\eta_{53} & 
\eta_{2} -\eta_{38}-\eta_{58}+\eta_{22} & 
\eta_{47} -\eta_{7}-\eta_{17}+\eta_{23} \\
\eta_{29} -\eta_{11}-\eta_{59}+\eta_{19} & 
\eta_{46} -\eta_{34}-\eta_{14}+\eta_{26} & 
\eta_{1} -\eta_{41}-\eta_{31}+\eta_{49} \\
\end{array} \right), \nonumber   
\end{eqnarray}
in the short notation $\eta_N$ defined in Eq.~(\ref{eq:notation-y})
with 
$M=M_1M_2M_3=56$.

\subsubsection{8-8-16 model}

Here we show the model with 
$(|I^{(1)}_{ab}|,|I^{(1)}_{ca}|,|I^{(1)}_{bc}|)=(8,8,16)$.
The following table shows zero-mode wavefunctions of left,
right-handed 
matter fields and Higgs fields.
\begin{center}
\begin{tabular}{c|c|c|c}
& $L_i (\lambda^{ab})$ & $R_j (\lambda^{ca})$ & $H_k (\lambda^{bc})$ 
\\ \hline
0 & 
$\frac{1}{\sqrt{2}}\left(\Theta^{1,8}-\Theta^{7,8}\right)$ & 
$\frac{1}{\sqrt{2}}\left(\Theta^{1,8}-\Theta^{7,8}\right)$ & 
$\Theta^{0,16}$ \\  
1 & 
$\frac{1}{\sqrt{2}}\left(\Theta^{2,8}-\Theta^{6,8}\right)$ & 
$\frac{1}{\sqrt{2}}\left(\Theta^{2,8}-\Theta^{6,8}\right)$ & 
$\frac{1}{\sqrt{2}}\left(\Theta^{1,16}+\Theta^{15,16}\right)$ \\  
2 & 
$\frac{1}{\sqrt{2}}\left(\Theta^{3,8}-\Theta^{5,8}\right)$ & 
$\frac{1}{\sqrt{2}}\left(\Theta^{3,8}-\Theta^{5,8}\right)$ & 
$\frac{1}{\sqrt{2}}\left(\Theta^{2,16}+\Theta^{14,16}\right)$ \\  
3 & - & - & 
$\frac{1}{\sqrt{2}}\left(\Theta^{3,16}+\Theta^{13,16}\right)$ \\  
4 & - & - & 
$\frac{1}{\sqrt{2}}\left(\Theta^{4,16}+\Theta^{12,16}\right)$ \\  
5 & - & - & 
$\frac{1}{\sqrt{2}}\left(\Theta^{5,16}+\Theta^{11,16}\right)$ \\  
6 & - & - & 
$\frac{1}{\sqrt{2}}\left(\Theta^{6,16}+\Theta^{10,16}\right)$ \\  
7 & - & - & 
$\frac{1}{\sqrt{2}}\left(\Theta^{7,16}+\Theta^{9,16}\right)$ \\  
8 & - & - & 
$\Theta^{8,16}$ \\  
\end{tabular}
\end{center}
This model has eight zero-modes for the Higgs fields.
Yukawa couplings $Y_{ijk} L_i R_j H_k$ are obtained as 
\begin{eqnarray}
Y_{ij}^k H_k &=& 
  y_{ij}^0 H_0 + y_{ij}^1 H_1 + y_{ij}^2 H_2 
+ y_{ij}^3 H_3 + y_{ij}^4 H_4 + y_{ij}^5 H_5
+ y_{ij}^6 H_6 + y_{ij}^7 H_7 + y_{ij}^8 H_8, 
\nonumber
\end{eqnarray}
where 
\begin{eqnarray}
y_{ij}^0 &=& 
\left( \begin{array}{ccc}
-y_g & 0 & 0 \\
0 & -y_e & 0 \\
0 & 0 & -y_g 
\end{array} \right), \qquad
y_{ij}^1 \ = \ \frac{1}{\sqrt{2}} 
\left( \begin{array}{ccc}
0 & -y_d & 0 \\
-y_d & 0 & -y_f \\
0 & -y_f & 0
\end{array} \right), 
\nonumber \\
y_{ij}^2 &=& \frac{1}{\sqrt{2}} 
\left( \begin{array}{ccc}
y_a & 0 &  -y_e  \\
0 & 0 & 0 \\ 
-y_e & 0 & y_i    
\end{array} \right), \qquad 
y_{ij}^3 \ = \ \frac{1}{\sqrt{2}} 
\left( \begin{array}{ccc}
0 & y_b  &  0 \\
y_b & 0 & y_h \\ 
0 & y_h & 0  
\end{array} \right), 
\nonumber \\
y_{ij}^4 &=& \frac{1}{\sqrt{2}} 
\left( \begin{array}{ccc}
0 & 0 & y_c+y_g \\ 
0 & y_a+y_i & 0 \\ 
y_c+y_g & 0 & 0 
\end{array} \right), \qquad 
y_{ij}^5 \ = \ \frac{1}{\sqrt{2}} 
\left( \begin{array}{ccc}
0 & y_h  & 0 \\
y_h & 0 & y_b \\ 
0 & y_b & 0 
\end{array} \right), 
\nonumber \\
y_{ij}^6 &=& \frac{1}{\sqrt{2}} 
\left( \begin{array}{ccc}
y_i & 0 &  -y_e  \\
0 & 0 & 0 \\ 
-y_e & 0 & y_a 
\end{array} \right), \qquad 
y_{ij}^7 \ = \ \frac{1}{\sqrt{2}} 
\left( \begin{array}{ccc}
0 & -y_f & 0 \\
-y_f & 0 & -y_d \\
0 & -y_d & 0
\end{array} \right), 
\nonumber \\
y_{ij}^8 &=& 
\left( \begin{array}{ccc}
-y_c & 0 & 0 \\
0 & -y_e & 0 \\
0 & 0 & -y_c 
\end{array} \right), 
\nonumber
\end{eqnarray}
and
\begin{eqnarray}
y_a &=& \eta_0 +2(\eta_{128} +2\eta_{256} +2\eta_{384})+\eta_{512}, 
\nonumber \\ 
y_b &=& \eta_{8} + \eta_{120} + \eta_{136} + \eta_{248} + \eta_{264}
          + \eta_{376} + \eta_{392} +\eta_{504}, 
\nonumber \\ 
y_c &=& \eta_{16} + \eta_{112} + \eta_{144} + \eta_{240} + \eta_{272}
          + \eta_{368} + \eta_{400} +\eta_{496}, 
\nonumber \\ 
y_d &=& \eta_{24} + \eta_{104} + \eta_{156} + \eta_{232} + \eta_{280}
          + \eta_{360} + \eta_{408} +\eta_{488}, 
\nonumber \\ 
y_e &=& \eta_{32} + \eta_{96} + \eta_{164} + \eta_{224} + \eta_{288}
          + \eta_{352} + \eta_{416} +\eta_{480}, 
\nonumber \\ 
y_f &=& \eta_{40} + \eta_{88} + \eta_{172} + \eta_{216} + \eta_{296}
          + \eta_{344} + \eta_{424} +\eta_{472}, 
\nonumber \\ 
y_g &=& \eta_{48} + \eta_{80} + \eta_{180} + \eta_{208} + \eta_{304}
          + \eta_{336} + \eta_{432} +\eta_{464}, 
\nonumber \\ 
y_h &=& \eta_{56} + \eta_{72} + \eta_{188} + \eta_{200}, + \eta_{312}
          + \eta_{328} + \eta_{440} +\eta_{456}, 
\nonumber \\
y_i &=& 2(\eta_{64} +\eta_{192} +\eta_{320}+\eta_{448}), 
\nonumber  
\end{eqnarray}
in the short notation $\eta_N$ defined in Eq.~(\ref{eq:notation-y})
with 
$M=M_1M_2M_3=1024$.

\section{Non-Abelian discrete symmetries}
In this appendix the group theoretical aspects of the some discrete
symmetries are explained.

\subsection{$D_4$}\label{sec:D4}

Here, we give a examples of $D_4$ which can appear in the models
containing at least two flavors.

The $D_4$ is the symmetry of a square, 
which is generated by the $\pi/4$ rotation $A$ and 
the reflection $B$, where they satisfy 
$A^4=e$, $B^2=e$ and $BAB=A^{-1}$.
Indeed, the $D_4$ consists of the eight elements, 
which are represented by $(-1)^t Z^r C^s$
with $t, r, s =0,1$. They are related each other by 
$A=ZC$, $B=Z$
where $Z$ and $C$ are defined by 
\begin{align}
Z= 
\begin{pmatrix}
1 & 0 \\ 0 & -1 
\end{pmatrix}, \ 
C= 
\begin{pmatrix}
0 & 1 \\ 1 & 0 
\end{pmatrix}.  
\end{align}
The $D_4$ has the following five conjugacy classes, 
\begin{eqnarray*}
\begin{array}{ccc}
 C_1:&\{ e \}, &  h=1\\
 C_1^{(1)}:&\{ -e \}, &  h=2\\
 C_2^{(0,1)}: &\{ C, -C \}, & h=2\\
 C_2^{(1,0)}: &\{ Z, -Z \}, &  h=2\\
 C_2^{(1,1)}: &\{ ZC, -ZC \}, & h=4,
\end{array}
\end{eqnarray*}
where $h$ denotes the order of the elements.

The $D_4$ has four singlets, ${\bf 1}_{++}$, 
${\bf 1}_{+-}$, ${\bf 1}_{-+}$ and ${\bf 1}_{--}$, 
and one doublet ${\bf 2}$.
The characters are shown in Table \ref{tab:D4-character}.
The tensor products are obtained as 
\begin{eqnarray*}
\left(
\begin{array}{c}
x_1 \\
x_2
\end{array}
\right)_{{\bf 2}} \otimes 
\left(
\begin{array}{c}
y_1 \\
y_2
\end{array}
\right)_{{\bf 2}} & =&  
\left( x_1 y_1 + x_2 y_2 \right)_{{\bf 1}_{++}}
+ \left( x_1 y_1 - x_2 y_2 \right)_{{\bf 1}_{+-}} \\
& & + \left(x_1 y_2 + x_2 y_1 \right)_{{\bf 1}_{-+}}
+ \left(x_1 y_2 - x_2 y_1 \right)_{{\bf 1}_{--}} ,
\end{eqnarray*}
\begin{eqnarray*}
 & & \left( x \right)_{{\bf 1}_{++}} \times 
\left(
\begin{array}{c}
y_1 \\
y_2
\end{array}
\right)_{{\bf 2}}= 
\left(
\begin{array}{c}
x y_1 \\
x y_2
\end{array}
\right)_{{\bf 2}}, \qquad 
\left( x \right)_{{\bf 1}_{+-}} \times 
\left(
\begin{array}{c}
y_1 \\
y_2
\end{array}
\right)_{{\bf 2}}= 
\left(
\begin{array}{c}
x y_1 \\
-x y_2
\end{array}
\right)_{{\bf 2}},  \\
 & & \left( x\right)_{{\bf 1}_{-+}} \times 
\left(
\begin{array}{c}
y_1 \\
y_2
\end{array}
\right)_{{\bf 2}}= 
\left(
\begin{array}{c}
x y_2 \\
x y_1
\end{array}
\right)_{{\bf 2}}, \qquad 
\left( x\right)_{{\bf 1}_{--}} \times 
\left(
\begin{array}{c}
y_1 \\
y_2
\end{array}
\right)_{{\bf 2}}= 
\left(
\begin{array}{c}
x y_2 \\
- xy_1
\end{array}
\right)_{{\bf 2}}.
\end{eqnarray*}

\begin{table}[t]
\begin{center}
\begin{tabular}{|c|c|c|c|c|c|c|}
\hline
   & $h$ & $\chi_{1_{++}}$ & $\chi_{1_{+-}}$ 
   & $\chi_{1_{-+}}$  & $\chi_{1_{--}}$ & $\chi_{2}$  \\ \hline
$C_1$ & 1 &  1 & 1 & 1 & 1 & 2 \\ \hline
$C_1^{(1)}$       & 2 &  1 & $1$  & $1$  & 1 & $-2$ \\ \hline
$C_2^{(0,1)}$ & 2 & 1 & -1 & $1$ & $-1$ & 0 \\ \hline
$C_2^{(1,0)}$ & 2 & 1 & 1 & $-1$ & $-1$ & 0 \\ \hline
$C_2^{(1,1)}$ & $4$ & 1 & $-1$ & $-1$ & $1$ & $0$ \\ \hline
\end{tabular}
\end{center}
\caption{Characters of $D_{4}$ representations}
\label{tab:D4-character}
\end{table}

\subsection{$\Delta(27)$}\label{sec:delta27}

The elements $g$ of $\Delta(27)$  are summarized as 
$g= \omega^t Z^r C^s,$ ($r, s, t=0,1,2$).
The elements $Z$ and $C$ are defined 
\begin{align}
Z= 
\begin{pmatrix}
1 & 0 & 0 \\ 0 & \omega & 0 \\ 0 & 0 & \omega^2
\end{pmatrix}, \ 
C= 
\begin{pmatrix}
0 & 1 & 0 \\ 0 & 0 & 1 \\ 1 & 0 & 0
\end{pmatrix},  
\end{align}
where $\omega$ is the cubic root of 1.
Therefore the order of this group is 27. 
The elements $Z$ and $C$ satisfy  the following algebra 
\begin{align}
C^3 = Z^3 = e, \ \  CZ=\omega ZC, 
\end{align}
where $e$ denotes the identity matrix. 
The conjugacy classes of $\Delta(27)$ are obtained as 
\begin{eqnarray*}
\begin{array}{ccc}
 C_1:&\{ e \}, &  h=1\\
 C_1^{(1)}:&\{ \omega \}, &  h=3\\
 C_1^{(2)}:&\{ \omega^2 \}, &  h=3\\
 C_3^{(0,1)}: &\{ C, \omega C, \omega^2 C \}, & h=3\\
 C_3^{(0,2)}: &\{ C^2, \omega C^2, \omega^2 C^2 \}, & h=3\\
 C_3^{(1,0)}: &\{ Z, \omega Z,  \omega^2 Z \}, & h=3\\
 C_3^{(1,1)}: &\{ ZC, \omega ZC,  \omega^2 ZC \}, & h=3\\
 C_3^{(1,2)}: &\{ ZC^2, \omega ZC^2,  \omega^2 ZC^2 \}, & h=3\\
 C_3^{(2,0)}: &\{ Z^2, \omega Z^2,  \omega^2 Z^2 \}, & h=3\\
 C_3^{(2,1)}: &\{ Z^2C, \omega Z^2C,  \omega^2 Z^2C \}, & h=3\\
 C_3^{(2,2)}: &\{ Z^2C^2, \omega Z^2C^2,  \omega^2 Z^2C^2 \}, & h=3\\
\end{array}
\end{eqnarray*}

The $\Delta(27)$ has nine singlets ${\bf 1}_{r,s}$ 
($r,s=0,1,2$) and two triplets, ${\bf 3}$ and 
$\bar{{\bf 3}}$.
The characters are shown in Table \ref{tab:delta-27}.

\begin{table}[t]
\begin{center}
\begin{tabular}{|c|c|c|c|c|}
\hline
&h&$\chi_{1_{(r,s)}}$&$\chi_{3}$&$\chi_{\bar{3}}$\\
\hline
$C_1$&1&1 & 3&3\\ 
\hline
$C_1^{(1)}$&3&1 & $3\omega$&$3\omega^2$\\ 
\hline
$C_1^{(2)}$&3&1 & $3\omega^2$&$3\omega$\\ 
\hline
$C_3^{(0,1)}$&$3$&$\omega^{s}$ & 
$ 0$ &$0$ \\ 
\hline
$C_3^{(0,2)}$&$3$&$\omega^{2s}$ & 
$ 0$ &$0$ \\ 
\hline
$C_3^{(1,p)}$&$3$&$\omega^{r+s p}$      &0&0 \\ 
\hline
$C_3^{(2,p)}$&$3$ & $\omega^{2r+s p}$      &0&0\\
\hline
\end{tabular}
\end{center}
\caption{Characters of $\Delta(27)$}
\label{tab:delta-27}
\end{table}

Tensor products between triplets are obtained as 
\begin{eqnarray}\label{eq:delta27-3-3}
\begin{pmatrix} x_1  \\ x_2  \\ x_3    \\
\end{pmatrix}_{\bf 3} &\otimes& 
\begin{pmatrix} y_1  \\ y_2  \\ y_3    \\
\end{pmatrix}_{\bf 3} =
 \begin{pmatrix} x_1 y_1 \\ x_2 y_2  \\ x_3 y_3   \\
 \end{pmatrix}_{\bar{{\bf 3}}} 
+
 \begin{pmatrix} x_1 y_2 \\ x_2 y_3  \\ x_3 y_1   \\
 \end{pmatrix}_{\bar{{\bf 3}}} 
+
 \begin{pmatrix} x_1 y_3 \\ x_2 y_1  \\ x_3 y_2   \\
 \end{pmatrix}_{\bar{{\bf 3}}} 
\end{eqnarray}

\begin{eqnarray}\label{eq:delta27-3-3}
\begin{pmatrix} x_1  \\ x_2  \\ x_3    \\
\end{pmatrix}_{\bar{\bf 3}} &\otimes& 
\begin{pmatrix} y_1  \\ y_2  \\ y_3    \\
\end{pmatrix}_{\bar{\bf 3}} =
 \begin{pmatrix} x_1 y_1 \\ x_2 y_2  \\ x_3 y_3   \\
 \end{pmatrix}_{{\bf 3}} 
+
 \begin{pmatrix} x_1 y_2 \\ x_2 y_3  \\ x_3 y_1   \\
 \end{pmatrix}_{{\bf 3}} 
+
 \begin{pmatrix} x_1 y_3 \\ x_2 y_1  \\ x_3 y_2   \\
 \end{pmatrix}_{{\bf 3}} 
\end{eqnarray}

\begin{eqnarray}\label{eq:delta27-3-3}
\begin{pmatrix} x_1  \\ x_2  \\ x_3    \\
\end{pmatrix}_{\bf 3} &\otimes& 
\begin{pmatrix} y_1  \\ y_2  \\ y_3    \\
\end{pmatrix}_{\bar{\bf 3}} =
 \begin{pmatrix} x_1 y_1 \\ x_2 y_2  \\ x_3 y_3   \\
 \end{pmatrix}_{{\bf 3}} 
+
 \begin{pmatrix} x_1 y_2 \\ x_2 y_3  \\ x_3 y_1   \\
 \end{pmatrix}_{{\bf 3}} 
+
 \begin{pmatrix} x_1 y_3 \\ x_2 y_1  \\ x_3 y_2   \\
 \end{pmatrix}_{{\bf 3}} 
\end{eqnarray}

\begin{eqnarray}\label{eq:delta27-3-3d}
\begin{pmatrix} x_1 \\ x_2 \\ x_3 
 \end{pmatrix}_{{\bf 3} } \otimes 
\begin{pmatrix} y_1 \\ y_2 \\ y_3 
 \end{pmatrix}_{\bar{\bf 3} } 
=&&
\sum_r(  x_1 y_1 +
+\omega^{r} x_2 y_2
+\omega^{2r} x_3 y_3
)_{{\bf 1}_{(0,r)}}
\nonumber\\
&+&
\sum_r(  x_1 y_3 +
+\omega^{r} x_2 y_1
+\omega^{2r} x_3 y_2
)_{{\bf 1}_{(1,r)}}
\nonumber\\
&+&
\sum_r(  x_1 y_2 +
+\omega^{r} x_2 y_3
+\omega^{2r} x_3 y_1
)_{{\bf 1}_{(2,r)}}
\end{eqnarray}

\subsection{$\Delta(54)$}\label{sec:delta54}

The elements $g$ of $\Delta(54)$  are summarized as 
$g= \omega^t Z^r C^s P^u,$ ($r, s, t=0,1,2, u=0,1$).
The elements $Z$ and $C$ are same as $\Delta(27)$ and 
$P$ is defined by
\begin{align}
P= 
\begin{pmatrix}
1 & 0 & 0 \\ 0 & 0 & 1 \\ 0 & 1 & 0
\end{pmatrix}.  
\end{align}
The order of $\Delta(54)$ is 54. 
These elements satisfy the following algebra as 
\begin{align}
P^2=e, \ \ PC= C^{-1}P, \ \ PZ= Z^{-1}P.
\end{align}
The conjugacy classes of $\Delta(54)$ are obtained as 
\begin{eqnarray*}
\begin{array}{ccc}
 C_1:&\{ e \}, &  h=1\\
 C_1^{(1)}:&\{ \omega \}, &  h=3\\
 C_1^{(2)}:&\{ \omega^2 \}, &  h=3\\
 C_6^{(0,1)+(0,2)}:&\{ Z, \omega Z, \omega^2 Z, Z^2, \omega Z^2,
 \omega^2 Z^2 \}, &  h=3\\
 C_6^{(1,0)+(2,0)}:&\{ C, \omega C, \omega^2 C, C^2, \omega C^2,
 \omega^2 C^2 \}, &  h=3\\ 
 C_6^{(1,2)+(2,1)}:&\{ C^2 Z, \omega C^2 Z, \omega^2 C^2 Z, C Z^2,
 \omega C Z^2, \omega^2 C Z^2 \}, &  h=3\\ 
 C_6^{(1,1)+(2,2)}:&\{ C Z, \omega C Z, \omega^2 C Z, C^2 Z^2,
 \omega C^2 Z^2, \omega^2 C^2 Z^2 \}, &  h=3\\ 
 C_9^{(1)}:&\{ P, Z P, Z^2 P, C P, C^2 P, \omega PCZ, \omega^2 PCZ^2,
 \omega^2 PC^2 Z, \omega PC^2 Z^2 \}, &  h=2\\ 
 C_9^{(2)}:&\{ \omega P, \omega Z P, \omega Z^2 P, \omega C P, \omega
 C^2 P, \omega^2  PCZ,  PCZ^2,  PC^2 Z, \omega^2 PC^2 Z^2 \}, &  h=6\\ 
 C_9^{(3)}:&\{ \omega^2 P, \omega^2 Z P, \omega^2 Z^2 P, \omega^2 C P,
 \omega^2 C^2 P,  PCZ,  \omega PCZ^2,  \omega PC^2 Z,  PC^2 Z^2 \}, &  h=6\\ 
\end{array}
\end{eqnarray*}

The $\Delta(54)$ has two singlets ${\bf 1}_1$, ${\bf 1}_2$ 
and four doublets ${\bf 2}_1$, ${\bf 2}_2$, ${\bf 2}_3$, ${\bf 2}_4$ and 
four triplets ${\bf 3}_1$, ${\bf 3}_2$, $\bar{{\bf 3}}_1$ and $\bar{{\bf 3}}_2$.
The characters are shown in Table \ref{tab:delta-54}.

\begin{table}[t]
\begin{center}
\begin{tabular}{|c|c|c|c|c|c|c|c|c|c|c|c|}
\hline
&h&$\chi_{1_1}$&$\chi_{1_2}$&$\chi_{2_1}$&$\chi_{2_2}$&$\chi_{2_3}$&$\chi_{2_4}$&
$\chi_{3_1}$&$\chi_{\bar{3}_1}$&$\chi_{3_2}$&$\chi_{\bar{3}_2}$\\
\hline
$C_1$&1&1&1&2&2&2&2&3&3&3&3\\ 
\hline
$C_1^{(1)}$&3&1&1&2&2&2&2&$3\omega$&$3\omega^2$&$3\omega$&$3\omega^2$ \\ 
\hline
$C_1^{(2)}$&3&1&1&2&2&2&2&$3\omega^2$&$3\omega$&$3\omega^2$&$3\omega$ \\ 
\hline
$C_6^{(0,1)+(0,2)}$&3&1&1&2&-1&-1&-1&0&0&0&0\\
\hline
$C_6^{(1,0)+(2,0)}$&3&1&1&-1&2&-1&-1&0&0&0&0\\
\hline
$C_6^{(1,2)+(2,1)}$&3&1&1&-1&-1&2&-1&0&0&0&0\\
\hline
$C_6^{(1,1)+(2,2)}$&3&1&1&-1&-1&-1&2&0&0&0&0\\
\hline
$C_9^{(1)}$&2&1&-1&0&0&0&0&1&1&-1&-1\\ 
\hline
$C_9^{(2)}$&6&1&-1&0&0&0&0&$\omega$&$\omega^2$&$-\omega$&$-\omega^2$\\ 
\hline
$C_9^{(3)}$&6&1&-1&0&0&0&0&$\omega^2$&$\omega$&$-\omega^2$&$-\omega$\\ 
\hline
\end{tabular}
\end{center}
\caption{Characters of $\Delta(54)$}
\label{tab:delta-54}
\end{table}

\ 
\


\begin{thebibliography}{99}



\bibitem{Dixon}
L.~J.~Dixon, J.~A.~Harvey, C.~Vafa and E.~Witten, Nucl.\ Phys.\ B\
{\bf 261}
, 678 (1985);
Nucl.\ Phys.\ B\ {\bf 274}, 285 (1986).
%



\bibitem{IMNQ}
L.~E.~Ib\'a\~nez, H.-P.~Nilles and F.~Quevedo, Phys.\ Lett.\ B\
{\bf 187}, 25 (1987);
L.~E.~Ib\'a\~{n}ez, J.~E.~Kim, H.-P.~Nilles and F.~Quevedo, Phys.\
Lett.\ B\ {\bf 191}, 282 (1987);
L.~E.~Ib\'a\~{n}ez, J.~Mas,
H.~P.~Nilles and F.~Quevedo, Nucl.\ Phys.\ B\ {\bf 301}, 157
(1988);
A.~Font, L.~E.~Ib\'a\~{n}ez, F.~Quevedo and A.~Sierra,
Nucl.\ Phys.\ B\ {\bf 331}, 421 (1990);
D.~Bailin, A.~Love and
S.~Thomas, Phys.\ Lett.\ B\ {\bf 194}, 385 (1987);
Y.~Katsuki, Y.~Kawamura, T.~Kobayashi, N.~Ohtsubo, Y. Ono and
K.~Tanioka, Nucl.\ Phys.\ B\ {\bf 341}, 611 (1990).
%


\bibitem{Kobayashi:2004ud}
T.~Kobayashi, S.~Raby and R.~J.~Zhang,
Phys.\ Lett.\ B {\bf 593}, 262 (2004)
[arXiv:hep-ph/0403065].
%




\bibitem{Forste:2004ie}
S.~F\"orste, H.~P.~Nilles, P.~K.~S.~Vaudrevange and A.~Wingerter,
Phys.\ Rev.\ D {\bf 70}, 106008 (2004);
%
  S.~F\"orste, H.~P.~Nilles and A.~Wingerter,
  Phys.\ Rev.\ D {\bf 72}, 026001 (2005)
  [arXiv:hep-th/0504117];
%
  Phys.\ Rev.\ D {\bf 73}, 066011 (2006)
  [arXiv:hep-th/0512270].
%


\bibitem{Buchmuller:2004hv}
W.~Buchm\"uller, K.~Hamaguchi, O.~Lebedev and M.~Ratz,
 Nucl.\ Phys.\ B {\bf 712}, 139 (2005)
[arXiv:hep-ph/0412318];
%
  K.~S.~Choi, S.~Groot Nibbelink and M.~Trapletti,
  JHEP {\bf 0412} (2004) 063
  [arXiv:hep-th/0410232];
%
  H.~P.~Nilles, S.~Ramos-Sanchez, P.~K.~S.~Vaudrevange and
  A.~Wingerter,
  JHEP {\bf 0604}, 050 (2006)
  [arXiv:hep-th/0603086];
%
  J.~E.~Kim and B.~Kyae,
  arXiv:hep-th/0608085.
%


\bibitem{Kobayashi:2004ya}
T.~Kobayashi, S.~Raby and R.~J.~Zhang,
Nucl.\ Phys.\ B {\bf 704}, 3 (2005)
  [arXiv:hep-ph/0409098].



\bibitem{Lebedev:2006kn}
  O.~Lebedev, H.~P.~Nilles, S.~Raby, S.~Ramos-Sanchez, M.~Ratz,
  P.~K.~S.~Vaudrevange and A.~Wingerter,
  Phys.\ Lett.\  B {\bf 645}, 88 (2007)
  [arXiv:hep-th/0611095];
%
  P.~K.~S.~Vaudrevange and A.~Wingerter,
  Phys.\ Rev.\  D {\bf 77}, 046013 (2008)
  [arXiv:0708.2691 [hep-th]].




\bibitem{Kim:2006hw}
  J.~E.~Kim and B.~Kyae,
  arXiv:hep-th/0608086.
%



\bibitem{Buchmuller:2005jr}
  W.~Buchmuller, K.~Hamaguchi, O.~Lebedev and M.~Ratz,
  Phys.\ Rev.\ Lett.\  {\bf 96}, 121602 (2006)
  [arXiv:hep-ph/0511035];
  %
  arXiv:hep-th/0606187.
%



\bibitem{Hamidi:1986vh}
  S.~Hamidi and C.~Vafa,
  Nucl.\ Phys.\  B {\bf 279}, 465 (1987);
  L.~J.~Dixon, D.~Friedan, E.~J.~Martinec and S.~H.~Shenker,
  Nucl.\ Phys.\  B {\bf 282}, 13 (1987).




\bibitem{Burwick:1990tu}
T.~T.~Burwick, R.~K.~Kaiser and H.~F.~M\"uller,
Nucl.\ Phys.\ B {\bf 355}, 689 (1991);
%
J.~Erler, D.~Jungnickel, M.~Spalinski and S.~Stieberger,
Nucl.\ Phys.\ B {\bf 397}, 379 (1993).
%



\bibitem{Kobayashi:2003vi}
T.~Kobayashi and O.~Lebedev,
Phys.\ Lett.\ B {\bf 566}, 164 (2003).
%



\bibitem{Kobayashi:2003gf}
T.~Kobayashi and O.~Lebedev,
Phys.\ Lett.\ B {\bf 565}, 193 (2003).
%



\bibitem{Ko:2004ic}
P.~Ko, T.~Kobayashi and J.~h.~Park,
Phys.\ Lett.\ B {\bf 598}, 263 (2004);
  Phys.\ Rev.\ D {\bf 71}, 095010 (2005)
  [arXiv:hep-ph/0503029].



%
\bibitem{Antoniadis:1989zy}
  H.~Kawai, D.~C.~Lewellen and S.~H.~H.~Tye,
  Phys.\ Rev.\ Lett.\  {\bf 57} (1986) 1832
  [Erratum-ibid.\  {\bf 58} (1987) 429],
  Nucl.\ Phys.\ B {\bf 288} (1987) 1;
  I.~Antoniadis, C.~P.~Bachas and C.~Kounnas,
  Nucl.\ Phys.\ B {\bf 289} (1987) 87;
%
I.~Antoniadis, J.~R.~Ellis, J.~S.~Hagelin and D.~V.~Nanopoulos,
  Phys.\ Lett.\ B {\bf 231} (1989) 65.



%
\bibitem{Faraggi:1991jr}
  A.~E.~Faraggi,
  Phys.\ Lett.\ B {\bf 278}, 131 (1992), 
%
  A.~E.~Faraggi, C.~Kounnas, S.~E.~M.~Nooij and J.~Rizos, 
  Nucl.\ Phys.\ B {\bf 695}, 41 (2004)
  [arXiv:hep-th/0403058];
A.~E.~Faraggi, C.~Kounnas and J.~Rizos,
  orbifold
  arXiv:hep-th/0606144;
A.~E.~Faraggi, E.~Manno and C.~Timirgaziu,
  arXiv:hep-th/0610118.
%


\bibitem{Faraggi:1993pr}
  A.~E.~Faraggi,
  Phys.\ Lett.\ B {\bf 326} (1994) 62
  [arXiv:hep-ph/9311312],
  Phys.\ Lett.\ B {\bf 544} (2002) 207
  [arXiv:hep-th/0206165];
  P.~Berglund, J.~R.~Ellis, A.~E.~Faraggi, D.~V.~Nanopoulos and
  Z.~Qiu,
  Phys.\ Lett.\ B {\bf 433} (1998) 269
  [arXiv:hep-th/9803262].



\bibitem{Polchinski:1995mt}
  J.~Polchinski,
  Phys.\ Rev.\ Lett.\  {\bf 75} (1995) 4724
  [arXiv:hep-th/9510017].




\bibitem{Berkooz:1996km}
  M.~Berkooz, M.~R.~Douglas and R.~G.~Leigh,
  Nucl.\ Phys.\  B {\bf 480}, 265 (1996)
  [arXiv:hep-th/9606139].

\bibitem{Blumenhagen:2000wh}
  R.~Blumenhagen, L.~Goerlich, B.~Kors and D.~Lust,
  JHEP {\bf 0010}, 006 (2000)
  [arXiv:hep-th/0007024].

\bibitem{Angelantonj:2000hi}
  C.~Angelantonj, I.~Antoniadis, E.~Dudas and A.~Sagnotti,
  Phys.\ Lett.\  B {\bf 489}, 223 (2000)
  [arXiv:hep-th/0007090].


\bibitem{Blumenhagen:2005mu}
  R.~Blumenhagen, M.~Cvetic, P.~Langacker and G.~Shiu,
  Ann.\ Rev.\ Nucl.\ Part.\ Sci.\  {\bf 55}, 71 (2005)
  [arXiv:hep-th/0502005];
%
  R.~Blumenhagen, B.~Kors, D.~Lust and S.~Stieberger,
  Phys.\ Rept.\  {\bf 445}, 1 (2007)
  [arXiv:hep-th/0610327].




\bibitem{Aldazabal:2000dg}
  G.~Aldazabal, S.~Franco, L.~E.~Ibanez, R.~Rabadan and A.~M.~Uranga,
  J.\ Math.\ Phys.\  {\bf 42}, 3103 (2001)
  [arXiv:hep-th/0011073];
%
  JHEP {\bf 0102}, 047 (2001)
  [arXiv:hep-ph/0011132].



\bibitem{Blumenhagen:2000ea}
  R.~Blumenhagen, B.~Kors and D.~Lust,
  JHEP {\bf 0102}, 030 (2001)
  [arXiv:hep-th/0012156].



\bibitem{Cvetic:2001tj}
  M.~Cvetic, G.~Shiu and A.~M.~Uranga,
  intersecting  brane
  Phys.\ Rev.\ Lett.\  {\bf 87}, 201801 (2001)
  [arXiv:hep-th/0107143];
%
  Nucl.\ Phys.\  B {\bf 615}, 3 (2001)
  [arXiv:hep-th/0107166].



\bibitem{Cremades:2003qj}
  D.~Cremades, L.~E.~Ibanez and F.~Marchesano,
  JHEP {\bf 0307} (2003) 038
  [arXiv:hep-th/0302105].



\bibitem{Froggatt:1978nt}
  C.~D.~Froggatt and H.~B.~Nielsen,
  Nucl.\ Phys.\  B {\bf 147} (1979) 277.


%
\bibitem{Kobayashi:2006wq}
  T.~Kobayashi, H.~P.~Nilles, F.~Pl\"oger, S.~Raby and M.~Ratz,
  arXiv:hep-ph/0611020.


\bibitem{Altarelli:2006kg}
  G.~Altarelli, F.~Feruglio and Y.~Lin,
  Nucl.\ Phys.\  B {\bf 775}, 31 (2007)
  [arXiv:hep-ph/0610165].


\bibitem{Conlon:2008qi}
  J.~P.~Conlon, A.~Maharana and F.~Quevedo,
  JHEP {\bf 0809}, 104 (2008)
  [arXiv:0807.0789 [hep-th]].


\bibitem{Marchesano:2008rg}
  F.~Marchesano, P.~McGuirk and G.~Shiu,
  JHEP {\bf 0904} (2009) 095
  [arXiv:0812.2247 [hep-th]].


\bibitem{Camara:2009xy}
  P.~G.~Camara and F.~Marchesano,
  arXiv:0906.3033 [hep-th].



\bibitem{Donagi:2008ca}
  R.~Donagi and M.~Wijnholt,
  arXiv:0802.2969 [hep-th];
%
%
  R.~Donagi and M.~Wijnholt,
  arXiv:0808.2223 [hep-th].



\bibitem{Beasley:2008dc}
  C.~Beasley, J.~J.~Heckman and C.~Vafa,
  JHEP {\bf 0901}, 058 (2009)
  [arXiv:0802.3391 [hep-th]];
%
%
  JHEP {\bf 0901}, 059 (2009)
  [arXiv:0806.0102 [hep-th]].


\bibitem{Curio:2000sc}
  G.~Curio, A.~Klemm, D.~Lust and S.~Theisen,
  Nucl.\ Phys.\  B {\bf 609}, 3 (2001)
  [arXiv:hep-th/0012213].

\bibitem{Kachru:2002he}
  S.~Kachru, M.~B.~Schulz and S.~Trivedi,
  JHEP {\bf 0310}, 007 (2003)
  [arXiv:hep-th/0201028].




\bibitem{Kachru:2003aw}
  S.~Kachru, R.~Kallosh, A.~D.~Linde and S.~P.~Trivedi,
  Phys.\ Rev.\  D {\bf 68}, 046005 (2003)
  [arXiv:hep-th/0301240].



\bibitem{Altarelli:2007cd}
See, e.g. , \\
G.~ Altarelli,
arXiv:0705.0860 [hep-ph];
E.~Ma,
 arXiv:0705.0327 [hep-ph]
 and references therein.


 \bibitem{dbkaplan}
D.~B.~Kaplan and M.~Schmaltz,
Phys. Rev. D {\bf 49}, 3741 (1994);
L.J.~Hall and H.~Murayama,
Phys. Rev. Lett. {\bf 75}, 3985  (1995)  ;
C.D.~Carone, L.J.~Hall and H.~Murayama,
Phys. Rev. D {\bf 53}, 6282 (1996).

\bibitem{babu}
K.S.~Babu, T.~Kobayashi and J.~Kubo, 
Phys. Rev. D {\bf 67}, 075018 (2003);
%
K.~Hamaguchi, M.~Kakizaki and M.~Yamaguchi, 
Phys. Rev. D {\bf 68}, 056007 (2003);
T.~Kobayashi, J.~Kubo and H.~Terao, 
Phys. Lett. B {\bf 568}, 83 (2003);
%
 G. G. Ross, L.~Velasco-Sevilla
 and  Oscar Vives, Nucl. Phys. B {\bf 692}, 50 (2004);
S. F.~King and G. G.~Ross, Phys. Lett. B {\bf 520}, 243 (2001);
B{\bf 574,} 239 (2003);
G. G.~Ross and L.~Velasco-Sevilla, Nucl. Phys. B {\bf 653}, 3 (2003);
Ki-Y.~Choi, Y.~Kajiyama,
J.~Kubo and H.M.~Lee,  Phys. Rev. D {\bf 70}, 055004 (2004);
%
N.~Maekawa and T.~Yamashita,
JHEP {\bf  0407,} 009 (2004);
K.S.~Babu and J.~Kubo, Phys. Rev. D {\bf 71}, 056006 (2005);
%
T.~Yamashita, hep-ph/0503265;
I.~de Medeiros Varzielas and  G.G.~Ross, hep-ph/0612220;
%
  P.~Ko, T.~Kobayashi, J.~h.~Park and S.~Raby,
  arXiv:0704.2807 [hep-ph].



\bibitem{murayama}
  L.~E.~Ibanez and G.~G.~Ross,
  Nucl.\ Phys.\  B {\bf 368}, 3 (1992);
%
  S.~P.~Martin,
  Phys.\ Rev.\  D {\bf 46}, 2769 (1992);
%
%
H.~Murayama and D.B.~Kaplan,  
Phys.~Lett. B {\bf 336},  221 (1994);
V.~Ben-Hamo and Y.~Nir,
Phys.~Lett.~B {\bf 339}, 77 (1994);
C.D.~Carone, L.J.~Hall and H.~Murayama,
Phys. Rev. D {\bf 53}, 6282 (1996).


\bibitem{kakizaki}  
M.~Kakizaki and M.~Yamaguchi,
JHEP~{\bf 0206,} 032 (2002);
R.~Harnik,~D.T.~Larson,~H.~Murayama and M.~Thormeier,
Nucl.~Phys.~B {\bf 706}, 372 (2005);
E.~Itou, Y.~Kajiyama and J.~Kubo,
Nucl.~Phys.~B {\bf 743}, 74 (2006).



\bibitem{Green:1984sg}
  M.~B.~Green and J.~H.~Schwarz,
  Phys.\ Lett.\  B {\bf 149}, 117 (1984).


\bibitem{Witten:1984dg}
  E.~Witten,
  Phys.\ Lett.\  B {\bf 149}, 351 (1984);
%
%
  M.~Dine, N.~Seiberg and E.~Witten,
  Nucl.\ Phys.\  B {\bf 289}, 589 (1987);
%
%
  W.~Lerche, B.~E.~W.~Nilsson and A.~N.~Schellekens,
  Nucl.\ Phys.\  B {\bf 289}, 609 (1987);
  J.~J.~Atick, L.~J.~Dixon and A.~Sen,
  Nucl.\ Phys.\  B {\bf 292}, 109 (1987);
%
  M.~Dine, I.~Ichinose and N.~Seiberg,
  Nucl.\ Phys.\  B {\bf 293}, 253 (1987).



\bibitem{Ibanez:1998qp}
  L.~E.~Ibanez, R.~Rabadan and A.~M.~Uranga,
  Nucl.\ Phys.\  B {\bf 542}, 112 (1999);
%
  Z.~Lalak, S.~Lavignac and H.~P.~Nilles,
  Nucl.\ Phys.\  B {\bf 559}, 48 (1999).



\bibitem{Schellekens:1986xh}
  A.~N.~Schellekens and N.~P.~Warner,
  Nucl.\ Phys.\  B {\bf 287}, 317 (1987).




\bibitem{Kobayashi:1996pb}
  T.~Kobayashi and H.~Nakano,
  Nucl.\ Phys.\  B {\bf 496}, 103 (1997).


\bibitem{Derendinger:1991hq}
  J.~P.~Derendinger, S.~Ferrara, C.~Kounnas and F.~Zwirner,
  Nucl.\ Phys.\  B {\bf 372}, 145 (1992).


\bibitem{Ibanez:1992hc}
  L.~E.~Ibanez and D.~L\"ust,
  Nucl.\ Phys.\  B {\bf 382}, 305 (1992).


\bibitem{Ibanez:1991zv}
  L.~E.~Ibanez, D.~L\"ust and G.~G.~Ross,
  Phys.\ Lett.\  B {\bf 272}, 251 (1991).


\bibitem{Kawabe:1993pz}
  H.~Kawabe, T.~Kobayashi and N.~Ohtsubo,
  Phys.\ Lett.\  B {\bf 325}, 77 (1994);
%
  Nucl.\ Phys.\  B {\bf 434}, 210 (1995).


\bibitem{Cremades:2004wa}
  D.~Cremades, L.~E.~Ibanez and F.~Marchesano,
  JHEP {\bf 0405}, 079 (2004)
  [arXiv:hep-th/0404229].


\bibitem{Abe:2009uz}
  H.~Abe, K.~S.~Choi, T.~Kobayashi and H.~Ohki,
  arXiv:0907.5274 [hep-th].


\bibitem{Abe:2010ii}
  H.~Abe, K.~S.~Choi, T.~Kobayashi and H.~Ohki,
  arXiv:1001.1788 [hep-th].




\bibitem{Choi:2009pv}
  K.~S.~Choi, T.~Kobayashi, R.~Maruyama, M.~Murata, Y.~Nakai, H.~Ohki
  and M.~Sakai,
  arXiv:0908.0395 [hep-ph].


\bibitem{Abe:2009dr}
  H.~Abe, K.~S.~Choi, T.~Kobayashi and H.~Ohki,
  arXiv:0903.3800 [hep-th].


\bibitem{Abe:2009vi}
  H.~Abe, K.~S.~Choi, T.~Kobayashi and H.~Ohki,
  arXiv:0904.2631 [hep-ph].


\bibitem{Abe:2008fi}
  H.~Abe, T.~Kobayashi and H.~Ohki,
  JHEP {\bf 0809}, 043 (2008)
  [arXiv:0806.4748 [hep-th]].



\bibitem{Abe:2008sx}
  H.~Abe, K.~S.~Choi, T.~Kobayashi and H.~Ohki,
  Nucl.\ Phys.\  B {\bf 814}, 265 (2009)
  [arXiv:0812.3534 [hep-th]].



\bibitem{Manton:1981es}
  N.~S.~Manton,
  Nucl.\ Phys.\  B {\bf 193}, 502 (1981);
%
%
  G.~Chapline and R.~Slansky,
  Nucl.\ Phys.\  B {\bf 209}, 461 (1982);
%
  S.~Randjbar-Daemi, A.~Salam and J.~A.~Strathdee,
  Nucl.\ Phys.\  B {\bf 214}, 491 (1983);
%
  C.~Wetterich,
  Nucl.\ Phys.\  B {\bf 222}, 20 (1983);
%
  P.~H.~Frampton and K.~Yamamoto,
  Phys.\ Rev.\ Lett.\  {\bf 52}, 2016 (1984);
%
  P.~H.~Frampton and T.~W.~Kephart,
  Phys.\ Rev.\ Lett.\  {\bf 53}, 867 (1984);
%
  K.~Pilch and A.~N.~Schellekens,
  Nucl.\ Phys.\  B {\bf 256}, 109 (1985).



\bibitem{Bachas:1995ik}
  C.~Bachas,
  arXiv:hep-th/9503030.




\bibitem{Troost:1999xn}
  J.~Troost,
  Nucl.\ Phys.\  B {\bf 568}, 180 (2000)
  [arXiv:hep-th/9909187].



\bibitem{'t Hooft:1979uj}
  G.~'t Hooft,
  Nucl.\ Phys.\  B {\bf 153}, 141 (1979).





\bibitem{Alfaro:2006is}
  J.~Alfaro, A.~Broncano, M.~B.~Gavela, S.~Rigolin and M.~Salvatori,
  JHEP {\bf 0701}, 005 (2007)
  [arXiv:hep-ph/0606070];
%
%
  D.~Hernandez, S.~Rigolin and M.~Salvatori,
  arXiv:0712.1980 [hep-ph].


\bibitem{vonGersdorff:2007uz}
  G.~von Gersdorff,
  Nucl.\ Phys.\  B {\bf 793}, 192 (2008)
  [arXiv:0705.2410 [hep-th]].


\bibitem{toron}
  G.~'t Hooft,
  Commun.\ Math.\ Phys.\  {\bf 81} (1981) 267;
  P.~van Baal,
  Commun.\ Math.\ Phys.\  {\bf 94} (1984) 397;
  Z.~Guralnik and S.~Ramgoolam,
  Nucl.\ Phys.\  B {\bf 521}, 129 (1998)
  [arXiv:hep-th/9708089].




\bibitem{Font:2008id}
  A.~Font and L.~E.~Ibanez,
  JHEP {\bf 0902}, 016 (2009)
  [arXiv:0811.2157 [hep-th]];
  arXiv:0907.4895 [hep-th].



\bibitem{Bourjaily:2009vf}
  J.~L.~Bourjaily,
  arXiv:0901.3785 [hep-th];
  arXiv:0905.0142 [hep-th].



\bibitem{Hayashi:2009ge}
  H.~Hayashi, T.~Kawano, R.~Tatar and T.~Watari,
  arXiv:0901.4941 [hep-th].



\bibitem{Marsano:2009ym}
  J.~Marsano, N.~Saulina and S.~Schafer-Nameki,
  arXiv:0904.3932 [hep-th];
%
  arXiv:0906.4672 [hep-th].



\bibitem{Blumenhagen:2009up}
  R.~Blumenhagen, T.~W.~Grimm, B.~Jurke and T.~Weigand,
  arXiv:0906.0013 [hep-th].




\bibitem{SYM}
  G.~F.~Chapline and N.~S.~Manton,
  Phys.\ Lett.\  B {\bf 120} (1983) 105;
  A.~H.~Chamseddine,
  Nucl.\ Phys.\  B {\bf 185} (1981) 403.




\bibitem{Mu} 
D.~Mumford, ``Tata Lectures on Theta, vol I, II, III,''
  In  Progress in Mathematics, Vol. 28 (1983) Birkhauser.


\bibitem{DiVecchia:2008tm}
  P.~Di Vecchia, A.~Liccardo, R.~Marotta and F.~Pezzella,
  arXiv:0810.5509 [hep-th].




\bibitem{Higaki:2005ie}
  T.~Higaki, N.~Kitazawa, T.~Kobayashi and K.~j.~Takahashi,
  Phys.\ Rev.\  D {\bf 72}, 086003 (2005)
  [arXiv:hep-th/0504019].



\bibitem{VSV}
  M.~A.~Virasoro,
  Phys.\ Rev.\  {\bf 177} (1969) 2309;
  J.~A.~Shapiro,
  Phys.\ Rev.\  {\bf 179} (1969) 1345;
  G.~Veneziano,
  Nucl.\ Phys.\  B {\bf 74} (1974) 365.




\bibitem{Koerber:2002zb}
  P.~Koerber and A.~Sevrin,
  JHEP {\bf 0210} (2002) 046
  [arXiv:hep-th/0208044].



\bibitem{Po}
J.~Polchinski, ``String Theory,'' vol. 1, Cambridge Univ. Press
(1998).



\bibitem{Atick:1987kd}
  J.~J.~Atick, L.~J.~Dixon, P.~A.~Griffin and D.~Nemeschansky,
  Nucl.\ Phys.\  B {\bf 298} (1988) 1.





\bibitem{AO}
  S.~A.~Abel and A.~W.~Owen,
  Nucl.\ Phys.\  B {\bf 682} (2004) 183
  [arXiv:hep-th/0310257].



\bibitem{CK}
  K.~S.~Choi and T.~Kobayashi,
  Nucl.\ Phys.\  B {\bf 797} (2008) 295
  [arXiv:0711.4894 [hep-th]].



\bibitem{CP}
  M.~Cvetic and I.~Papadimitriou,
  Phys.\ Rev.\  D {\bf 68} (2003) 046001
  [Erratum-ibid.\  D {\bf 70} (2004) 029903]
  [arXiv:hep-th/0303083];
  S.~A.~Abel and A.~W.~Owen,
  Nucl.\ Phys.\  B {\bf 663} (2003) 197
  [arXiv:hep-th/0303124];
  V.~Braun, Y.~H.~He and B.~A.~Ovrut,
  JHEP {\bf 0604} (2006) 019
  [arXiv:hep-th/0601204];
  C.~M.~Chen, T.~Li, V.~E.~Mayes and D.~V.~Nanopoulos,
  Phys.\ Rev.\  D {\bf 78}, 105015 (2008)
  [arXiv:0807.4216 [hep-th]].



\bibitem{BKM}
  T.~T.~Burwick, R.~K.~Kaiser and H.~F.~Muller,
  Nucl.\ Phys.\  B {\bf 355} (1991) 689;
\bibitem{Stieberger:1992bj}
  S.~Stieberger, D.~Jungnickel, J.~Lauer and M.~Spalinski,
  Mod.\ Phys.\ Lett.\  A {\bf 7}, 3059 (1992)
  [arXiv:hep-th/9204037].




\bibitem{Brignole:1997dp}
  A.~Brignole, L.~E.~Ibanez and C.~Munoz,
  arXiv:hep-ph/9707209.




\bibitem{Bianchi:2005yz}
  M.~Bianchi and E.~Trevigne,
  JHEP {\bf 0508} (2005) 034
  [arXiv:hep-th/0502147].



\bibitem{Antoniadis:2004pp}
  I.~Antoniadis and T.~Maillard,
  Nucl.\ Phys.\  B {\bf 716} (2005) 3
  [arXiv:hep-th/0412008].



\bibitem{Branco:1983tn}
  G.~C.~Branco, J.~M.~Gerard and W.~Grimus,
  Phys.\ Lett.\  B {\bf 136}, 383 (1984);
%
  C.~Luhn, S.~Nasri and P.~Ramond,
  J.\ Math.\ Phys.\  {\bf 48}, 073501 (2007)
  [arXiv:hep-th/0701188];
%
  I.~de Medeiros Varzielas, S.~F.~King and G.~G.~Ross,
  Phys.\ Lett.\  B {\bf 648}, 201 (2007)
  [arXiv:hep-ph/0607045];
%
  E.~Ma,
  Mod.\ Phys.\ Lett.\  A {\bf 21}, 1917 (2006)
  [arXiv:hep-ph/0607056];
%
  Phys.\ Lett.\  B {\bf 660}, 505 (2008)
  [arXiv:0709.0507 [hep-ph]].




\bibitem{Grimus} 
W.~Grimus and L.~Lavoura,
  Phys.\ Lett.\  B {\bf 572}, 189 (2003);
%
%
%
  W.~Grimus, A.~S.~Joshipura, S.~Kaneko, L.~Lavoura and M.~Tanimoto,
  JHEP {\bf 0407}, 078 (2004);
  A.~Blum, R.~N.~Mohapatra and W.~Rodejohann,
  Phys.\ Rev.\  D {\bf 76}, 053003 (2007);
  A.~Blum, C.~Hagedorn and M.~Lindner,
  Phys.\ Rev.\  D {\bf 77}, 076004 (2008)
  [arXiv:0709.3450 [hep-ph]];
%
W.~Grimus, A.~S.~Joshipura, S.~Kaneko, L.~Lavoura, H.~Sawanaka and
M.~Tanimoto,
  Nucl.\ Phys.\  B {\bf 713}, 151 (2005).
%
  H.~Ishimori, T.~Kobayashi, H.~Ohki, Y.~Omura, R.~Takahashi and
  M.~Tanimoto,
  Phys.\ Lett.\  B {\bf 662}, 178 (2008)
  [arXiv:0802.2310 [hep-ph]];
  H.~Ishimori, T.~Kobayashi, H.~Ohki, Y.~Omura, R.~Takahashi and
  M.~Tanimoto,
  Phys.\ Rev.\  D {\bf 77}, 115005 (2008)
  [arXiv:0803.0796 [hep-ph]];
%
  A.~Adulpravitchai, A.~Blum and C.~Hagedorn,
  JHEP {\bf 0903}, 046 (2009)
  [arXiv:0812.3799 [hep-ph]].




\bibitem{Ishimori:2008uc}
  H.~Ishimori, T.~Kobayashi, H.~Okada, Y.~Shimizu and M.~Tanimoto,
  arXiv:0811.4683 [hep-ph]; 
%
  H.~Ishimori, T.~Kobayashi, H.~Okada, Y.~Shimizu and M.~Tanimoto,
  JHEP {\bf 0912} (2009) 054
  [arXiv:0907.2006 [hep-ph]];  
%
  J.~A.~Escobar and C.~Luhn,
  J.\ Math.\ Phys.\  {\bf 50}, 013524 (2009)
  [arXiv:0809.0639 [hep-th]].




%



\bibitem{Ishimori:2010au}
  H.~Ishimori, T.~Kobayashi, H.~Ohki, H.~Okada, Y.~Shimizu and
  M.~Tanimoto,
  arXiv:1003.3552 [hep-th].





\bibitem{Blumenhagen:2005tn}
  R.~Blumenhagen, M.~Cvetic, F.~Marchesano and G.~Shiu,
  JHEP {\bf 0503}, 050 (2005)
  [arXiv:hep-th/0502095].



\bibitem{Haba:2006dz}
  N.~Haba, A.~Watanabe and K.~Yoshioka,
  Phys.\ Rev.\ Lett.\  {\bf 97}, 041601 (2006)
  [arXiv:hep-ph/0603116].



\bibitem{Kobayashi:2008ih}
  T.~Kobayashi, Y.~Omura and K.~Yoshioka,
  Phys.\ Rev.\  D {\bf 78}, 115006 (2008)
  [arXiv:0809.3064 [hep-ph]].



\bibitem{Seidl:2008yf}
  G.~Seidl,
  arXiv:0811.3775 [hep-ph].



\bibitem{Blumenhagen:2002gw}
  R.~Blumenhagen, L.~Gorlich and T.~Ott,
  JHEP {\bf 0301}, 021 (2003)
  [arXiv:hep-th/0211059];
%
  M.~Cvetic and I.~Papadimitriou,
  Phys.\ Rev.\  D {\bf 67}, 126006 (2003)
  [arXiv:hep-th/0303197];
%
%
  M.~Cvetic, T.~Li and T.~Liu,
  Nucl.\ Phys.\  B {\bf 698}, 163 (2004)
  [arXiv:hep-th/0403061].



\bibitem{Green:1987mn}
  M.~B.~Green, J.~H.~Schwarz and E.~Witten,
  ``Superstring Theory. Vol. 2: Loop Amplitudes, Anomalies And
  Phenomenology,''
{\it  Cambridge, Uk: Univ. Pr. ( 1987) 596 P. ( Cambridge Monographs
On Mathematical Physics)}



\bibitem{Ibanez:1987xa}
  L.~E.~Ibanez, H.~P.~Nilles and F.~Quevedo,
  Phys.\ Lett.\  B {\bf 192}, 332 (1987).




\bibitem{Fujikawa:1979ay}
K.~Fujikawa, Phys. Rev. Lett. \textbf{42} (1979), 1195.



\bibitem{Fujikawa:1980eg}
K.~Fujikawa, Phys. Rev. \textbf{D21} (1980), 2848.




\bibitem{AlvarezGaume:1983ig}
L.~Alvarez-Gaume and E.~Witten, Nucl. Phys. \textbf{B234} (1984), 269.

\bibitem{AlvarezGaume:1984dr}
L.~Alvarez-Gaume and P.~H. Ginsparg, Ann. Phys. \textbf{161} (1985),
423.

\bibitem{Fujikawa:1986hk}
K.~Fujikawa, S.~Ojima, and S.~Yajima, Phys. Rev. \textbf{D34} (1986),
3223.



\bibitem{Rohlin:1959}
V.~Rohlin, Dokl. Akad. Nauk. \textbf{128} (1959), 980 --983.



\bibitem{Csaki:1997aw}
C.~Csaki and H.~Murayama, Nucl. Phys. \textbf{B515} (1998), 114--162,
  [hep-th/9710105].





\bibitem{Ibanez:1991hv}
  L.~E.~Ibanez and G.~G.~Ross,
  Phys.\ Lett.\  B {\bf 260} (1991) 291;
  T.~Banks and M.~Dine,
  Phys.\ Rev.\  D {\bf 45}, 1424 (1992);
  L.~E.~Ibanez,
  Nucl.\ Phys.\  B {\bf 398}, 301 (1993);
  K.~Kurosawa, N.~Maru and T.~Yanagida,
   Phys. Lett. B{\bf 512}, 203 (2001);
  J.~Kubo and D.~Suematsu, 
  Phys. Rev. D{\bf 64}, 115014 (2001);
K. S.~Babu, I.~Gogoladze and K.~Wang, 
  Nucl. Phys. B{\bf 660}, 332 (2003);
  M.~Dine and M.~Graesser,
  JHEP {\bf 0501}, 038 (2005);
  T.~Araki,
  arXiv:hep-ph/0612306;
H.~Dreiner and M.~Thormeier, 
Phys. Rev. D{\bf 69}, 053002 (2004);
  H.~Dreiner, H.~Murayama and M.~Thormeier,
     Nucl. Phys. B{\bf 729}, 278 (2005);
  A.~H.~Chamseddine and H.~K.~Dreiner,
  Nucl.\ Phys.\  B {\bf 458}, 65 (1996);
  H.~K.~Dreiner, C.~Luhn and M.~Thormeier,
  Phys.\ Rev.\  D {\bf 73}, 075007 (2006);
  H.~K.~Dreiner, C.~Luhn, H.~Murayama and M.~Thormeier,
  arXiv:hep-ph/0610026.



\bibitem{Friedan:1985ge}
  D.~Friedan, E.~J.~Martinec and S.~H.~Shenker,
  Nucl.\ Phys.\ B {\bf 271}, 93 (1986).




\bibitem{Font:1988mk}
  A.~Font, L.~E.~Ib\'{a}\~{n}ez and F.~Quevedo,
  Phys.\ Lett.\ B {\bf 217} (1989) 272.



\bibitem{Faraggi:2006bs}
  A.~E.~Faraggi, S.~Forste and C.~Timirgaziu,
  JHEP {\bf 0608}, 057 (2006).
%



\bibitem{Katsuki:1989cs}
  Y.~Katsuki, Y.~Kawamura, T.~Kobayashi, N.~Ohtsubo, Y.~Ono and
  K.~Tanioka,
  DPKU-8904.




\bibitem{Araki:2008ek}
  T.~Araki, T.~Kobayashi, J.~Kubo, S.~Ramos-Sanchez, M.~Ratz and 
P.~K.~S.~Vaudrevange,
  Nucl.\ Phys.\  B {\bf 805}, 124 (2008)
  [arXiv:0805.0207 [hep-th]].



\bibitem{jones}
  D.R.T~Jones and L.~Mezincescu,
  Phys. Lett. B {\bf 136}, 242 (1984);
  B {\bf 138}, 293 (1984);
   P.~West, Phys. Lett. B {\bf 137}, 371 (1984);
A.J.~Parkes and P.C~West, Phys. Lett. B {\bf 138}, 99 (1984);
Nucl. Phys. B {\bf 256}, 340 (1985);
A.J.~Parkes,
  Phys. Lett. B {\bf 156}, 73 (1985);
D.R.T~Jones and A.J.~Parkes,
  Phys. Lett. B {\bf 160}, 267 (1985).



\bibitem{piguet1}
O.~Piguet and K.~Sibold, Int. Mod. Phys. A {\bf 1},
913 (1986);
Phys. Lett. B {\bf 177}, 373 (1986).





\bibitem{ferrara}
S.~Ferrara and B.~Zumino, Nucl. Phys. B {\bf 87}, 207 (1975).




\bibitem{piguet2}
O.~Piguet and K.~Sibold,
Nucl. Phys. B {\bf 196}, 428 (1982); B {\bf 196}, 447 (1982).



\bibitem{Ibanez:1992uh}
  L.~E.~Ibanez and D.~L\"ust,
  Phys.\ Lett.\  B {\bf 302}, 38 (1993).



\bibitem{Dixon:1990pc}
  L.~J.~Dixon, V.~Kaplunovsky and J.~Louis,
  Nucl.\ Phys.\  B {\bf 355}, 649 (1991);


\bibitem{Kobayashi:1991rp}
  T.~Kobayashi and N.~Ohtsubo,
  Int.\ J.\ Mod.\ Phys.\ A {\bf 9}, 87 (1994).



\bibitem{Araki:2007ss}
  T.~Araki, K.~S.~Choi, T.~Kobayashi, J.~Kubo and H.~Ohki,
  Phys.\ Rev.\  D {\bf 76}, 066006 (2007)
  [arXiv:0705.3075 [hep-ph]].






\end{thebibliography}
\end{document}